\documentclass{aa}
\usepackage{graphicx}
\usepackage{rotating}
\usepackage{txfonts}
\usepackage{epsf}
\usepackage{epstopdf}

\begin{document}

\title{IMAGES\thanks{intermediate-mass Galaxy Evolution Sequence}
 II. A surprisingly low fraction of undisturbed rotating spiral disks at $z\sim$0.6.\\
 \Large{The morpho-kinematical relation 6 Gyrs ago} }

\author{  
           B. Neichel\inst{1}                  
  \and F. Hammer\inst{1}                   
  \and M. Puech\inst{2,1}                  
  \and H. Flores\inst{1}                   
  \and M. Lehnert\inst{1}                  
  \and A. Rawat\inst{1,3}                  
  \and Y. Yang\inst{1}                     
  \and R. Delgado\inst{1,4}                     
  \and P. Amram\inst{5}                    
  \and C. Balkowski\inst{1}                
  \and C. Cesarsky\inst{2}                 
  \and H. Dannerbauer\inst{6}              
  \and I.  Fuentes-Carrera\inst{1}          
  \and B. Guiderdoni\inst{7}               
  \and A. Kembhavi\inst{3}                 
  \and Y. C. Liang\inst{8}                 
  \and N. Nesvadba\inst{1}                 
  \and G. {\"O}stlin\inst{9}              
  \and L. Pozzetti\inst{10}                 
  \and C. D. Ravikumar\inst{11}            
  \and S. di Serego Alighieri\inst{12}      
  \and D. Vergani\inst{13}                 
  \and J. Vernet\inst{2}                   
  \and H. Wozniak\inst{7}                  
}                                          
\offprints{benoit.neichel@obspm.fr}
\authorrunning{B. Neichel et al.}
\titlerunning{IMAGES II. Morpho-kinematic}
\institute{
GEPI, Observatoire de Paris, CNRS, University Paris Diderot; 5 Place Jules Janssen, 92195 Meudon, France 
\and 
ESO, Karl-Schwarzschild-Strasse 2, D-85748 Garching bei M\"unchen, Germany
\and 
Inter-University Centre for Astronomy and Astrophysics, Post Bag 4, Ganeshkhind, Pune 411007, India 
\and 
IFARHU-SENACYT, Technological University of Panama, 0819-07289 Panama, Rep. of Panama
\and 
Laboratoire d'Astrophysique de Marseille, Observatoire Astronomique de 
Marseille-Provence, 2 Place Le Verrier, 13248 Marseille, France
\and 
MPIA, K{\"o}nigstuhl 17, D-69117 Heidelberg, Germany
\and 
Centre de Recherche Astronomique de Lyon, 9 Avenue Charles Andr, 69561 Saint-Genis-Laval 
Cedex, France
\and 
National Astronomical Observatories, Chinese Academy of Sciences, 20A Datun Road, Chaoyang District, Beijing 100012, PR China 
\and 
Stockholm Observatory, AlbaNova University Center, Stockholms Center for Physics, Astronomy and Biotechnology, Roslagstullsbacken 21, 10691 Stockholm, Sweden 
\and 
 INAF - Osservatorio Astronomico di Bologna, via Ranzani 1, 40127 Bologna, Italy
\and 
Department of Physics, University of Calicut, Kerala 673635, India
\and 
INAF, Osservatorio Astrofisico di Arcetri, Largo Enrico Fermi 5, I-50125, Florence, Italy
\and 
IASF-INAF - via Bassini 15, I-20133, Milano, Italy
}
\date{Received ...... ; accepted ...... }

\abstract{ We present a first combined analysis of the morphological and dynamical properties for the intermediate-mass Galaxy Evolution Sequence (IMAGES) sample. It is a representative sample of  52 $z\sim$0.6 galaxies with $M_{\rm stell}$ from 1.5 to 15 $\!\times\!10^{10}M_{\sun}$  that possesses 3D resolved kinematics and HST deep imaging in at least two broad band filters. We aim at evaluating the evolution of rotating spirals robustly since $z\sim$0.6, as well as at testing the different schemes for classifying galaxies morphologically. We used all the information provided by multi-band images, color maps, and 2D light fitting to assign a morphological class to each object. We divided our sample into spiral disks, peculiar objects, compact objects, and mergers. 
Using our morphological classification scheme, 4/5 of the identified spirals are rotating disks, and more than 4/5 of identified peculiar galaxies show complex kinematics, while automatic classification methods such as concentration-asymmetry and GINI-M20 severely overestimate the fraction of relaxed disk galaxies. Using this methodology, we find that the fraction of undisturbed  rotating spirals has increased by a factor $\sim$ 2 during the past 6 Gyrs, a much higher fraction than was found previously based on morphologies alone. These rotating spiral disks are forming stars very rapidly, even doubling their stellar masses over the past 6 Gyrs, while most of their stars were formed a few Gyrs earlier, which reveals a large gas supply. Because they are the likely progenitors of local spirals, we can conjecture how their properties are evolving. Their disks show some evidence of inside-out growth, and the gas supply/accretion is not random since the disk needs to be stable in order to match the local disk properties.

\keywords{
Morphology -- 
3D spectroscopy --
kinematics and dynamics -- 
galaxy evolution} }

\maketitle
%


\section{Introduction} 
\label{intro}

In the local universe, galaxy morphologies can be organized along the Hubble sequence. Both the physical and kinematic properties of galaxies vary systematically with the Hubble type (\cite{Roberts94}). In this scheme, rotating disk galaxies constitute the majority of the galaxy population. They represent 70\% of the intermediate-mass galaxy population, which itself includes at least 2/3 of the present-day stellar mass (see \cite{Hammer05} and references therein).

In the distant universe, morphological investigations based on HST imaging have brought observational evidence that a large fraction of galaxies have peculiar morphologies that do not fit into the standard Hubble sequence (e.g. \cite{Abraham96}; \cite{Brinchmann98}; \cite{VandenBergh01}; \cite{Zheng05}).
For instance, Zheng et al. (2005) conclude that at least 30\% of the intermediate-mass galaxies at $z$=0.4-1 have peculiar morphologies, whereas they represent less than few percent in the local universe. Over this fraction, more than 2/3 are luminous compact galaxies (LCGs - \cite{Rawat07}), an enigmatic population that has almost completely vanished in the local universe, with up to a factor $\sim$10 decrease in comoving number density (\cite{Werk04}).\\
While different observations suggest that most of the massive elliptical galaxies were in place prior to $z$=1 (\cite{Jimenez06}; \cite{Bernardi06}), the formation and evolution of spiral galaxies is still under debate.
Different physical processes can modify the galaxy properties over cosmic time, and the main hypothesis processes are: \\
(i) the secular evolution with slow and continuous matter accretion through the inter-galactic medium (e.g., \cite{Semelin05}, \cite{Birnboim07}), \\
(ii) minor mergers and accretion of small satellites (\cite{Somerville01}), \\
(iii) more violent evolution through hierarchical merging (e.g., Hammer et al. 2005) as a function of lookback time.

To help establish the relative importance of each process, both high-resolution imaging and integral field spectroscopy are required. Morphological studies can offer substantial clues to the merging rate (\cite{Conselice03}, \cite{Bell06}, \cite{Lotz06a}), but kinematics studies using 3D spectroscopy appear to be a prerequisite to sampling the whole velocity field of individual galaxies and to directly distinguishing between interacting and non-interacting galaxies. The internal kinematics of galaxies are of primary interest because they probe the evolutionary state of these objects.\\

In this context, we are pursuing an ESO Large Program using the integral field capability of GIRAFFE (LP: IMAGES) to gather a complete and representative sample of velocity fields and dispersion maps of intermediate-mass galaxies at intermediate redshift. Galaxies are selected in different fields by their absolute J band magnitude (M$_J$(AB) $<$ -20.3 - see \cite{Ravikumar07} for more details). A first part of this sample was observed during the FLAMES/GIRAFFE guaranteed time and is described in Flores et al. (2006), Puech et al. (2006a), Puech et al. (2006b), and Puech et al. (2007). The other part was observed in the frame of the large program IMAGES and is described in  Yang et al. (2007) (hereafter Paper I). 
Based on this unique sample, Flores et al. (2006) and Paper I have developed a robust classification scheme to divide the sample into three distinct classes based on dynamical characteristics.
Paper I finds that the whole sample can be distributed between these three kinematical classes because 43$\pm$12\% of the galaxies have complex kinematics, 25$\pm$12\% have perturbed rotation, and only 32$\pm$12\% are consistent with pure rotation. Interestingly, objects supported by pure rotation all lie in the local Tully-Fisher relation (\cite{Puech08} - hereafter Paper III), whereas objects identified as having perturbed or complex kinematics possess dynamical properties (specific angular momentum, V/$\sigma$) that are statistically different from local disks and whose dispersions in properties are described by simulations of mergers (\cite{Puech07}). In a first approach, Puech et al. (2007) suggest that galaxies with pure rotation could be associated with secular evolution, perturbed rotation with minor mergers (\cite{Puech07b}), and galaxies with complex kinematics could be the result of more violent events (e.g. major mergers). \\
In the first part of this paper, we try to understand to what extent the morphological appearance of a galaxy reflects the underlying dynamical state. By combining the high-resolution, multi-band HST imaging with the 3D spectroscopy, we explore whether a correlation between morphology and kinematics exists, like in the local universe. 
After describing the sample in Sect. 2,  we present our derivation of a morphological classification of the sample in Sect. 3 and compare this classification with the kinematical classification in Sect. 4. In Sect 5 we compare the morphological method developed in this paper with automatic morphological classifiers. This morpho-kinematical analysis will lead us to focus on one specific group: the rotating spiral disks. In Sect. 6, we discuss the properties of these objects that dynamically and morphologically correspond to the local class of rotating spirals. In this paper we adopt the ``Concordance" cosmological parameters of
$H_0\!=\!70$ km s$^{-1}$ Mpc$^{-1}$, $\Omega_M\!=\!0.3$ and $\Omega_\Lambda\!=\!0.7$. Magnitudes are given in the AB-system.

\section{Data}
\label{ssac}
The whole sample is composed of 63 galaxies that were observed with the integral field units of FLAMES/GIRAFFE on the ESO-VLT. This data set comes from Paper I, and it combines the Flores et al. (2006) sample with the IMAGES large program  sample, using the same selection criteria: $M_J(AB)<$-20.3 and EW([OII])$>$15$\AA$.
The Flores et al. (2006) sample includes 28 galaxies. Originally, spectra from the CFRS (Hammer et al. 1997) and from FORS1 and FORS2 data for the HDFS (\cite{Vanzella02}, \cite{Rigopoulou05}) were used to select galaxies with the [OII] emission lines detected. This first run of observations demonstrates that the FLAMES/GIRAFFE instrument was able to detect all galaxies with EW([OII])$>$15$\AA$ and I$_{AB} \le$23.5 after 8 to 13hrs of integration time (see Flores et al. 2006 for more details). 
In the preparation of the IMAGES large program, Ravikumar et al. (2007) obtained spectra of 580 galaxies in the GOODS-South field from VIMOS observations. From this catalog, combined with VVDS (Vimos VLT Deep Survey, \cite{LeFevre04}) and FORS2 (\cite{Vanzella06}) spectroscopic data, Ravikumar et al. (2007) were able to draw a representative sample of 640 galaxies at $z\leq$1 with $I_{AB}<23.5$. They show that, within the redshift range of 0.4$<z<$0.75, $I_{AB}<23.5$ galaxies include almost all intermediate-mass galaxies (e.g. at least 95\% of $M_J(AB)<$-20.3 galaxies). The 35 galaxies of paper I come from this sample.\\
The Flores et al. (2006) sample and the IMAGES sample are merged by applying the $M_J(AB)<$-20.3 and EW([OII])$>$15$\AA$ selection criteria. This merged sample of 63 galaxies is representative of the luminosity function in the z=0.4-0.75 range, with a confidence level $>$99.99\% (see Fig. 1 in Paper I).\\
Because our morphological classification use color information (see Sect. \ref{colormap}), we chose to restrict ourselves to a smaller sample (52 objects) with multi-band HST data (see Sect. \ref{im}). In the following we refer to the 63 galaxies as the ``original sample" and to the sub-sample of 52 galaxies having multi-band data as our ``working sample". The working sample follows the $z\sim$0.6 luminosity function with very similar confidence level.

\subsection{Kinematic observations}
\label{kin}
A first part of the original sample (28 objects) was observed during FLAMES/GIRAFFE guaranteed time (ESO runs 071.B-0322(A), 072.A-0169(A), and 75.B-0109(A)). The other part (35 objects) was observed in the frame of the large program IMAGES (ESO run 174.B-0328 - PI: F. Hammer). The [OII]$\lambda \lambda$3726,3729 doublet was used to derive both velocity fields and velocity dispersion maps. These maps were used to divide the sample into three distinct classes based on their dynamical characteristics:\begin{itemize}
\item Rotating disks (RD): when the velocity field shows a rotation pattern that follows the optical major axis and the dispersion map show a peak near the dynamical center.
 \item Perturbed rotators (PR): when the velocity field shows a rotation pattern but the dispersion map shows a peak not located at the dynamical center, or does not show any peak. 
\item Complex kinematics (CK): when velocity field and dispersion map show complex distributions, very different from that expected for simple rotating disks.
\end{itemize}
A complete description of the methods and analyses for reducing the data and classifying the galaxies is given in Flores et al. (2006) and Paper I. The final kinematic classification for each galaxy can be found in Table \ref{toto} (col. 10).

\subsection{Complementary imaging data}
\label{im}
To carry out our morphological classification, we make use of all the HST data publicly available.
Galaxies were selected in 4 different fields, namely: the Canada France Redshift Survey 03h and 22h, the Hubble Deep Field South and the Chandra Deep Field South. In the following, we describe each field and the corresponding HST data.

\subsubsection{Chandra Deep Field South.} 
The CDFS (GOODS project) was observed with ACS in F435W (3 orbits), F606W (2.5 orbits), F775W (2.5 orbits), and F850W (5 orbits) bands (0.05"/pix - \cite{Giavalisco04}). We used the publicly available version v1.0 of the reduced, calibrated, stacked, mosaicked and drizzled images (drizzled pixel scale = 0.03"). Thirty-five objects of the original sample are located in the CDFS. All these galaxies are included in our working sample. 

\subsubsection{Hubble Deep Field South.} 
The HDFS was undertaken in 1998 (\cite{Williams00}). It consists of a deep principal field observed during 150 orbits and 9 contiguous flanking fields observed for 2 orbit each (\cite{Lucas03}). 
For the principal field, WFPC2 imaging with the F300W, F450W, F606W, and F814W broadband filters was performed (drizzled pixel scale = 0.04"  - \cite{Casertano00}). 
The flanking fields were only observed with WFPC2 - F814W band (drizzled pixel scale = 0.05"/pix). Nine galaxies of the original sample are embedded in the HDFS: 5 are in the principal field, 2 are in FF2, and 2 in FF5. The 5 galaxies located in the principal field are included in the working sample.

\subsubsection{Canada-France Redshift Survey.}
The CFRS consist in 4 different fields located at 03h, 10h, 14h, and 22h. It was partially observed with the WFPC2 camera (0.1"/pix). A first campaign was in the 03h and 14h fields with F450W and F814W filters (\cite{Schade95}). These observations were completed with F814W filter orbits, including the 10h and 22h field (\cite{Brinchmann98}). Finally, F606W and completion of F814W were carried out in the 03h and 14h fields (\cite{Zheng04}).\\
The original sample is divided between the 03h and 22h fields: 13 objects in the 03h field, 6 objects in the 22h one. 
As mentioned above, the CFRS was only partially observed by HST and only low-resolution, ground-based images (taken at CFHT, 0.2"/pix, \cite{Schade96}) are available for 4 objects in the 22h field and 1 in the 03h.
We do not include these objects in the working sample, as low-resolution data could lead to morphological misidentifications. 
The 2 remaining objects in the 22h field only have F814W images and are also not included in the working sample. The 12 remaining objects located in the 03h have either F606W and/or F450W images in addition to the F814W images and are included in the working sample (typical exposure time $\sim$6400s). 
Finally, it is noteworthy that one object (CFRS031032) in the 03h field has been observed with ACS (F814W and F555W filters - 0.05"/pix) during a gravitational lense program (Prop. 9744 - PI:C. Kochanek).\\

Table \ref{obj} summarizes the data available for the original sample and the working sample. Middle row lists all the objects included in the original sample. Last row lists the objects with HST multi-band data included in the working sample. 
Despite the fact that the working sample includes galaxies observed at different resolutions, spatial samplings, and depths (ACS compared to WFPC2 and observed at different wavelengths), we find that none of these effects restricts our ability to homogeneously classify all the galaxies morphologically.  Moreover, note that the majority of the galaxies have high-resolution data taken with ACS with its better-sampled data and depth.

\begin{table}[h!]
\begin{center}
 \caption{Number of objects per field. From the 63 galaxies that have been observed with GIRAFFE, we restrict ourselves to the 52 objects with multi-band HST data.}
\begin{tabular}{c|ccccc}
\hline\hline
\# of obj. & 03h & 22h & HDFS  & CDFS  & Total  \\ \hline
Original Sample & 13 & 6 & 9 & 35 & 63 \\ \hline
Working Sample & 12 & 0 & 5 & 35 & 52 \\  \hline\hline
 \end{tabular}
 \label{obj}
\end{center}
\end{table}

\section{Morphological classifications}

We chose to follow a morphological analysis in three steps: (i) a surface brightness profile analysis was carried out to quantify structural parameters, (ii) we constructed a set of color maps for the whole sample, (iii) a morphological label was  assigned to each object based on visual inspection of the images and color maps, and a detailed analysis of the structural parameters and physical properties derived from the first two steps. \\

\subsection{Surface brightness analysis}

The surface brightness profile analysis was essentially performed to derive the half-light radius and the bulge fraction (B/T). 

\subsubsection{Half-light radius}
\label{hlr}
The half-light radius can be used as a compactness parameter. We consider that, under a half-light radius cutoff, objects are too compact to be clearly classified. These objects are fairly common at high $z$, and they represent a specific morphological category (\cite{Rawat07}). We chose to follow the compactness criteria defined in Melbourne et al. (2005), which classify as compact all objects with a half-light radius lower than 3kpc. Making a detailed morphological analysis for galaxies smaller than this size threshold, we found that our morphological analysis became unreliable.\\
Two different techniques are used to derive the half-light radius. The first one is based on the IRAF task ``Ellipse" (see \cite{Hammer01} for a complete description of the procedure) and the second one on our own IDL procedure. In both cases, we adopted the method detailed in Bershady et al. (2000), which defines the total aperture as twice the Petrosian radius. Both methods give consistent results within an error of one pixel, which represents $\sim$0.2kpc ($\sim$0.7kpc) at $z$=0.6 for ACS drizzled images (respectively WFPC2 in CFRS).\\

\subsubsection{Bulge fraction: B/T}
\label{bt}
\begin{figure}[!ht]
   \begin{center}
   \begin{tabular}{cc}
  \includegraphics[height=3.3cm]{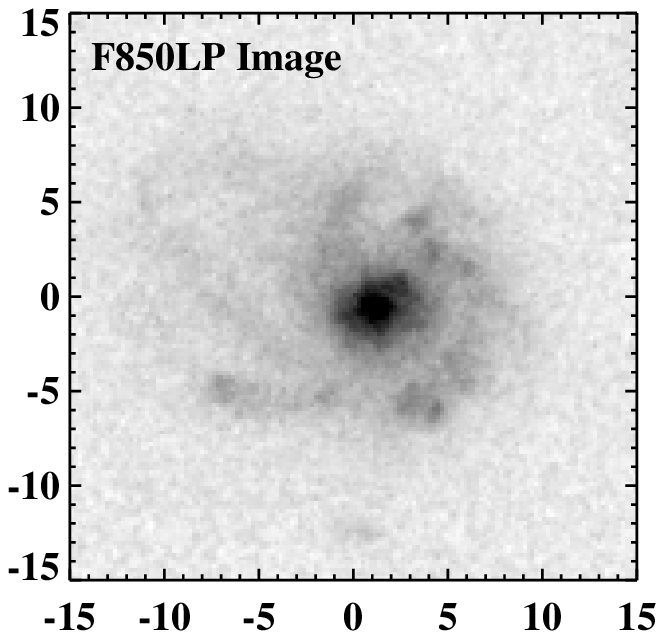} &
   \includegraphics[height=3.3cm]{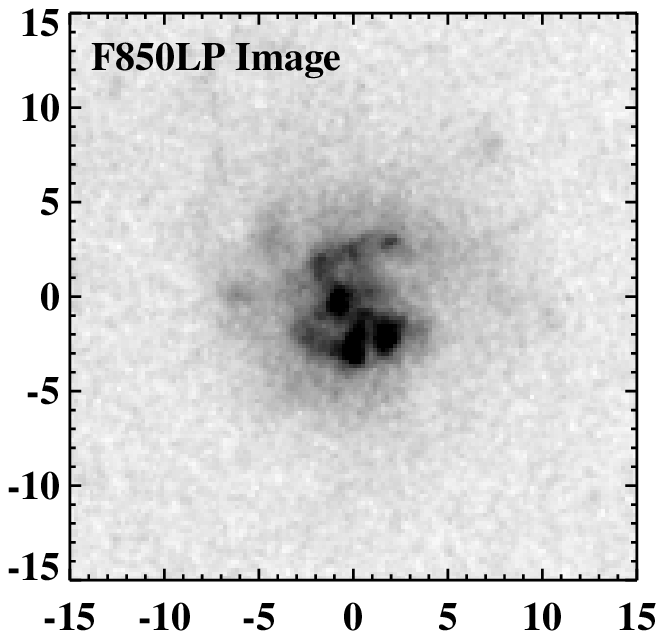} \\   
   \includegraphics[height=3.3cm]{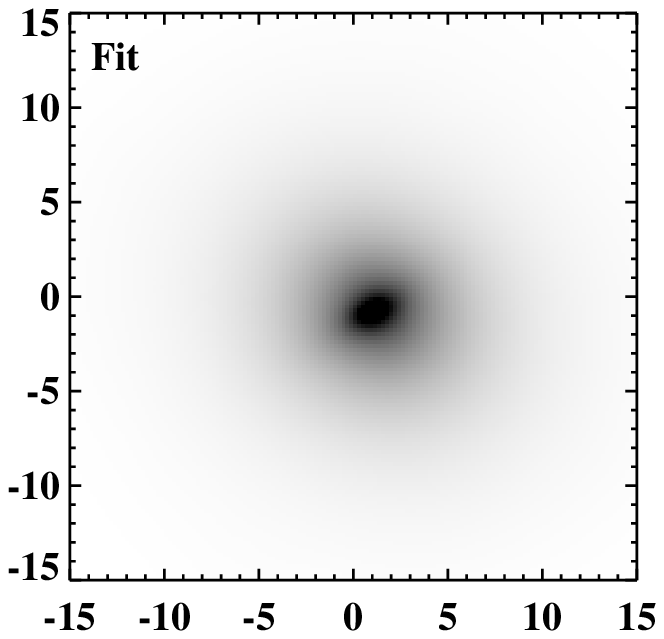} &
 \includegraphics[height=3.3cm]{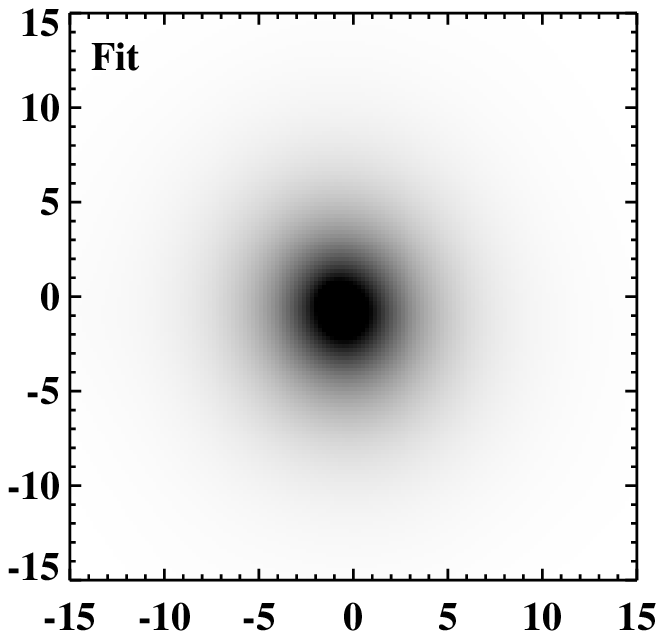} \\ 
  \includegraphics[height=3.3cm]{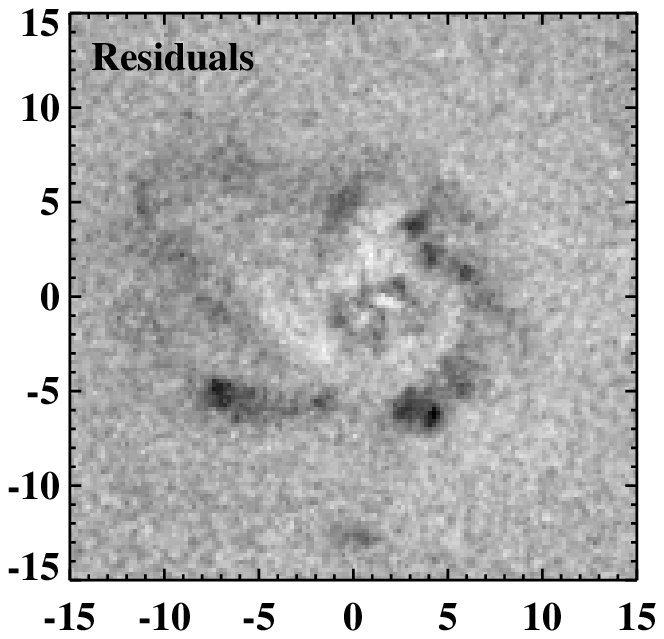} &
 \includegraphics[height=3.3cm]{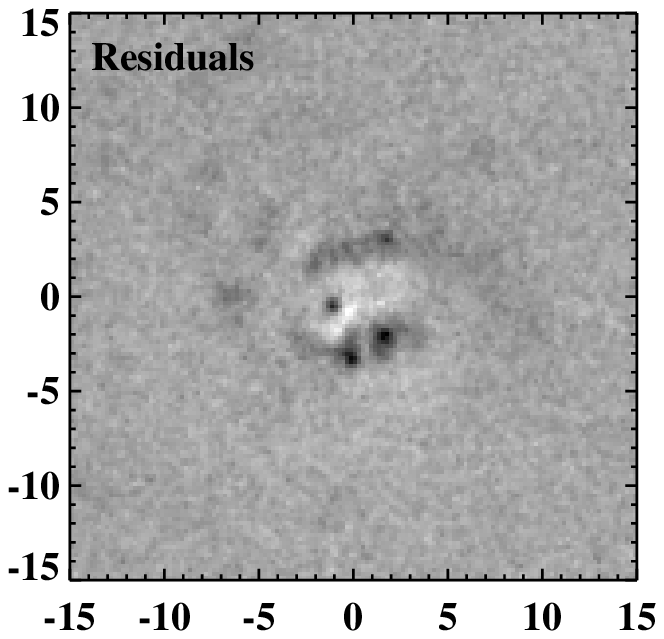} \\  
 \end{tabular}
   \end{center}
   \caption{F850LP image, Galfit model image, residual image. For these two objects (J033237.54-274848.9 on the left and J033214.97-275005.5 on the right), the software gives a similar confidence level in the fitting process ($\chi^2\simeq$1.4). For J033237.54-274848.9 (left), the $\chi^2$ is high due to spiral arms, whereas it is due to the two clumps located southeast of the center for J033214.97-275005.5 (right).}
   \label{galfit} 
   \end{figure}

The B/T parameter is broadly correlated with the traditional Hubble type as early type galaxies have high B/T, whereas late types are expected to have low B/T. Then, it provides a valuable first guess for the morphological type of an object (\cite{Kent85}). 
To derive the B/T for each galaxy, we used the Galfit software (\cite{Peng02}), which performs a 2D modeling of the galaxy flux assuming predefined light distributions. 
We followed the same procedure as described in detail in Rawat et al. (2007). Briefly, each galaxy is modeled as a combination of a bulge and a disk: the intensity profile of the bulge is modeled with a Sersic law and the disk as an exponential function. All the structural parameters (including Sersic index) are allowed to be free during the fitting process, except for the sky, which was held fixed at the estimate made by SExtractor. All neighboring objects within 5.0 arcsec of the target are simultaneously fitted with a single Sersic component to avoid any flux contamination. Finally, during the fitting process, the star closest to each galaxy is used as the PSF for convolution (see Rawat et al. (2007) for more details).\\ 
The results of the fitting procedure are a list of structural parameters, an image of the modeled galaxy, and a residual map constructed by the difference between real and modeled light distribution. This last map is used to compute a $\chi^2$ that  measures how far the modeled image is compared to the original light distribution. This parameter is generally used to estimate the quality of the fit. However, we find that this parameter alone is not sufficient for deciding whether a fit is reliable or not. To illustrate this point, Fig. \ref{galfit} shows the fitting procedure for two galaxies with their z-band images, fitting light models, and residuals. For these two objects, the software gives a similar level of confidence ($\chi^2=1.38$ and $\chi^2=1.46$), whereas in the first case (left column) the $\chi^2$ is due to spiral arms and to real irregularities in the second (right column).
In fact, the main limitation of parametric methods is that they use an ``a priori" assumption on the light distribution, making them unsuitable for treating patchy light distribution often found in spiral galaxies, such as spiral arms. A spiral galaxy with prominent arms unavoidably produces a $\chi^2 > 1$ because the spiral pattern is taken as if it was noise during the fitting. However, we can guage the appropriateness of the fit by carefully examining the residual maps in all of the high resolution images. A careful examination of each residual map allows us to define those that are relatively symmetric due to spiral arm patterns that show the same spiral-like symmetry in all images (e.g., right column) from asymmetric ones (e.g. left column) that show asymmetries in all the images.

Following this recommendation, we define a quality factor for each object to represent our confidence in the fitting result:\\
\begin{list}{}
\item Q=1: Secure fit: if the visual inspection of the fitting result and residual image shows that the Galfit solution is correct. 
\item Q=2: Possibly secure fit: if the visual inspection of the fitting result and residual image shows that the Galfit solution could be correct. 
\item Q=3: Unreliable fit or fit failed: if the visual inspection shows that the Galfit type is not consistent with the original light distribution or if the fitting procedure crashes.\\
\end{list}
We find that, for 40\% of the galaxies in our sample, the 2D fit failed or seems unreliable (Q = 3) because light distribution presents overly distorted features. All these galaxies are morphologically peculiar objects. 
For galaxies that can be modeled by bulge+disk, we considered a segregation of galaxies into different Hubble types following \cite{Rawat07}: early type are for 0.8 $<$ B/T$<$1, S0 for 0.4 $<$ B/T $<$ 0.8 and late type for 0.0 $<$ B/T $<$ 0.4.\\
Both B/T and half-light radius are evaluated in each available filter. However, to minimize any effects of k-correction, we only keep those derived from F850LP (for CDFS) or F814W (for HDFS/CFRS) as it corresponds to the rest frame V band for all the objects, due to the narrow redshift range of the sample. 
Final results for the rest-frame V band are presented in Table \ref{toto} (Cols. 4, 5, 6, \& 9). \\

\subsection{Color maps}
\label{colormap}

We made use of the available multi-band images to construct two sets of color maps.
The first set was obtained by combining, pixel by pixel, the magnitude in two observed bands (see \cite{Abraham99}, \cite{Menanteau01}, \cite{Menanteau04}, \cite{Zheng05} for examples). 
In the classification process, they are principally used to distinguish regular spirals with red bulges and blue disks (see Sect. \ref{method}). 
Moreover, this color information can be compared with evolutionary synthesis models and then used to investigate the distribution of the stellar populations within the galaxies (see Sect. \ref{colorbis}). Physical properties, including whether some regions are dusty or starforming, can be derived from these color maps. This method provides a very direct way to identify the physical nature of objects and compare distant galaxies with local galaxies in the Hubble sequence.  
The B$_{435}$-z$_{850}$ color maps are produced for the 35 galaxies observed with ACS in the CDFS. For the objects located in the HDFS/CFRS, we obtained either B$_{450}$-I$_{814}$ or V$_{606}$-I$_{814}$ color maps. The pixels assigned to the color maps were derived following the method described in Zheng et al. (2004). This method allows quantitative measurements of the reliability of each pixel in the color map and then better constraints of the low signal-to-noise regions.
Figure \ref{allimages} (second row) shows the color map images for the 52 galaxies with multi-band data available (see caption for details).\\
As a complement to these quantitative color maps, we constructed another set of ``false three-color images" (see Figs. \ref{FigZheng} and \ref{herbier} for examples).
These maps were constructed by combining three flux-weighted band images (B -V-z or B-V-I as blue-green-red).
These maps cannot be used to derive physical parameters, however, in the morphological classification process, they are useful for deriving the global appearance of the galaxies.

\subsection{Visual classification}
\label{method}

\begin{figure*}
   \begin{center}
  \includegraphics[height=9cm]{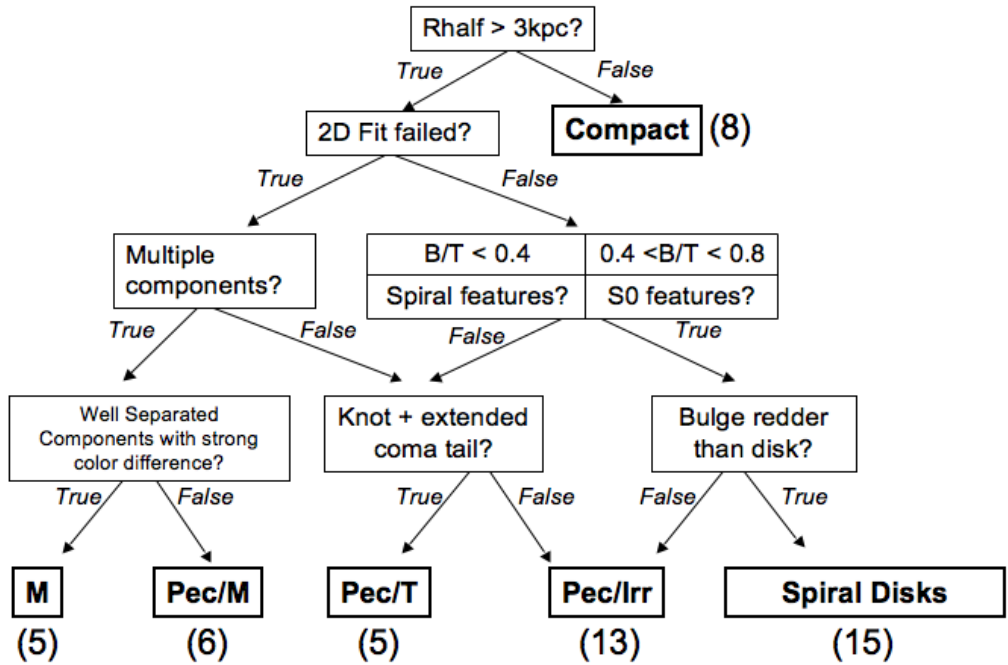}
   \end{center}
   \caption{Decision tree used to morphologically classify the sample. We assign each object a morphological label following step-by-step the arms of the tree. For each morphological class, we also report the number of galaxies.}
   \label{tree} 
   \end{figure*}

A classification only based on B/T is difficult, principally because of the $\chi^2$ problem illustrated in Fig. \ref{galfit}. Moreover, it would not include the information provided by the color maps. To account for all the available information, we chose to visually classify all the galaxies. However, to reduce the subjectivity of such a task, we constructed a decision tree presented in Fig. \ref{tree}. 
This decision tree was built to make the morphological classification simpler to understand. All the information provided by structural parameters, multi-band images, and color maps are included in the tree. At each step, a simple and unique criterion is considered. Following the arms of the tree step-by-step provides a reproductive and quantitative tool for performing the morphological classification.\\

Galaxies were separated into four morphological classes as follows:
\begin{enumerate}
\item Sp: the spiral disks, i.e., all objects with regular structures (arms) and a highly symmetric disk. Moreover, these spiral disk galaxies must show a bulge redder than the disk.
\item Pec: the peculiar galaxies, i.e., objects with asymmetric features in the image or in the color map. As the peculiar class includes objects with different characteristics, we chose to divide this class into three subcategories: tadpole like (Pec/T), which are objects showing a knot at one end plus an extended tail ;  suspected mergers (Pec/M), which are peculiar objects for which the irregularities could be associated with merger/interaction events; and irregulars (Pec/Irr) for objects similar to local irregulars.
\item C: the compact galaxies, i.e., all objects barely resolved and too concentrated to be decomposed into a bulge and a disk. We chose to classify as compact all objects with a half-light radius lower than 3kpc (see Sect. \ref{hlr}).
\item M: the obvious merging/interacting systems, i.e., objects showing tidal tails, multiple cores, two components, etc. They generally also present very perturbed color maps, or extreme colors.
\end{enumerate}

\begin{figure}[!ht]
   \begin{center}
  \includegraphics[width=8cm]{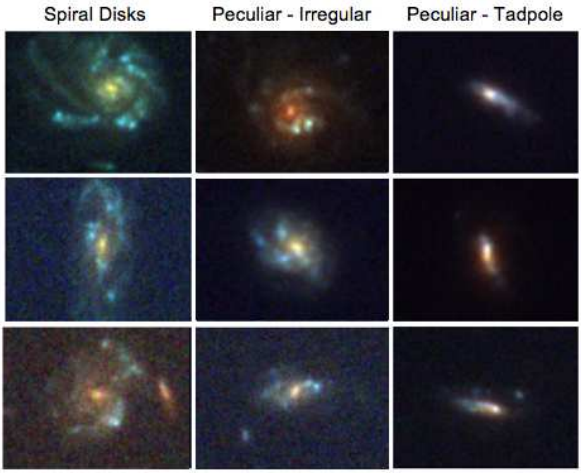}
   \end{center}
   \caption{B-V-z color images of nine galaxies selected in the CDFS sample with B/T$<$0.4. The three galaxies in the first column are classified as spiral disks, those in the second column are Peculiar/Irregulars, and those in the third column are Peculiar/Tadpoles.}
   \label{FigZheng} 
   \end{figure}

To distribute our sample into the four classes, we followed the decision tree shown in Fig. \ref{tree}. 
We first identified all the compact galaxies with the half-light radius information. 
Then, we investigated all the objects with B/T$<$0.4 to classify whether or not each galaxy is a disk. We made use of the color maps to classify the galaxies with a red bulge surrounded by a bluer symmetric disk. The remaining objects with B/T$<$0.4, but not satisfying the spiral disks criterion, were distributed between the Pec/Tadpoles and Pec/Irregular categories. 
To illustrate this process, Fig. \ref{FigZheng} shows a three-color image of nine galaxies with  B/T $<$ 0.4 selected in the CDFS. The three galaxies in the first column satisfy the spiral disk criteria. The three galaxies in the second column clearly show more irregular features than those of the galaxies presented in the first column. These kinds of objects do not pass the ``Spiral features?" test in the third row of the decision tree, and are then classified as Peculiar/Irregular galaxies. Similarly, objects like those in the third column do not pass the ``Spiral features?" test, and as they show a knot plus an extended tail, they are assigned to the Peculiar/Tadpole galaxy class.

\begin{figure}[h!]
   \begin{center}
  \begin{tabular}{c}
  \includegraphics[height=4.3cm]{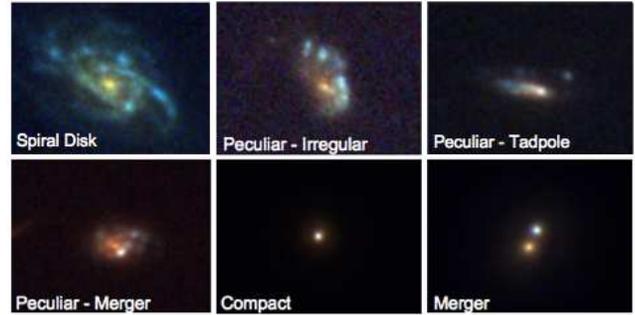}
  \end{tabular}
   \end{center}
   \caption{B-V-z color images of six galaxies representative of the six morphological classes used in this study. From top to bottom and left to right: spiral disks, Peculiar/Irregular, Peculiar/Tadpole, Peculiar/Merger, compact galaxies, and major mergers.}
   \label{herbier} 
   \end{figure}

Following the same procedure, we also investigated the 0.4$<$B/T$<$0.8 range to look for regular S0s. We find no galaxies satisfying this criterion.
Finally, all the remaining objects for which B/T is unreliable or fit failed were distributed between the peculiar and merger class. At this step, we identified the major mergers as galaxies with well-separated components with a strong color difference.
Fig. \ref{herbier} shows, for each morphological class, a three-color image of a representative galaxy and Table \ref{toto} (Col. 7) summarize the morphological classification assigned to each object.\\
Of course this method is not fully objective, and the distinction between two classes may be ambiguous for some objects.
The third object of the first column in Fig. \ref{FigZheng} is an example of such a case, and it illustrates the limitations of our method. Nevertheless, these ambiguous objects only represent a low percentage of the whole sample, and the uncertainties are reduced because 3 authors (BN, FH, AR) have independently classified the sample using the same set of criteria as outlined in Figure \ref{tree}.

\section{Comparing the morphological and kinematical classifications}
\label{comp}

\begin{figure}[h!]
   \begin{center}
  \includegraphics[height=10cm]{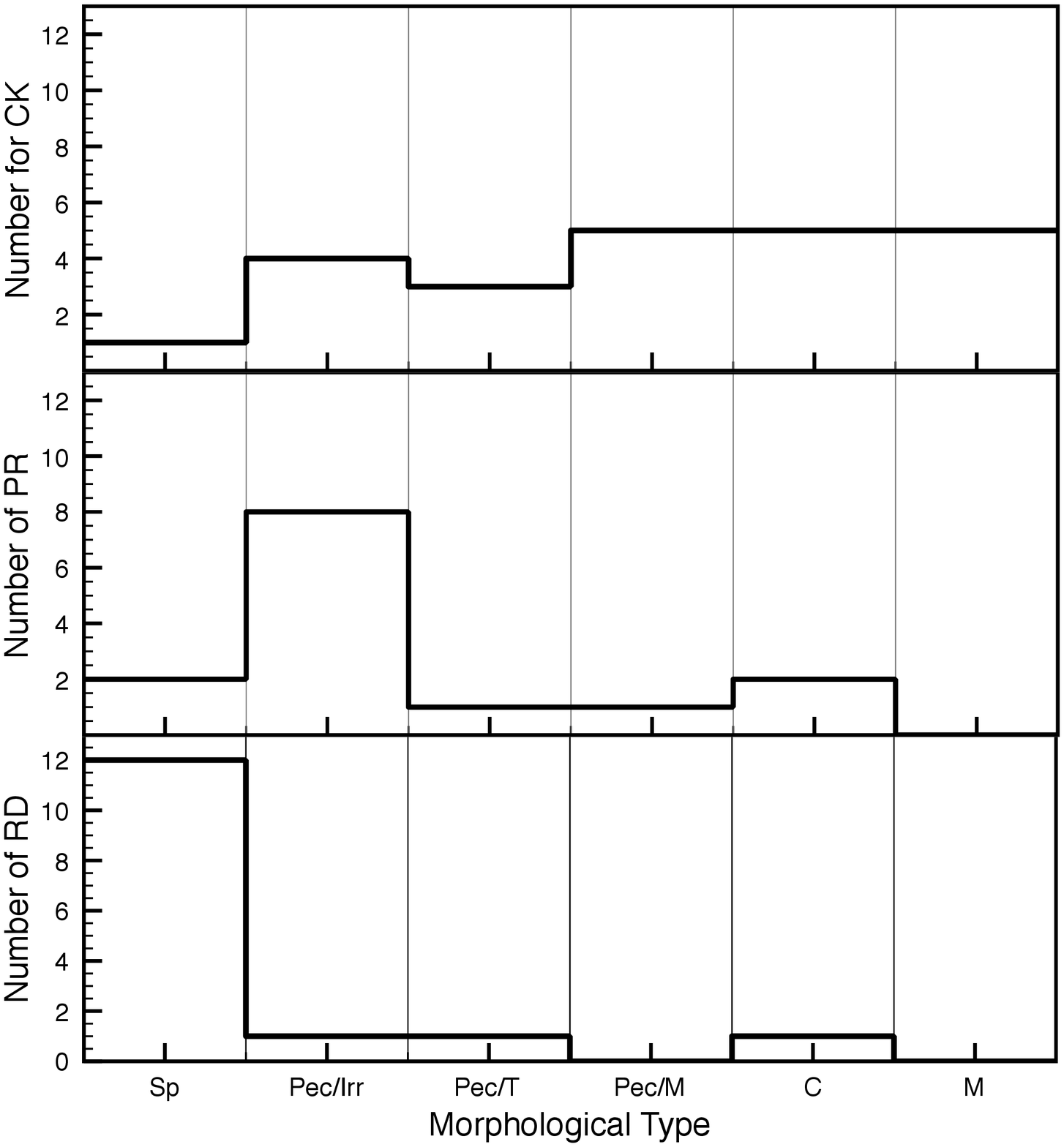}
   \end{center}
   \caption{Comparison between the morphological classification and the kinematical classification. Morphological classes are quoted in the bottom of X-coordinate, kinematical classes in Y-coordinate. RD refers to rotating disks, PR to perturbed rotations and CK to complex kinematics.}
   \label{histo} 
   \end{figure}

We compared the morphological classification derived in the previous section with the dynamical state of our objects as classified in Paper I.
As described in Sect. \ref{kin}, galaxies were classified into three kinematical class: RD, PR and CK (see Paper I for the kinematical classification). This kinematic classification was derived with a completely independent method, so it provides a unique test of whether the morphological classification is consistent with the kinematic state of the galaxies.
In Fig. \ref{histo} we report the comparison between the morphological classification derived with our method and the kinematic one. 

We find good agreement between morphological and kinematical classifications: 80\% of rotating disks are classified as spiral disks and more than 70\% of complex kinematics are peculiar or merging galaxies. 
Among the galaxies classified as spiral disks, only one has complex kinematics and two have perturbed rotation. The spiral, whose kinematics were classified as complex (J033213.06-274204.8), is a galaxy for which the velocity field is not aligned with the optical axis. The two galaxies that show perturbed kinematics include one galaxy (J033248.28-275028.9) that shows an elongated $\sigma$ peak possibly due to a large bar and one (J033226.23-274222.8) that shows a sigma peak shifted from the dynamical center but located on a minor merger event (\cite{Puech07b}). In this last case, the morphology alone would be insufficient
for detecting such situations.
Peculiar galaxies are mainly distributed between PR and CK. Interestingly, galaxies identified as possible mergers or tadpoles are mainly CK (8/11), whereas Pec/Irregulars are mainly PR (9/13). Not surprisingly, all the galaxies identified as major merger are CK. Although the sample is small, this may mean that our morphological classification correlates well with kinematical properties.
Finally, compact galaxies are mostly CK (5/8) and only one is RD (see also \cite{Puech06b}). This last object is morphologically ambiguous as it possibly shows spiral arms, and has a very complex color distribution. \\
To test our morphological classification against the kinematical one, we made a $\chi^2$ test between the two independent categories of rotating disks and complex kinematics. The $\chi^2$ test is used to determine the significance of differences between two independent groups. The hypothesis is that the two groups (RD and CK) differ with respect to the relative frequency with which their galaxies fall into different morphological classes. 
We find that the morphological classification of the two populations of RD and CK are inconsistent with a confidence level over 99.98\%. We also made the same test by considering the peculiar class as a single class, and we found a confidence level of 99.99\%. This suggests that our methodology is comparatively robust in classifying galaxies.

\section{Comparison with automatic classification}
\label{automatic}
We have shown in the previous section that it was possible to recover a good correlation between morphology and kinematics following our classification scheme. More particularly, our method at least allows accurate separation of rotation disks from complex kinematics.
Nevertheless, our method is not fully automatic and requires visual inspection of each galaxy. This human interaction unavoidably introduces subjectivity. 
In this section, we explore whether simple morphological parameters could efficiently distinguish RD from CK in a fully automatic way.
As a first indication, we can use the B/T parameter derived by Galfit. For instance, we can assume a classification where all the objects with B/T$<$0.4 are considered as spirals. We find 28 galaxies satisfying such a criteria. However, beyond this number, only 14 are RD and 7 are CK. We then conclude that a  classification only based on B/T is not efficient at clearly distinguishing the RD from CK and would unavoidably lead to overestimating the number of rotating spirals. As illustrated in Fig. \ref{FigZheng}, a visual check of each object is necessary to clearly disentangle regular objects from more perturbed ones and to isolate only spiral disks. In fact, it appears that a full automatization of the decision tree would not be easy to produce.\\
We also explore the ``non-parametric" methods to find whether a correlation between these parameters and the kinematical state of a galaxy could be derived. To do so, we compute four parameters which are often used in morphological analyses: Concentration, Asymmetry, Gini and M20.
The Concentration (C) parameter (\cite{Abraham94}) roughly correlates with the bulge over disk ratio (B/D) and the Asymmetry (A) parameter (\cite{Abraham96}) divides the sample between irregulars and more symmetric objects. More recently two other parameters have been introduced: the Gini coefficient (\cite{Abraham03}) which is a measure of the relative distribution of galaxy pixel flux values, and the M20 parameter (\cite{Lotz04}), which is the relative second-order moment of the brightest 20\% of a galaxy's pixels. \\
For C and A, pixels assigned to each galaxy are defined at 1.5$\sigma$ above background. C and A are measured following Abraham et al. (1996) method. Given the narrow redshift range in the sample, we do not attempt to correct the concentration as described in Brinchmann et al. (1998) (Appendix A).
Gini and M20 are measured following Lotz et al. (2004). \\

\begin{figure}[h!]
   \begin{center}
  \includegraphics[height=6cm]{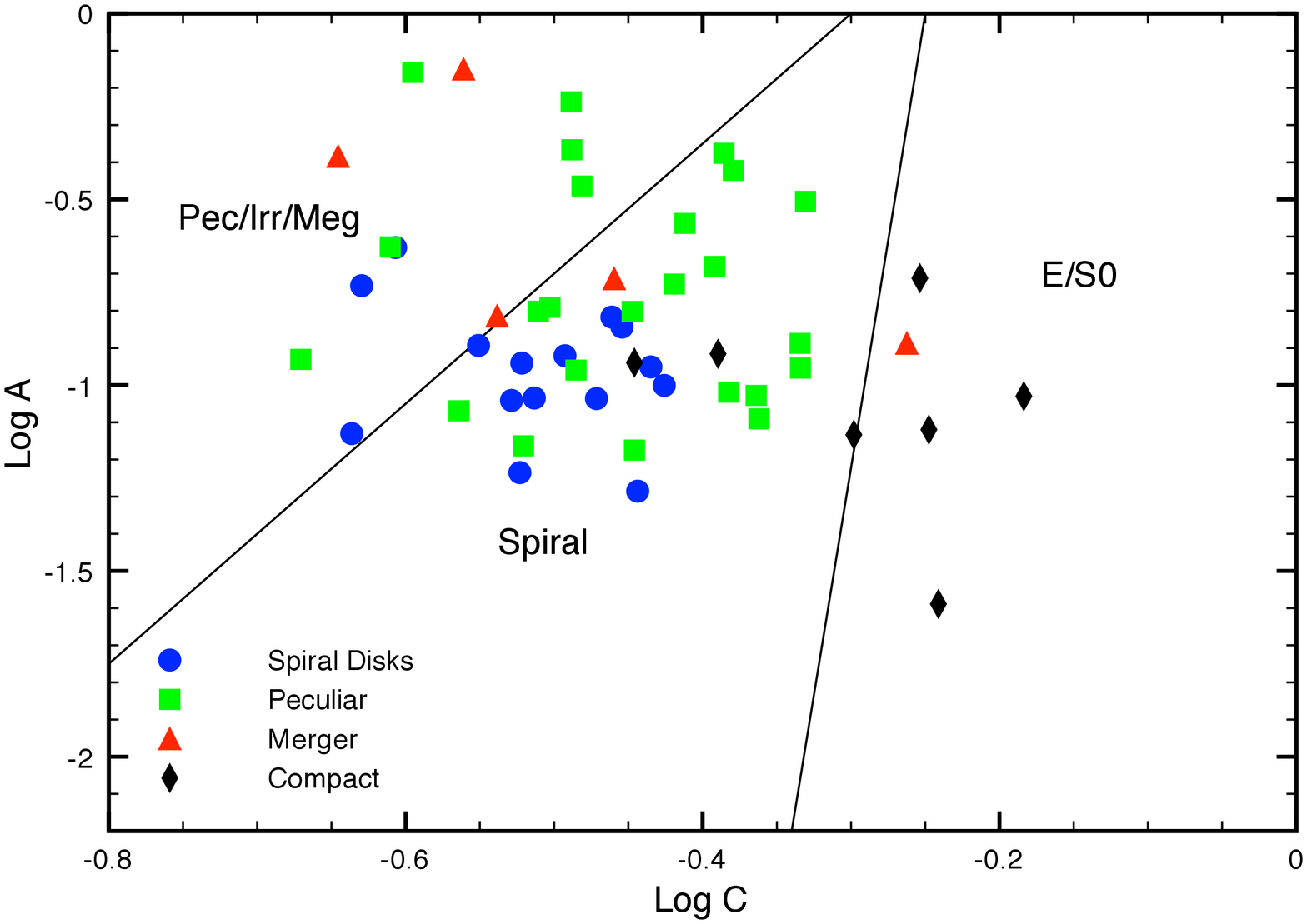}
    \includegraphics[height=6cm]{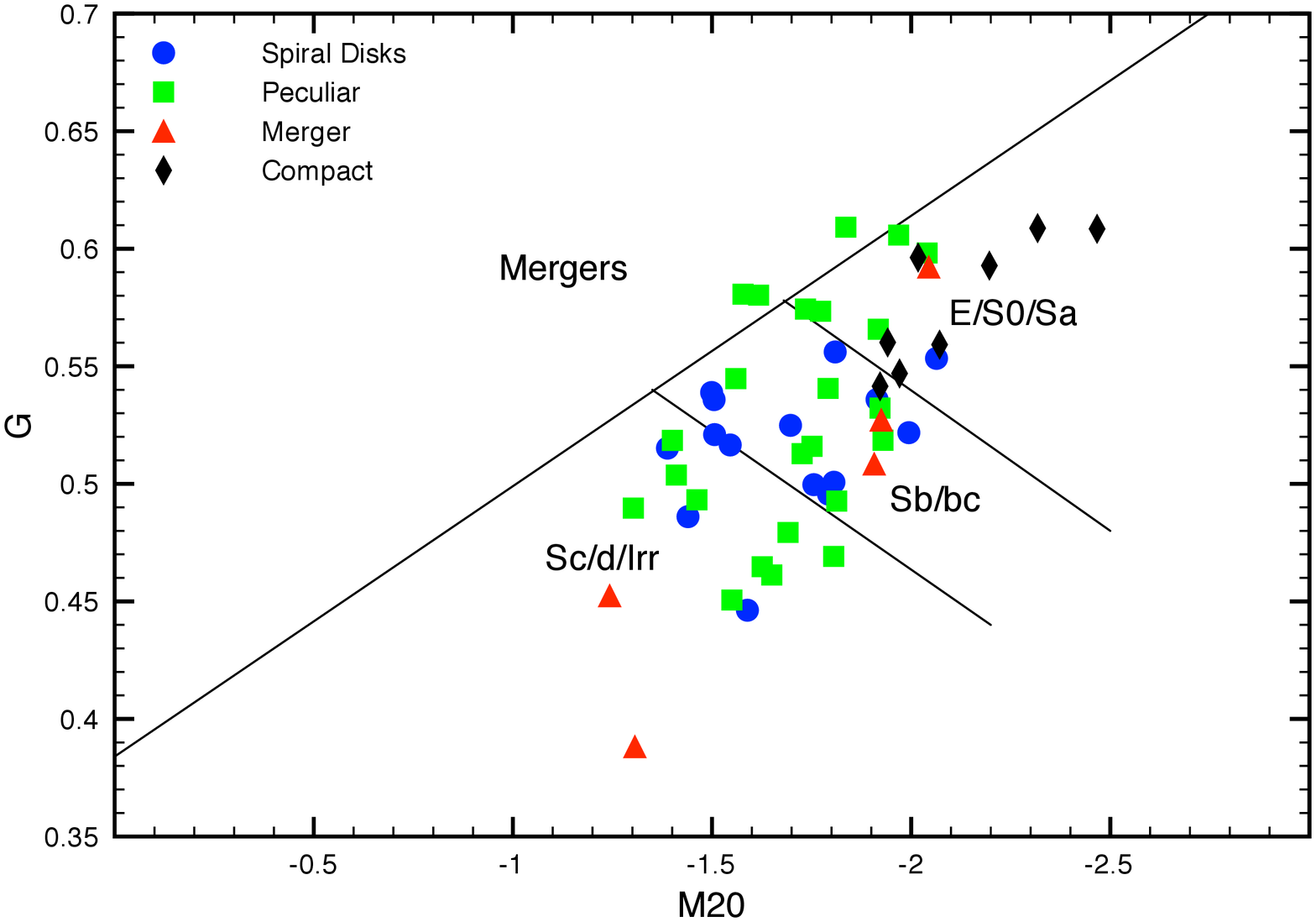}
   \end{center}
   \caption{{\bf top:} log(A)-log(C) diagram. {\bf Bottom:} Gini-M20 diagram. Different symbols refer to the morphological classes: spiral disks  are represented as blue circles, peculiars as green squares, compact galaxies as black diamonds, and mergers as red triangles.}
   \label{cas2} 
   \end{figure} 
   
In Fig. \ref{cas2} we show how the sample is distributed in a log(A)-log(C) plane (top) and in a Gini-M20 plane (bottom). We find that the segregation between different morphological classes is partially consistent with our classification. 
For instance, there is a clear offset between the compact galaxies (black diamonds) and the spiral disks (blue circles), whereas the distinction between peculiar (green squares) and spiral disks is less obvious. All the spiral disks indeed lie in specific regions in both diagrams. However, several objects that have been identified by our method as peculiar (or even merger (red triangles)) galaxies are mixed with the spiral disks. 
These last peculiar galaxies indeed show symmetric features or smooth light distribution, which explain why they lie near spiral disks. However, as illustrated in Fig. \ref{FigZheng}, they are clearly inconsistent with our spiral disk criterion.
It could be then difficult to define a criteria to isolate the spiral disks from other morphological types. 
As an example, we add the limits defined in Abraham et al. (1996) and Lotz et al. (2008) in both diagrams. These limits have been calibrated on local samples and are generally used to distinguish three or more rough morphological types.
With these limits, we find that the agreement for disks is good, since most of the objects designed as spiral disks are lying in the spiral (respectively Sb/bc) domain of both diagrams. The agreement is also good for the compact galaxies, as they almost all lie in or near the early type domain, which is consistent with the findings of Abraham et al. (1996). Note, however, that these objects could not be directly classified as early type objects because they show a relatively complex and/or blue color distribution (see sect. \ref{colorbis} and Fig. \ref{allimages}). For peculiar objects, the assignments are not very accurate, and around half of these objects they do not lie in the peculiar region: Pec/Irr/Merg for C-A and Sd/d/Irr and Merger for Gini-M20.
These results are consistent with those of Cassata et al. (2005), who find that these methods (even if efficient in disentangling early-type from late-type) generally failed to resolve the different classes contributing to late-type galaxies (see also \cite{Conselice03}, \cite{Yagi06}, \cite{vanderWel08}).

\begin{figure}[h!]
   \begin{center}
  \includegraphics[height=9cm]{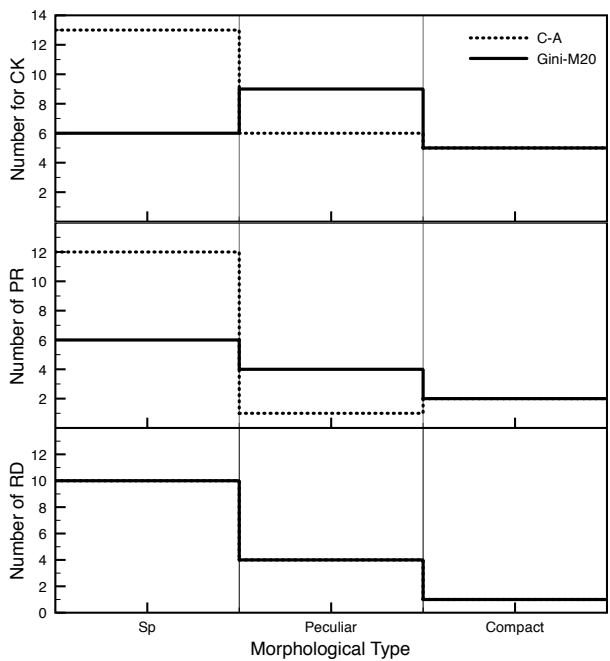}
   \end{center}
   \caption{Comparison between the concentration-assymetry (dotted filled histogram), Gini-M20 (solid histogram) morphological classification, and the kinematical classification. Morphological classes are quoted in the bottom of X-coordinate, kinematical classes in Y-coordinate. RD refers to rotating disks, PR to perturbed rotations and CK to complex kinematics. The line for the CA and Gini/M20 exactly overlap for the RD
types.}
   \label{histo_comp} 
   \end{figure}

To test the impact of this limitation on the morpho-kinematical correlation, we assume a morphological classification based only on the boundaries defined in Abraham et al. (1996) and Lotz et al. (2008). All the objects lying in the spiral region for the log(A)-log(C) plane (respectively Sb/bc for the Gini-M20 plane) are assumed to be spirals, and all objects lying in the Pec/Irr/Merg  (respectively Sd/d/Irr and Merger for Gini-M20) are assumed to be peculiar. The compact galaxies are included independently as they are identified when using the half-light radius. All the remaining galaxies lying in the E/S0 domain are not included, because the working sample does not contain early type galaxies.
In Fig. \ref{histo_comp} we report the comparison between these classifications and the kinematical one. 

In Fig. \ref{histo_comp}, it appears that the morpho-kinematical correlation is less obvious with the automatic methods. The same $\chi^2$ test as in Sect. \ref{comp} gives 50\% and 90\%  confidence for the log(A)-log(C) and Gini-M20 planes, respectively. For comparison, our classification considering the peculiars and mergers as a single peculiar class gives 99.9\%.
Whereas these methods almost recover the number of rotating spirals correctly, they mix the spiral disks with some peculiar galaxies: several galaxies lying in the spiral region have complex kinematics.
This result indeed confirms the finding of Conselice et al. (2003),  Conselice et al. (2005), and more recently Atkinson et al. (2007),  and Kassin et al. (2007). For example, Conselice et al. (2003) find that the asymmetry index is not sensitive to all phases of the merging process, particularly those in the beginning or at the end of the mergers. They argue that the total number of mergers  would be underestimated by a factor of two, which is similar to what we find with our analysis. Kassin et al. (2007) report that some galaxies that were fairly normal according to Gini/M20 looked more disturbed or compact by visual inspection. These objects generally present larger scatter in the Tully-Fisher relation (TFR) than visually identified disks. Combined with the result of Flores et al. (2006) showing that perturbed and complex kinematics are driving the scatter of the TFR, this also suggests that automatic methods mix the spiral disks with complex kinematics.
Similar to B/T, a method only based on a few morphological parameters will unavoidably overestimate the number of rotating spirals, and it fails to uniquely identify all the CKs.

\section{Rotating spiral disks}
\label{color}

Our classification scheme is efficient in isolating different morpho-kinematical classes. We therefore use this method to
focus on a particular subsample defined by the rotating spiral disks. This subsample includes all the galaxies sharing the
spiral disk morphology and a relatively relaxed dynamical state (rotating disks and perturbed rotations). In doing so we have assumed that a disk, even with a slight kinematic peculiar structure (i.e. PR), may be considered as a disk. This includes two galaxies, J033226.23-274222.8 (a minor merger, see Puech et al, 2007) and J033248.28-275028.9.
Over the 52 galaxies of our sample, we find 14 objects sharing the spiral disk morphology and a rotating velocity field. These galaxies are the 14 first objects of Table \ref{toto} (see also Table \ref{SFR} and Fig. \ref{allimages}), they are particularly interesting because they can be directly compared with the local rotating spirals.


\subsection{Sizes and B/T}
In terms of size and B/T distributions, these galaxies are consistent with the local late type spirals.
The disk scale lengths derived from Galfit range from 2.5kpc to 6.5kpc with a mean value of 4.1kpc.
For comparison, the disk scale length of the Milky Way is estimated to be between 2kpc and 3kpc, whereas the disk scale length of M31 ranges between  5kpc and 6kpc (see Sect. 2.1 in \cite{Hammer07} and references therein).
This indicates that the rotating spiral disks are roughly in the same size range compared to the local spirals.
In terms of B/T, all these rotating spirals are distributed between 0.01 and 0.17. 
Around half of these objects have B/T$<$0.1 consistent with Sc type or later, and the other half are in the range 0.1$<$B/T$<$0.2, consistent with Sb type.

\subsection{Fractional number of rotating spiral disks}
\label{fraction}
Over the 52 galaxies of the working sample, we find 14 objects (27\%) with both a spiral disk morphology and a simple rotational velocity field.
As discussed in Sect. \ref{ssac}, the sample is representative of intermediate-mass galaxies {\em{with}} EW([OII])$>$15$\AA$ at intermediate redshift. Let us now consider the general population of galaxies at $z$=0.4 - 0.8, with $M_{stell}>$ 1.5x$10^{10}M_{\odot}$, including those with EW([OII])$<$15$\AA$. Following Hammer et al. (1997), the fraction of intermediate-mass galaxies with EW([OII])$>$15$\AA$ represents 60\% of intermediate-mass galaxies. 
Then, the fraction of rotating spiral disks depends on how we distribute the remaining 40\% of those galaxies with EW([OII])$<$15$\AA$ between the different morphological classes. To do so, we first assume that these quiescent galaxies are mainly E/S0 and evolved spirals (see for instance Table 1. in \cite{Cassata05}). We then assume a population of 23\% of E/S0 consistent with a similar morphological study based on a sample of 111, 0.4$<z<$1 intermediate-mass galaxies (\cite{Zheng05}). This fraction is also consistent with other several studies (e.g. \cite{VandenBergh01}, \cite{Brinchmann98}, \cite{Cassata05}, \cite{Conselice05}, \cite{Lotz08}). This leads to a fraction of  17\% of evolved spirals with EW([OII])$<$15$\AA$.
Accounting for these corrections, we find that 33\% (= 27x0.6 + 17)\% of intermediate-mass galaxies are rotating spiral disks at $z\sim$0.6. 
In Table \ref{fraction} we report the distribution of all the morpho-kinematical classes. RspD is for rotating spiral disks, C for compact galaxies and M for mergers. The morphological distribution for galaxies with  EW([OII])$<$15$\AA$ is from (\cite{Zheng05}).
\begin{table}[h!]
\begin{center}
\caption{Fraction number of the different morphological classes.}
\begin{tabular}{cc|cccc}
\hline \hline
\multicolumn{2}{c|}{with EW([OII])$<$15$\AA$} & \multicolumn{4}{c}{with EW([OII])$>$15$\AA$} \\ 
E/S0 & RSpD  &  RSpD & Pec & C & M\\ \hline
23\% & 17\% & 16\%  & 28\% & 10\% & 6\% \\ \hline\hline
\end{tabular}
 \label{fraction}
\end{center}
\end{table}

\subsection{Colors}
\label{colorbis}
\begin{figure}[h!]
   \begin{center}
   \begin{tabular}{c}
  \includegraphics[height=9cm]{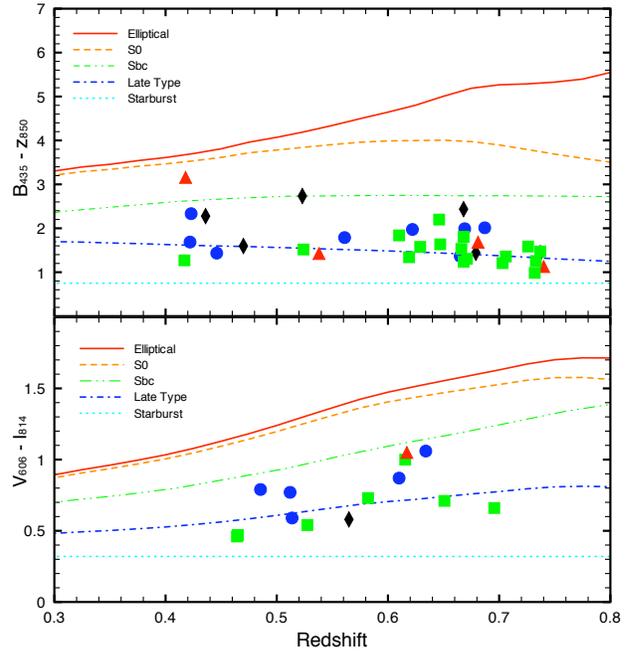}
  \end{tabular}
   \end{center}
   \caption{B$_{435}$-z$_{850}$ (top) and V$_{606}$-I$_{814}$ (bottom) observed color as a function of redshift. Three solar metallicity, beginning their evolution at $z$=5 galaxy models, including an elliptical galaxy (single burst - filled red curve), S0 galaxy ($\tau$=1Gyr - orange dotted line), and  late type galaxy ($\tau$=7Gyr - blue dotted line) are plotted for comparison. A young starburst (assuming $L_v \propto v^{-1}$ - cyan dotted line) is also plotted. Different morphological classes are represented by blue circles for spiral disks, green squares for peculiars, black diamonds for compacts, and red triangles for mergers. The mean error for B$_{435}$-z$_{850}$ plot is 0.03mag  (0.01mag for V$_{606}$-I$_{814}$ respectively), much smaller than the symbol size.}
   \label{ssp1} 
   \end{figure}

Based on the color information provided by the color maps, we now investigate the distribution of the stellar populations within each galaxy. 
To do so, we compare the observed colors of our objects with modeled color-redshift evolution curves as predicted by Bruzual-Charlot stellar population synthesis code GALAXEV (\cite{Bruzual03}). 
We can compute 3 models of galaxies providing B$_{435}$-z$_{850}$ (for galaxies located in the CDFS) and V$_{606}$-I$_{814}$ (for CFRS and HDFS) observed colors at different redshift. The 3 spectral template are (i) an instantaneous burst corresponding to elliptical type objects; (ii) an exponentially decaying star formation rate (SFR) with e-folding time scale of 1Gyr, corresponding to S0 like objects; 
(iii) an exponentially decaying SFR with e-folding time scale of 7Gyr, corresponding to late type like object. All these models assume a solar metallicity and an epoch of formation at $z$=5. We also compute the theoretical color of a young starburst assuming a power law spectrum $L_v \propto v^{-1}$ and add the evolution curve of an Sbc galaxy from Benitez et al. (2004) templates.
Figure \ref{ssp1} shows these evolution curves, as well as the integrated color of each galaxy. For comparison, we include all the morphological classes.

In Fig. \ref{ssp1}, almost all our objects lie between the starburst and the Sbc evolution model. No specific trends between different morphological classes or between different redshifts can be identified. This suggests that our objects are mainly composed of a mix of stellar populations including old, intermediate, and young populations. This is indeed confirmed by spectroscopic studies (see e.g. Hammer et al. 2005). This mix clearly appears on the color maps where the resolved colors allow us to investigate the distribution of stellar populations within each object (see Fig. \ref{allimages}). 
As described above, the spiral disks share an old and red central population with a younger and blue population located in the disk. In the following, we try to characterize these populations quantitatively.

\subsubsection{Central color}
\begin{figure}[h!]
   \begin{center}
   \begin{tabular}{c}
  \includegraphics[height=9cm]{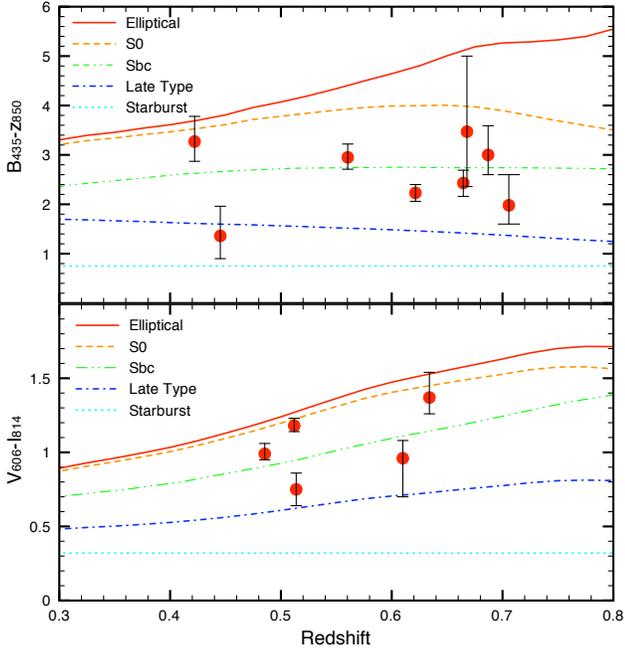}
  \end{tabular}
   \end{center}
   \caption{B$_{435}$-z$_{850}$ ({\bf top}) and V$_{606}$-I$_{814}$ ({\bf bottom}) observed color of the bulges for the rotating spiral disks  sub-sample. The different evolutionary models are the same as in Fig. \ref{ssp1}.  Error bars are used to show the maximal pixel-by-pixel color range within each bulge.}
   \label{bulbos} 
   \end{figure}

To derive the central color of the rotating spiral disks, we follow the procedure defined in Ellis et al. (2001). We construct a large elliptical aperture using second-order moments, and then a circular aperture with a radius of 5\% of the semi-major axis of the large aperture is used to integrate the inner colors.
Ellis et al. (2001) demonstrate that, following this method, the contamination from disk is negligible, and then, that these inner colors can be associated with the color of the bulges. We show these 'bulge'-integrated colors as well as the modeled evolution curves in Fig. \ref{bulbos} (see red filled dots). As an estimation of the uniformity of the color inside each bulge, we overplot the maximal pixel-by-pixel range of colors.

The modeled evolution curves are mainly used to compare the bulge integrated colors with the passively evolving system formed in a single burst at $z$=5 (elliptical model).
Indeed, bulges as red as the elliptical model indicate that their populations have comparatively similar ages and metallicities (\cite{Peletier99}), and then that these bulges were formed at high redshift, whereas bluer bulges probably host a younger population.
 Moreover, dust extinction may significantly redden the color, and the galaxy sample certainly suffers from reddening effects because the mean inclination is particularly high ( $\sim60^{\circ}$). This would make the bulges intrinsically bluer than observed.
Nevertheless, even with the effect of dust extinction, we find that only 3 bulges are as red as the elliptical model, whereas most of them lie near the Sbc model evolution curve.
This clearly suggests that these bulges host a significant population of young/intermediate age stars.

\subsubsection{Color of disks}
The disks of these distant rotating spirals show very blue outskirts with the color consistent with young star-forming regions (see Fig. \ref{allimages}). 
Accounting for the uncertainties, we identify as a pure starburst region all those pixels where the observed B-z color (observed V-I ) is lower than 1mag (0.4 mag) for CDFS (respectively HDFS and CFRS) galaxies (see Fig. \ref{ssp1}). 
By doing this, we are able to (i) compute the fraction of light coming from pure starburst regions, and (ii) find where this starburst activity is located in the disk.

To derive the relative strength of the starburst activity, we first computed the ratio between the UV (observed B or V band) flux coming from the pure starburst regions over the total UV flux. The total UV flux were defined to include all the pixels that are 4$\sigma$ above the background, following the method described in Zheng et al. (2004). We find that between 5\% and 40\% of the global UV flux can be associated with pure star-forming regions (mean value = 15\%). 

  \begin{figure}[h!]
   \begin{center}
   \begin{tabular}{c}
  \includegraphics[height=2.8cm]{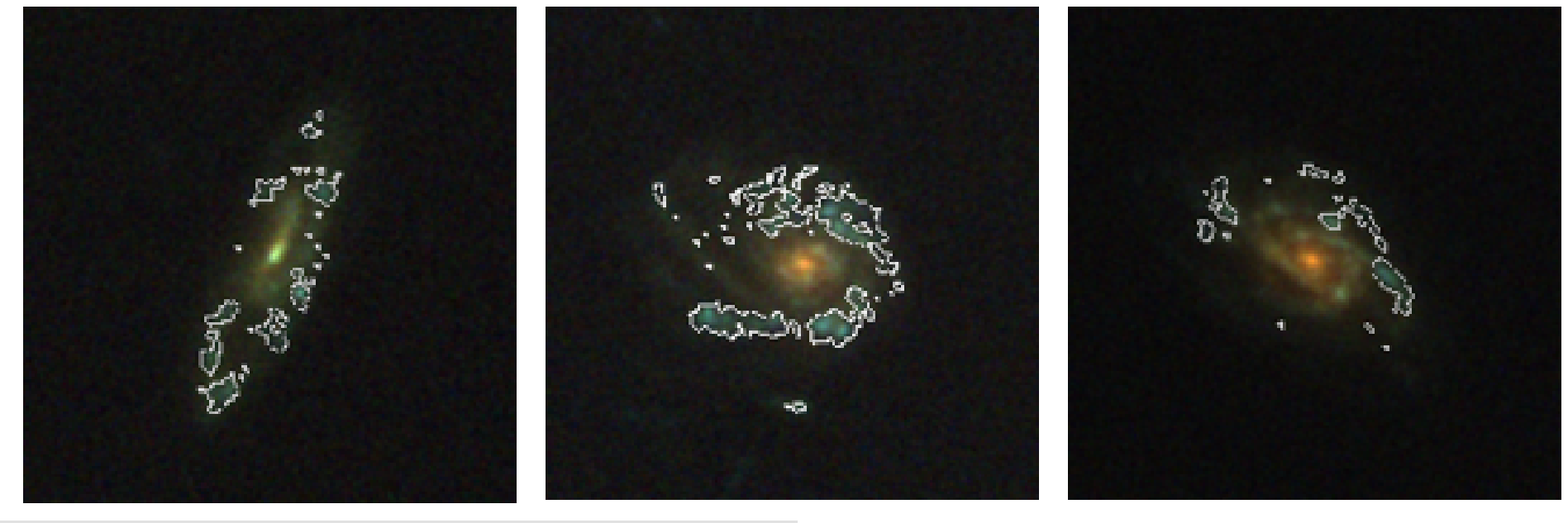} \\
    \includegraphics[height=4.5cm]{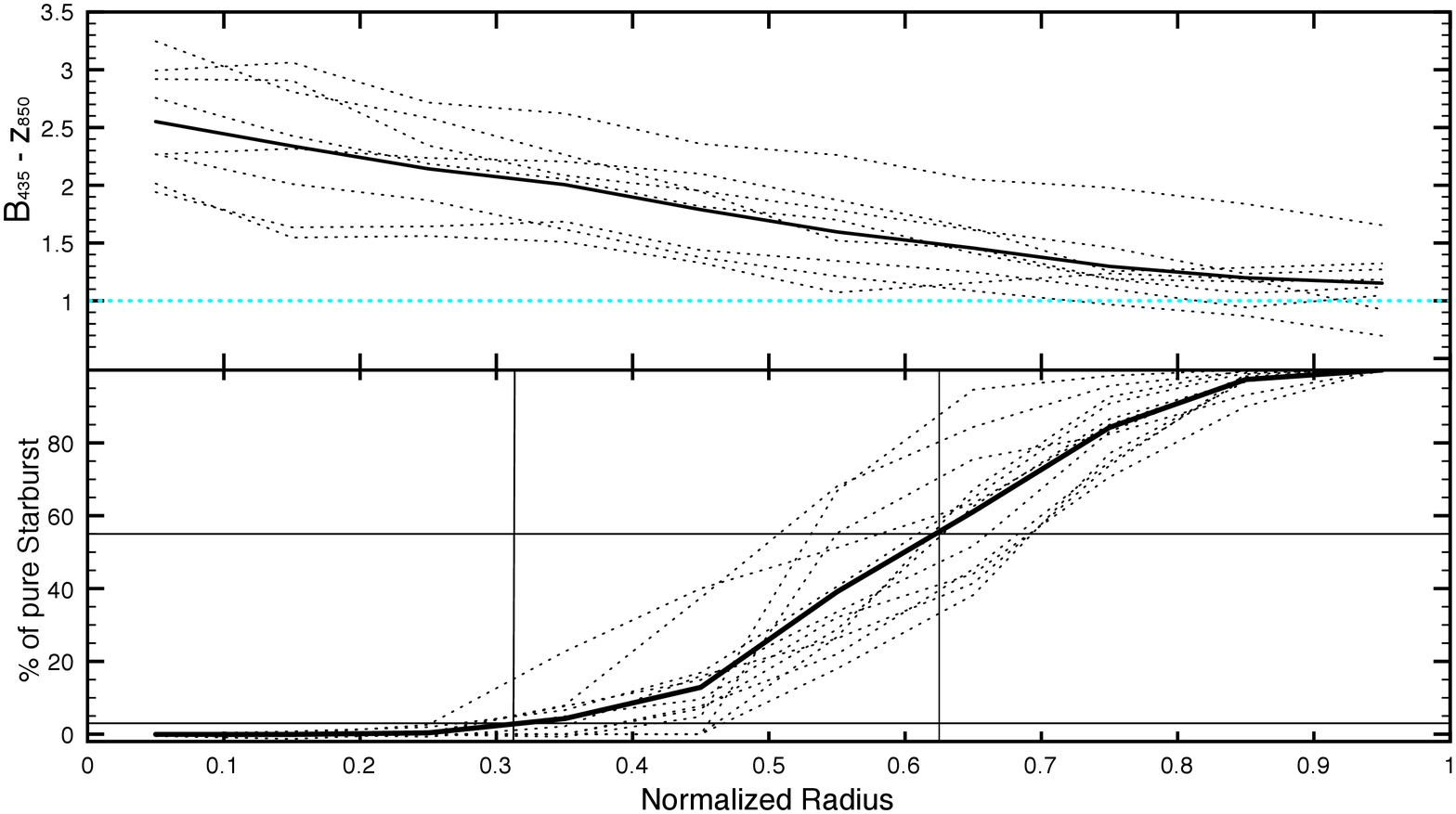} 
  \end{tabular}
   \end{center}
   \caption{{\bf Top:}  Three color images of three rotating spiral disks for which we have identified the star-bursting regions (B-z $<$1 or V-I$<$0.4). {\bf Middle:} Color profiles for the CDFS galaxies.  {\bf Bottom:} Repartition of the starburst light for a growing elliptical aperture. For each aperture, it represents the integrated UV light coming from pure starburst regions normalized by the UV flux coming from all star-bursting regions.}
   \label{delta3} 
   \end{figure} 
   
In Fig. \ref{delta3} (top) we show a three-color image of three galaxies for which we overplot a contour around the starbursting regions. It clearly appears that the starbursting regions are located towards the disk outskirts. 
More precisely, in Fig. \ref{delta3} (bottom), we define several elliptical apertures growing from the center of each object to an outer radius. The maximal radius being defined to include all the galaxy pixels that are 4$\sigma$ above the background. This allows us to define a normalized radius and compare together galaxies with different sizes. Notice that this maximal radius roughly corresponds to the optical radius, as we find it to be $\sim$ 3.2 times the disk scale length (\cite{Persic91}). In each aperture, we integrate the UV flux coming from the pure starburst regions and normalize this value by the UV flux coming from {\it all} star-bursting regions (see above). These results are shown in Fig. \ref{delta3} (bottom) for each galaxy (dotted lines), as well as the mean trend (solid line). This plot also clearly indicates that the starburst activity is increasing towards the disk outskirts. Assuming the maximal radius to be 3.2 times the disk radius, we find that more than 95\% (respectively 45\%) of the starburst activity is located at a greater radial distance than Rd (respectively 2Rd). \\

\subsection{Stellar mass and SFR}
\label{DT}
Table \ref{SFR} shows the stellar mass and SFR for the rotating spiral disks sample.
The stellar masses have been derived by Ravikumar et al. (2007) for galaxies located in the CDFS and by Flores et al. (2006) for the galaxies in the CFRS and HDFS. \\
For the SFR, we report the SFR$_{UV}$ and SFR$_{IR}$ from Hammer et al. (2005) for the CFRS galaxies. SFR$_{UV}$ are derived from $H_{\alpha}$ luminosities and SFR$_{IR}$ from ISOCAM observations. 
For the CDFS galaxies, we derived the SFR$_{UV}$ using the Kennicutt (1998) calibration based on the rest frame 2800$\AA$ and the SFR$_{IR}$ from Spitzer observations.
To derive the SFR$_{IR}$ we used the 24$\mu$m MIPS data. At $z\sim$0.6, this flux roughly corresponds to the rest-frame 15$\mu$m flux so we can use the Chary \& Elbaz calibration to derive the total IR luminosity, and then use the classical Kennicutt calibration to derive the SFR$_{IR}$. 
Although the SFR for CFRS and CDFS galaxies are derived with different methods, they give consistent results. Moreover, the majority of the galaxies come from the CDFS that have the best quality data.
For the galaxies that have not been detected by Spitzer or ISOCAM, only an upper limit of the SFR$_{IR}$ can be derived from the detection limits. \\
Finally, in the last column of Table \ref{SFR} we report the ratio of the stellar mass to the SFR$_{tot}$, where SFR$_{tot}$=SFR$_{IR}$+SFR$_{UV}$ (and only a lower limit for galaxies not detected in IR). This ratio provides a time-scale that can be roughly associated with a mass-doubling time (if SFR is constant).

\begin{figure}[h!]
   \begin{center}
    \includegraphics[height=3.3cm]{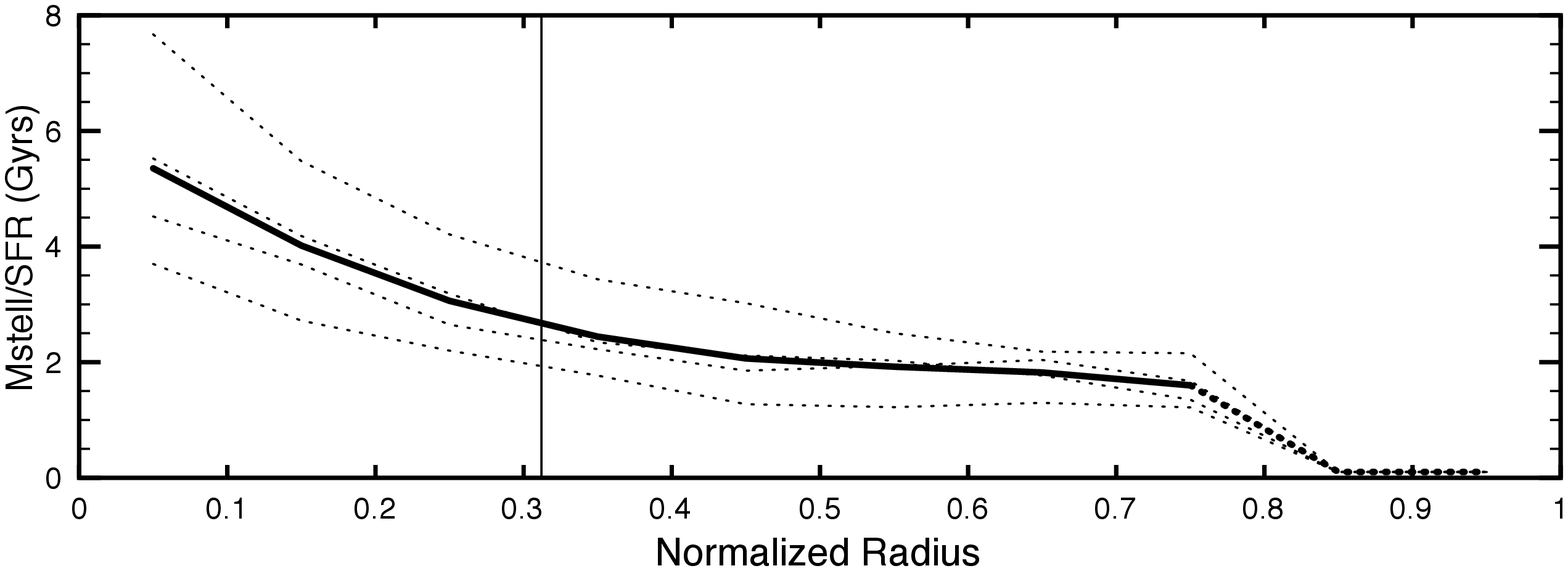} 
   \end{center}
   \caption{Ratio between stellar mass and SFR$_{tot}$ as a function of galaxy radius. SFR$_{tot}$ is defined as the sum of SFR$_{UV}$ and SFR$_{IR}$. Only galaxies which have been detected by Spitzer are used in this analysis. The vertical line is at 1/3.2 of Rmax which roughly corresponds to the disk scale length limit.}
   \label{dt} 
   \end{figure}

\begin{table*}[t]
\begin{center}
\caption{Stellar masses, star forming rates and M$_{stell}$/SFR for the rotating spiral disks. }
\begin{tabular}{ccccc}
\hline\hline
ID & $Log_{10}$(M$_{stell}$) & SFR$_{UV}$&  SFR$_{IR}$ & M$_{stell}$/SFR\\ 
      &  $Log_{10}$(M/$M_{\sun}$)  &   $M_{\sun}yr^{-1}$   &   $M_{\sun}yr^{-1}$  & Gyr \\ \hline 
J033212.39-274353.6 & 10.61 & 1.0 & 16.2 & 2.4 \\
J033219.68-275023.6 & 10.88 & 5.1 & 18.6 & 3.2 \\
J033230.78-275455.0 & 10.66 & 2.9 & $<$10  & $>$3.5 \\ 
J033231.58-274121.6 & 10.16 & 3.2 & $<$10.5 & $>1$  \\
J033237.54-274838.9 & 10.70 & 7.9 & 22.3 & 1.7 \\ 
J033238.60-274631.4 & 10.53 & 2.1 & $<$7.5 & $>$3.5 \\
J033226.23-274222.8 & 10.72 & 3.2 & 17.5 & 2.5 \\
J033248.28-275028.9 & 10.09 & 2.0 & $<$3.5 & $>$2.2 \\
HDFS4020 &  9.82 & - & - &- \\
CFRS030046 & 10.64 & 3.9 & $<$15 & $>$2.3 \\
CFRS030085 & 10.21 & 2.3 & 50.9 & 0.3 \\
CFRS030619 & 10.63 & 1.6 & $<$13 & $>$2.9 \\
CFRS031353 & 10.85 & 4.1 & $<$25 & $>$2.4 \\
CFRS039003 & - & 12.12 & 94.95 & - \\ \hline\hline
\end{tabular}
 \label{SFR}
\end{center}
\end{table*}
In the previous section, we have shown that starburst regions are mostly located in the disk outskirts, suggesting very active and recent  star formation in the outer parts of the disks. 
In a crude analysis, we can compare the M$_{stell}$/SFR ratio between these outer regions and the inner parts of the disks. We restrict this analysis for CDFS galaxies that have been detected by Spitzer, because they represent the most robust available data. Our purpose is not to quantify these values precisely, but rather to derive general trends. 
We first assume that the spatial distribution of the SFR2800$\AA$ follows the observed B-band flux distribution: for $z\sim$0.6 galaxies, the observed B-band indeed roughly corresponds to the 2800$\AA$ flux. To do so, some cruder assumptions are required, such as
 (i) the SFR$_{IR}$ follows the same distribution as the SFR2800$\AA$, and (ii) the stellar mass is sampled by the observed z-band flux. These last two assumptions are probably very approximate, since we have no idea how the dust is distributed within the galaxies.
However, these assumptions are needed to distribute the SFR and the stellar mass at different radii and to evaluate the distribution of the M$_{stell}$/SFR time scale. We computed this time-scale in different annuli, from the center to the maximal radius. Results are shown in Fig. \ref{dt}. All the galaxies show a negative gradient from large time scale near the center (between 4 and 8 Gyr) to shorter scales near the disk outskirts (between 1 and 2 Gyr). In the regions even farther out, the stellar populations are dominated by young starbursts (see Fig. \ref{delta3}). For these short-lived young stellar populations (typically $<$100 Myrs), a significant correction of the mass estimation should be applied (see \cite{Bell03}). Rather than try to correct for this effect, we arbitrarily decrease the M$_{stell}$/SFR time scale toward 0.1 Gyr at the larger radius.\\
Under all the assumptions made above, these observations clearly suggest an inside-out formation process in which the disk is gradually built from the outskirts. 

\section{Discussion}

\subsection{A low fraction of rotating spiral disks at $z\sim$0.6}
\label{lowfraction}

It is tempting to define morpho-kinematical classes from the relatively good coincidence between our morphological classification and the kinematic classes (see  Fig. \ref{histo}). Most galaxies can be distributed between three main distinct classes: (i) the rotating spiral disks for which both kinematics and morphology are relaxed, (ii) the galaxies with irregular morphology and perturbed kinematics, and (iii) the fully unrelaxed objects for which both morphology and kinematics are complex.

It is useful to compare this distribution with what we know from the local universe. There, the overwhelming majority of objects are dominated by rotation or dispersion (e.g. GASPH: \cite{Garrido02} - SAURON: \cite{Emsellem04}), whereas galaxies with anomalous kinematics are generally ones in close pairs or interactions (e.g. \cite{Kannappan04}). Disk galaxies constitute the majority of local objects with $\sim$ 70\% of the intermediate-mass galaxies (\cite{Nakamura04}, \cite{Hammer07}) and we can reasonably assume that these disks are almost all dynamically relaxed as they lie on the tight Tully-Fisher relationship. 
Indeed, Kannappan et al. (2004) show that, at low $z$, large TF offsets are produced by kinematically anomalous objects.
At higher redshift, Paper III find a similar result as they observe that the dispersion around the TF relation increases from RDs (0.31 mag) to PRs (0.80 mag) and CKs (2.08 mag). Restricting the distant TF relation to dynamically well-relaxed RDs, the local and distant relations have comparable dispersion (e.g. \cite{Pizagno07} for the local reference). It suggests that both distant RD and local spirals are comparably kinematically relaxed. Of course, some kinematically distorted or even complex velocity fields may also fall on the TFR (see for instance Fig. 8/9 in Flores et al. 2006), but they represent a very small fraction.
Then, a significant change between distant and local galaxies concerns this fraction of rotating spirals. Indeed, we found that at $z\sim$0.6, $\sim$1/3 of galaxies are regular rotating disks (see Section \ref{fraction} and Table \ref{fraction}), which implies that the number of rotating spiral disks has increased by a factor $\sim$2 during the last 6 Gyrs.

However, one could argue that our classification scheme (either morphological and kinematical) is too restrictive and that perturbed rotating Pec/Irr galaxies should be included in the rotating spiral disk sample. 
Even if we include this category among the rotating spiral disks, this would lead to a fraction of $\sim$3/7 of ``regular galaxies" that is still less than the local fraction.
Moreover, Paper I convincingly showed that the kinematic classification is robust and that discrepancies seen in velocity fields or sigma maps correspond to the morphological perturbation seen in the optical images. Given the good correlation between kinematics and morphology, it is unlikely that some rotating spiral disks may have been missed.
Only two galaxies that are morphologically classified as spiral disks are PR (J033226.23-274222.8 and J033248.28-275028.9) and one is CK (J033213.06-274204.8 - see also Sect. 4). 
These particular cases point out that morphology alone is insufficient for detecting a slight kinematic peculiar
structure, such as is caused by minor mergers (e.g. Puech et al. 2007b for J033226.23-274222.8).
On the other hand, over the 15 RD galaxies, three are not morphologically classified as spiral disks (J033234.04-275009.7 (``Peculiar - Irregular" object in Fig. \ref{herbier}), J033241.88-274853.9 (first object in the third column of Fig. \ref{FigZheng}), J033245.11-274724.0 a compact galaxy). \\
These exceptions show that some perturbed morphology can nonetheless have relaxed dynamics and may require further analysis (e.g. Puech et al. 2008b for a detailed analysis of J033241.88-274853.9). In each instance, to verify whether k-correction may have influenced our classification, we re-examined the morphology on the sole basis of the F850LP/F814W filter. Doing so, we find that all the morphological disturbed features that prompt us to classify an object as peculiar (and which are generally blue) are always present in these 'red' filters. This suggests that k-correction does not strongly affect our conclusions.\\\\
 Finally, the main limitation of this study is the still small number of objects, leading to possible non negligible statistical uncertainties. 
 However, each of the 4 different fields taken independently give the same distribution in morphological as in kinematical numbers. A simple $\chi^2$ test between the different fields confirms this trend. Moreover, the sample follows the luminosity function in the $z$ = 0.4 - 0.75 range (see Fig. 1 in Paper I). This indicates that, even if the number of objects is small, our result, a small number of rotating spiral disks at z= 0.6, is robust.\\

 \subsection{Comparison with previous works}

The evolution in the number fraction of regular spirals is surprising and in contradicts former studies (e.g. \cite{Lilly98} ;  \cite{Ravindranath04}, \cite{Sargent07}) that found that most of the spirals were in place at $z$=1. 
The differences observed with these former studies certainly come from the different methodology used to identify disk galaxies. 

Lilly et al. (1998) used a 2D light-fitting procedure, approximatively similar to the one presented in Sect. \ref{bt}. The main  differences are that (i) they used a de Vaucouleur light profile for the bulge components instead of a Sersic profile, and (ii) for each galaxy,  they subtracted a version of itself rotated by 180$^{\circ}$ from it. This last step is performed to ``symmetrize" the galaxies, but it would not modify the global underlying light distribution. They then classified as disk-dominated objects all those galaxies with B/T$<$0.5.
Even if their method is slightly different from our derivation of B/T, we have seen in Sect. \ref{automatic} and in Fig. \ref{cas2} that a method only based on this structural parameter is not able to clearly distinguish RD from CK so it unavoidably overestimates the number of spirals.

In their analysis, Ravindranath et al. (2004) use a single Sersic component to model the brightness profile of the galaxies. They then classify as disk-dominated all objects with n$<$2. 
In a crude approach, we reproduced their procedure for the galaxies located in the CDFS, following the method described in Sect. \ref{bt} but with only one Sersic component per galaxy. By doing this, we find 30 galaxies with n$<$2, over which 8 are RD but 13 are CK (9 are PR). This is agree with Cassata et al. (2005), who find that ``the sersic index alone provides just a broad and not unequivocal  indication of the morphological type".

The method developed in the present study is not fully automatic but instead based on a detailed analysis of each galaxy in order to assign them a morphological label. Structural parameters (R$_{half}$ and B/T), color information, and visual inspection are used in a complementary way. In this process, only the more regular galaxies (in their morphology and color distribution) are identified as spirals. This method indeed leads to a smaller number of spirals compared to previous studies; however, the classification seems robust because, on one hand, almost all these spirals are rotating disks, and on the other, almost all the rotating disks are spirals. This indicates that our classification is, at least, efficient in correctly recovering the rotating spirals.
We then suspect that, among the objects identified as spirals by former studies, a large fraction of them are dynamically perturbed or complex and have indeed peculiar morphologies. As an illustration of this, 12\% of the objects classified as large disks galaxies by Lilly et al. (1998) are included in the list of visually identified mergers presented by Le F\`{e}vre et al. (2000). As discussed in Sect. \ref{automatic}, automatic methods unavoidably mix rotating spirals with dynamically complex objects. Strikingly, we precisely find that only about 50\% of the objects identified as spirals by these methods are in fact rotating spiral disks.
It is then probable that former studies have overestimated the number of spirals, even if only a braoder 3D spectroscopic survey could definitely resolve this issue.


\subsection{Are rotating spiral disks forming stars inside out?}

In this section we consider the subsample of rotating spiral disks with emission lines, e.g. 16\% of the whole sample of $M_{stell}>$ 1.5x$10^{10}M_{\odot}$, galaxies at z=0.4-0.8. Because it is probable that these galaxies are evolving into present-day spirals, they vary likely represent one fourth of today's spiral progenitors (see Sect. \ref{lowfraction}). Based on their morphology, kinematics, and color distribution, the rotating spiral disks identified in this paper share many properties with local spirals. They all show regular patterns, with a red central bulge surrounded by a blue disk, and all have integrated colors between the late type and Sbc models.

However, those disks are certainly not passively evolving systems because they are actively forming stars (Table \ref{SFR}), and their bulges are bluer by $\sim$1 magnitude than elliptical stellar population models (Fig. \ref{bulbos}). The latter result is not new, since its discovery by Ellis et al. (2001), who concluded that secondary star formation superimposed on preexisting old populations (rejuvenation) have occurred in the inner regions of spirals galaxies (\cite{Thomas06}). Alternatively, Hammer et al. (2005) suggest that bulges at $z\sim$0.6 may have been formed by mergers, few Gyrs prior to the redshift epoch. 

Let us assume that the subsample of rotating spiral disks is made of galaxies that are likely unaffected by galaxy interactions, at least since z=1. It is quite likely that they have not undergone a major merger recently, and the probability that a galaxy at z=0.6 experiences another merger event is quite low. Table \ref{SFR} shows that most of them (if not all) present a doubling time ranging from 0.2 to 4 Gyrs. In the absence of interactions, it may be reasonable to assume an exponential decay of their SFRs, or a constant SFR as a lower limit of their past SFR. Because all of them but one are disk-dominated, it means that most of the stars in their disks have been formed rapidly, e.g. in a few Gyrs or less. Thus it implies quite a rapid building of the stellar disk and, also assuming a pre-existing old population in the bulge, a significant decrease in their B/T values. 

What could be the subsequent evolution of the rotating spiral disks at z=0.4-0.8? Sect. \ref{DT} provides evidence that recent starbursts are occurring in their outskirts. We also see decreasing mass-doubling time from the center to the edge, which suggests an inside-out disk growth; however, this conclusion is severely dependent on the assumption that IR light follows the UV light,  which is difficult to verify given the limited spatial resolution of Spitzer.  
An independent argument is provided by their location along the Tully Fisher relation (TFR, see Paper III). They delineate a TFR that is 0.34 dex shifted towards lower stellar masses than are local spirals. This is in remarkable agreement with the evolution of the stellar mass metallicity (0.3 dex on M-Z, \cite{Liang06}). It may mean that the rotating spiral disks have to double their stellar masses to reach the local TFR and the M-Z relationship.  Combined with the rapid formation of their stellar disks prior to their redshift epochs and with the fact that they show recent starbursts in their outskirts, it seems likely that their stellar disks are still growing inside-out (and their B/T values are decaying) during the 6 Gyr elapsed time since z=0.6. Such a conclusion is also in good agreement with results from Trujillo et al. (2005).

\subsection{Testing various scenarios of spiral formation}

 As these galaxies are dynamically relaxed, we expect the SFR to be self-regulated by the disk (e.g. \cite{Silk03}) and star formation to decay monotonically as the gas supply is exhausted. In contrast with theoretical predictions, we find that one quarter of the local spiral progenitors are sustaining a high SFR, which necessarily implies massive gas accretion or, alternatively, pre-existing large amounts of gas in an already relaxed gaseous disk. What is the origin of this gas and which scenario of spiral formation is able to reproduce our observations?
 
Despite the still small number of observations, these distant disks seem to have systematically lower V/$\sigma$ values than their local counterparts (\cite{Puech07}). They seem to be heated, which could possibly be associated with turbulence induced by processes related to star formation, and unlikely to minor mergers given that our observations are quite good at identifying minor mergers (\cite{Puech07b}). At the same time that rotating disk spirals are transforming their gas to stars to reach the local relationships (TFR and M-Z), they have to evolve towards thin disks, e.g. higher values of V/$\sigma$. 

We may infer that the necessary gas that feeds the star formation has a specific orientation to stabilize their disks during the past 6 Gyrs. Processes based on isotropic gas accretion (\cite{Dekel06}) may have difficulty reproducing such an evolution, while gas accretion from clumps or filaments (e.g. \cite{Keres05}) requires some preferential alignment to stabilize the disk. Thus it may be necessary for the gas to already pre-exist in a gaseous thin disk, and we may be simply seeing the gas transformation into stars in such systems. 

An alternative is that this gas reservoir is provided by re-accretion following a past major merger (\cite{Barnes02}; \cite{Hammer05};\cite{Governato07}). This post-merger gas accretion will constantly deposit gas at the outer edge of a spheroidal remnant, which will naturally give rise to the inside-out formation of a new disk surrounding the newly formed bulge. Extrapolating Barnes (2002) simulations at higher redshift, where the gas fraction is expected to be higher, may 
result in remnants with even larger and more massive gas disks.
Following this scenario, the morphological expectations are a relatively undisturbed disk (since it is assembled progressively by smooth re-accretion), with a clear inside-out signature with the youngest stars located at the outer regions of the disk, which closely matches what is observed in the rotating spiral disks (see above). From the viewpoint of the kinematics, we expect a relatively high velocity dispersion of the gas caused by shock-heating between the infalling gas material with the already accreted gaseous disk, and a progressive smooth increase in the angular momentum of the disk with time. Strikingly, these kinematical properties have already been observed in these galaxies by Puech et al. (2007).  Besides this, the bulge colors shown in Fig. \ref{bulbos} are also easy to reproduce if these bulges have been assembled a few Gyrs prior to the epoch at which we are observing them.

Since the disk growth time scales range from less than 1 Gyr to 4 Gyr, this suggests that these rotating spiral disks would have undergone their last major merger at $z$=1 or slightly later. At these redshifts, the gas fraction in galaxies is expected to be much higher (\cite{Liang06}, \cite{Law07}), which is more than two times higher at z=0.65 than today. This gives even more credit to the post-merger gas accretion scenario, because the gas reservoir available for the disk rebuilding phase would be fairly large. From $z$=1 to $z$=0, the merger rate in this mass range is expected to be between 50 to 75\% (see introduction in \cite{Hammer07}). Then, between 25 and 50\% of local intermediate-mass galaxies would have escaped such an event since $z$=1, which is, given the large uncertainties associated, in relatively good agreement with the estimation of the fraction  of intermediate-mass rotating spiral disks, i.e., 33\%. Given the much higher merger rate expected at $z>$1 (\cite{Conselice03}, \cite{Lotz06a}), our results are reasonably consistent with these spiral disks possibly undergoing a major merger before $z$=1.

\section{Conclusion}

We have presented a morphological analysis of a representative sample of intermediate-mass galaxies at $z\sim$0.6.
To derive our morphological classification, both structural parameters and color maps were used, combined with a visual inspection of each object. 
The method used in the present study is slightly different from previous studies, as we made used of the color information to distinguish only those spiral disks with properties comparable to local spirals (bulge redder than disk). \\
This method was tested against an independent kinematical classification, and we find good agreement between the morphological and the dynamical states of the galaxies (see Fig. \ref{histo}).
There is a definite offset in morphology between RD and CK. The major overlap between the different kinematical classes are for irregular galaxies known to span a wide range of kinematical properties.\\
We also computed the concentration, asymmetry, GINI and M20 parameters and find they are not able to recover the same correlation. It appears that, when applied automatically, these methods will miss around half of the complex kinematics and unavoidably overestimate the fraction of spirals. These results show that, when derived properly, the morphological information can be representative of the underlying kinematical properties, even at $z\sim$0.6. This correlation represents an interesting new tool for understanding the mechanisms in the formation and evolution of galaxies.\\

Our classification scheme allows us to isolate a particular morpho-kinematical class: the rotating spiral disks.
This class includes objects sharing the spiral disk morphology and rotating disks properties.
We find that the number of these rotating spiral disks is much lower (by a factor of $\sim$2) at $z\sim$0.6 than today. The radial distribution of color along their disks and their high SFR suggest that their stellar disks have been formed relatively recently before $z\sim$0.6 through inside-out processes. These findings are consistent with scenarios for which the large gas supply stabilizes the disk, e.g. leading ultimately to present-day thin disks by increasing V/$\sigma$ values with time. It opens up two possibilities, either that z=0.6 rotating spiral disks are forming their stars from a pre-existing gas reservoir already distributed in a thin disk or, alternatively, a post-merger accretion scenario where a disk is rebuilt thanks to gas accretion left over from the merging event (\cite{Barnes02} ; \cite{Hammer05}).


\begin{onecolumn}
\begin{table}[h!]
\begin{center}
 \caption{Catalog of derived parameters for the working sample.}
\begin{tabular}{cccccccccc}
\hline\hline
IAU ID & Our ID & $z^a$ &  B/T$^b$ & $\chi ^2$ & Q$^c$ & Type$^d$ & $R_{1/2}$$^e$ & $R_D^f$ & Cl$^g$ \\ 
(1)                &     (2)        & (3) & (4) &  (5)          &  (6)     & (7)    & (8)& (9) & (10)  \\ \hline 
J033212.39-274353.6 & 3400803 & 0.4213 & 0.12$\pm$0.01 & 1.45 & 1 & Sp & 6.0 & 3.8 & RD \\
J033219.68-275023.6 & 3202670 & 0.5596 & 0.02$\pm$0.00 & 1.52 & 1 &  Sp & 5.4 & 3.2 & RD \\
J033230.78-275455.0 & 2102060 & 0.6857 & 0.04$\pm$0.01 & 1.18 & 1 & Sp  & 7.6 & 4.2 & RD \\ 
J033231.58-274121.6 & 2500322 & 0.7057 & 0.02$\pm$0.01 & 1.32 & 1 & Sp & 7.6 & 3.5 &  RD \\
J033237.54-274838.9 & 2300477 & 0.6638 & 0.06$\pm$0.13 & 1.38 & 1 & Sp & 6.0  & 3.4 & RD \\ 
J033238.60-274631.4 & 2301047 & 0.6201 & 0.02$\pm$0.01 & 1.26 & 1  & Sp & 5.7 & 3.5 & RD \\
J223256.07-603148.8 & HDFS4020 & 0.5138 & 0.17$\pm$0.03 & 1.32 & 1 &  Sp & 6.1 & 3.2 & RD \\
J030228.72+001333.9 & CFRS030046 & 0.5120 & 0.12$\pm$0.06  &  1.56 & 1    & Sp & 8.3 & 6.3 &RD \\
J030225.28+001325.1 & CFRS030085 & 0.6100 & 0.02$\pm$0.03 &  1.27 & 1 &  Sp & 7.6 & 4.5 & RD \\
J030246.94+001032.6 & CFRS030619 & 0.4854 & 0.17$\pm$0.09  & 1.15 & 1 & Sp  & 3.9 & 2.5 & RD \\
J030248.41+000916.5 & CFRS031353 & 0.6340 & 0.13$\pm$0.03 &  1.38 & 1 & Sp & 6.4 & 4.5 & RD \\
J030232.16+000639.1& CFRS039003 & 0.6189 &  0.08$\pm$0.03 &  1.23 & 1 & Sp & 4.9 & 3.2 & RD \\

J033226.23-274222.8 & 3401338 & 0.6671 & 0.05$\pm$0.01 & 1.15 & 1 & Sp & 10.2 &  6.5 & PR \\
J033248.28-275028.9 & 1202537 & 0.4446 & 0.04$\pm$0.03 & 1.43 & 1 &  Sp & 7.2 & 4.4 & PR \\
J033213.06-274204.8 & 3500001 & 0.4215 & 0.16$\pm$0.05 & 1.32 & 1 & Sp & 6.6 & 3.9 & CK \\ \hline

J033234.04-275009.7 & 2300055 & 0.7024 &   -   &  -     & 3 & Pec/Irr & 4.5 & - & RD \\
J033210.25-274819.5 & 4301297 & 0.6087 &   -   &  -     & 3 & Pec/Irr & 4.6 & - & PR \\
J033214.97-275005.5 & 3300063 & 0.6665 & 0.01$\pm$0.00 & 1.46 & 2  & Pec/Irr & 5.8 & 2.8 & PR \\
J033233.90-274237.9 & 2401349 & 0.6180 &   -   & -      & 3 & Pec/Irr & 3.1 & - & PR \\ 
J033239.04-274132.4 & 2500233 & 0.7319 &  -  & - & 3 & Pec/Irr & 3.3 & - & PR \\
J033249.53-274630.0 & 1302369 & 0.5221 & 0.02$\pm$0.00 & 1.18 & 2 & Pec/Irr & 3.1 & 2.1 & PR \\
J033250.53-274800.7 & 1301018 & 0.7360 & 0.04$\pm$0.02 & 1.17 & 1  & Pec/Irr & 4.3 & 2.3 & PR \\
J223252.74-603207.3 & HDFS4040 & 0.4650 & 0.12$\pm$0.02 & 1.54 & 1 & Pec/Irr & 5.3 & 3.9 & PR \\
J030249.10+001002.1 & CFRS031349 & 0.6155 & 0.39$\pm$0.09 & 1.27 & 1 &  Pec/Irr & 3.8 & 1.8 & PR \\
J033250.24-274538.9 & 1400714 & 0.7310 &  - & - &  3 & Pec/Irr & 4.8 & - & CK \\
J030242.19+001324.3& CFRS030488 & 0.6069 & -                        & -        & 3      & Pec/Irr & 6.4 & - & CK \\
J030239.38+001327.1 & CFRS030523 & 0.6508 & 0.82$\pm$0.01 & 1.1 &  2 & Pec/Irr  & 3.6 & 0.8 & CK \\
J030245.67+001027.9 & CFRS030645 & 0.5275 & 0.42$\pm$0.09 &  1.62 & 2 & Pec/Irr & 4.6 & 0.8 & CK \\

J033241.88-274853.9 & 2300404 & 0.6670 & 0.07$\pm$0.01 & 1.70 & 2 & Pec/T & 3.5 & - & RD \\
J223302.45-603346.5 & HDFS5150 & 0.6956 & -                           &     -    & 3 & Pec/T & 4.1 & - & PR \\
J033217.62-274257.5 & 3401109 & 0.6457 & 0.16$\pm$0.02 & 1.16 &  2 & Pec/T  & 3.2 & 1.45 & CK \\
J033219.32-274514.0 & 3400329 & 0.7241 & 0.20$\pm$0.04 & 1.14 &  2 & Pec/T & 3.9 & - & CK \\
J033225.26-274524.0 & 3400279 & 0.6648 & 0.23$\pm$0.02 & 1.20 & 2 & Pec/T & 3.6 & 2.3 & CK \\

J033219.61-274831.0 & 3300651 & 0.6699 & - & - & 3  & Pec/M & 3.1 & - & PR \\
J033230.43-275304.0 & 2200433 & 0.6453 & - & - & 3  & Pec/M & 4.7 & - & CK \\
J033234.12-273953.5 & 2500971 & 0.6273 &  -                        &   -       & 3   & Pec/M & 5.0 & - & CK\\
J033239.72-275154.7 & 2200829 & 0.4151 &   -     &     -     & 3 & Pec/M  & 3.5 & - & CK \\
J223257.52-603305.9 & HDFS5030 & 0.5821 & 0.26$\pm$0.01 & 1.46  & 2 & Pec/M & 4.3 & 2.7 & CK \\
J030240.45+001359.4& CFRS030508 & 0.4642 & - & - & 3 & Pec/M & 3.3 & - & CK \\  \hline

J033245.11-274724.0 & 2300800 & 0.4346 &  0.21$\pm$0.01& 1.77 & 2  & C  & 1.8 & 0.5 & RD \\
J033232.96-274106.8 & 2500425 & 0.4681 & -  & - & 3 & C & 1.5 & - & PR \\
J030238.74+000611.5 & CFRS031032 & 0.6180 & - & - & 3 & C & 1.8 & - & PR \\
J033220.48-275143.9 & 3202141 & 0.6778 & 0.01$\pm$0.03 & 1.10 & 2 & C & 2.6 & 1.2 & CK \\
J033228.48-274826.6 & 3300684 & 0.6686 & - & - & 3  & C & 2.3 & - & CK \\
J033240.04-274418.6 & 2400536 & 0.5220 & - & - &  3  & C & 2.4 & - & CK \\
J033244.20-274733.5 & 2300750 & 0.7360 & -  & -  &  3    &  C & 1.7 & - & CK \\
J223256.08-603414.1 & HDFS5140 & 0.5649 & 0.42$\pm$0.01 &  1.87 &  2  &  C & 2.8 & 2.1 & CK \\ \hline

J033210.76-274234.6 & 4402679 & 0.4169 &    -     &   -                        &   3  & M & 4.4 & - & CK \\
J033224.60-274428.1 & 3400618 & 0.5368 & 0.09$\pm$0.03 & 1.19 & 2 & M & 3.8 & 2.5 & CK \\ 
J033227.07-274404.7 & 3400743 & 0.7381 & - & - & 3 & M & 4.5 & - & CK \\
J033230.57-274518.2 & 2400243 & 0.6799 &  - & -   & 2 & M & 9.2 & - & CK \\
J030252.03+001033.4& CFRS031309 & 0.6170 & -         & -       &    3  & M & 9.6 & - & CK \\ \hline\hline
 \multicolumn{10}{l}{} \\
 \multicolumn{10}{l}{$^a$ Redshift measured by [OII] emission.}\\ 
  \multicolumn{10}{l}{$^b$ Bulge fraction derived from Galfit.}\\ 
 \multicolumn{10}{l}{$^c$ Quality factor: 1 = secure; 2 = possibly secure; 3 = fit failed or unreliable}\\ 
 \multicolumn{10}{l}{$^d$ Morphological galaxy type: Sp = Spiral Disk; Pec/Irr = Peculiar-Iregular; Pec/T = Peculiar-Tadpole}\\ 
  \multicolumn{10}{l}{Pec/M = Peculiar-Merger; C = Compact; M = Merger.}\\ 
 \multicolumn{10}{l}{$^e$ half-light Radius (kpc).}\\ 
 \multicolumn{10}{l}{$^f$ Disk radius (kpc) derived from Galfit.}\\ 
 \multicolumn{10}{l}{$^g$ Kinelatical type: RD = Rotating Disk; PR = Perturbed rotation; CK = Complex kinematics.}\\ 
  \multicolumn{10}{l}{} \\

 \end{tabular}
 \label{toto}
\end{center}
\end{table}

%
\newpage

 \begin{figure}[h!]
   \begin{center}
   \begin{tabular}{cccc}
      \includegraphics[height=2.6cm]{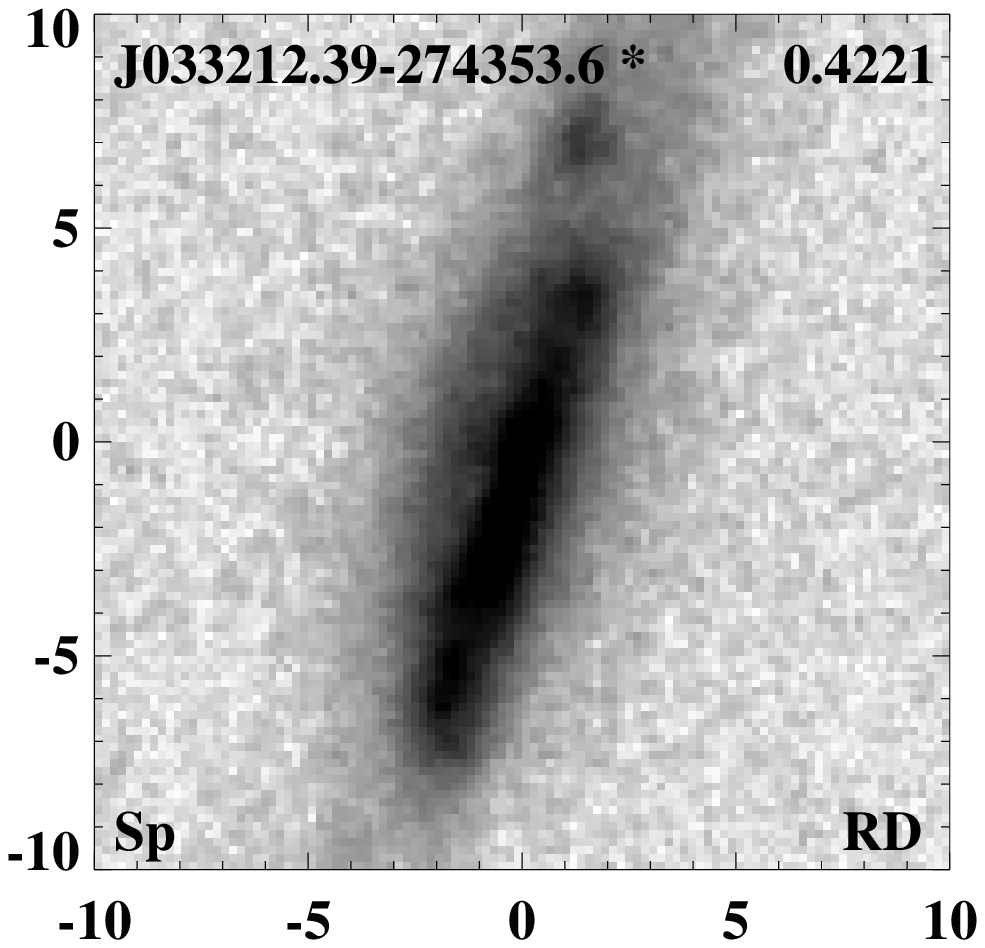} &
 \includegraphics[height=2.6cm]{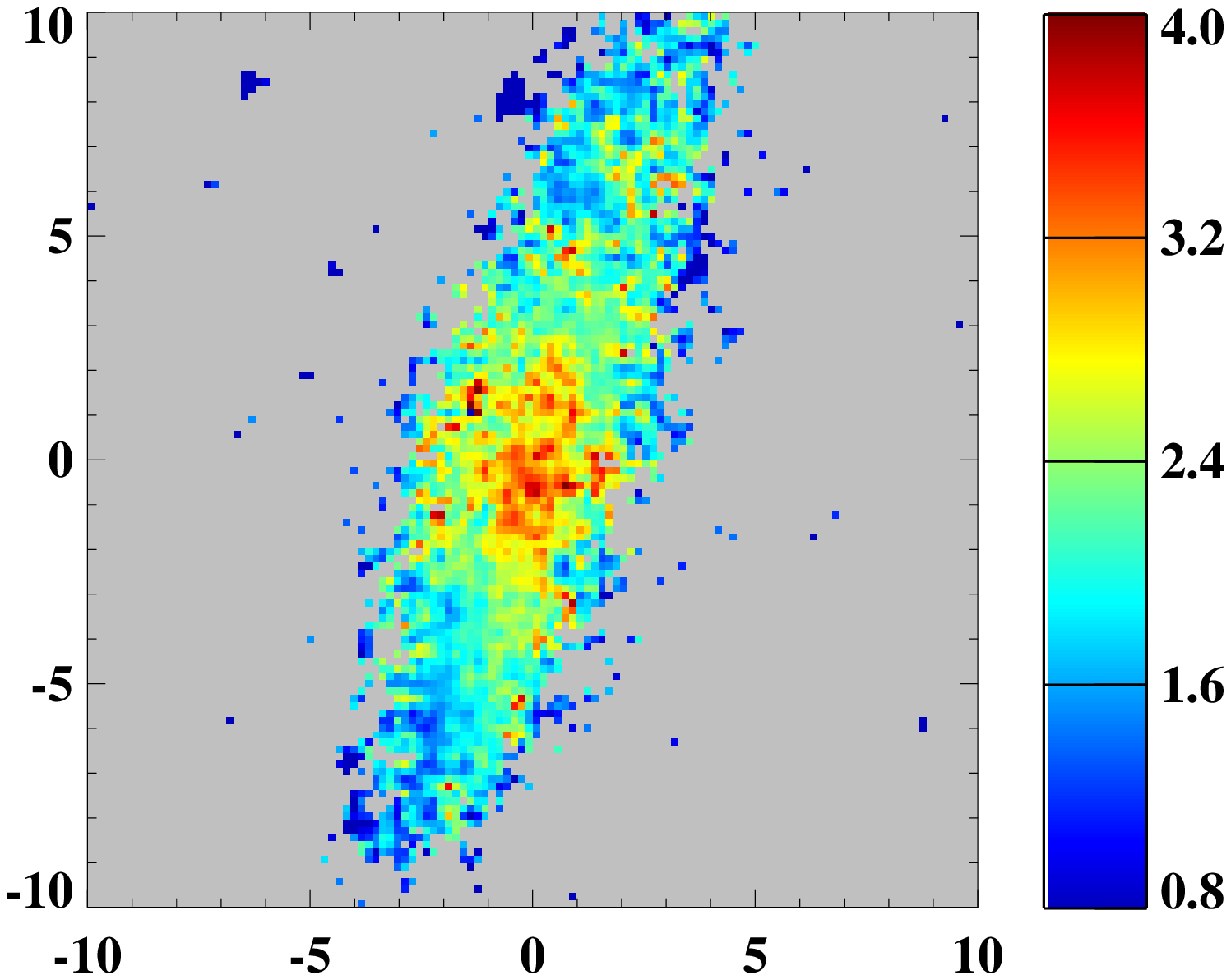} &
 \includegraphics[height=2.6cm]{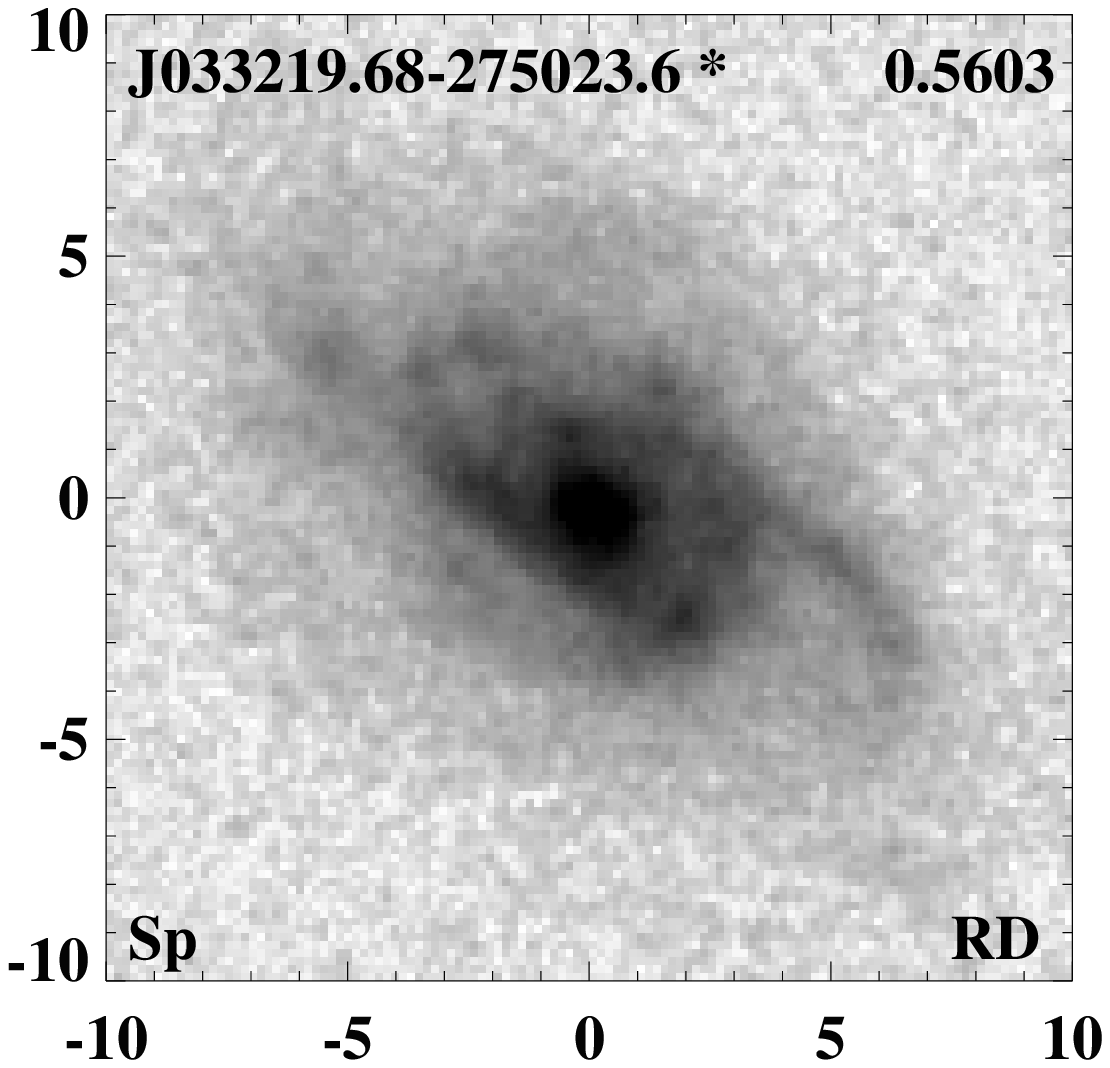} &
 \includegraphics[height=2.6cm]{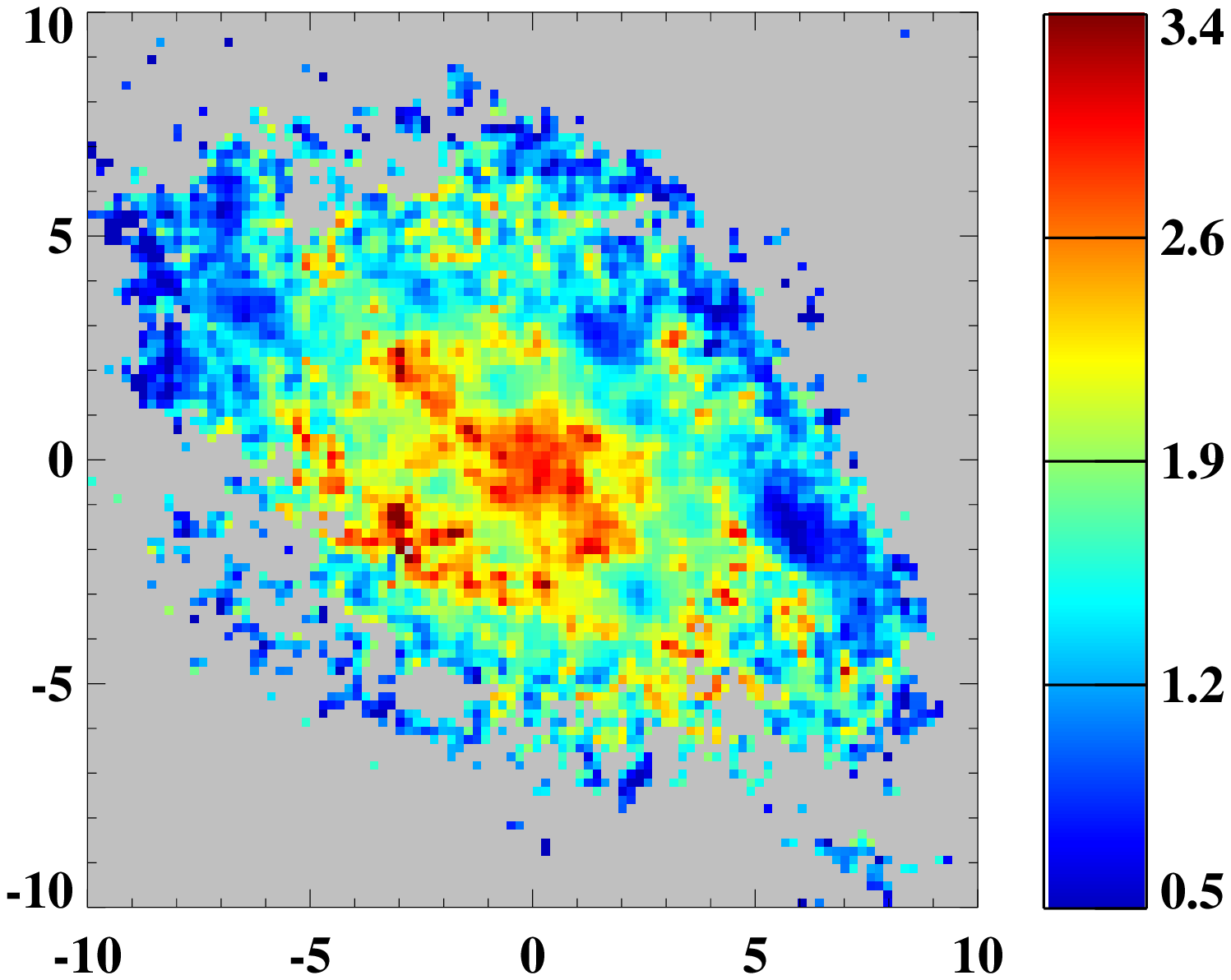} \\

  \includegraphics[height=2.6cm]{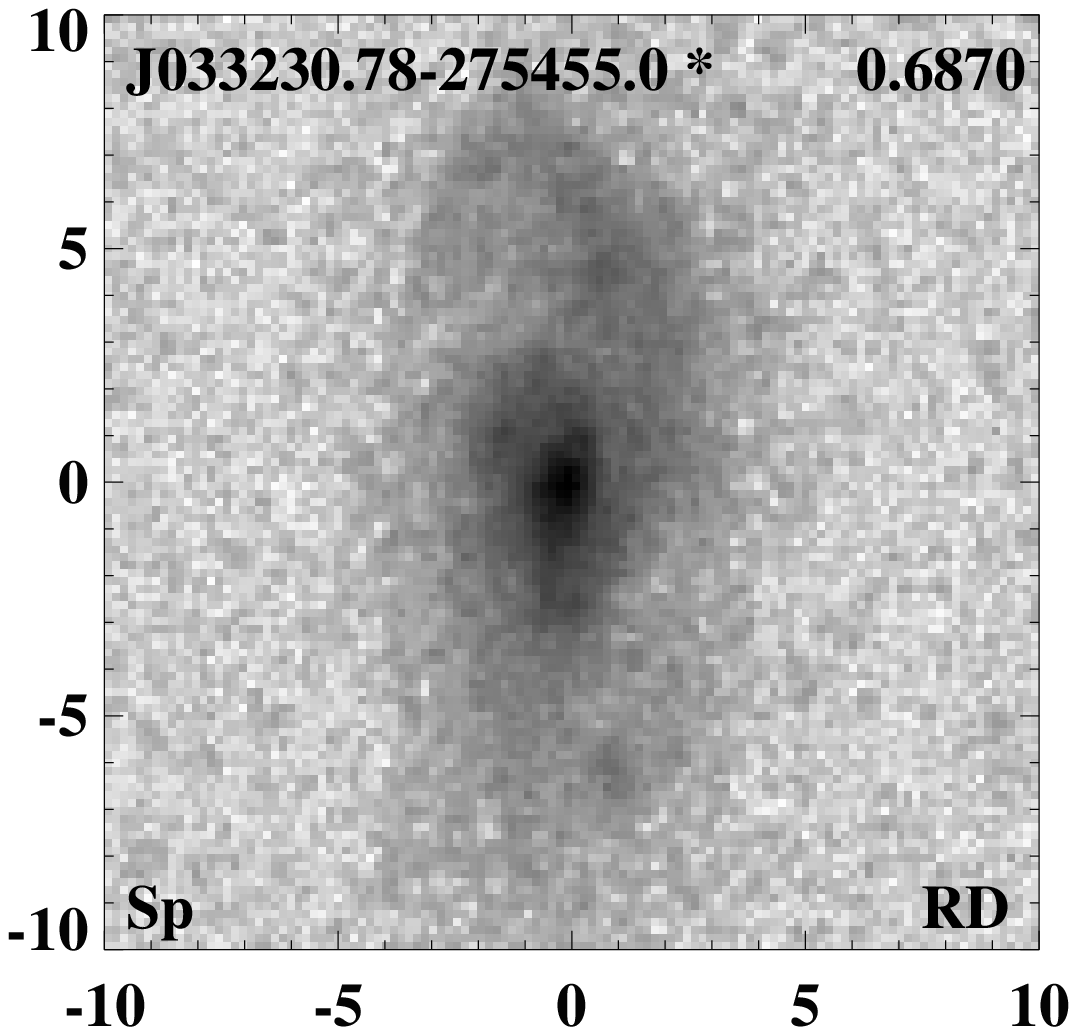} &
 \includegraphics[height=2.6cm]{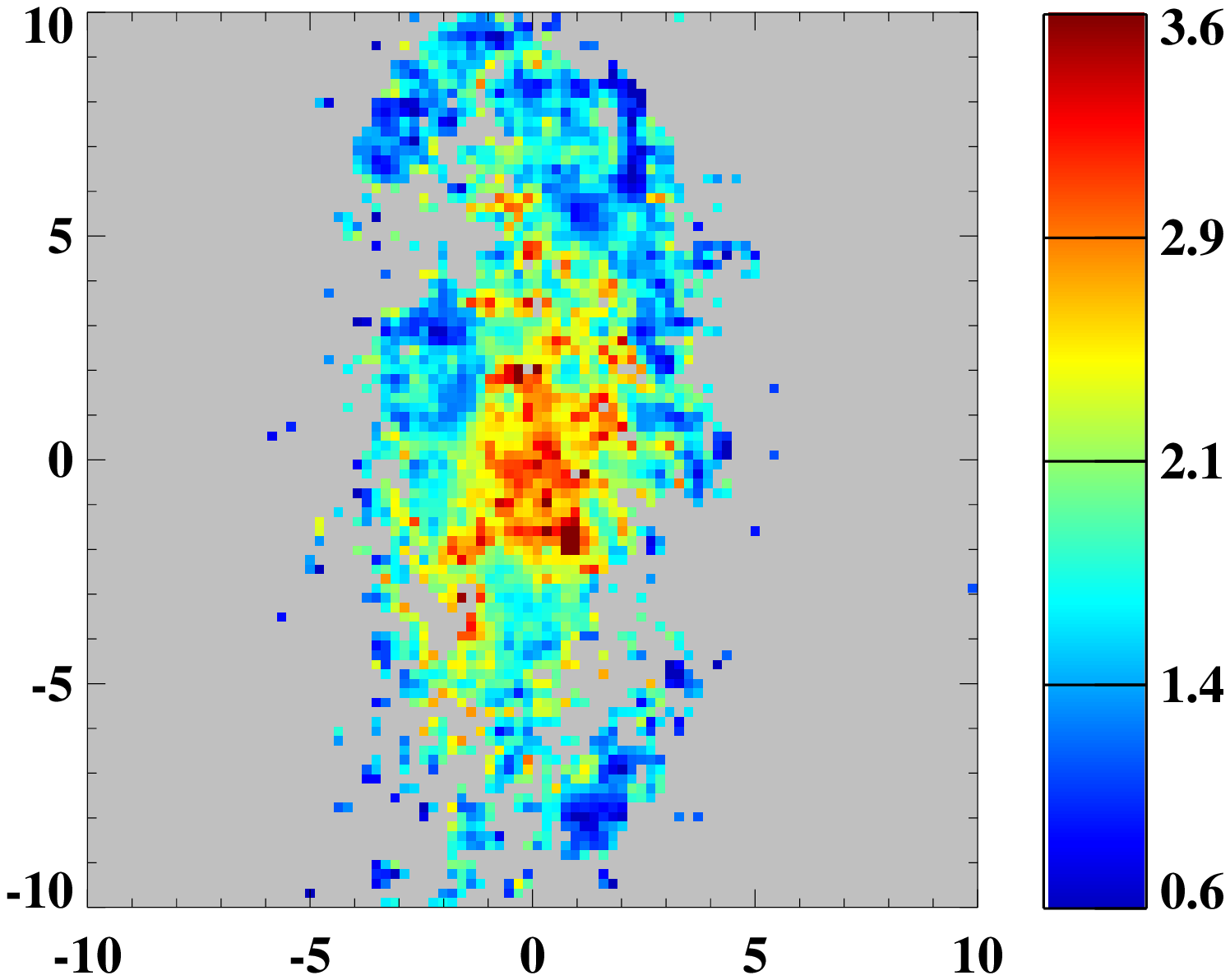} &
   \includegraphics[height=2.6cm]{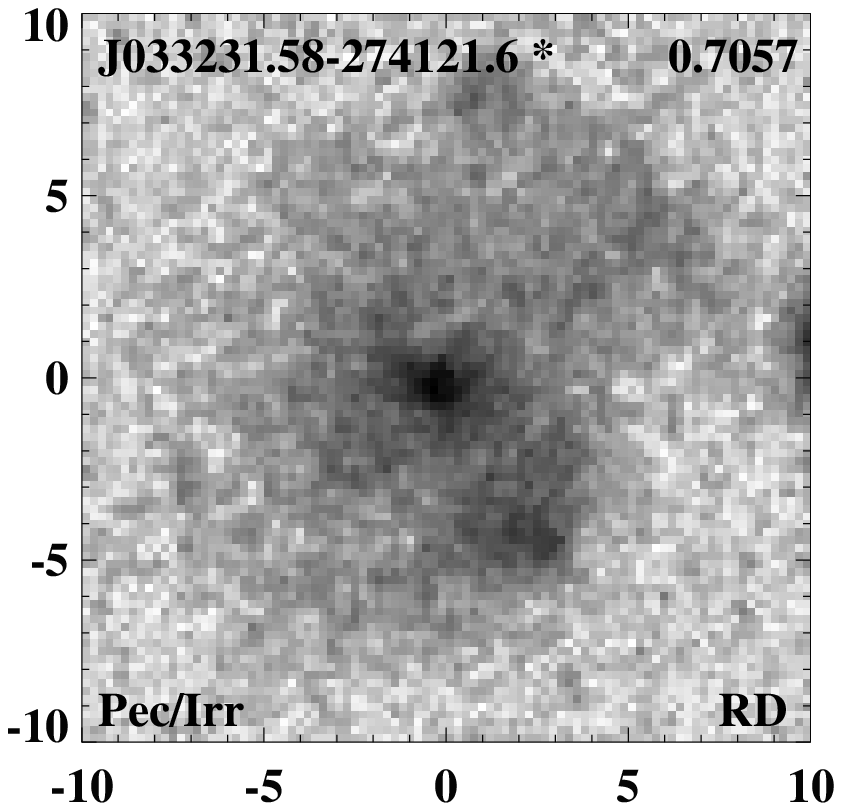} &
 \includegraphics[height=2.6cm]{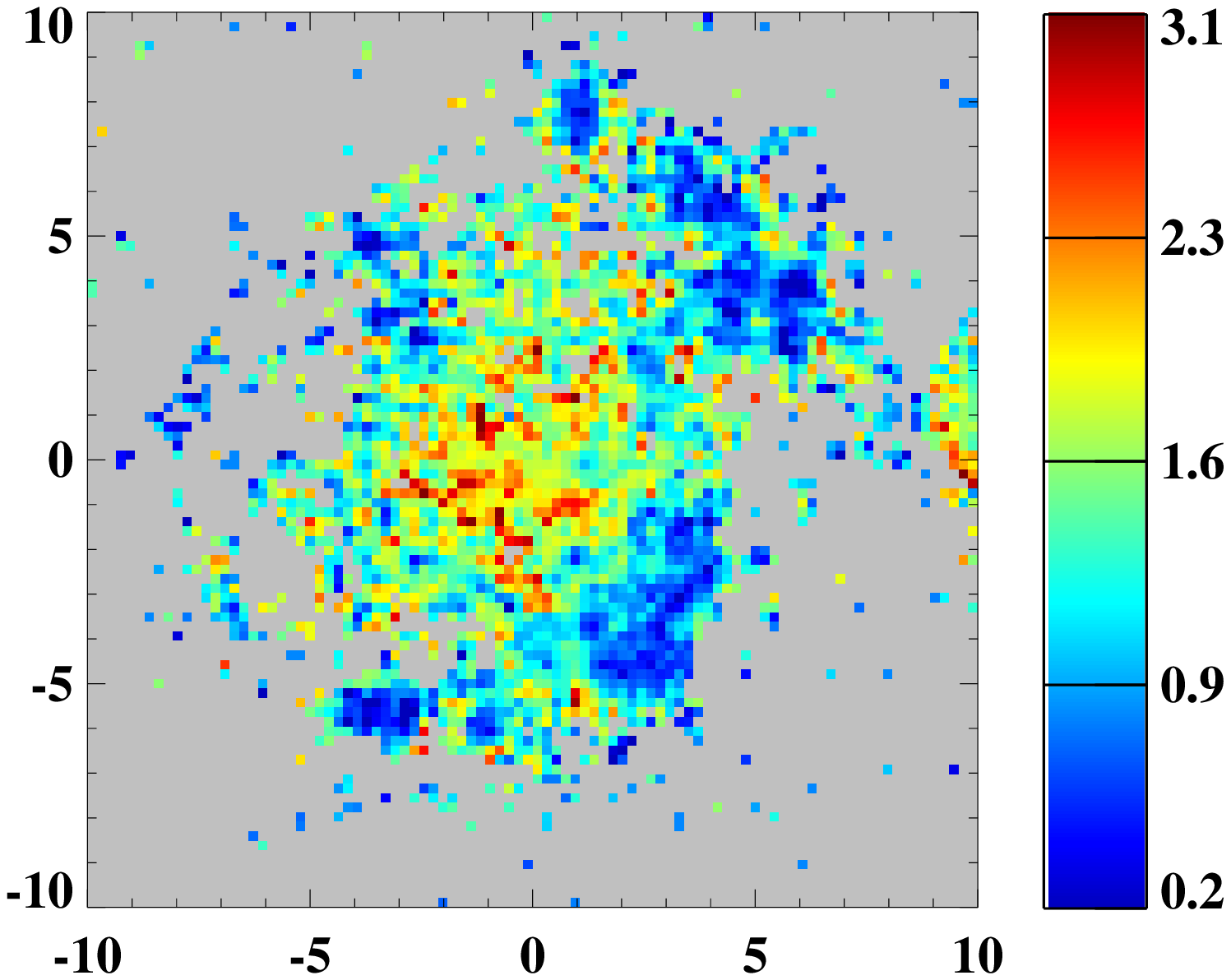} \\
 
       \includegraphics[height=2.6cm]{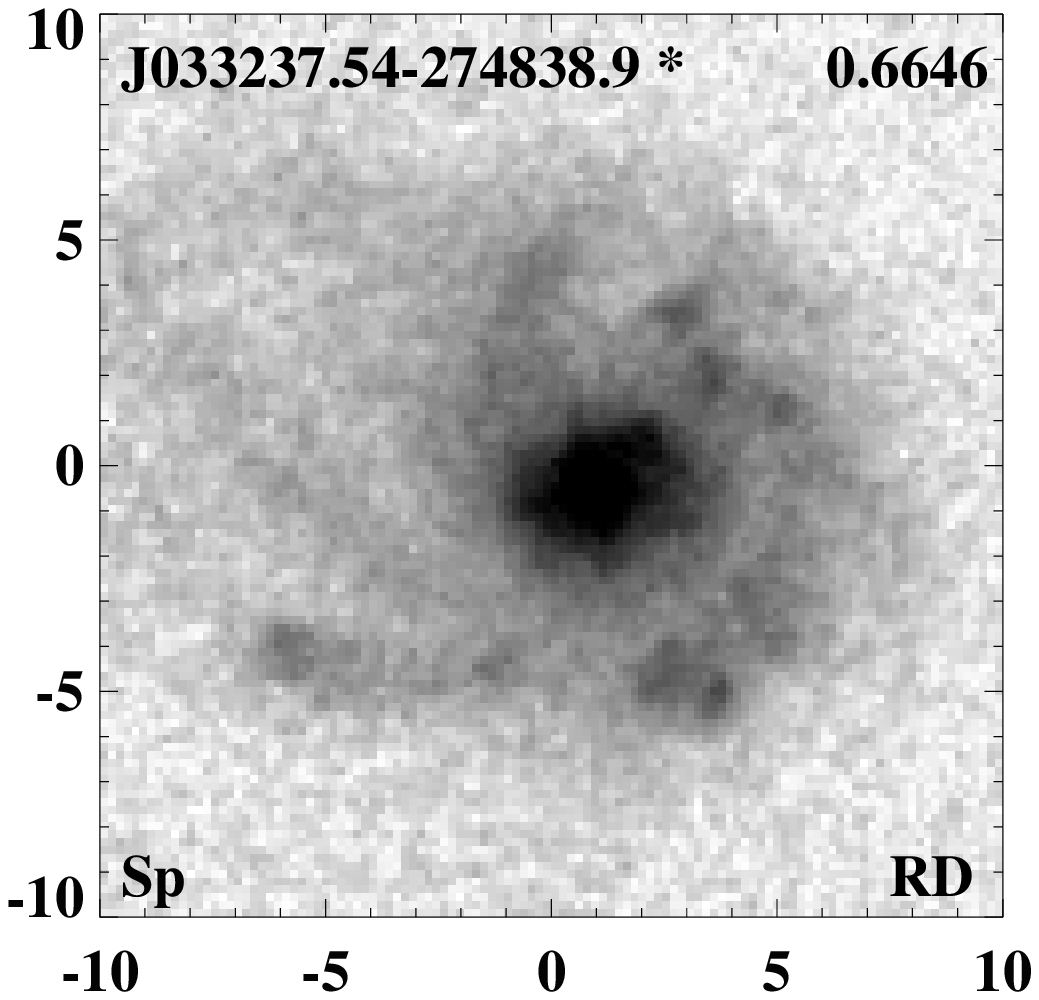} &
 \includegraphics[height=2.6cm]{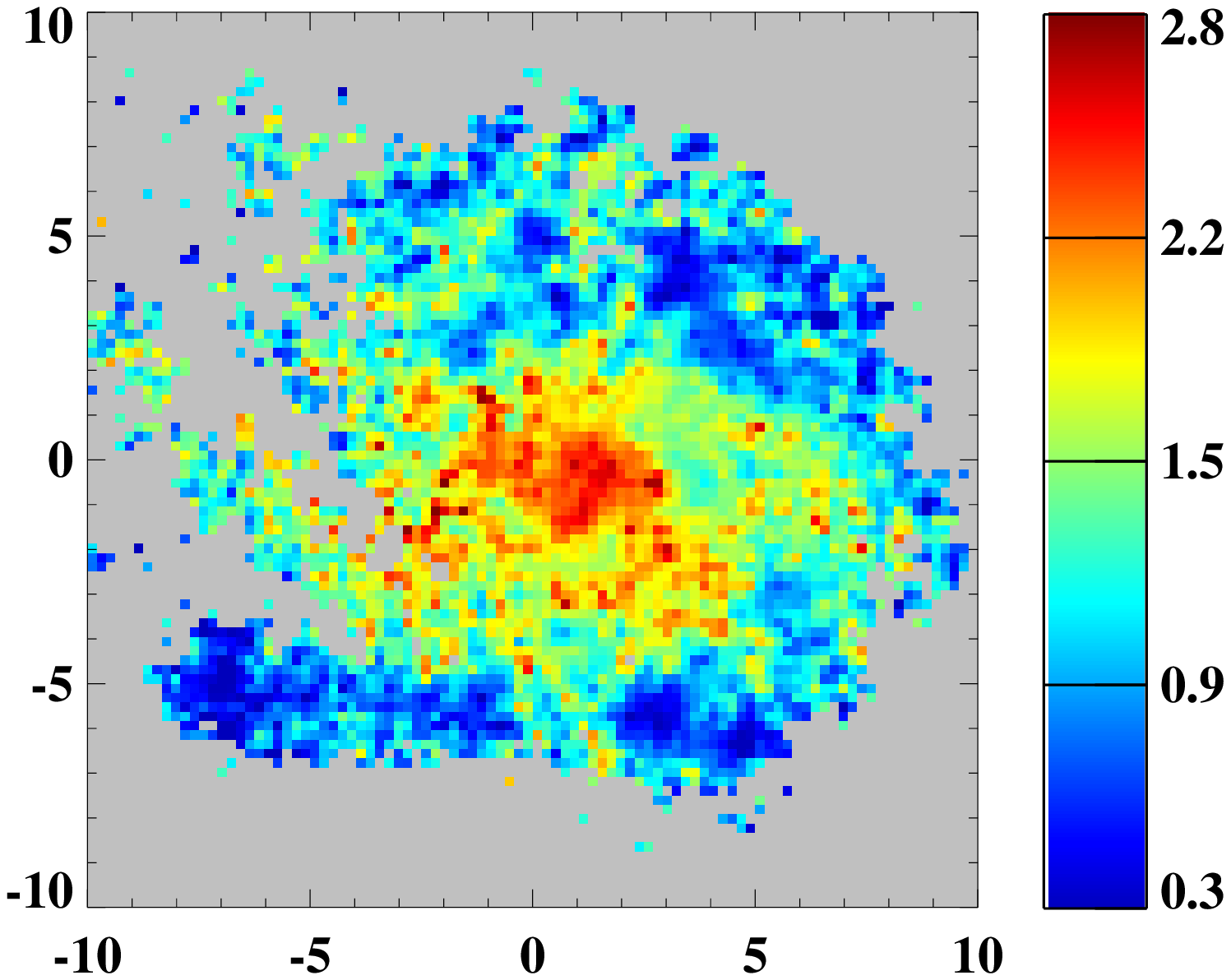} &
\includegraphics[height=2.6cm]{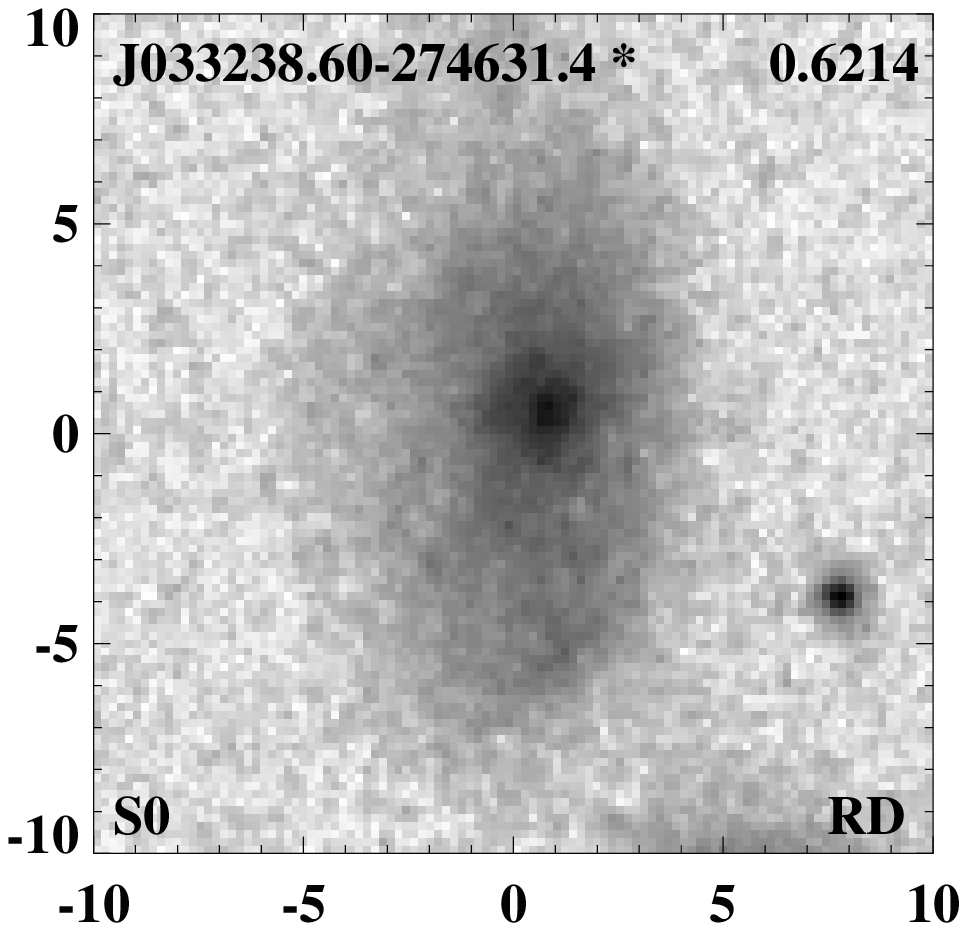} &
 \includegraphics[height=2.6cm]{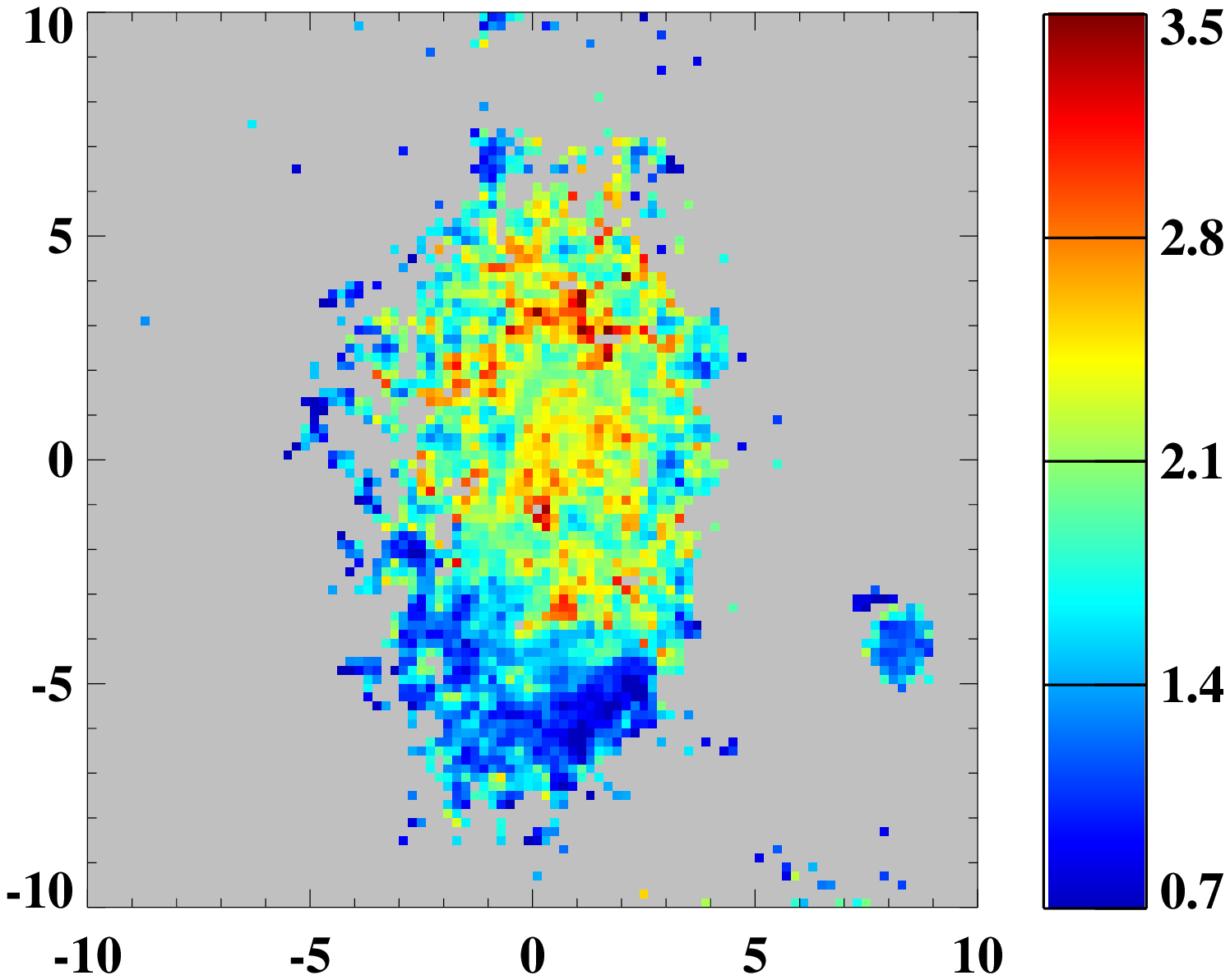} \\
 
   \includegraphics[height=2.6cm]{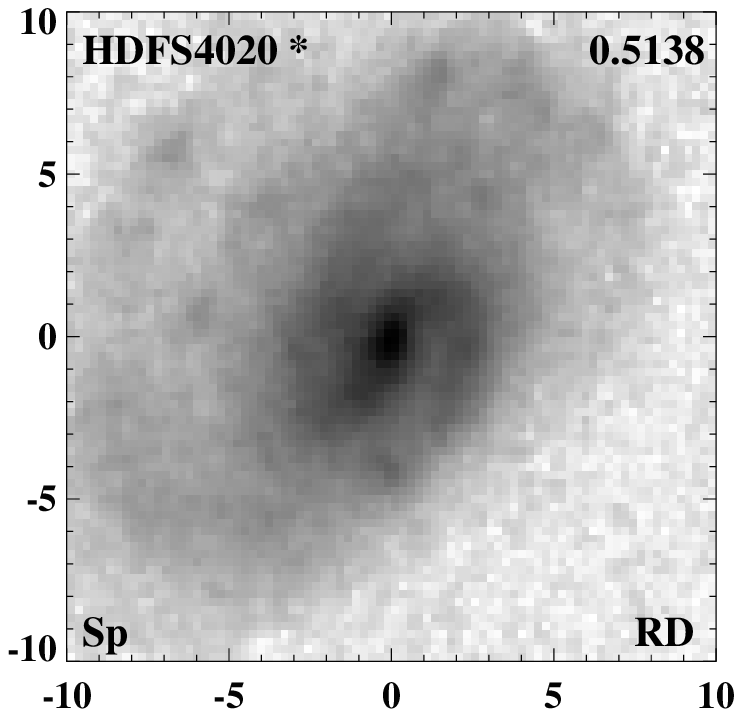} &
 \includegraphics[height=2.6cm]{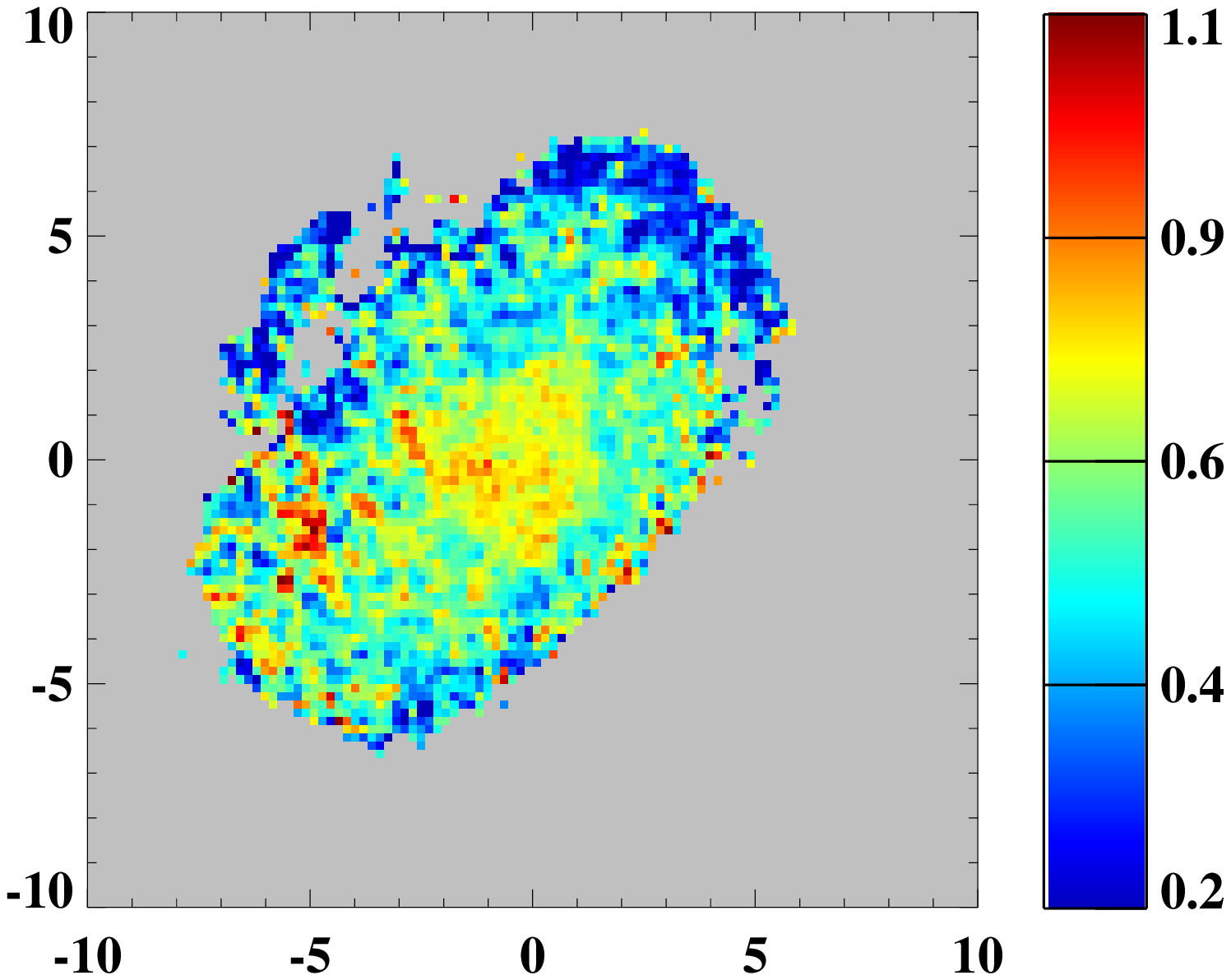} &
 \includegraphics[height=2.6cm]{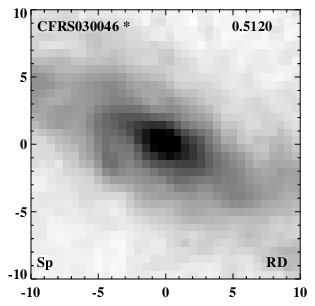} &
 \includegraphics[height=2.6cm]{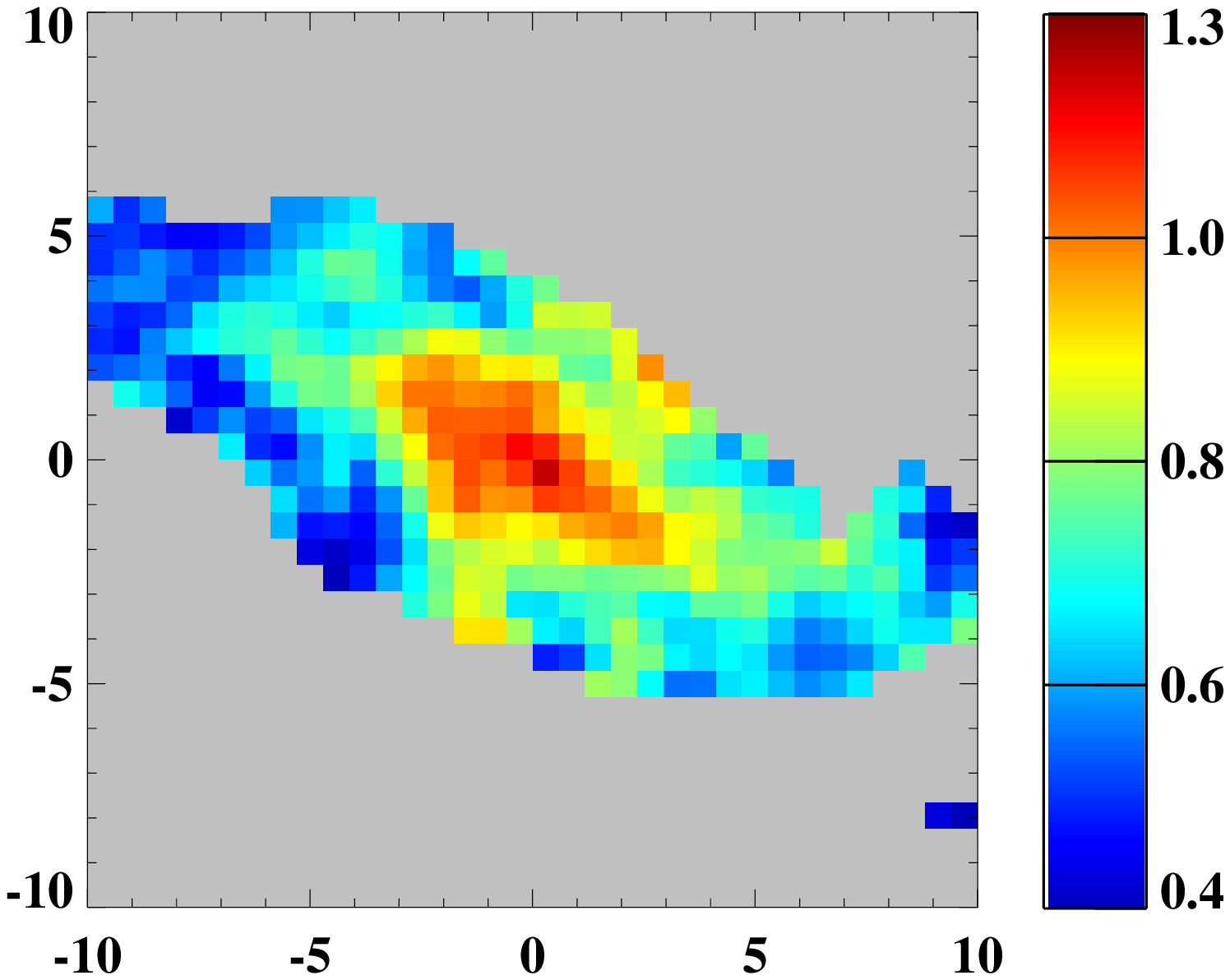} \\
 
  \includegraphics[height=2.6cm]{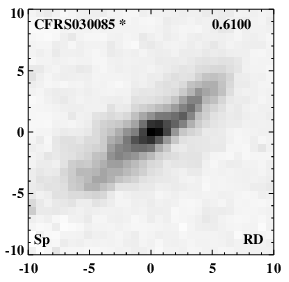} &
 \includegraphics[height=2.6cm]{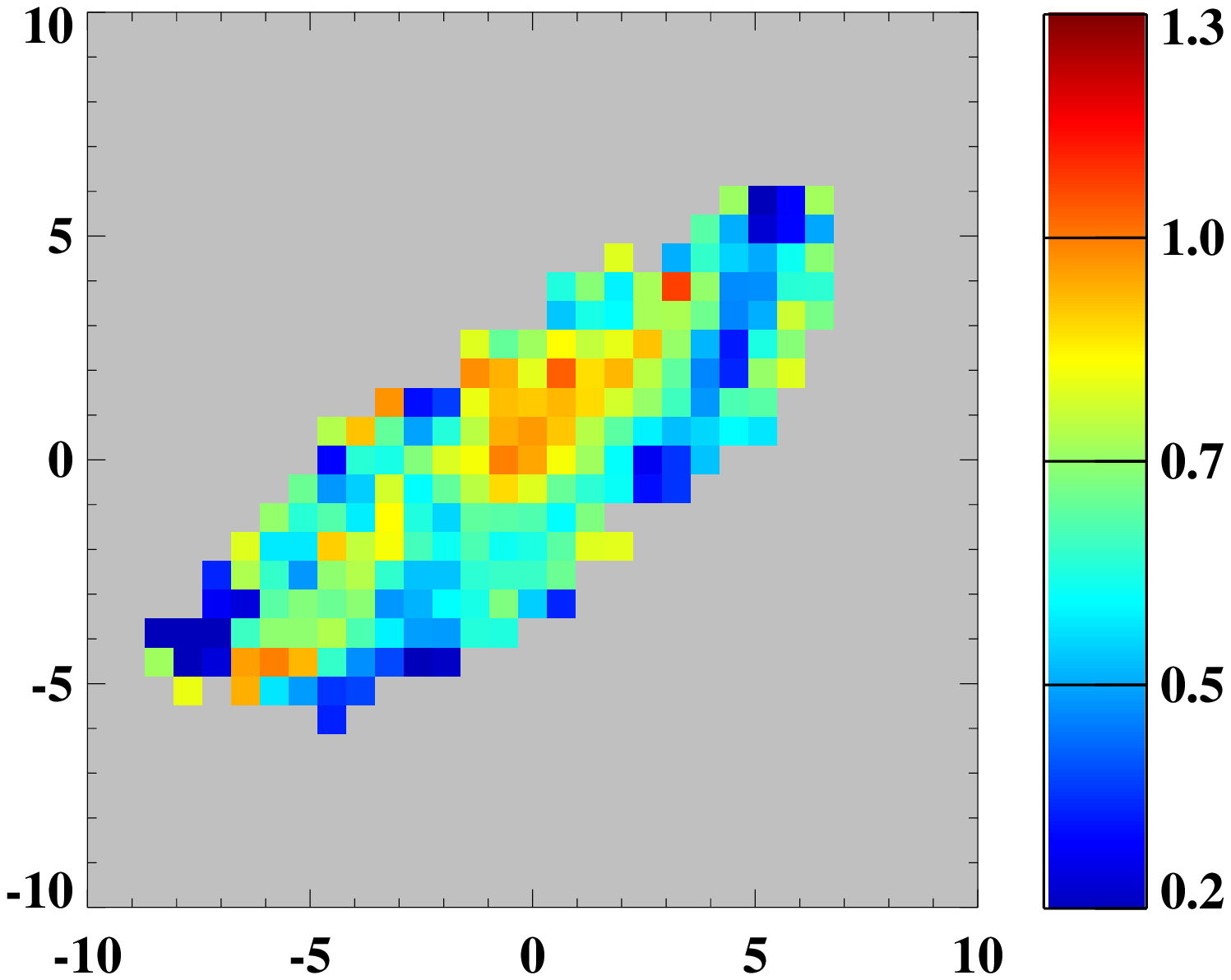} &
    \includegraphics[height=2.6cm]{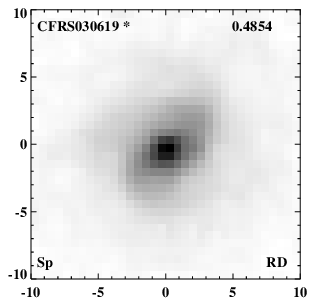} &
 \includegraphics[height=2.6cm]{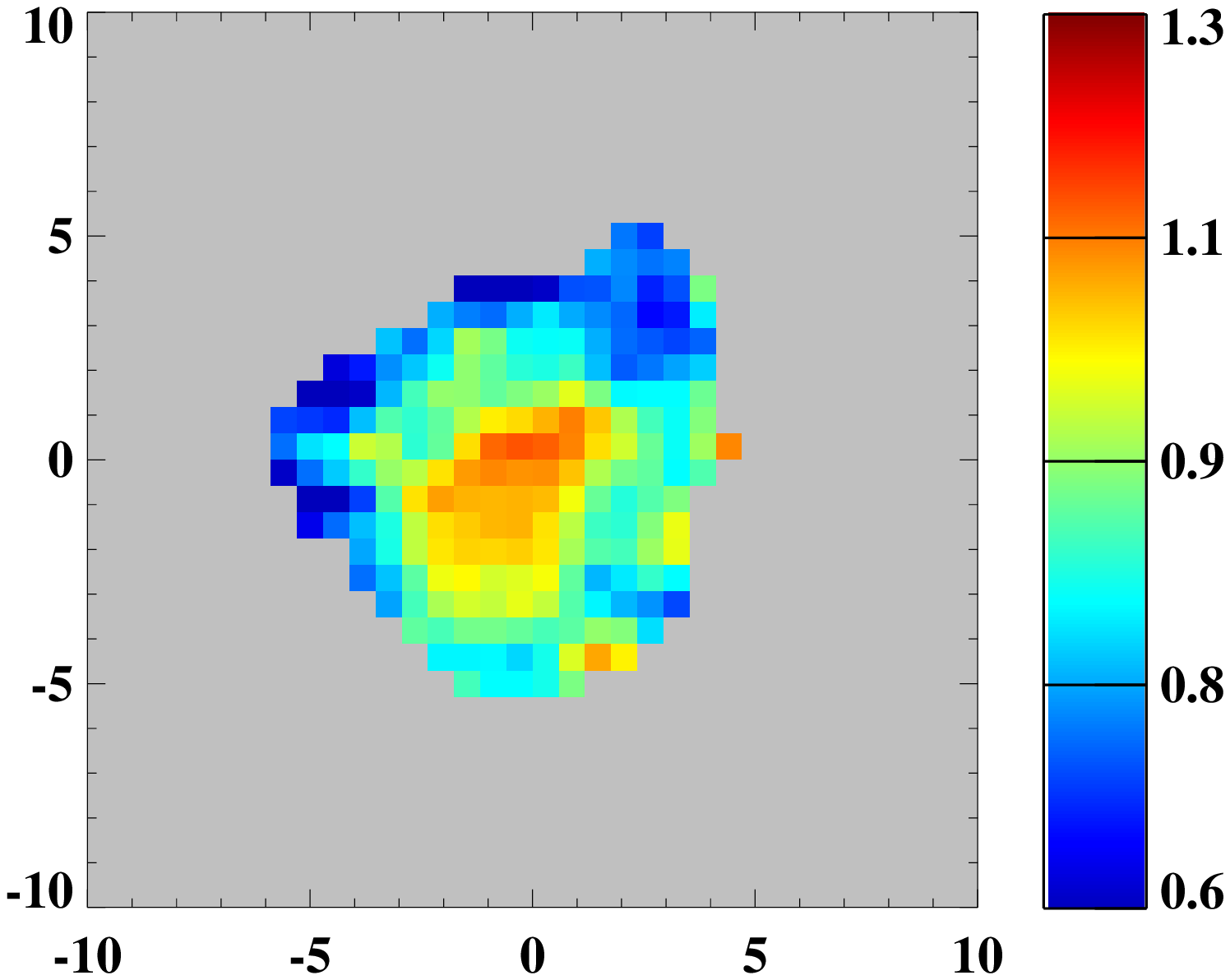} \\
 
    \includegraphics[height=2.6cm]{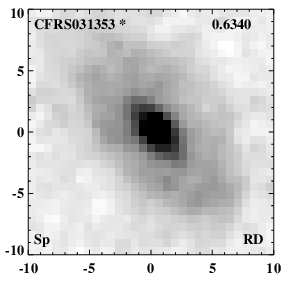} &
 \includegraphics[height=2.6cm]{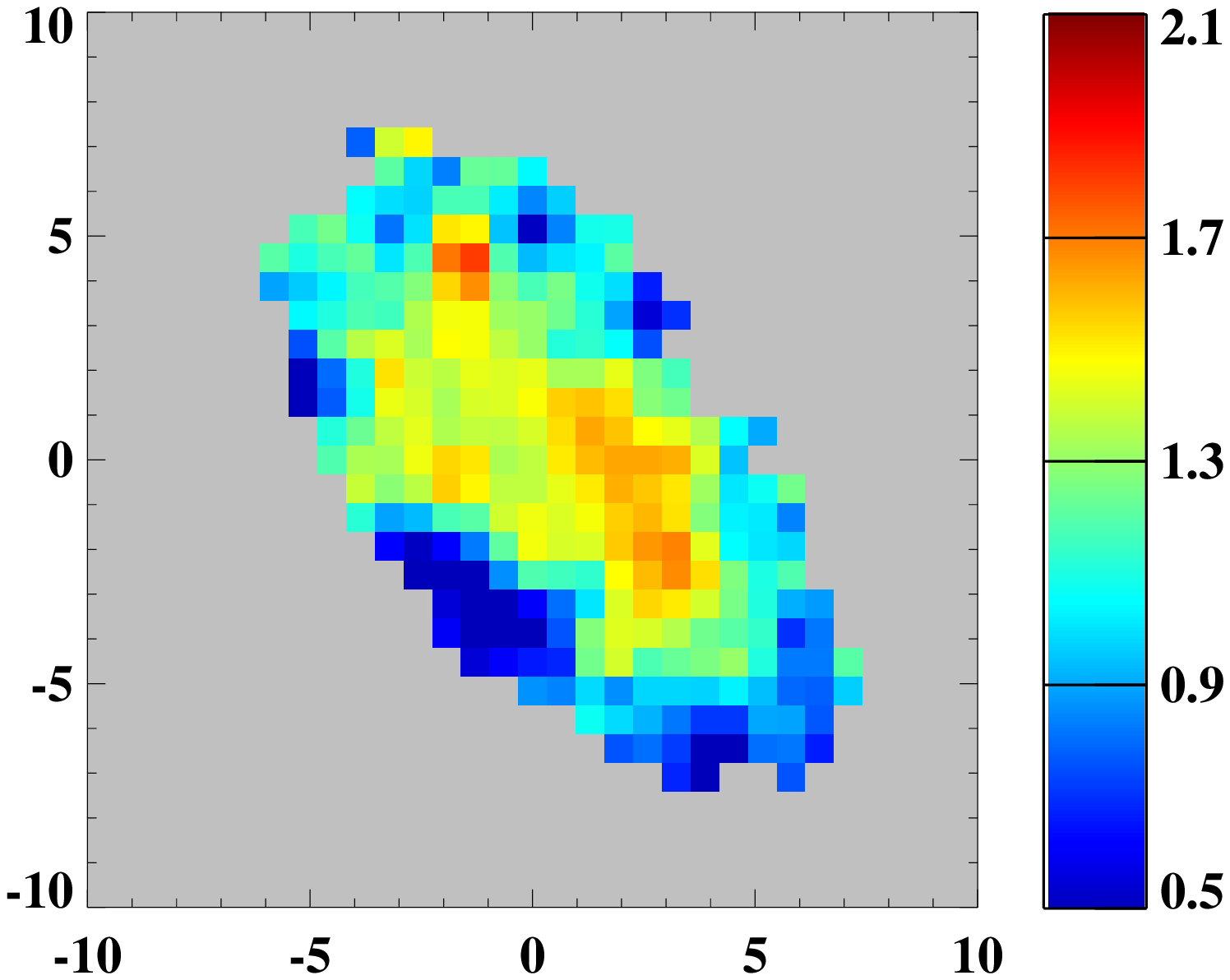}  &
 \includegraphics[height=2.6cm]{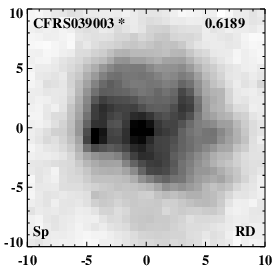} &
 \includegraphics[height=2.6cm]{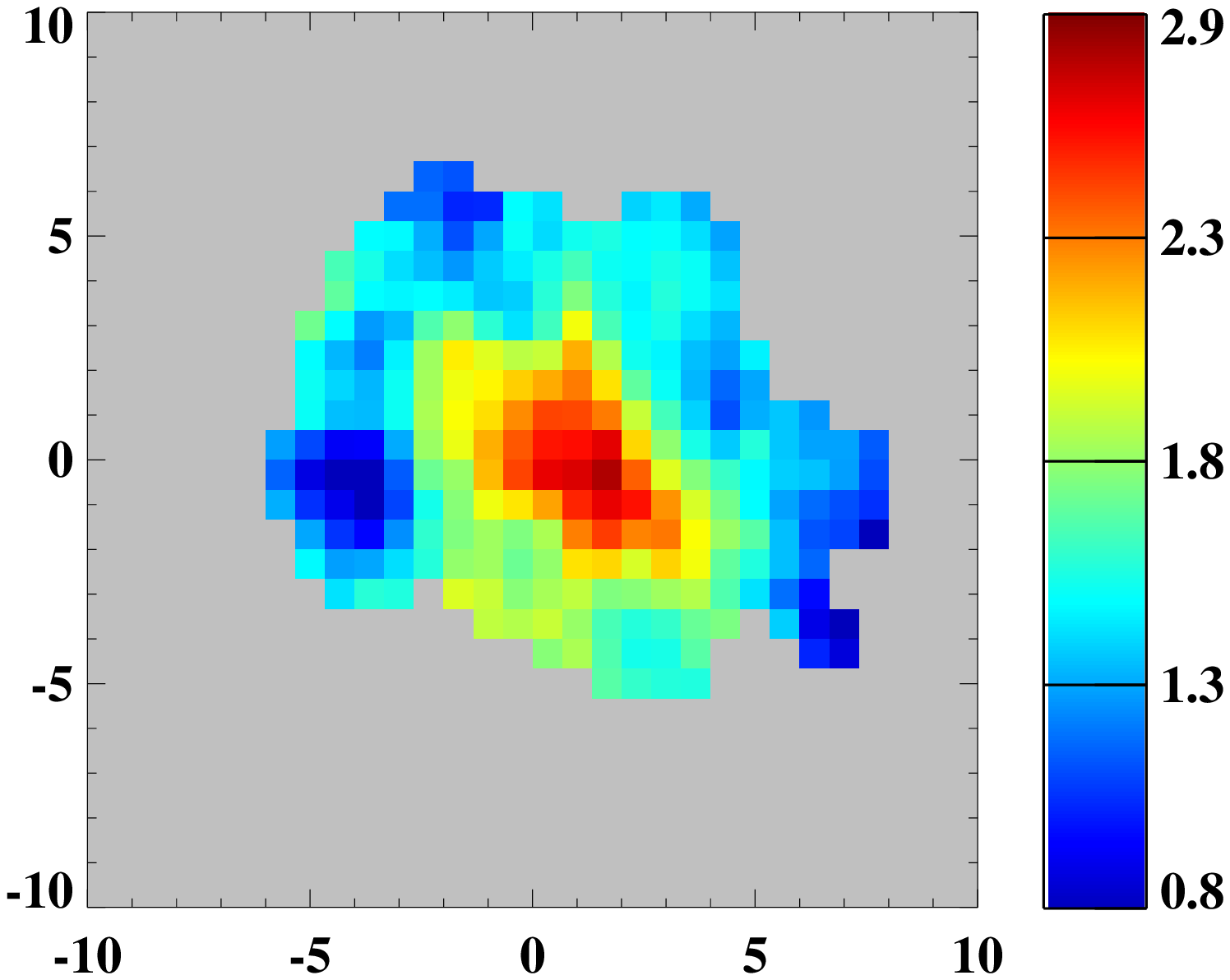} \\
 
   \includegraphics[height=2.6cm]{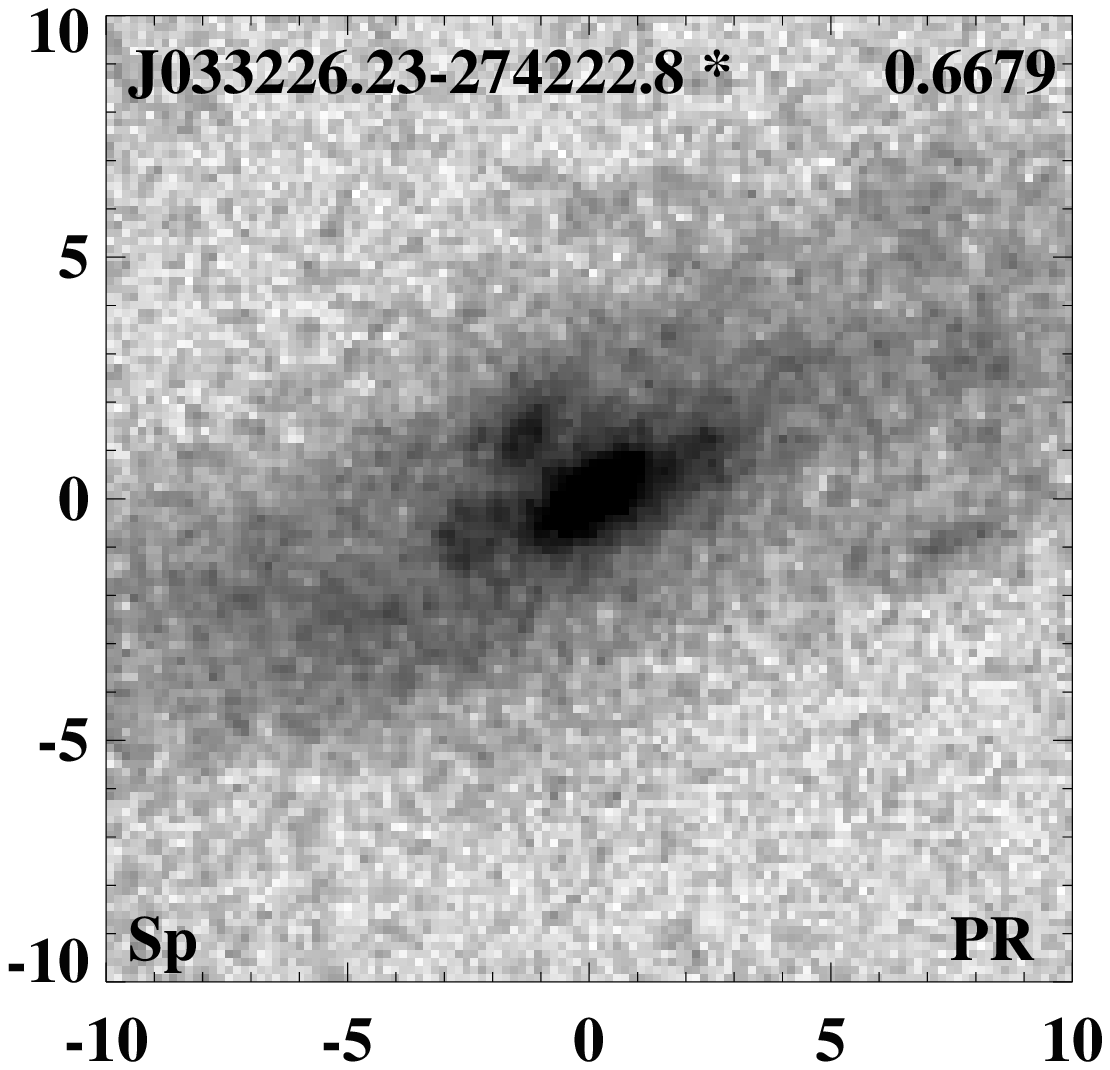} &
 \includegraphics[height=2.6cm]{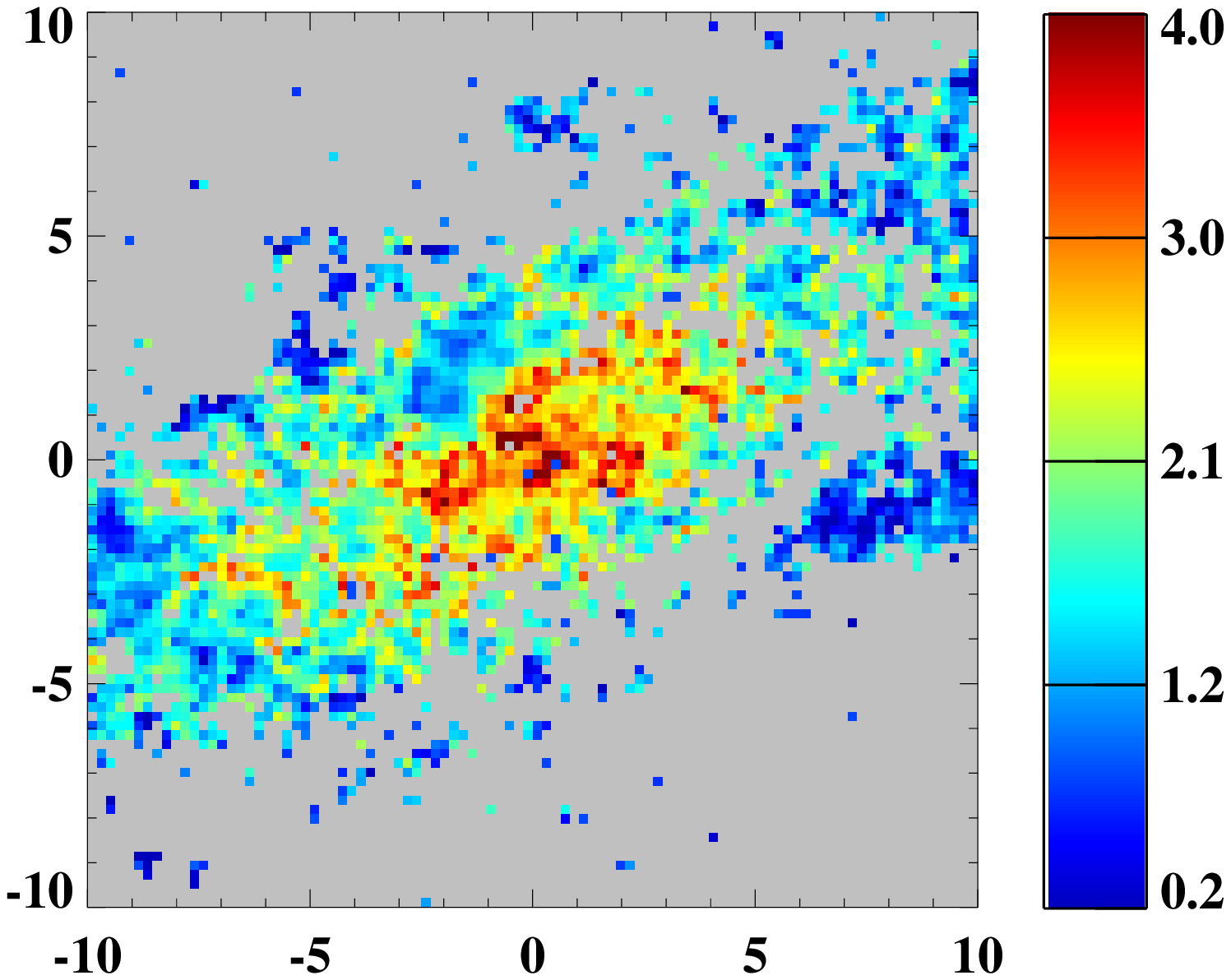} &
   \includegraphics[height=2.6cm]{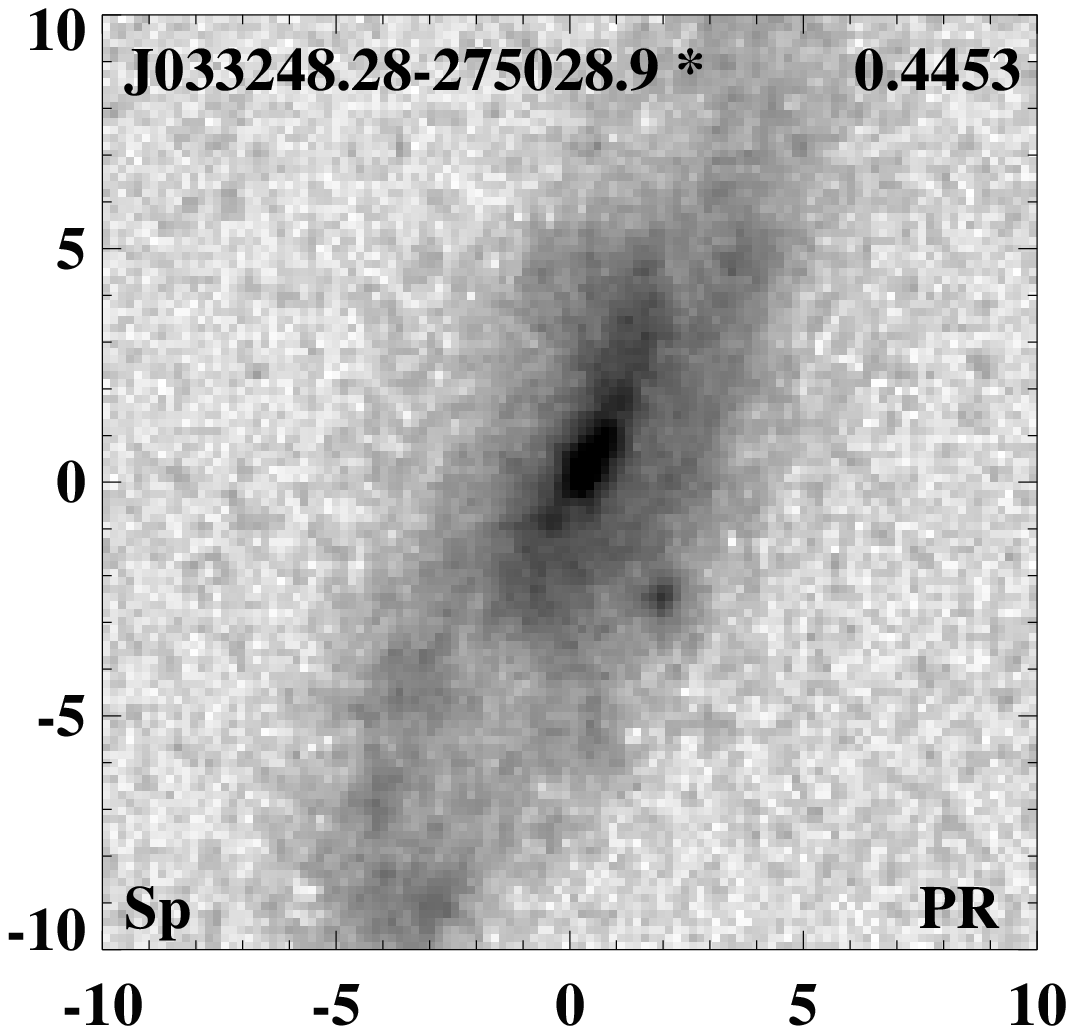} &
 \includegraphics[height=2.6cm]{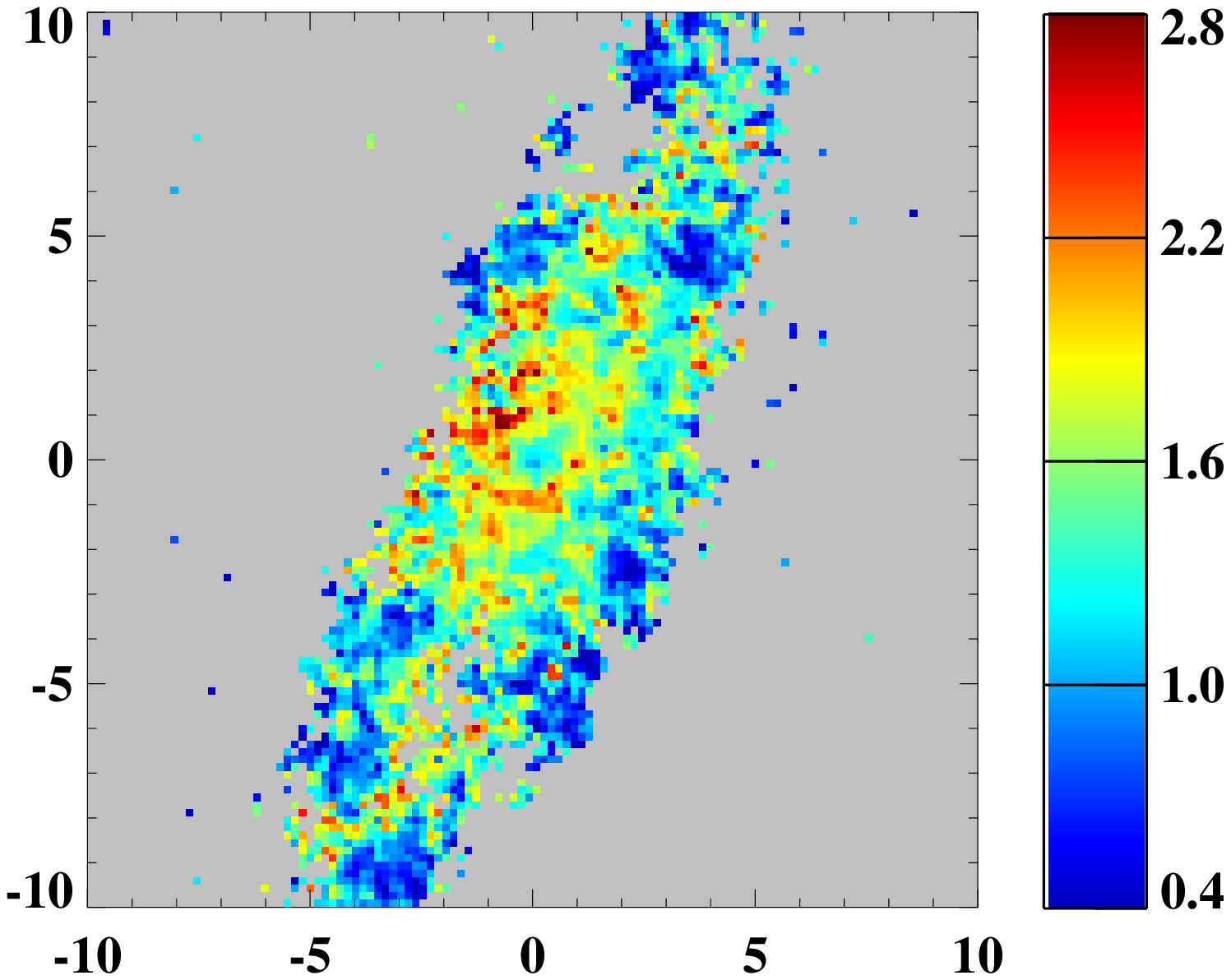} \\
 
   \includegraphics[height=2.6cm]{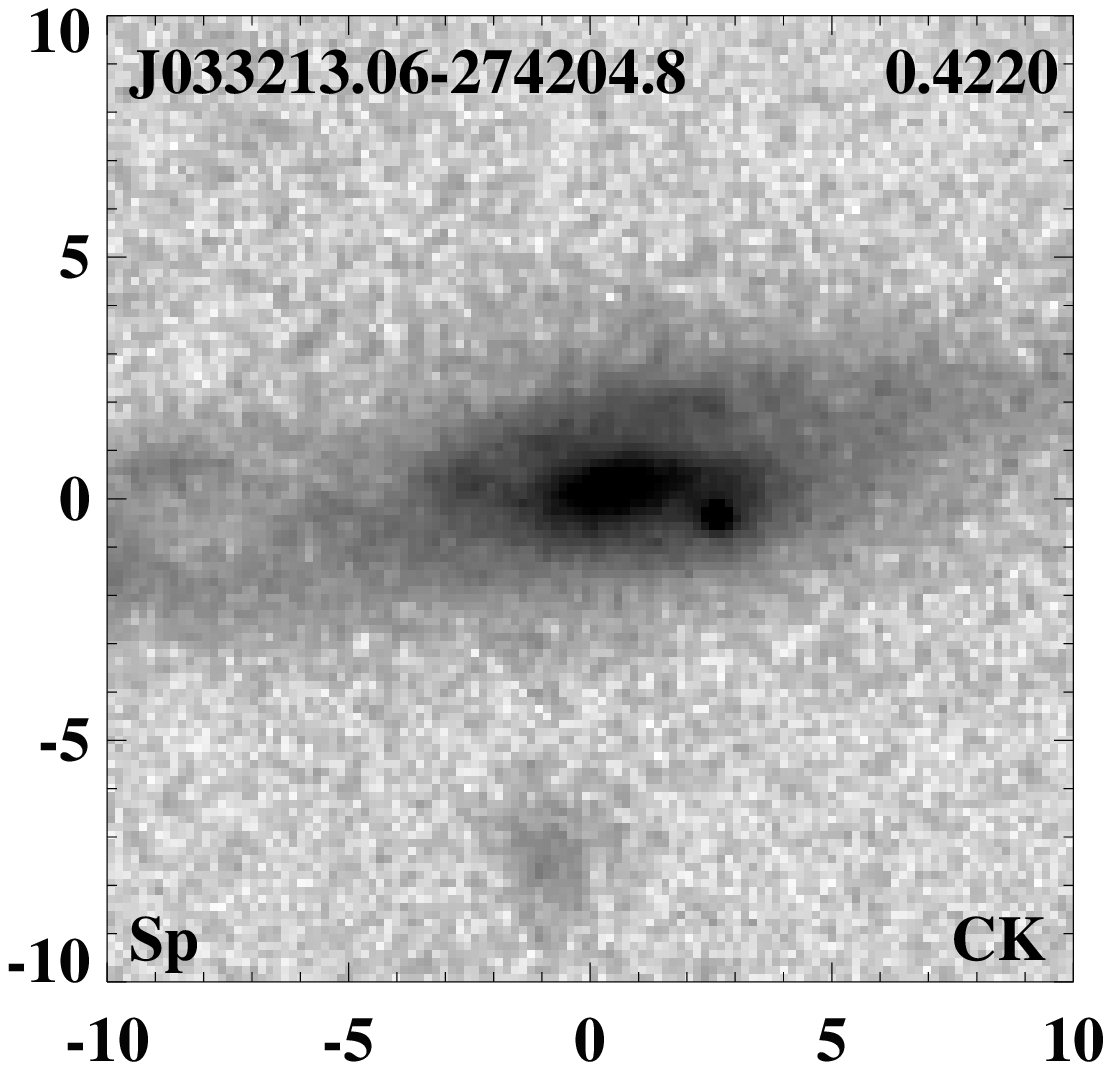} &
 \includegraphics[height=2.6cm]{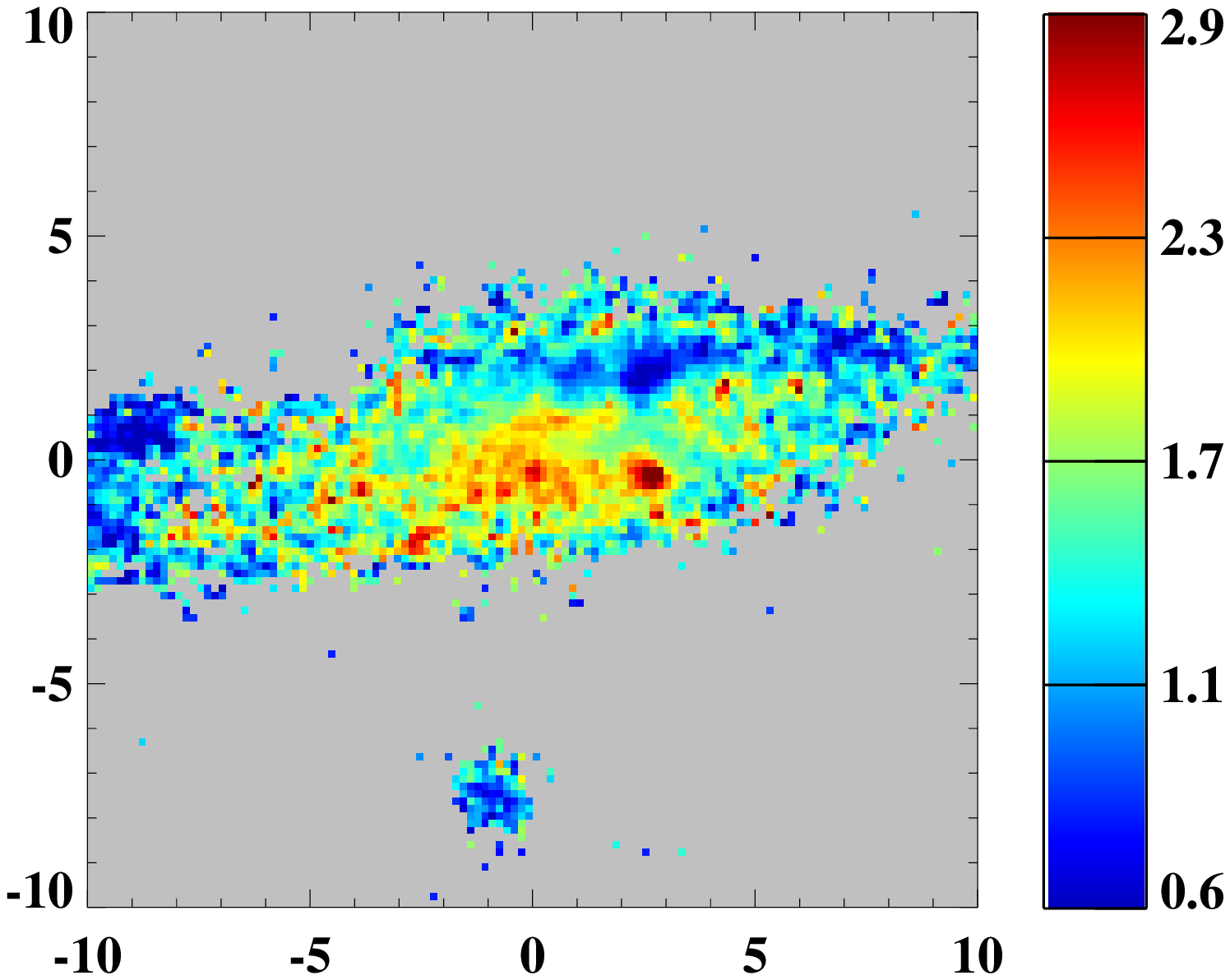} &
   \includegraphics[height=2.6cm]{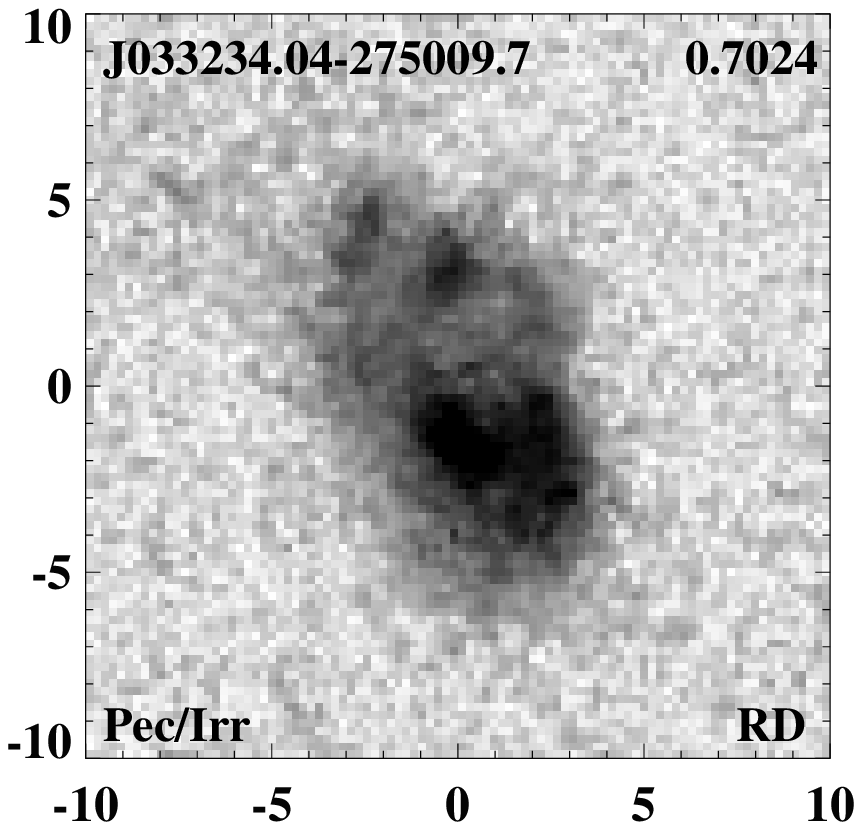} &
 \includegraphics[height=2.6cm]{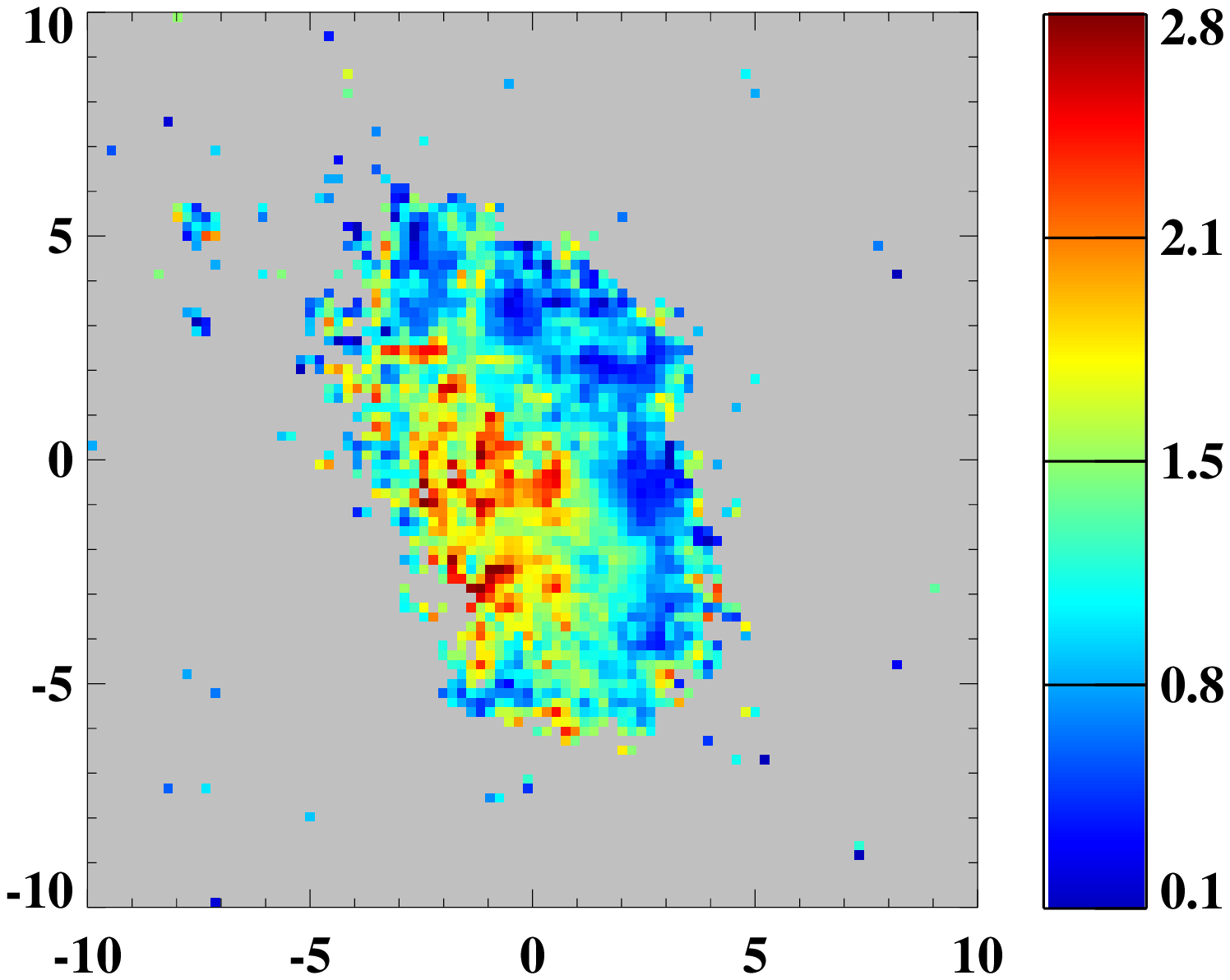} \\
 
    \includegraphics[height=2.6cm]{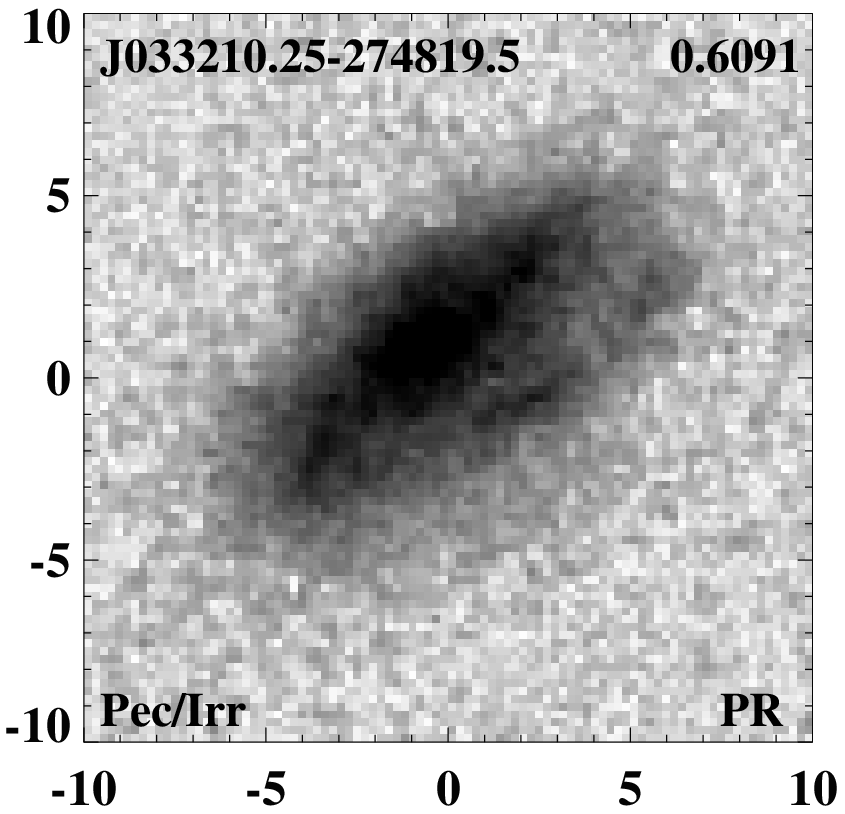} &
 \includegraphics[height=2.6cm]{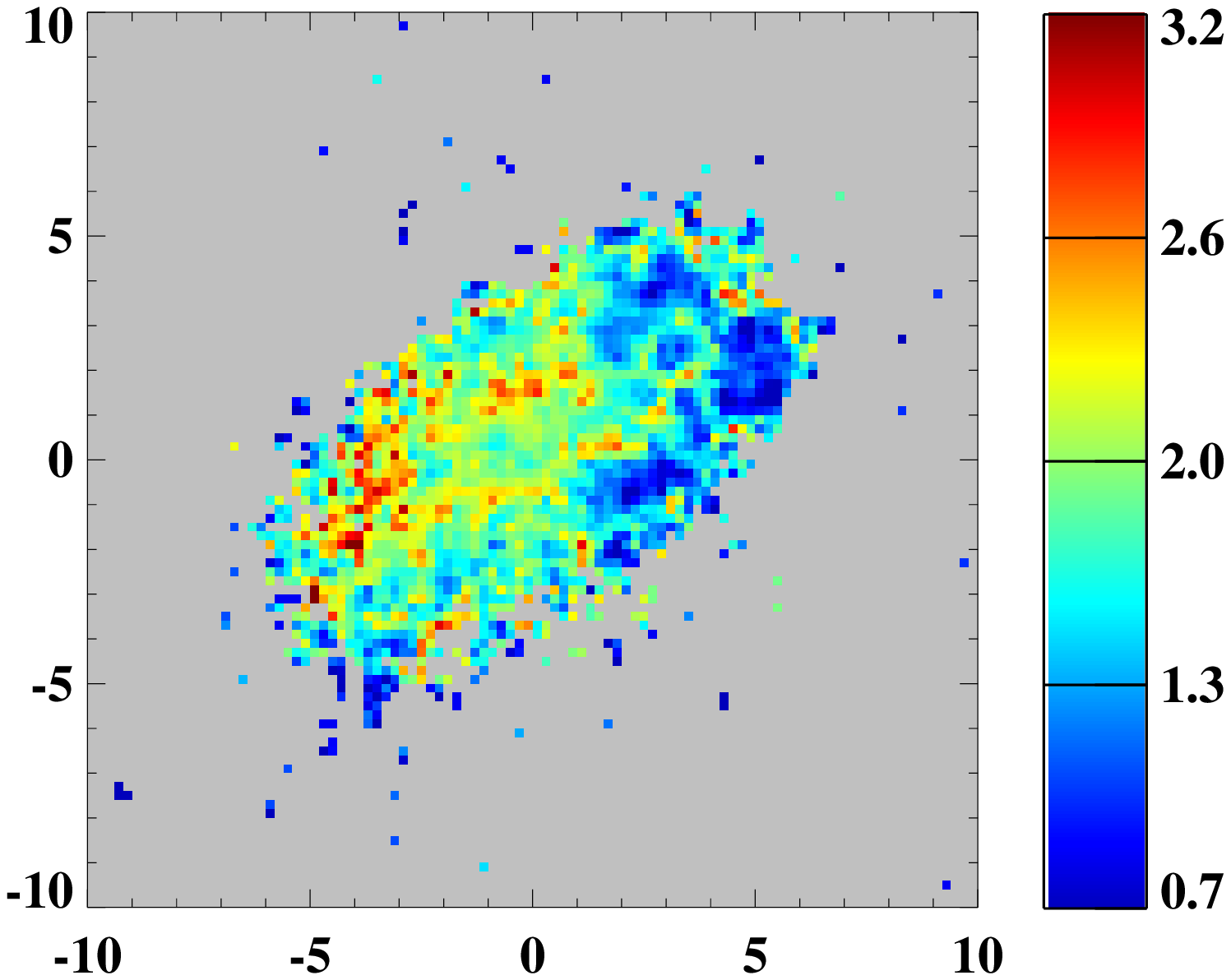} &
    \includegraphics[height=2.6cm]{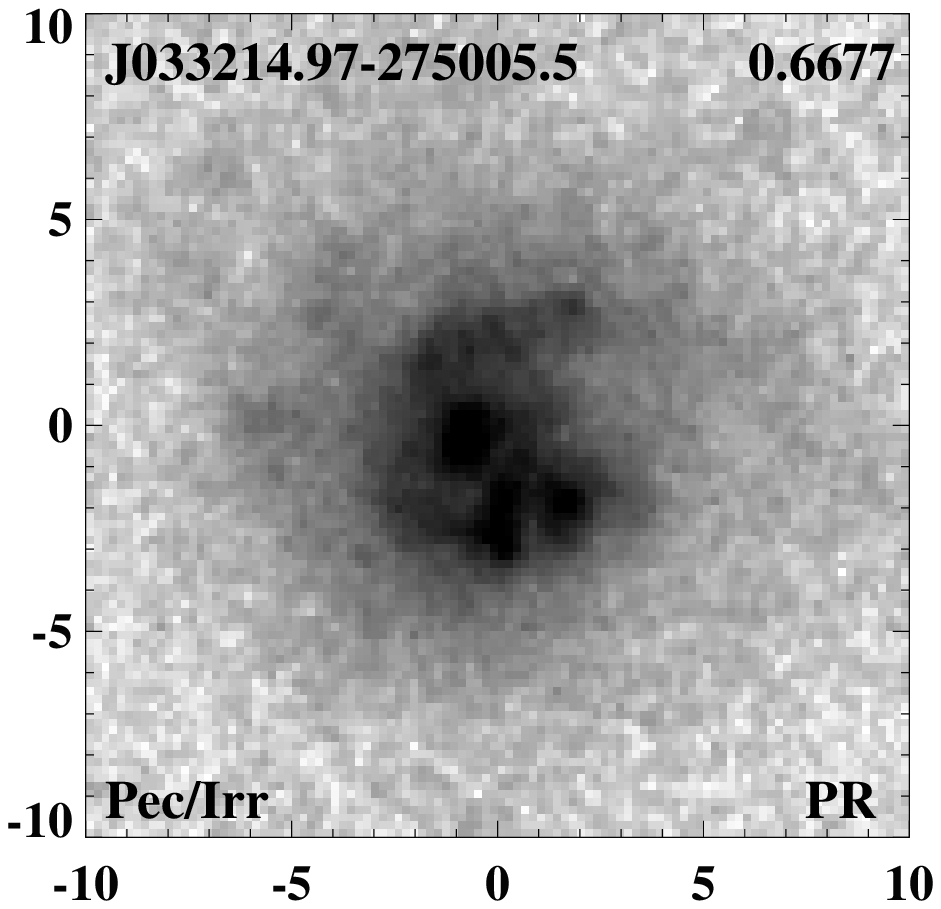} &
 \includegraphics[height=2.6cm]{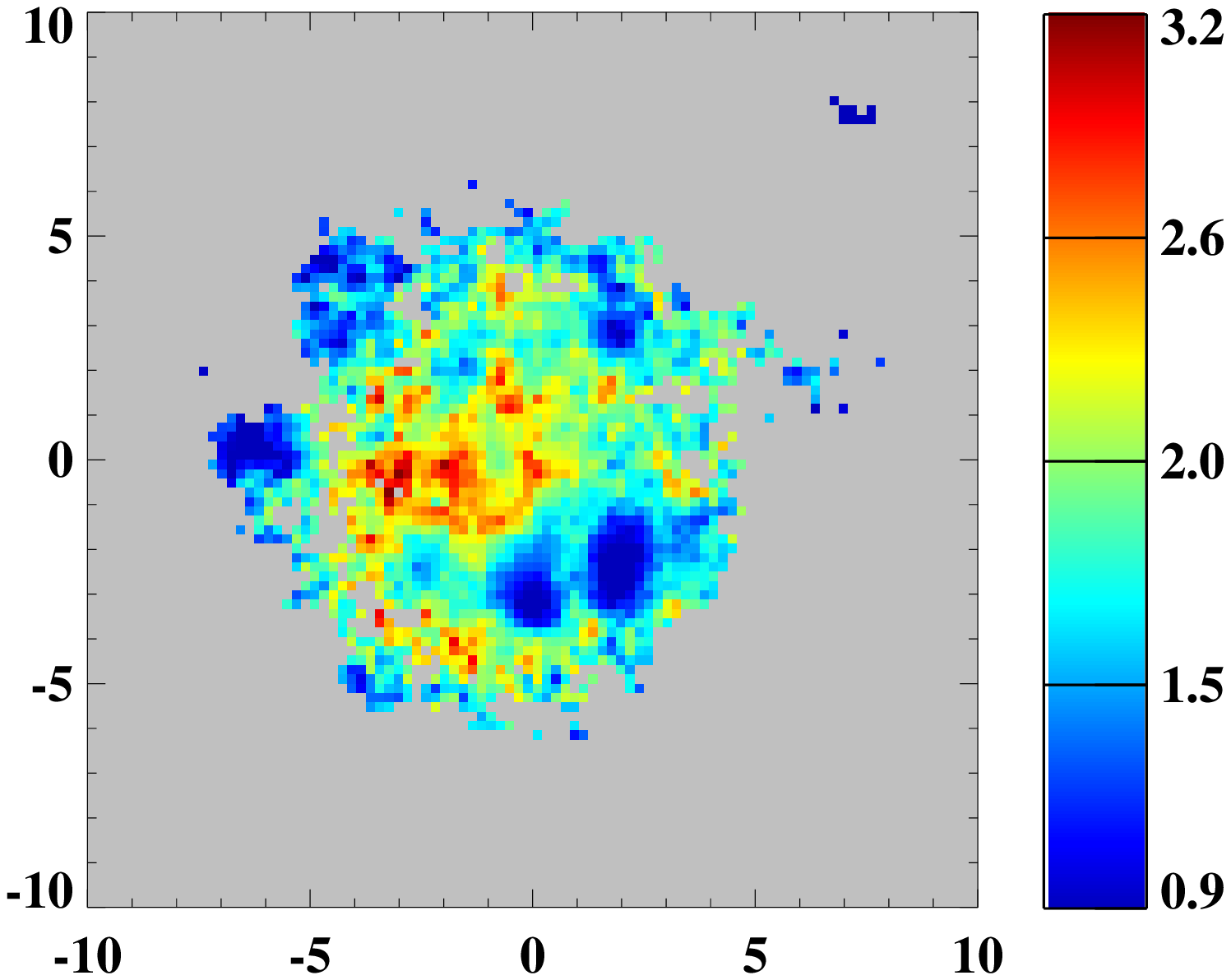} \\
  
    & & & \\
  \multicolumn{4}{l}{{\bf Fig. 12.}  F850LP/F814W image, color map. Explanations are given at the end of this figure.} \\
  \multicolumn{4}{l}{Rotating spiral disks are marked with an asterisk near their name.}

        \end{tabular}
   \end{center}
   \label{fig1} 
   \end{figure}

   \newpage

    \begin{figure}[h!]
   \begin{center}
   \begin{tabular}{cccc}        
   
   \includegraphics[height=2.6cm]{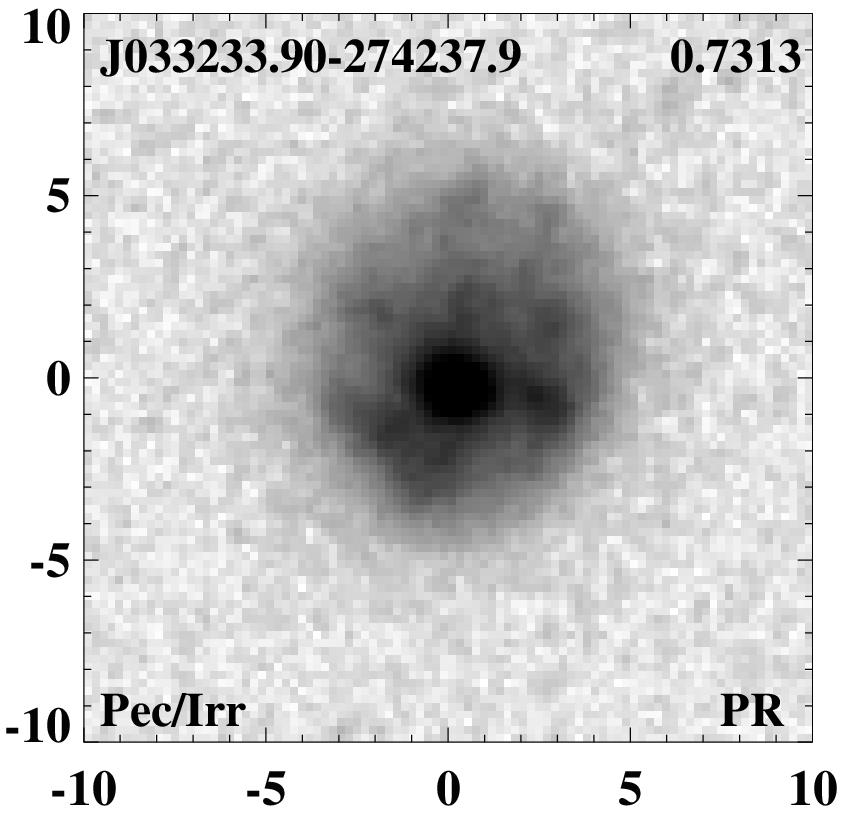} &
 \includegraphics[height=2.6cm]{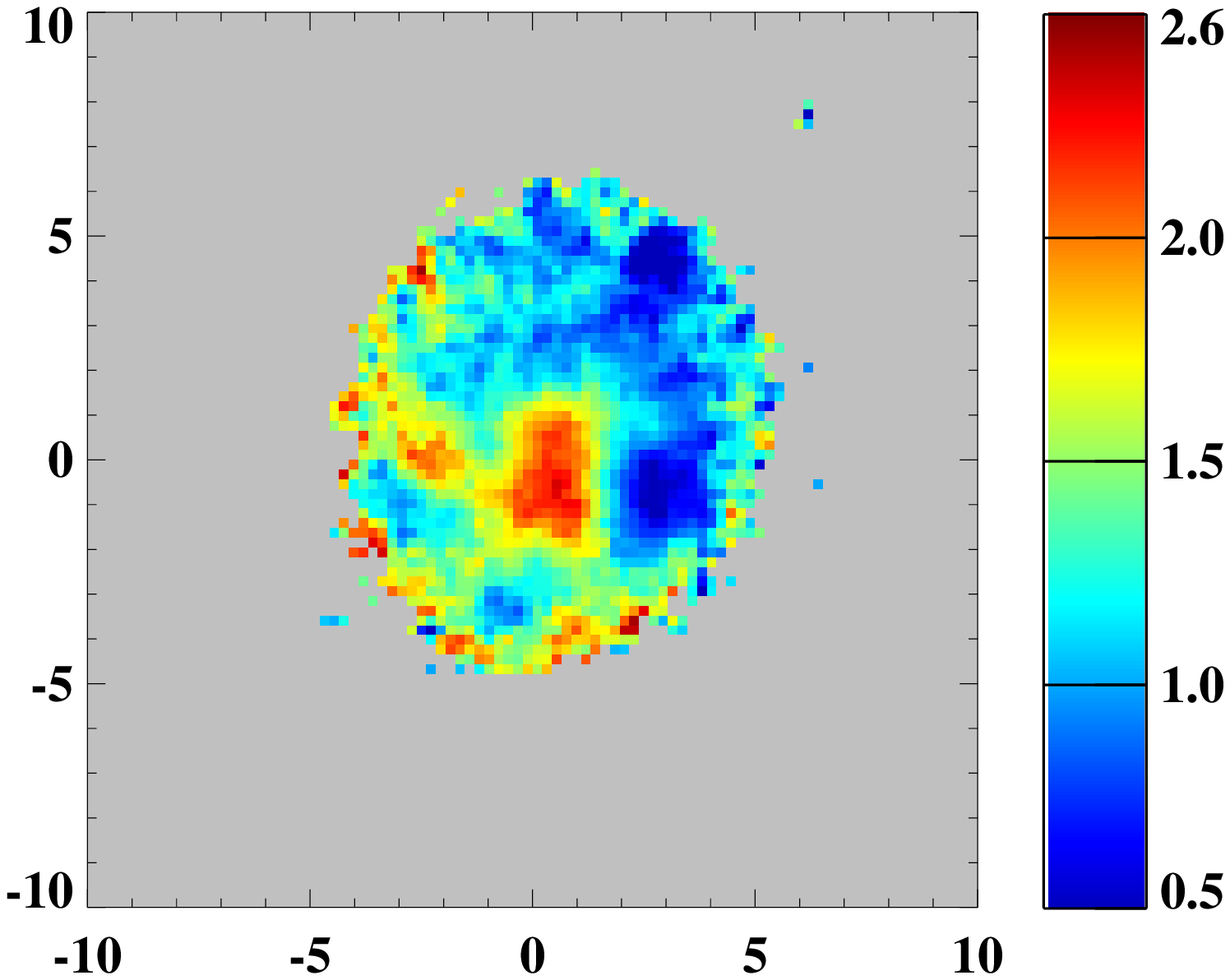} &
 \includegraphics[height=2.6cm]{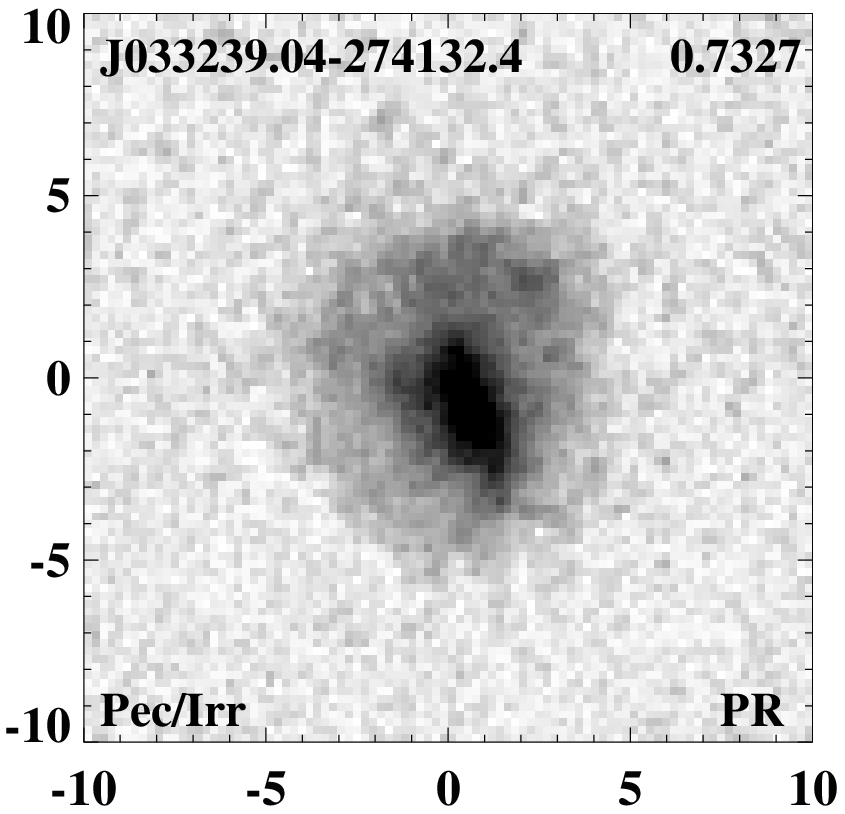} &
 \includegraphics[height=2.6cm]{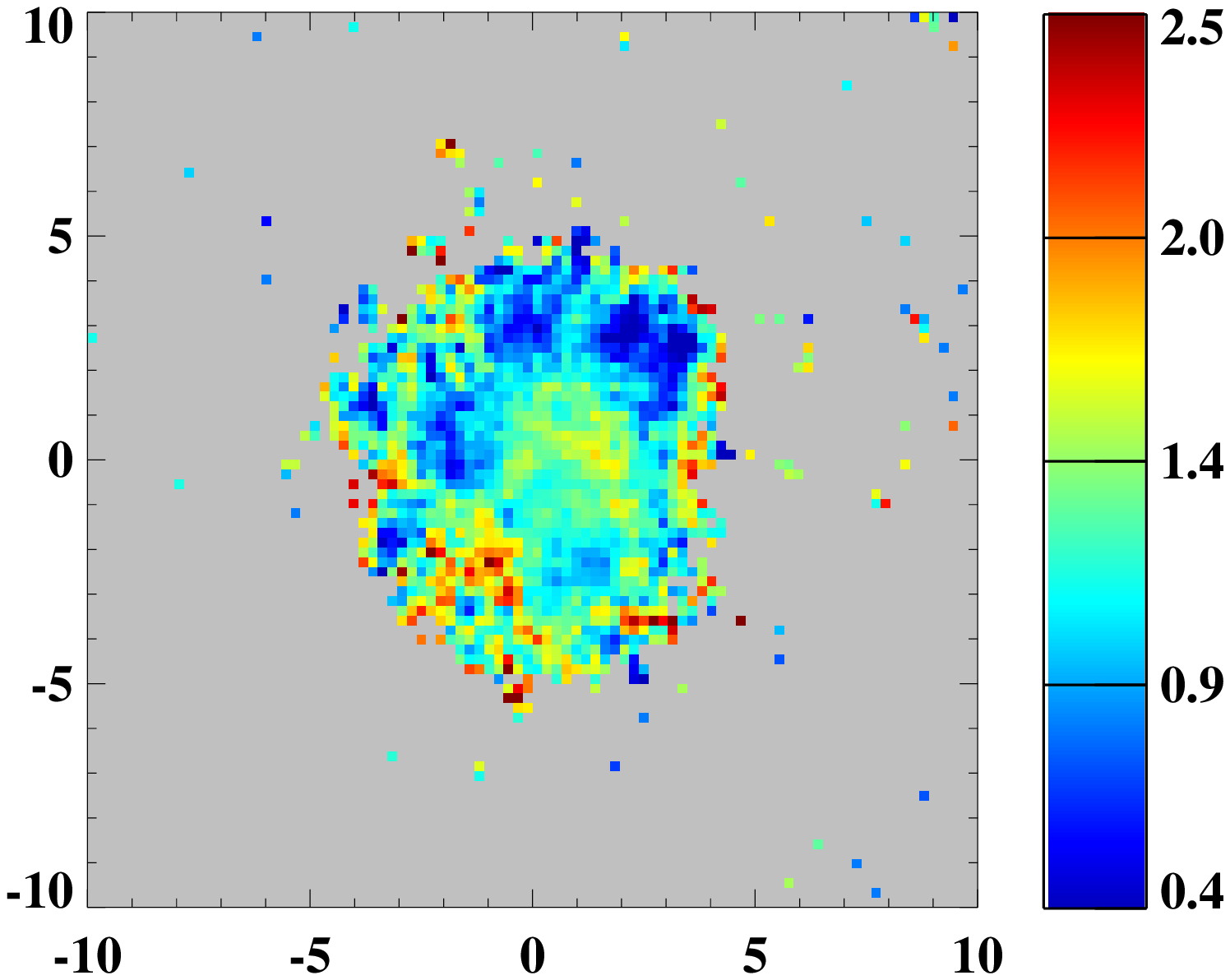} \\
 
 \includegraphics[height=2.6cm]{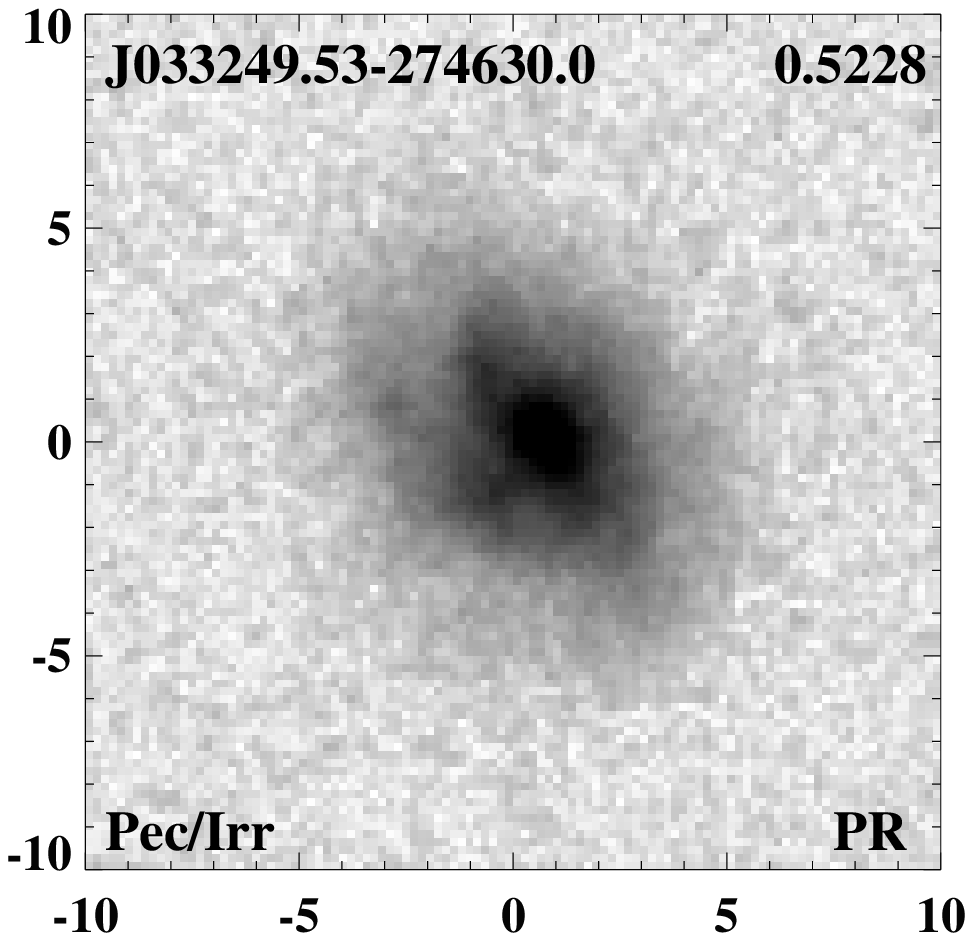} &
 \includegraphics[height=2.6cm]{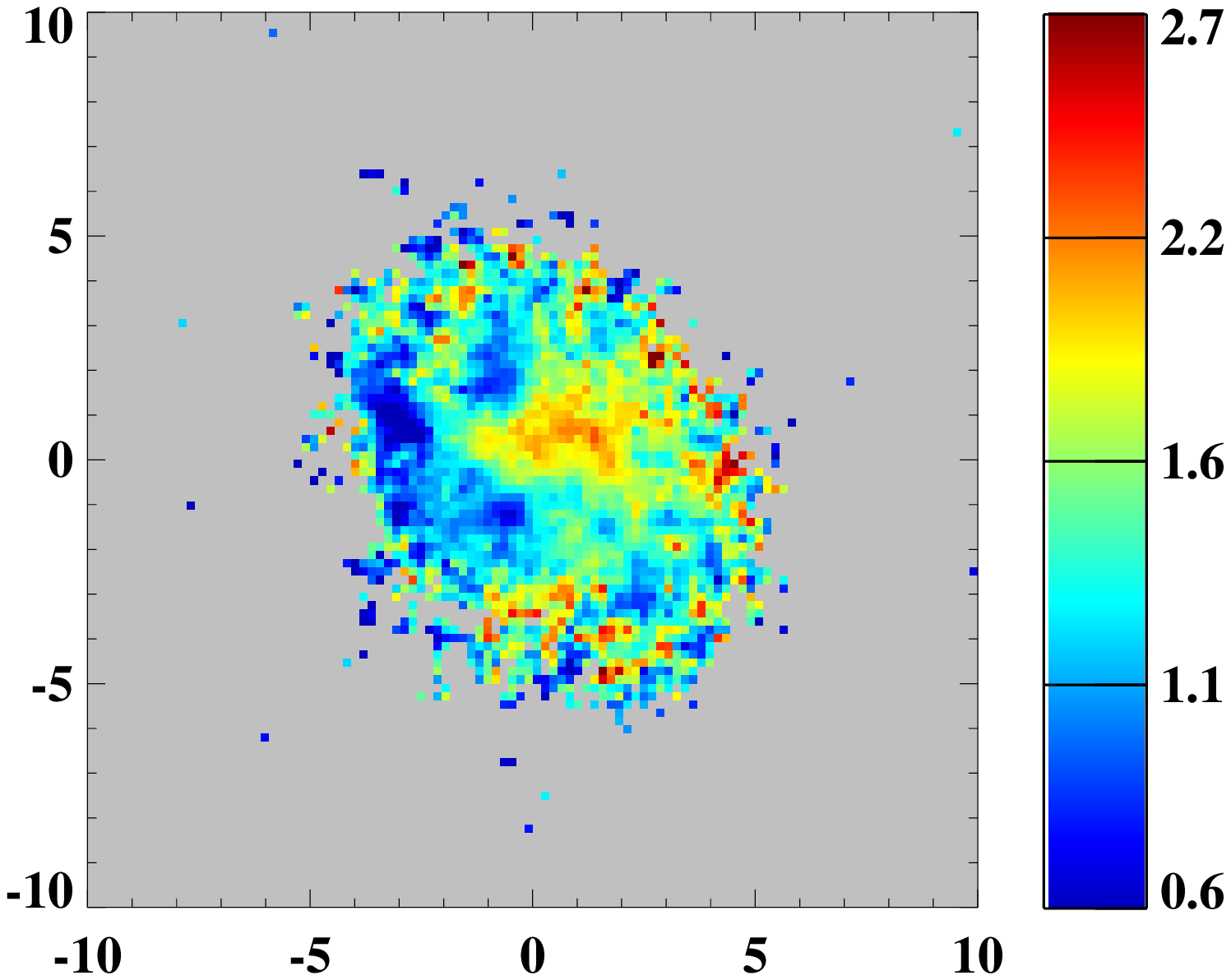} &
  \includegraphics[height=2.6cm]{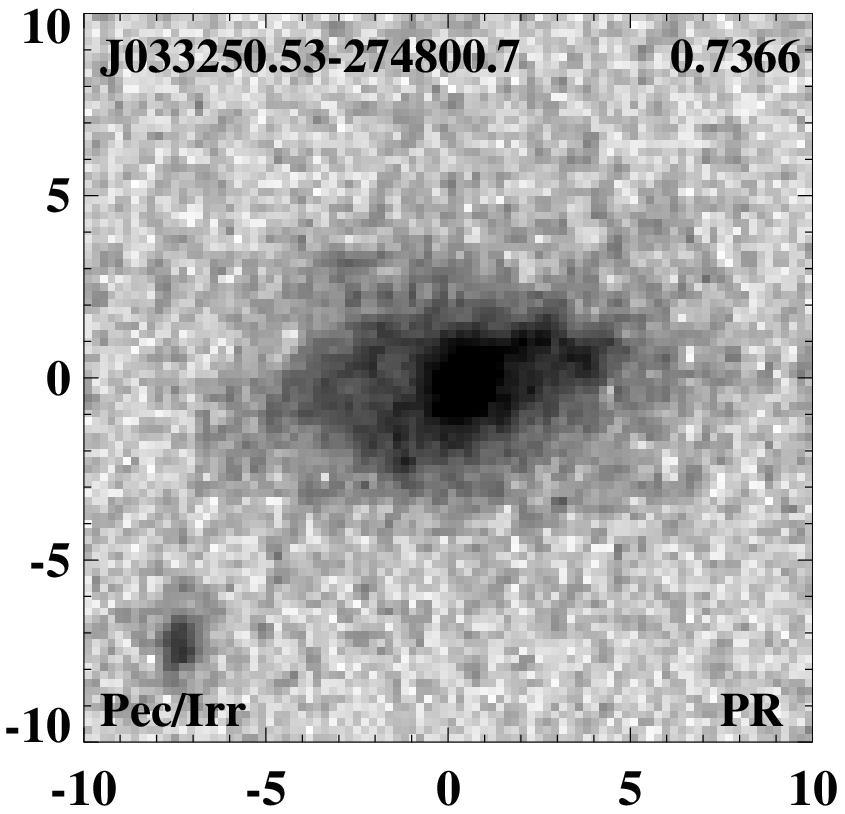} &
 \includegraphics[height=2.6cm]{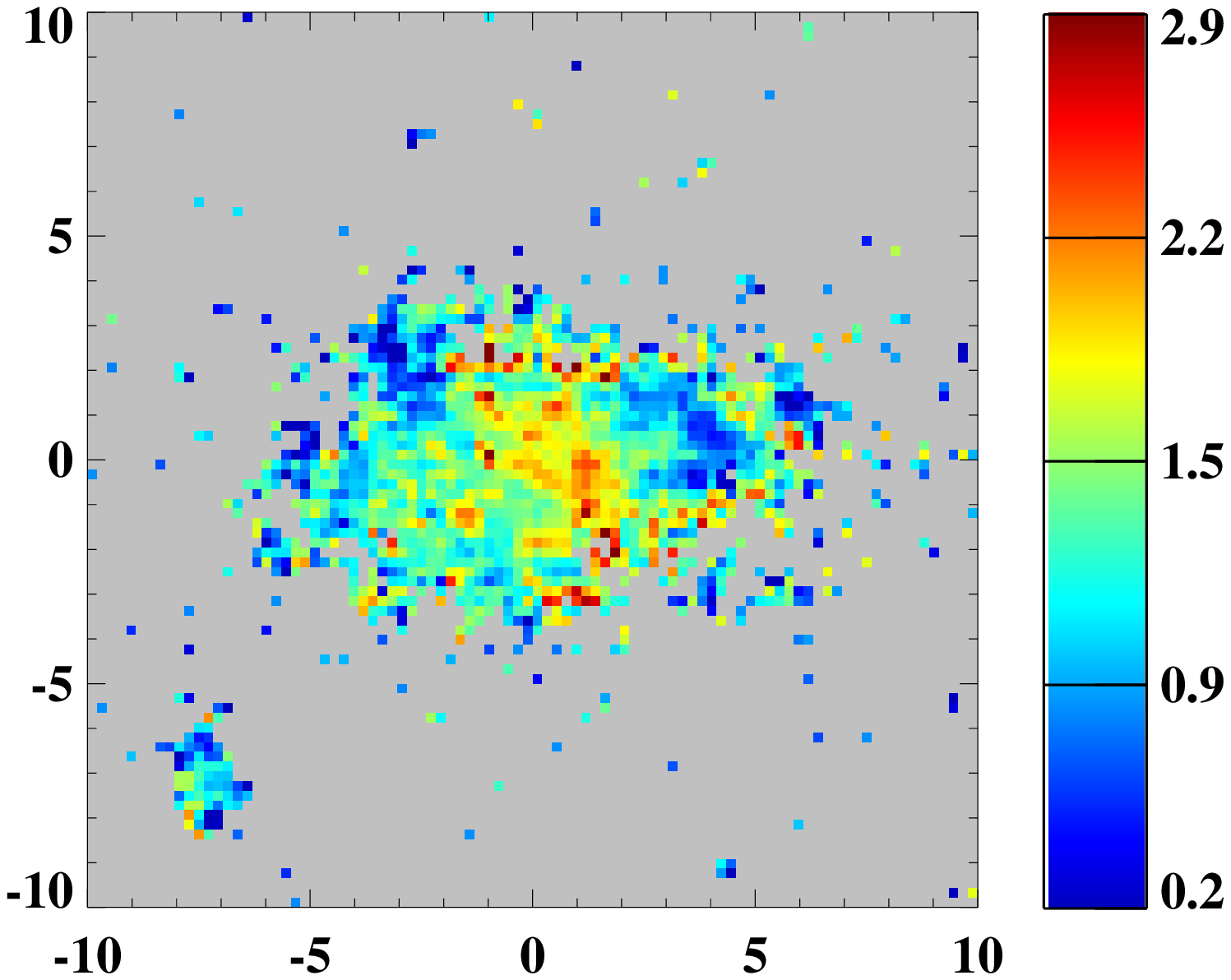} \\
 
\includegraphics[height=2.6cm]{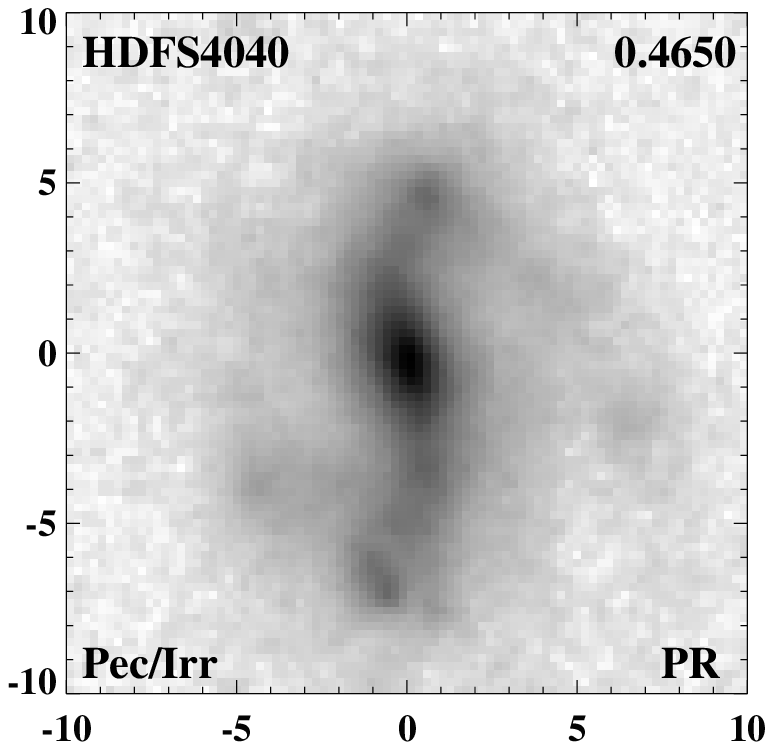} &
 \includegraphics[height=2.6cm]{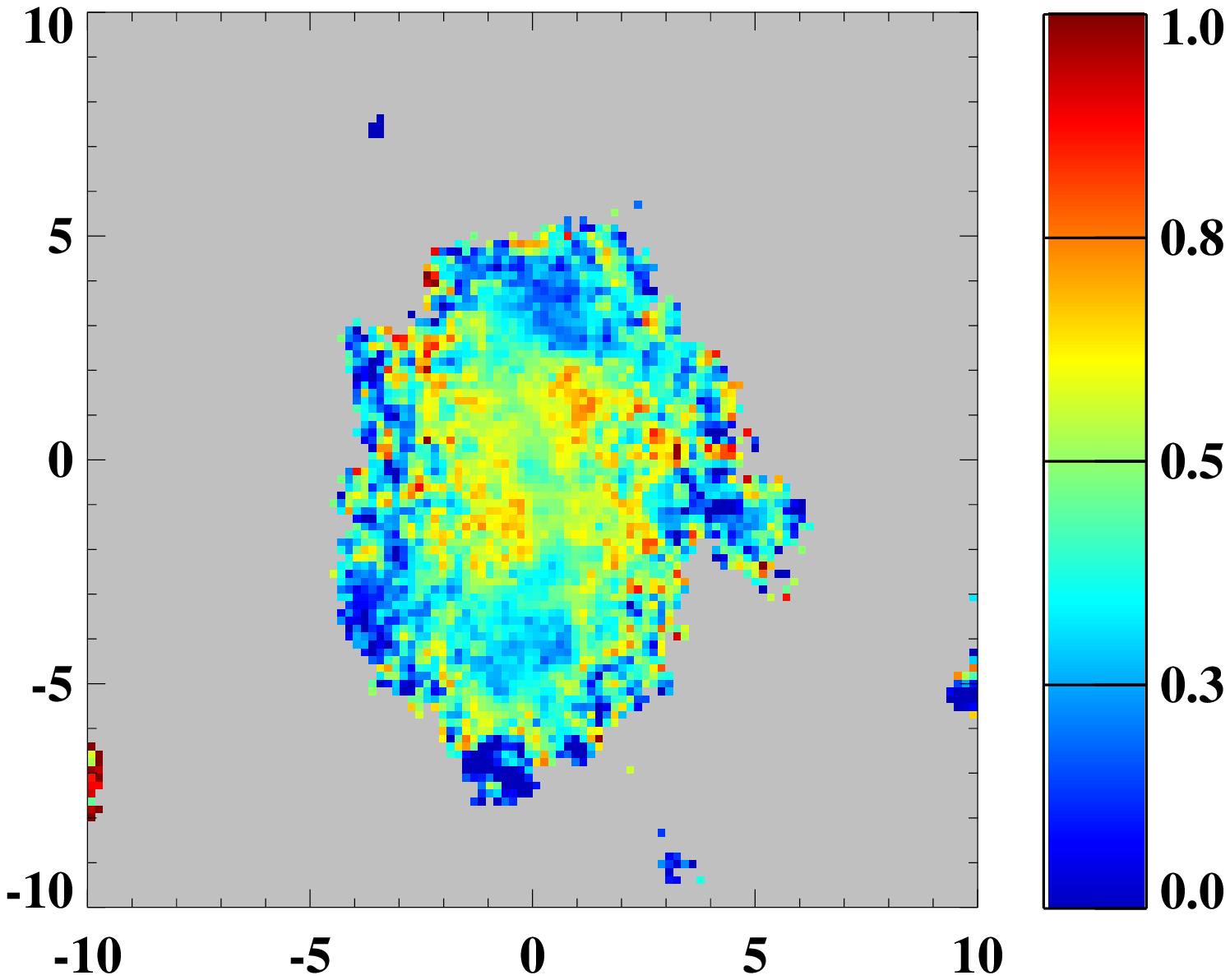} &
  \includegraphics[height=2.6cm]{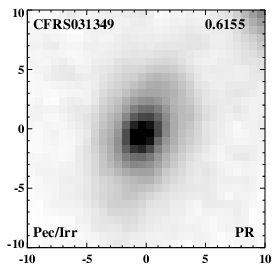} &
 \includegraphics[height=2.6cm]{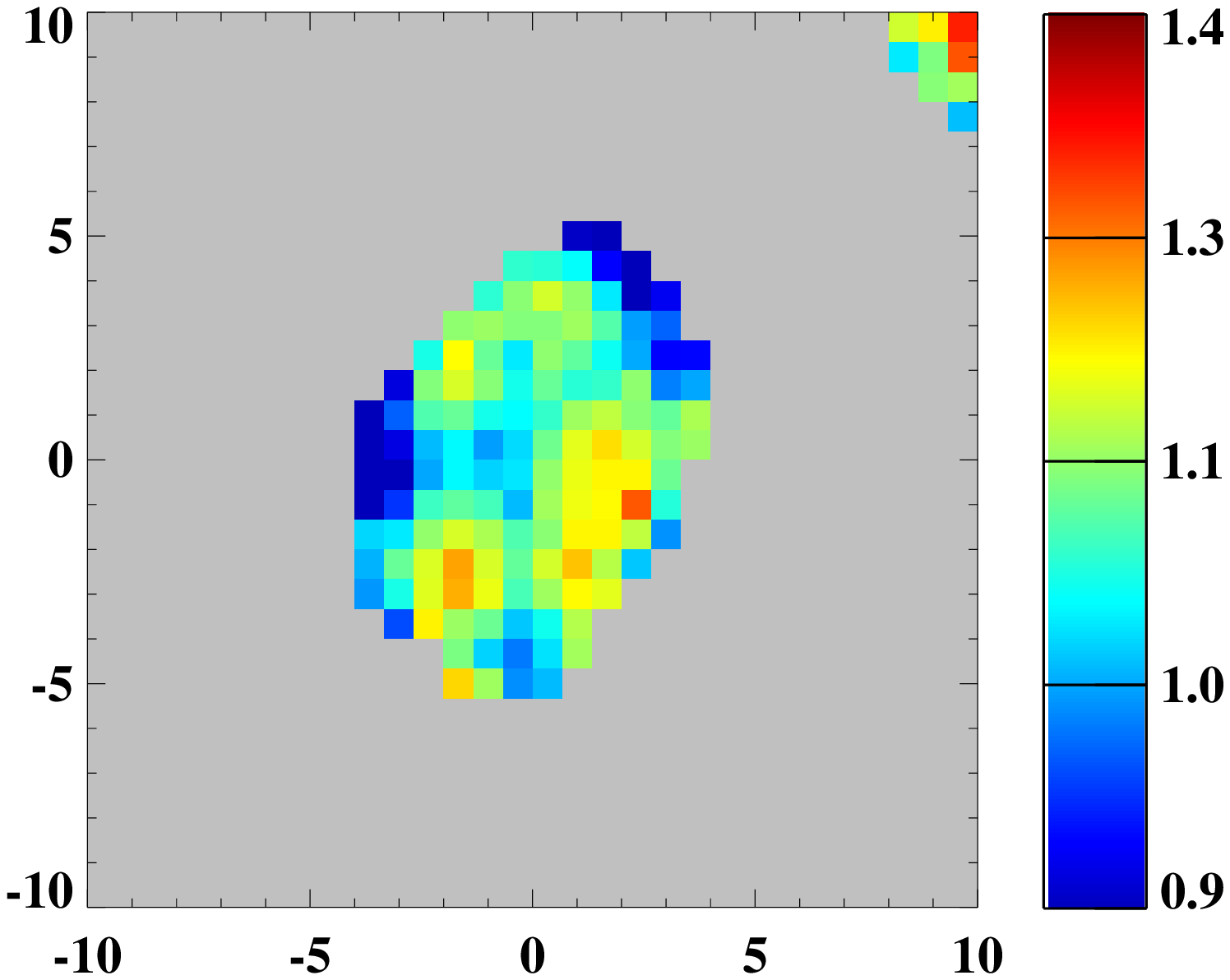} \\

 \includegraphics[height=2.6cm]{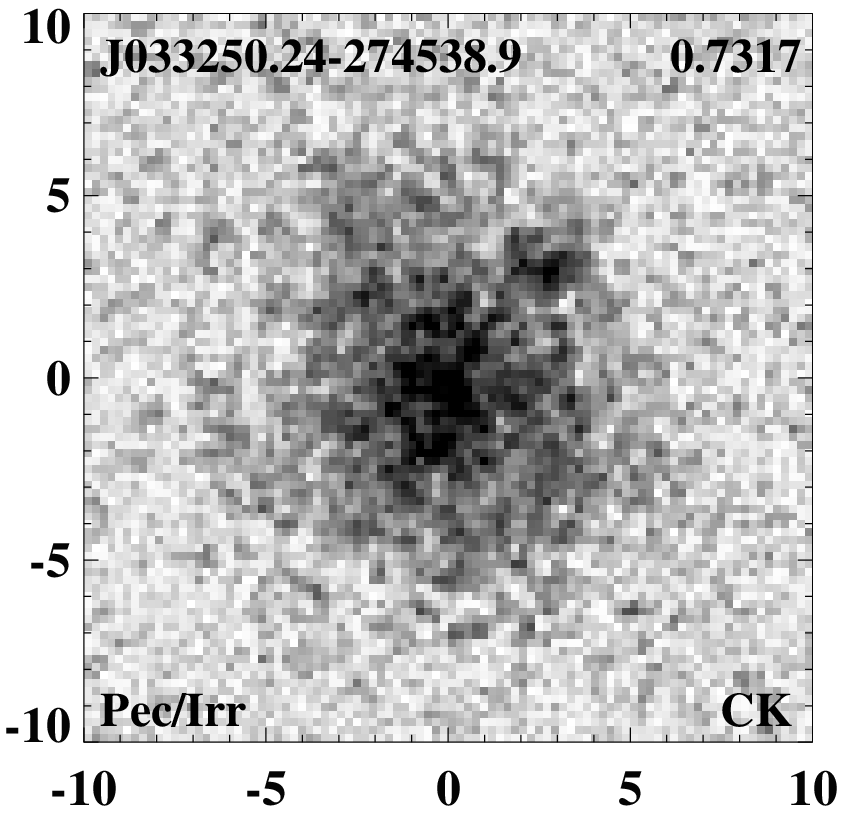} &
 \includegraphics[height=2.6cm]{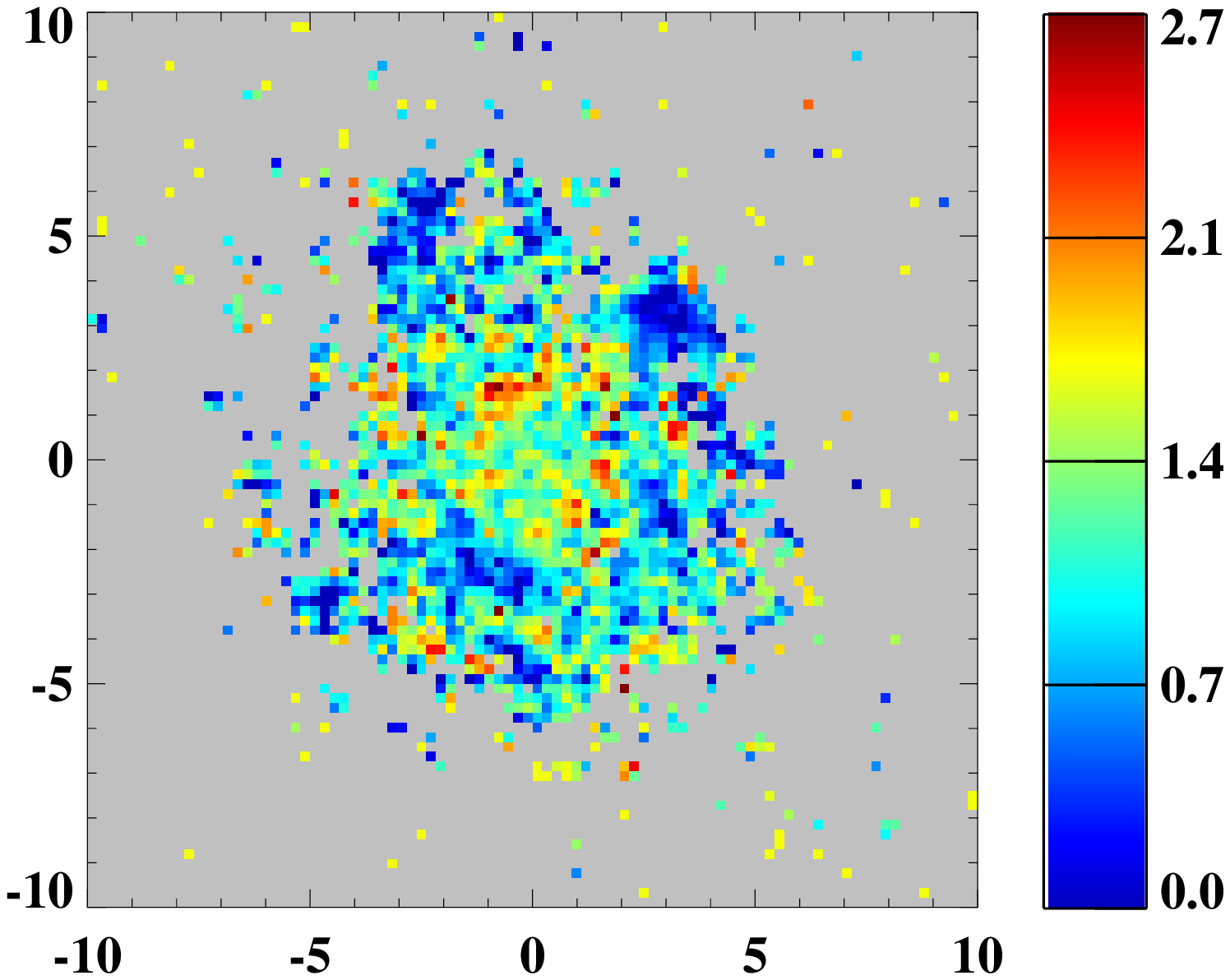} &
 \includegraphics[height=2.6cm]{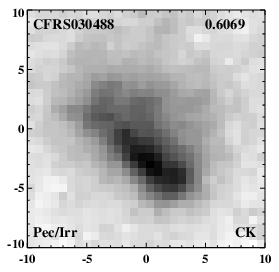} &
 \includegraphics[height=2.6cm]{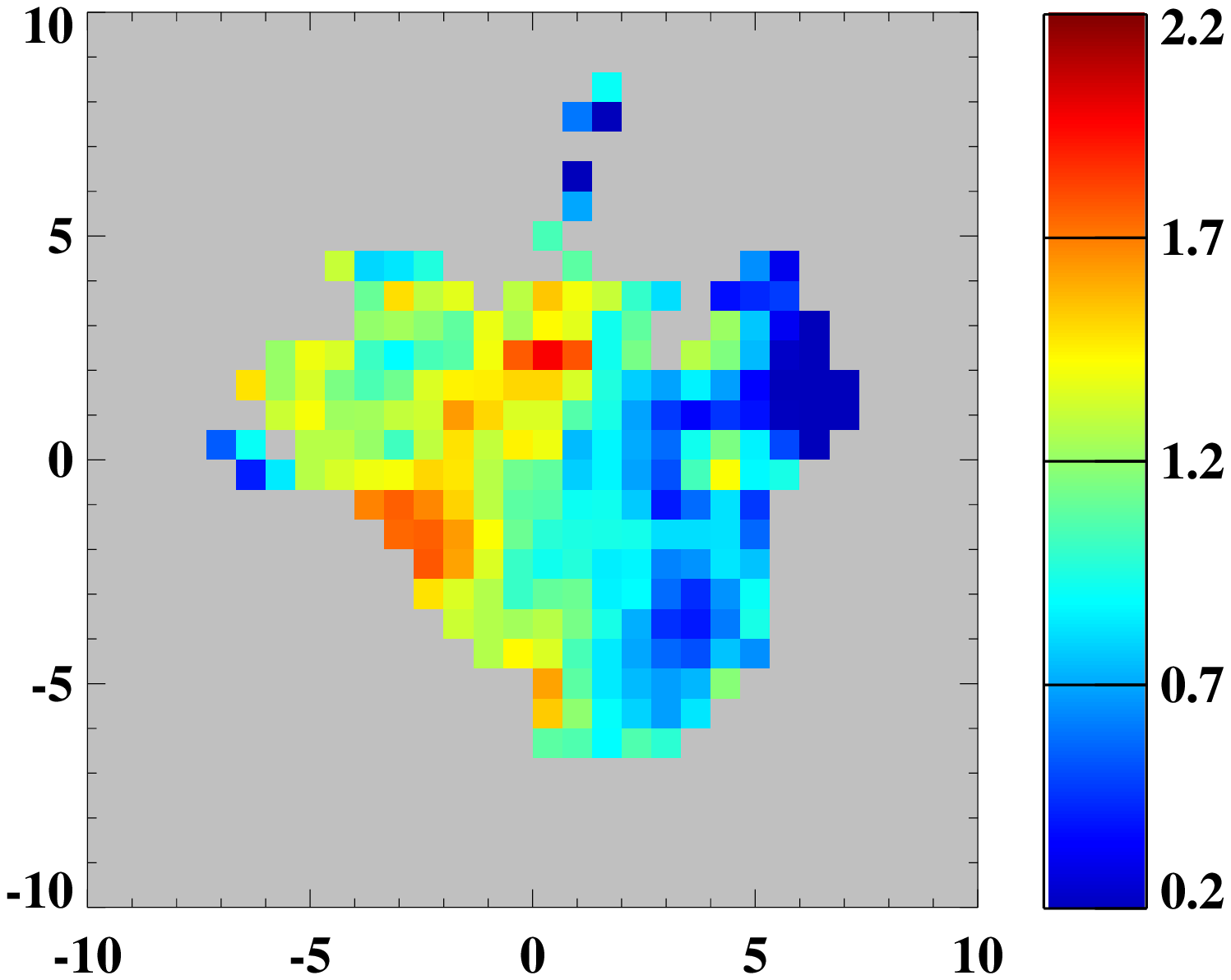} \\
 
  \includegraphics[height=2.6cm]{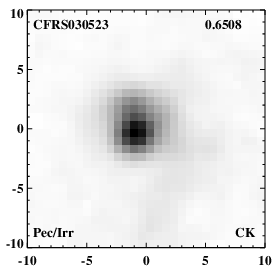} &
 \includegraphics[height=2.6cm]{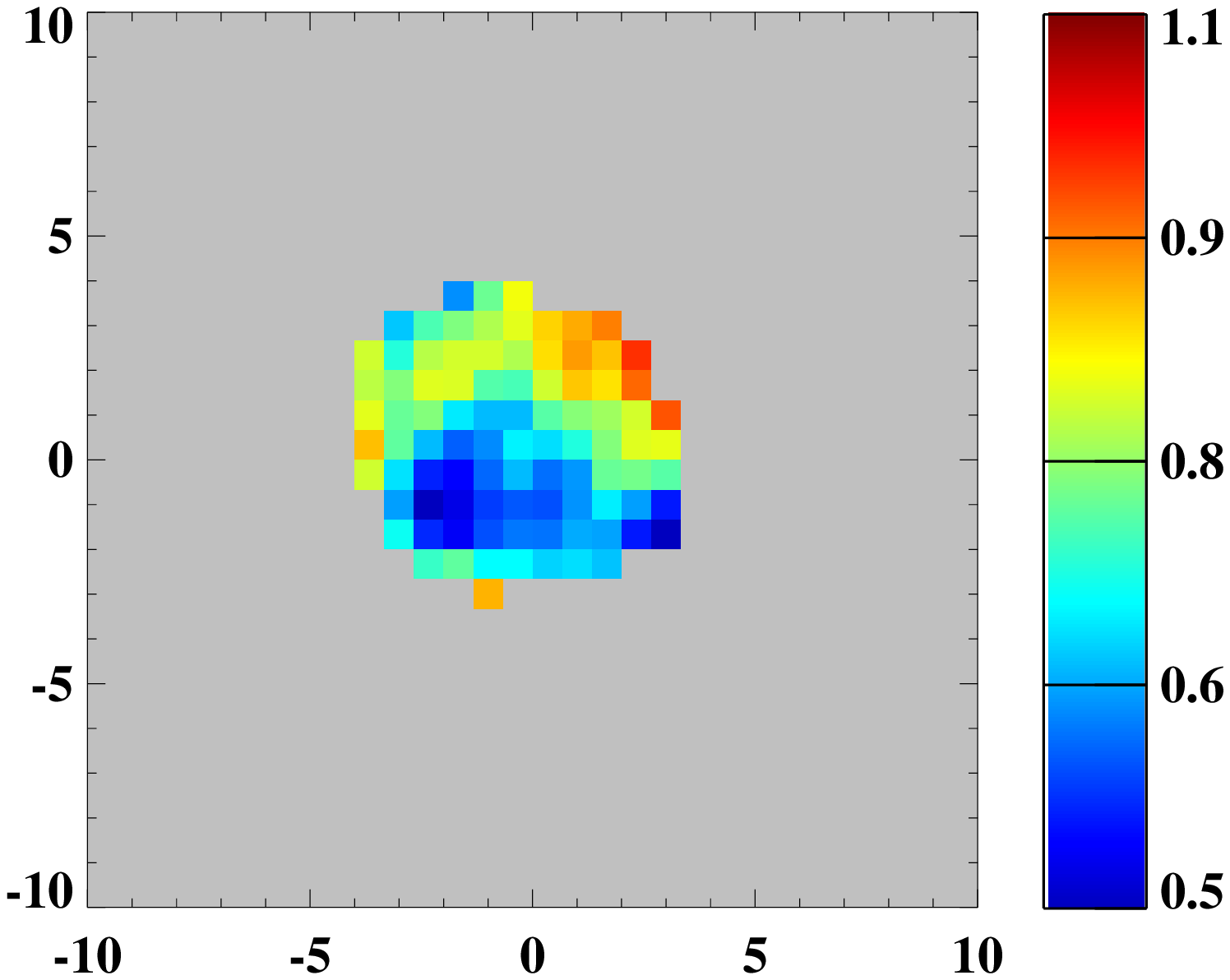} &
 \includegraphics[height=2.6cm]{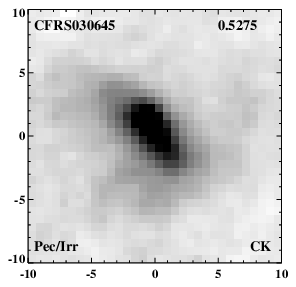} &
 \includegraphics[height=2.6cm]{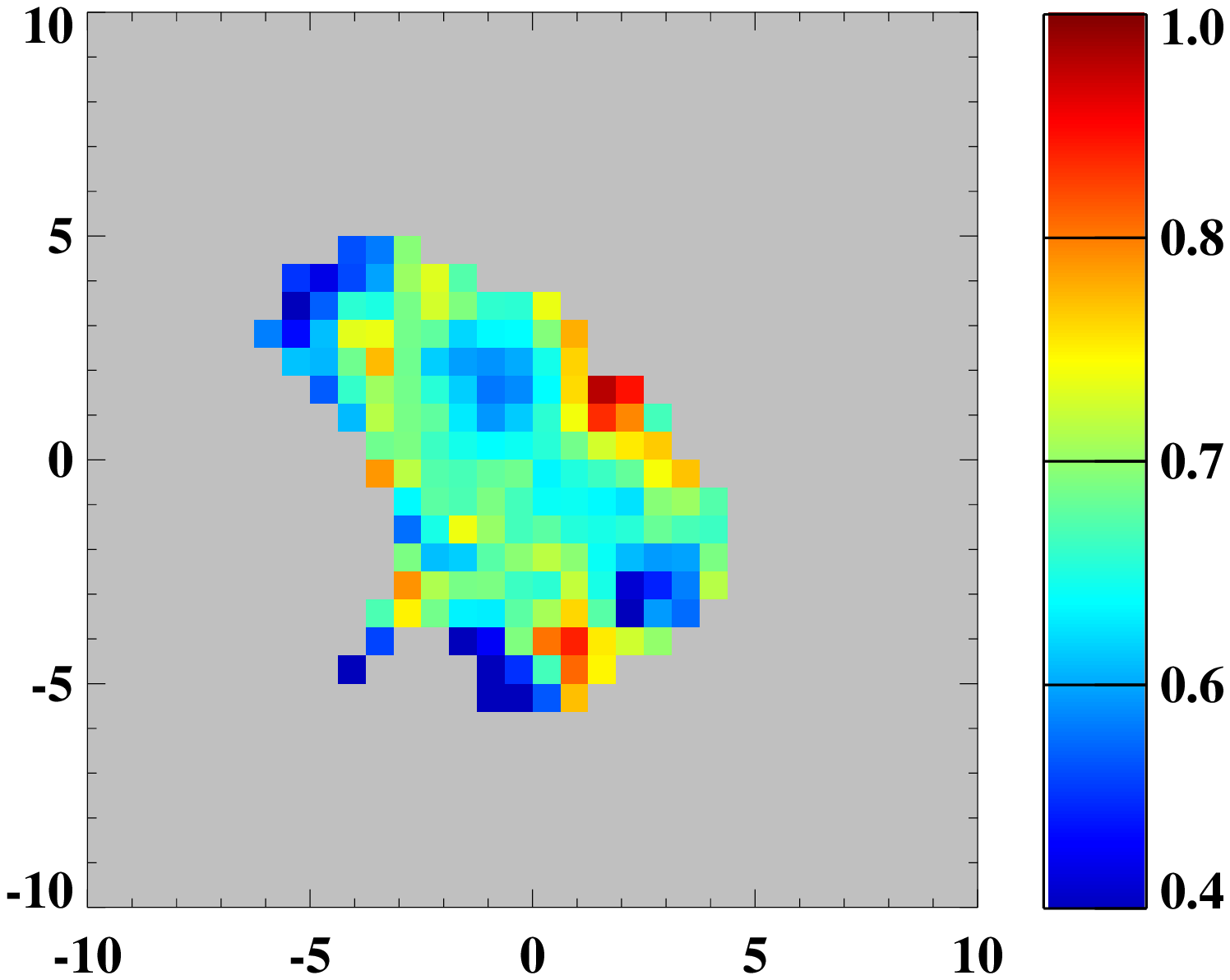} \\
 
  \includegraphics[height=2.6cm]{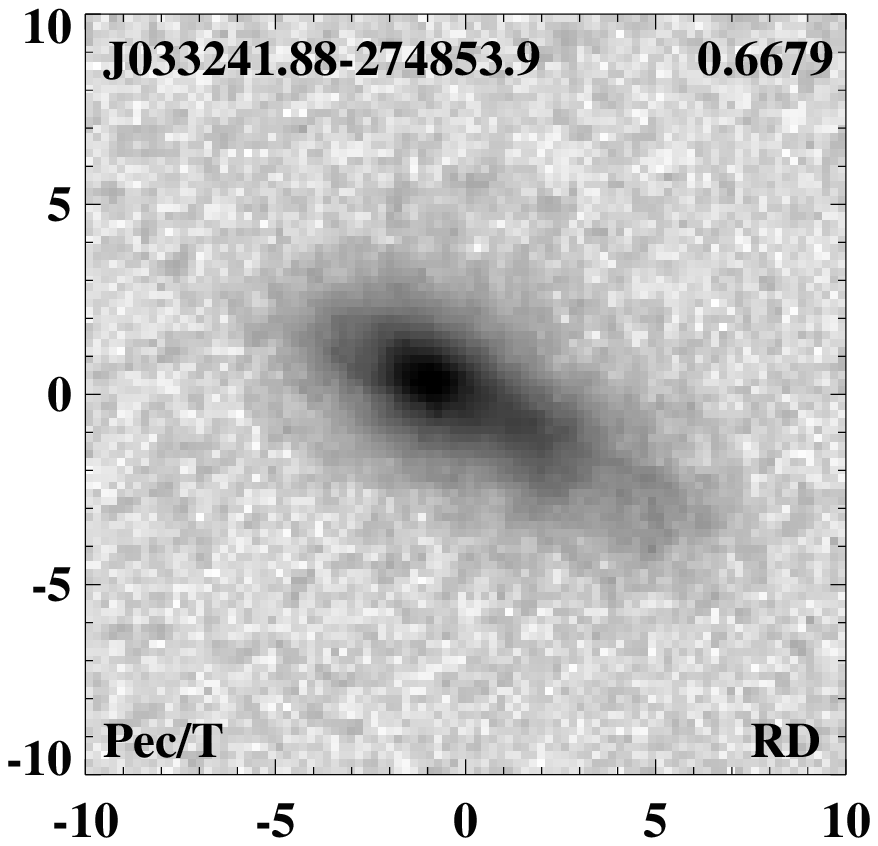} &
 \includegraphics[height=2.6cm]{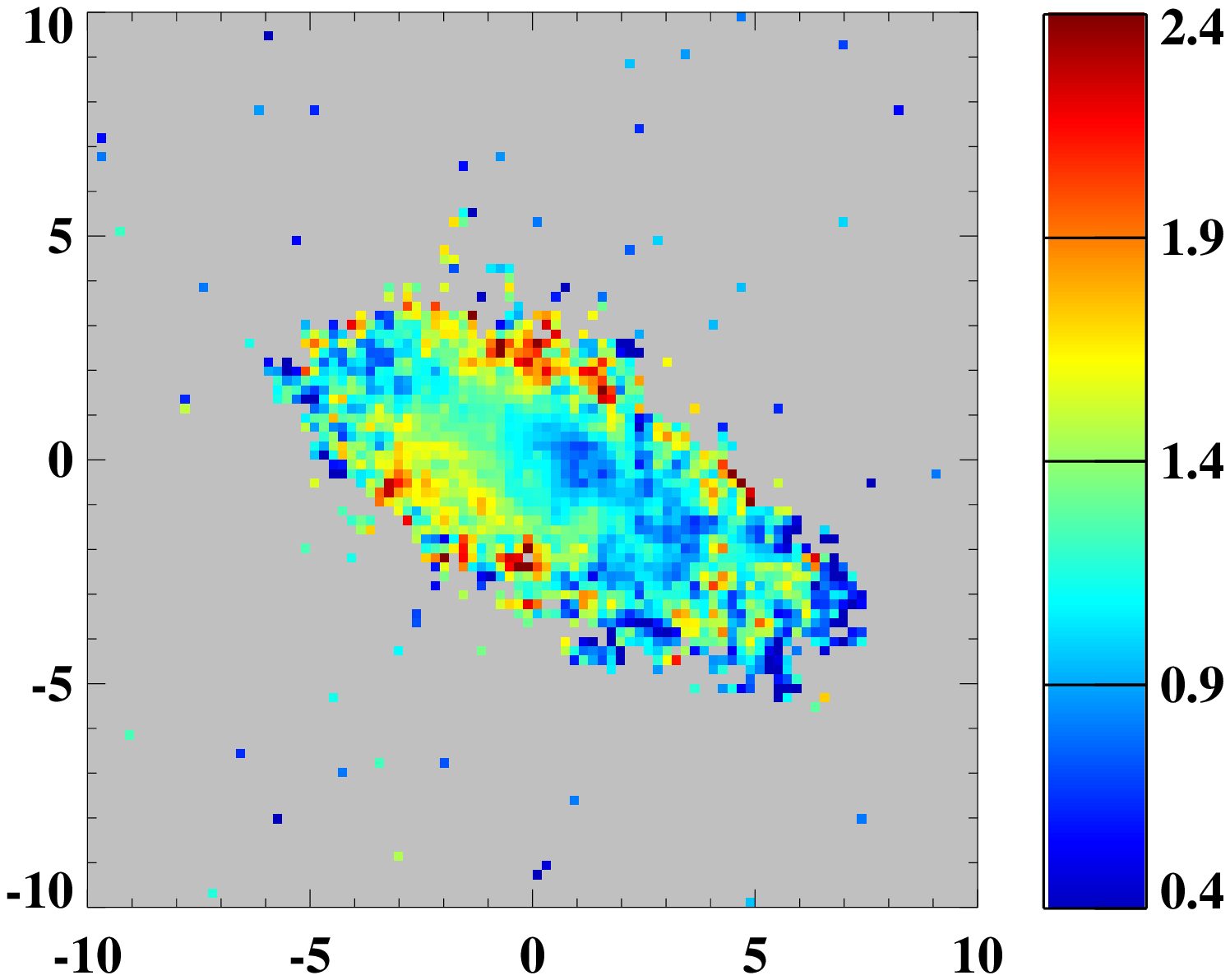} &
 \includegraphics[height=2.6cm]{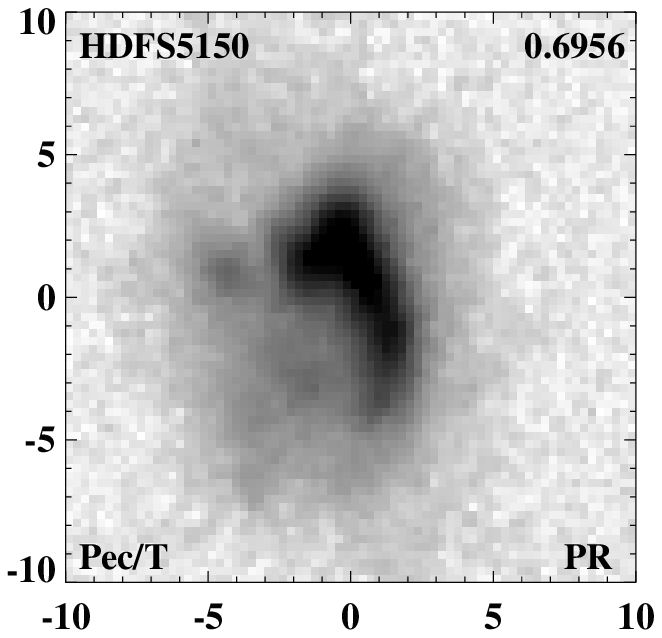} &
 \includegraphics[height=2.6cm]{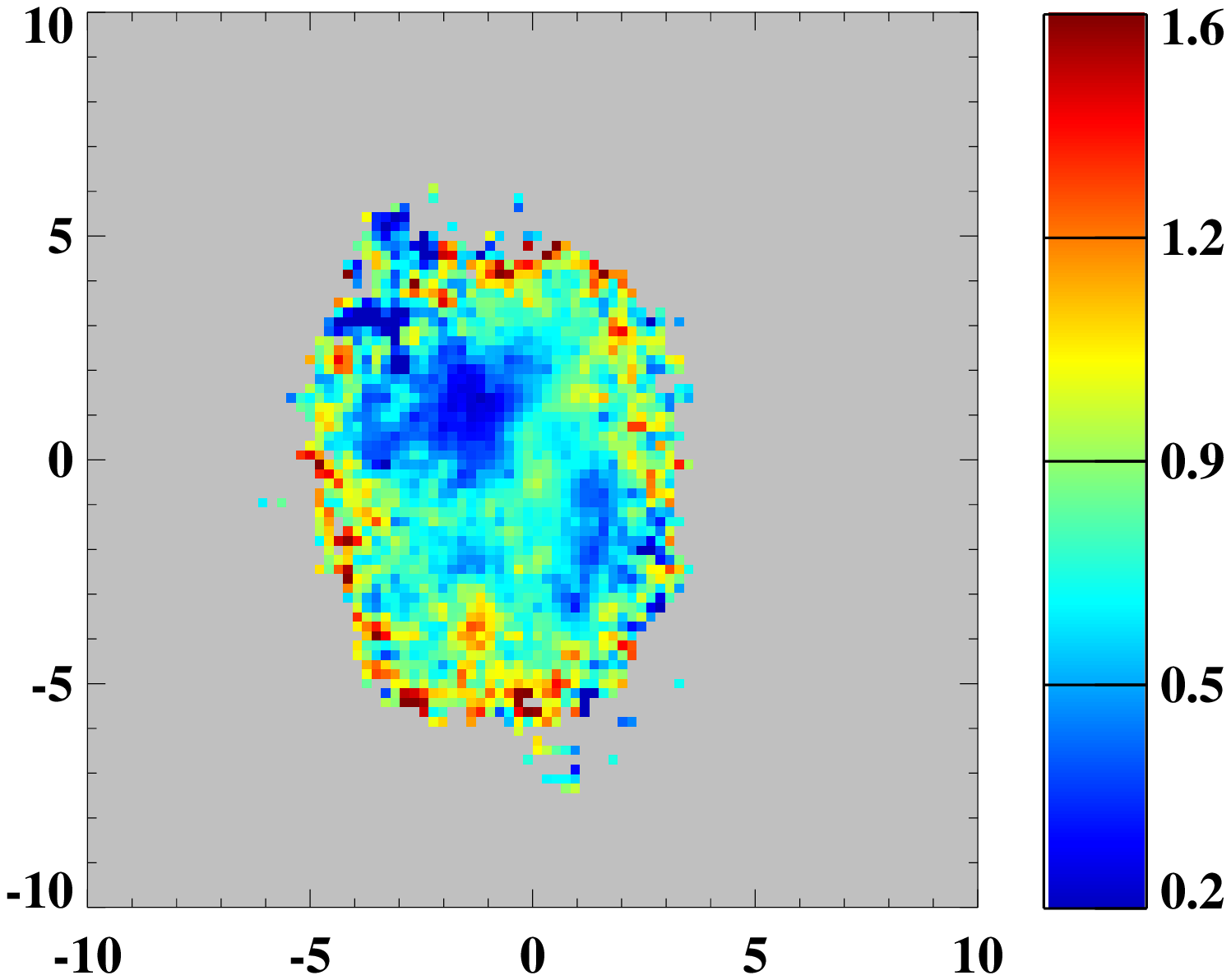} \\
 
    \includegraphics[height=2.6cm]{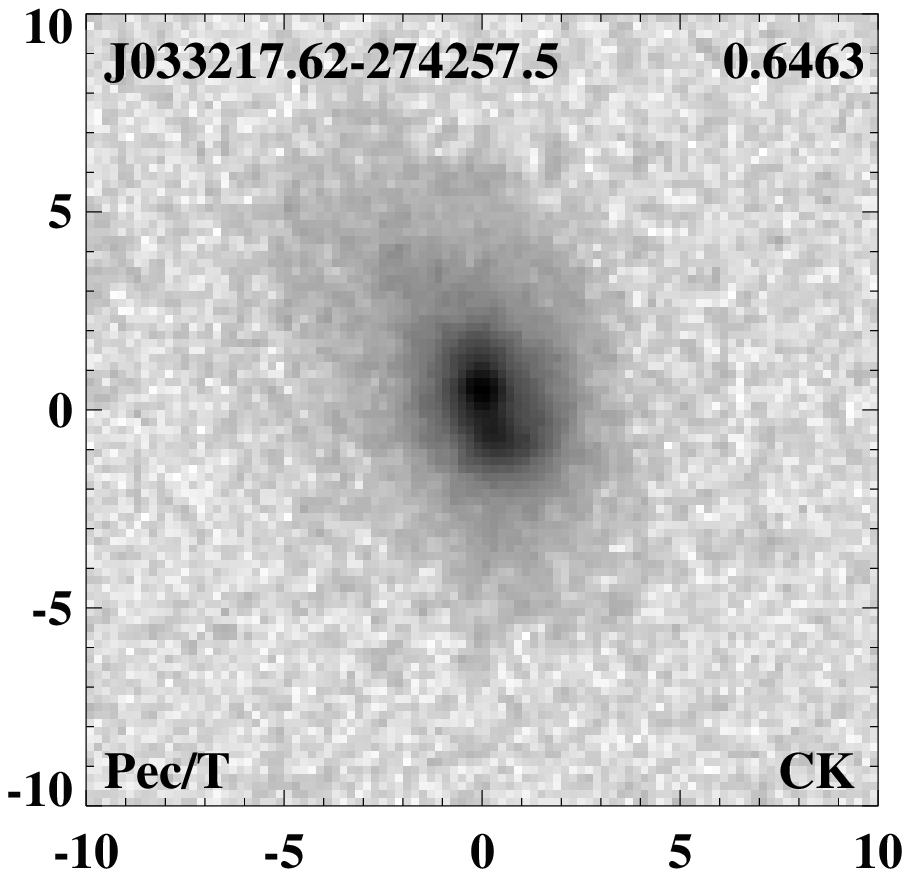} &
 \includegraphics[height=2.6cm]{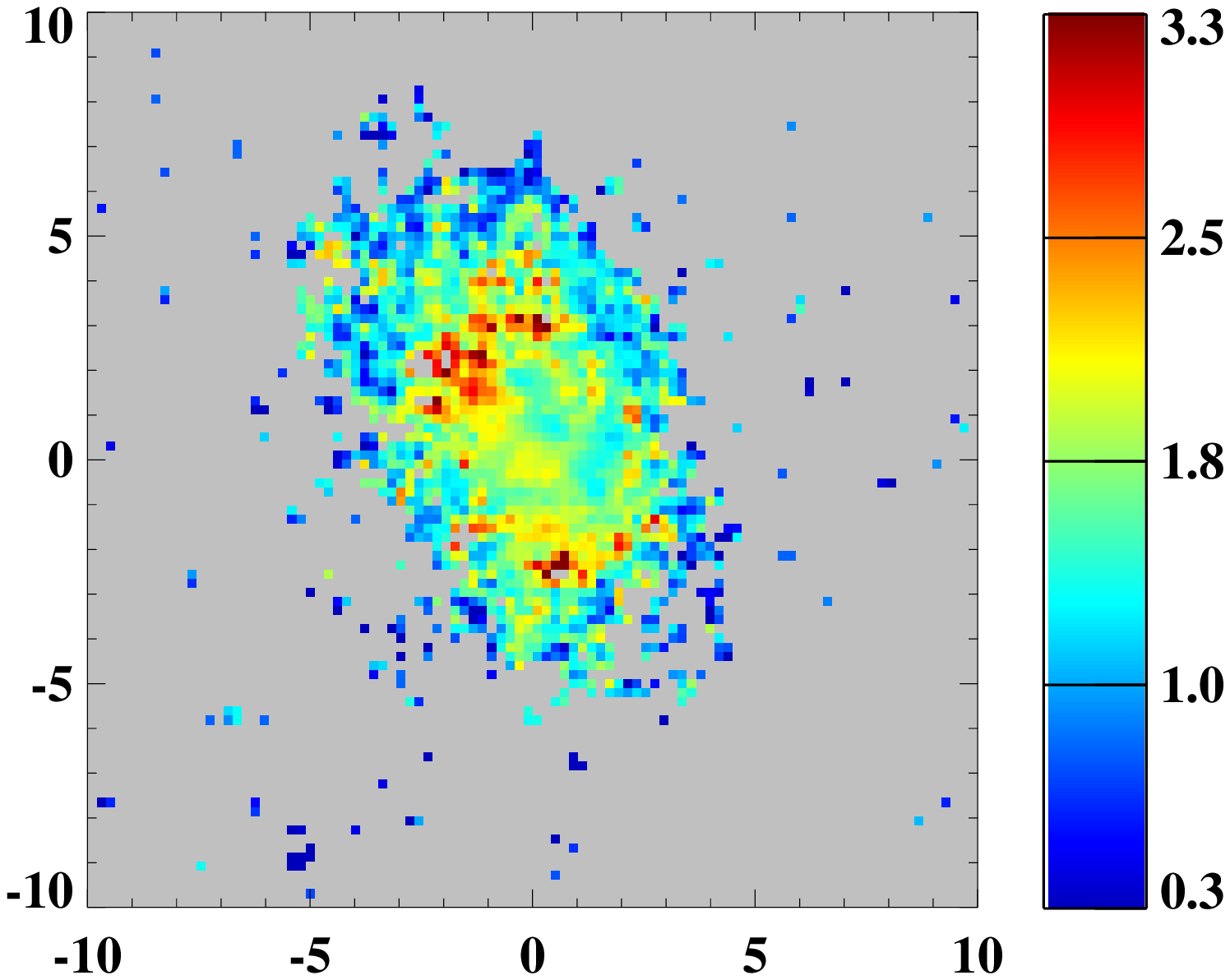} &
  \includegraphics[height=2.6cm]{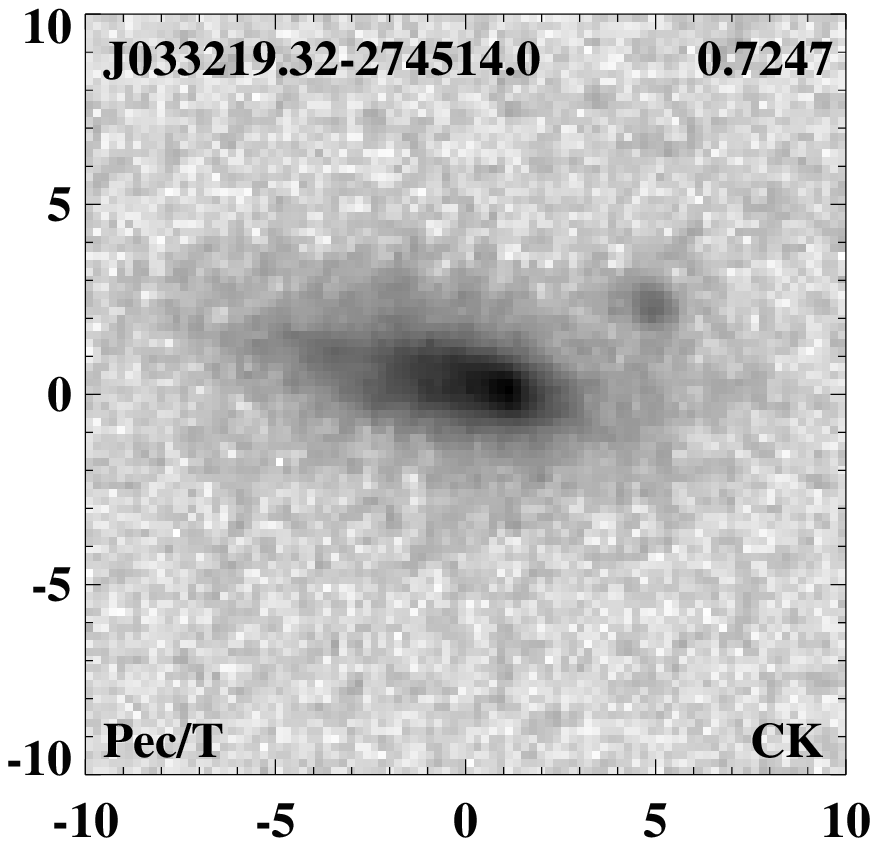} &
 \includegraphics[height=2.6cm]{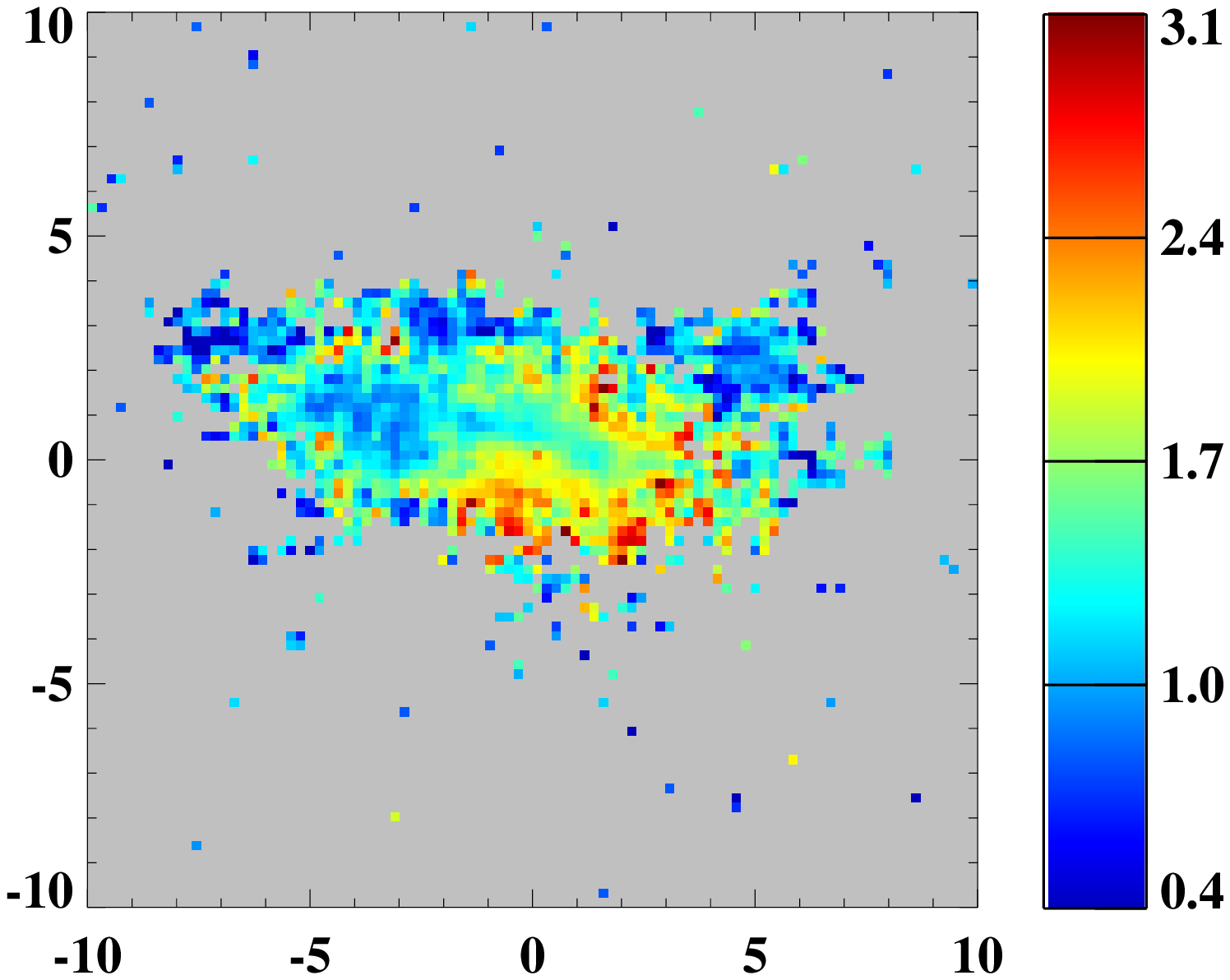} \\
 
   \includegraphics[height=2.6cm]{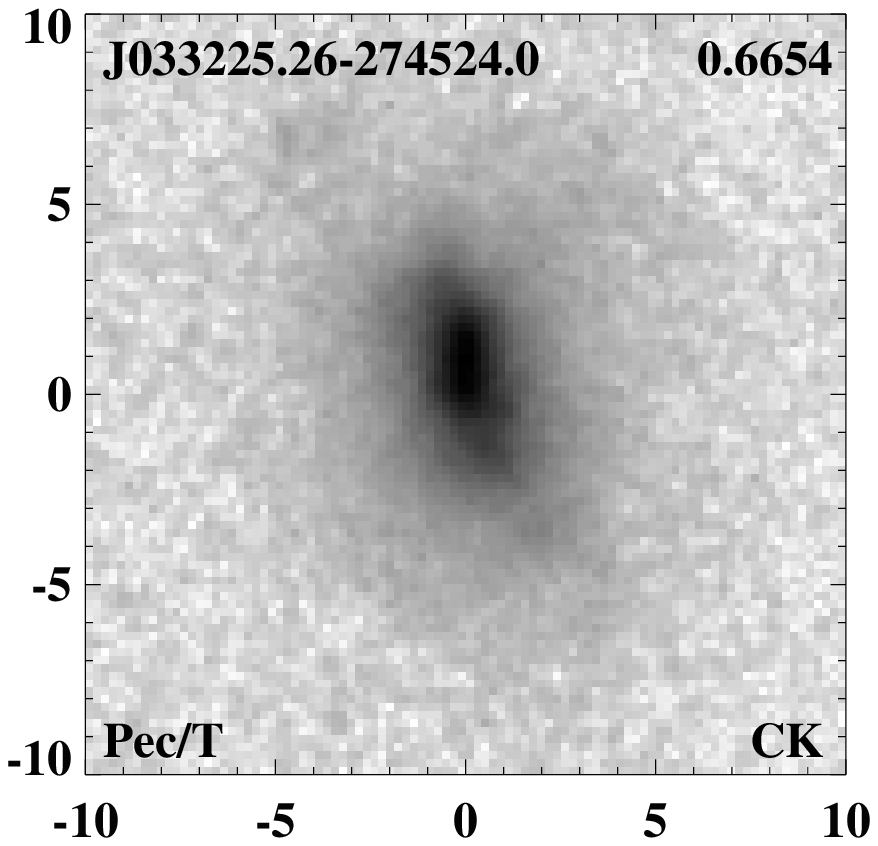} &
 \includegraphics[height=2.6cm]{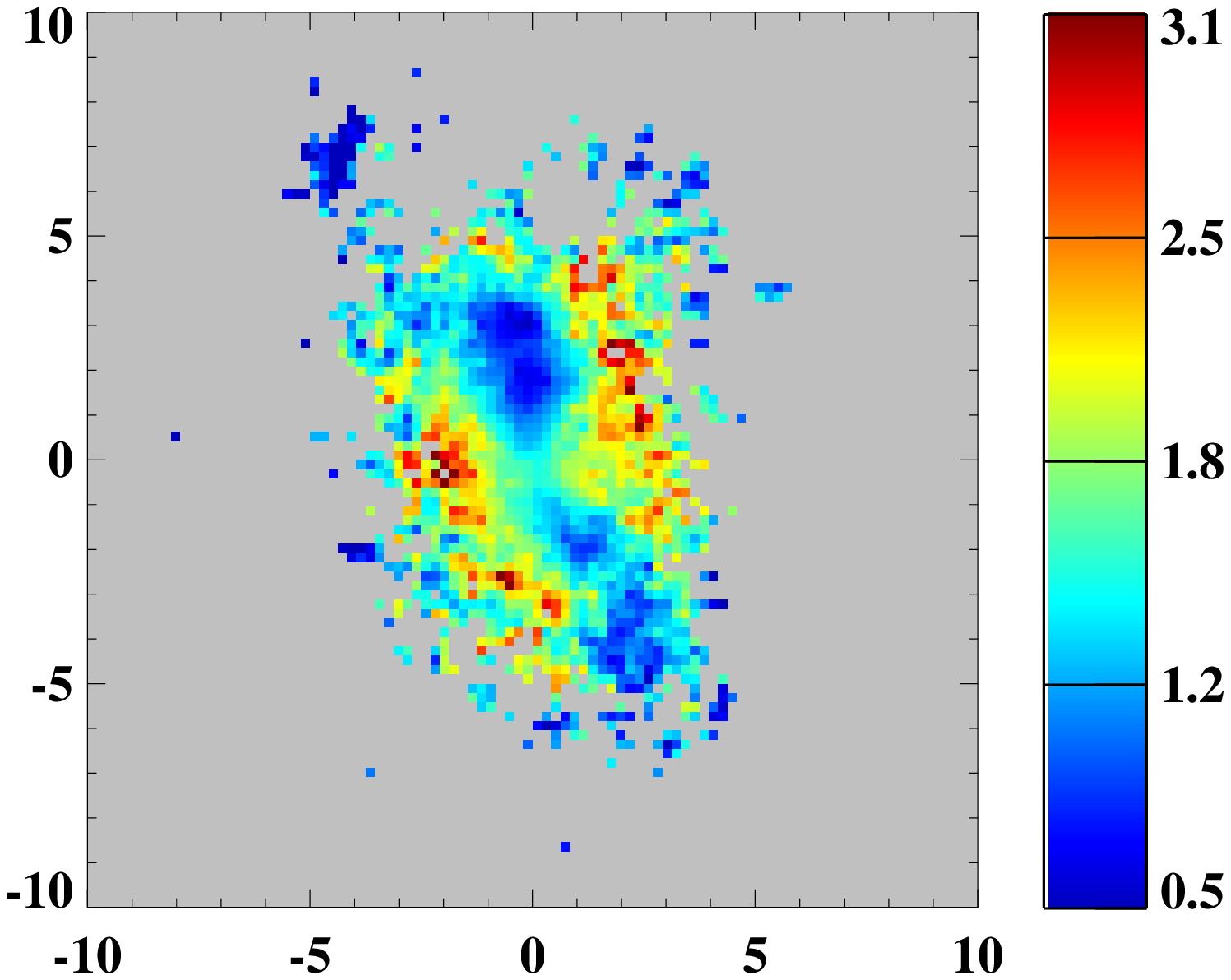} &
 \includegraphics[height=2.6cm]{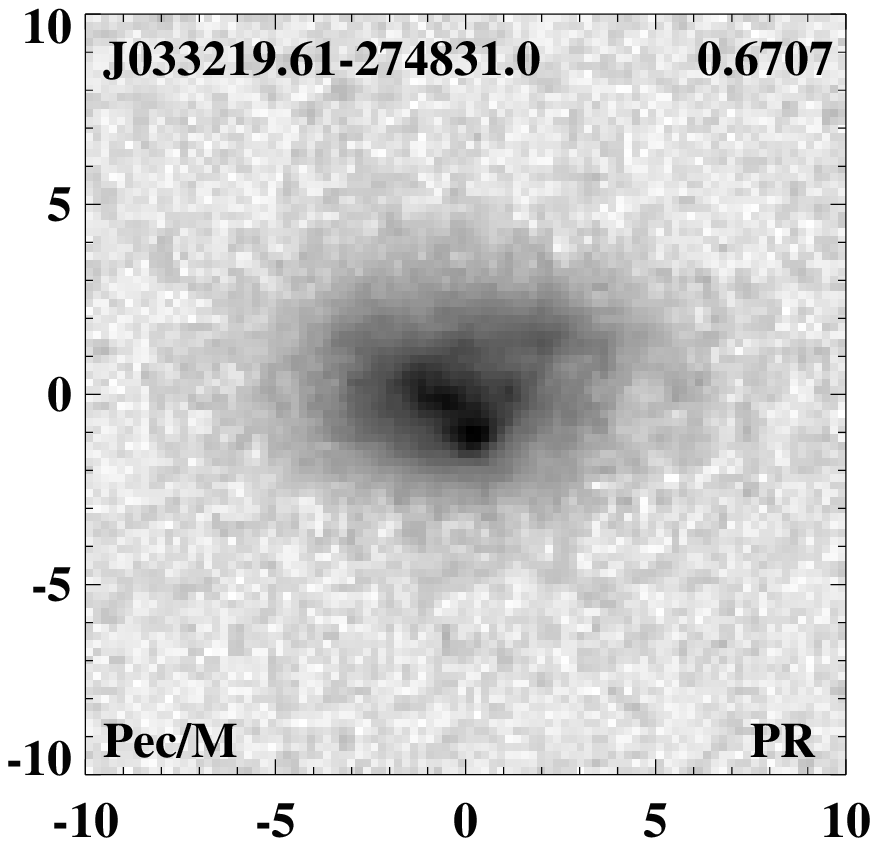} &
 \includegraphics[height=2.6cm]{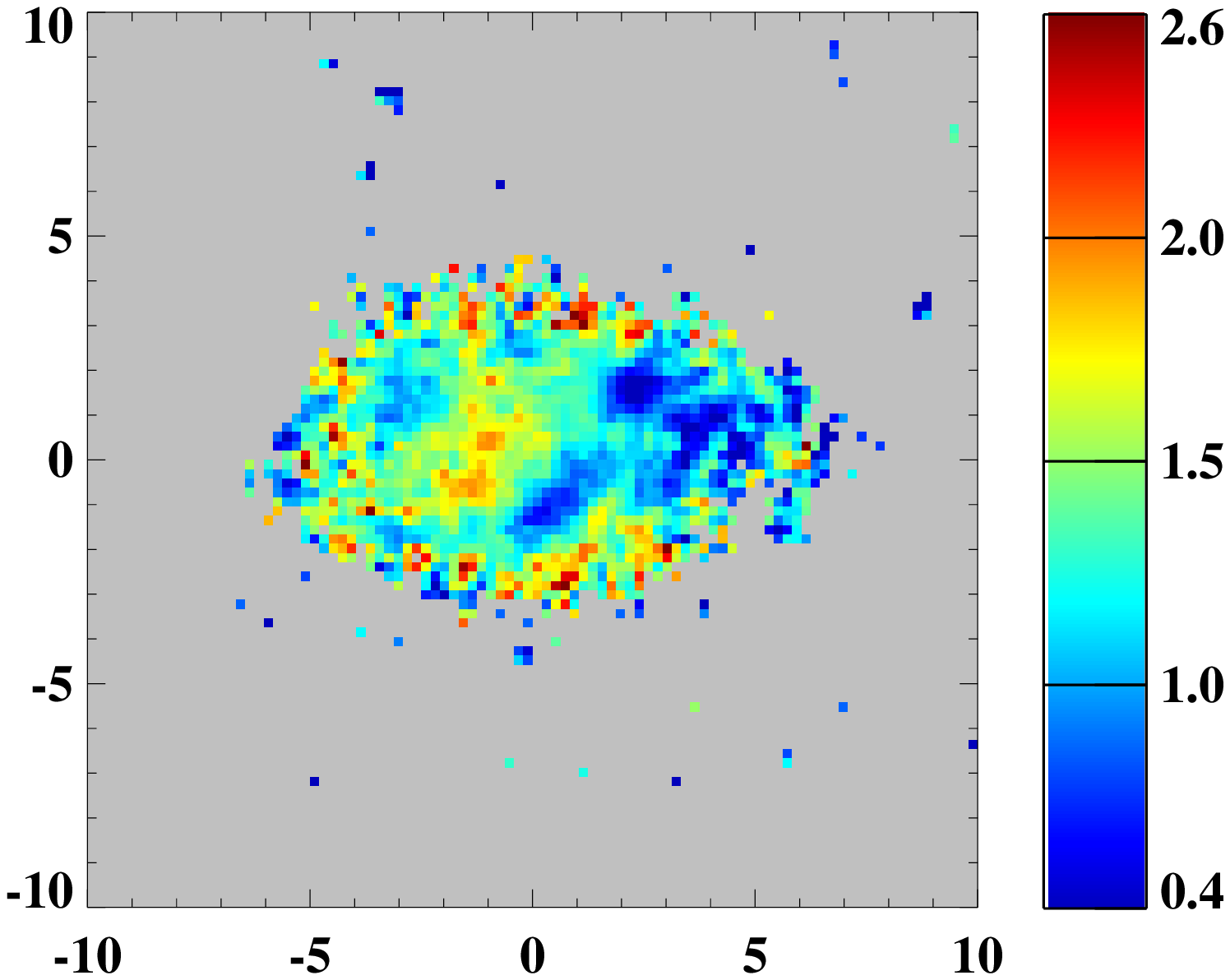} \\
 
 \includegraphics[height=2.6cm]{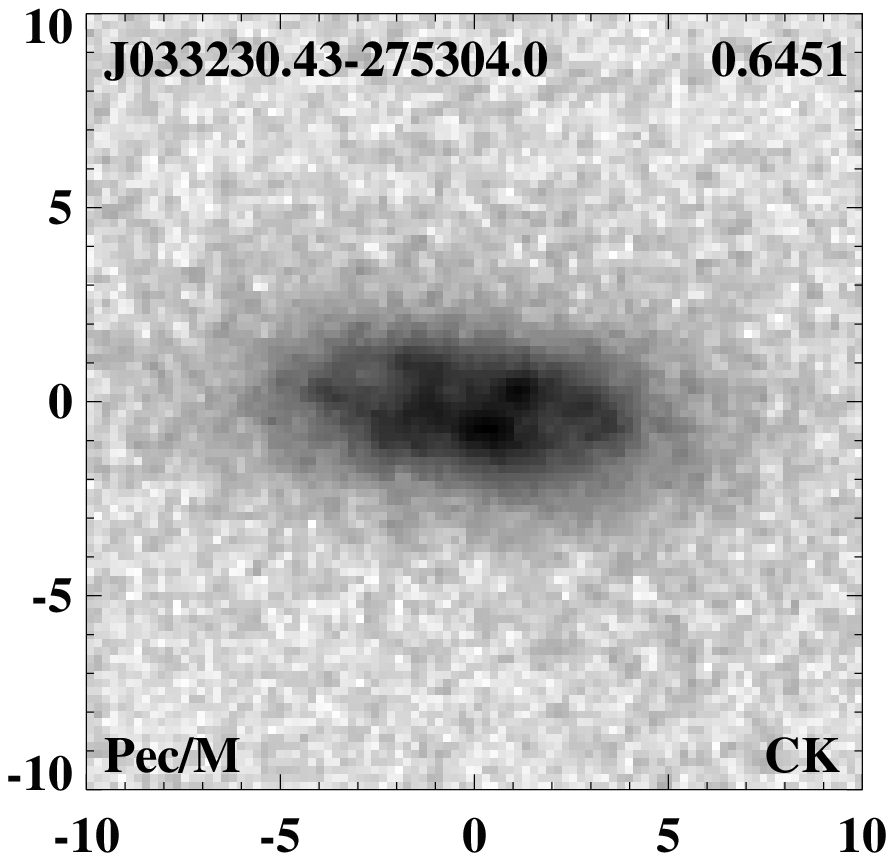} &
 \includegraphics[height=2.6cm]{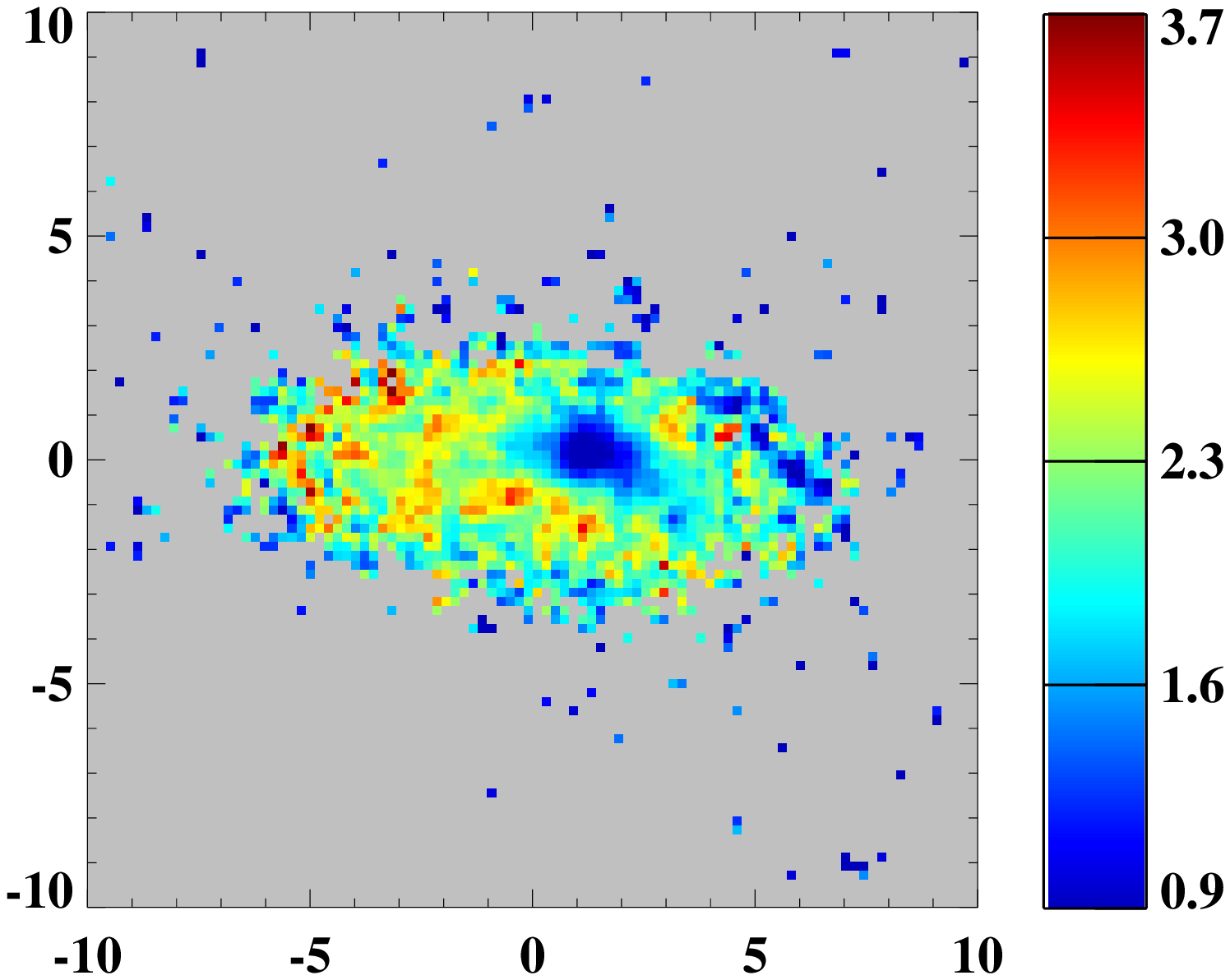} &
    \includegraphics[height=2.6cm]{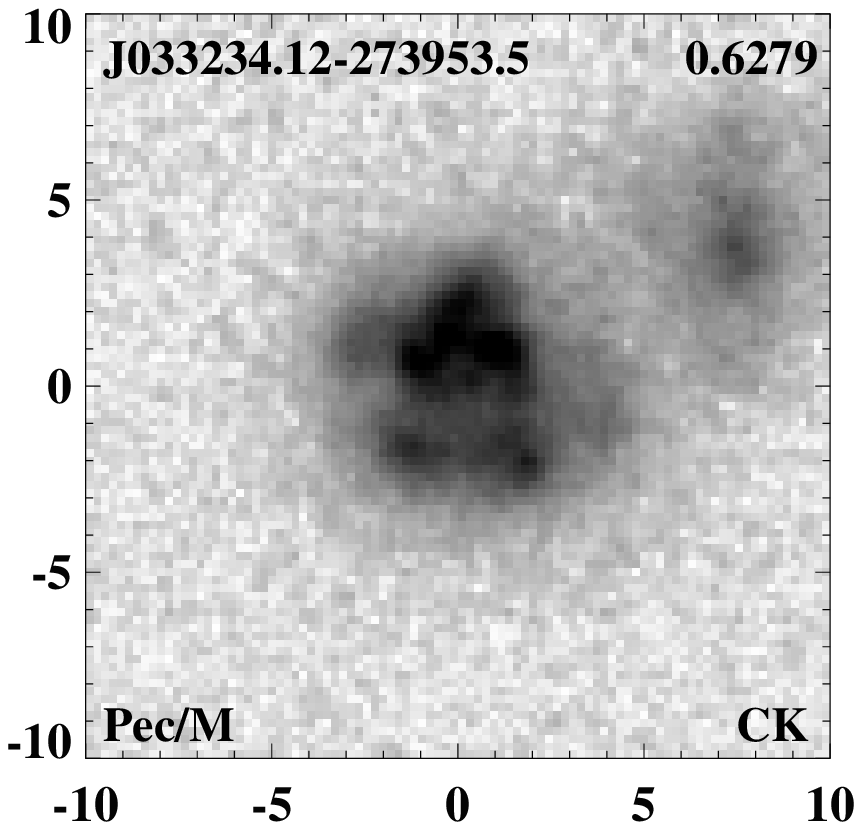} &
 \includegraphics[height=2.6cm]{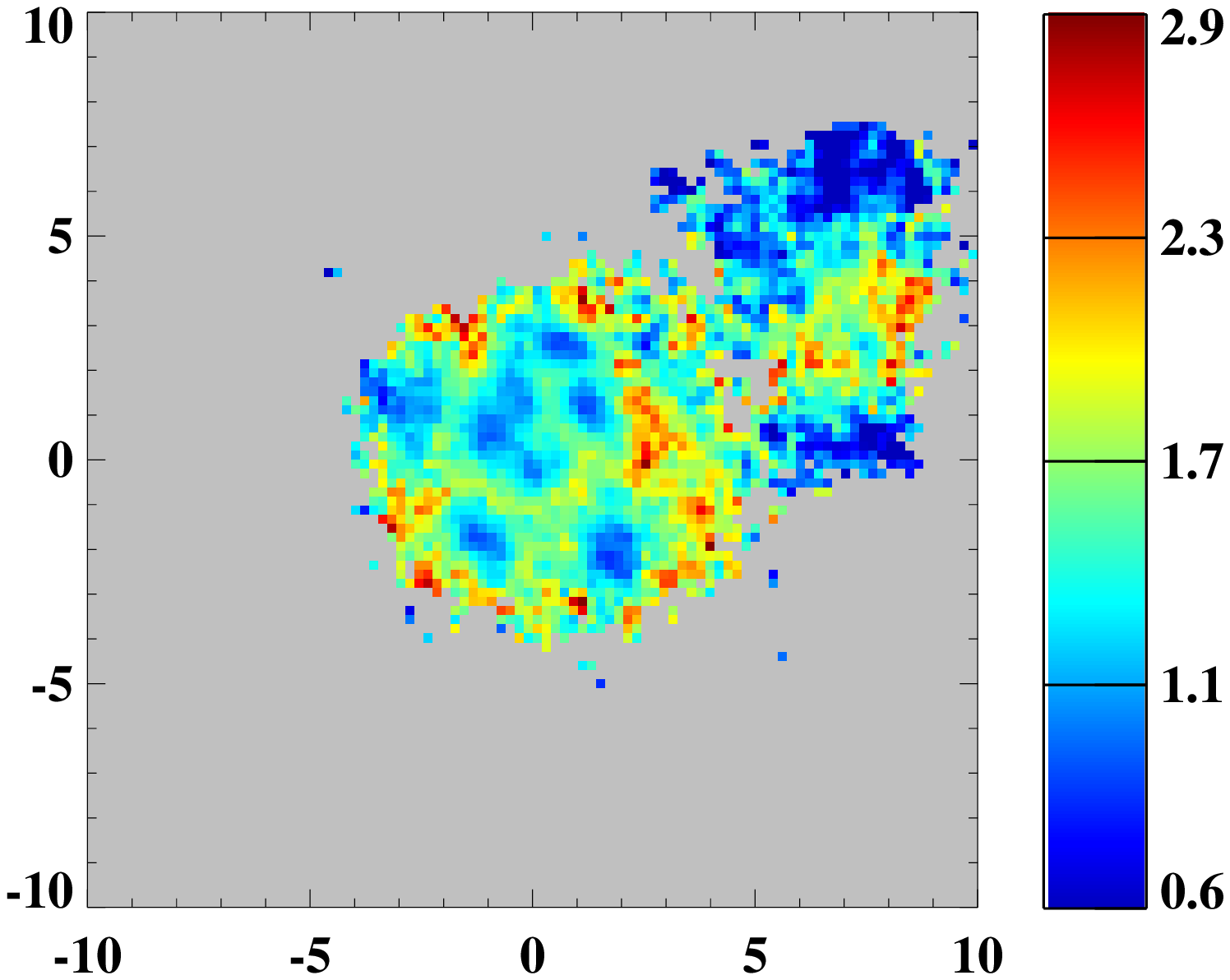} \\
 
     & & & \\
  \multicolumn{4}{l}{{\bf Fig. 12.}  F850LP/F814W image, color map. Explanations are given at the end of this figure.}
        \end{tabular}
   \end{center}
   \label{fig1} 
   \end{figure} 

  \newpage
  
   \begin{figure}[h!]
   \begin{center}
   \begin{tabular}{cccc}      
   
   \includegraphics[height=2.6cm]{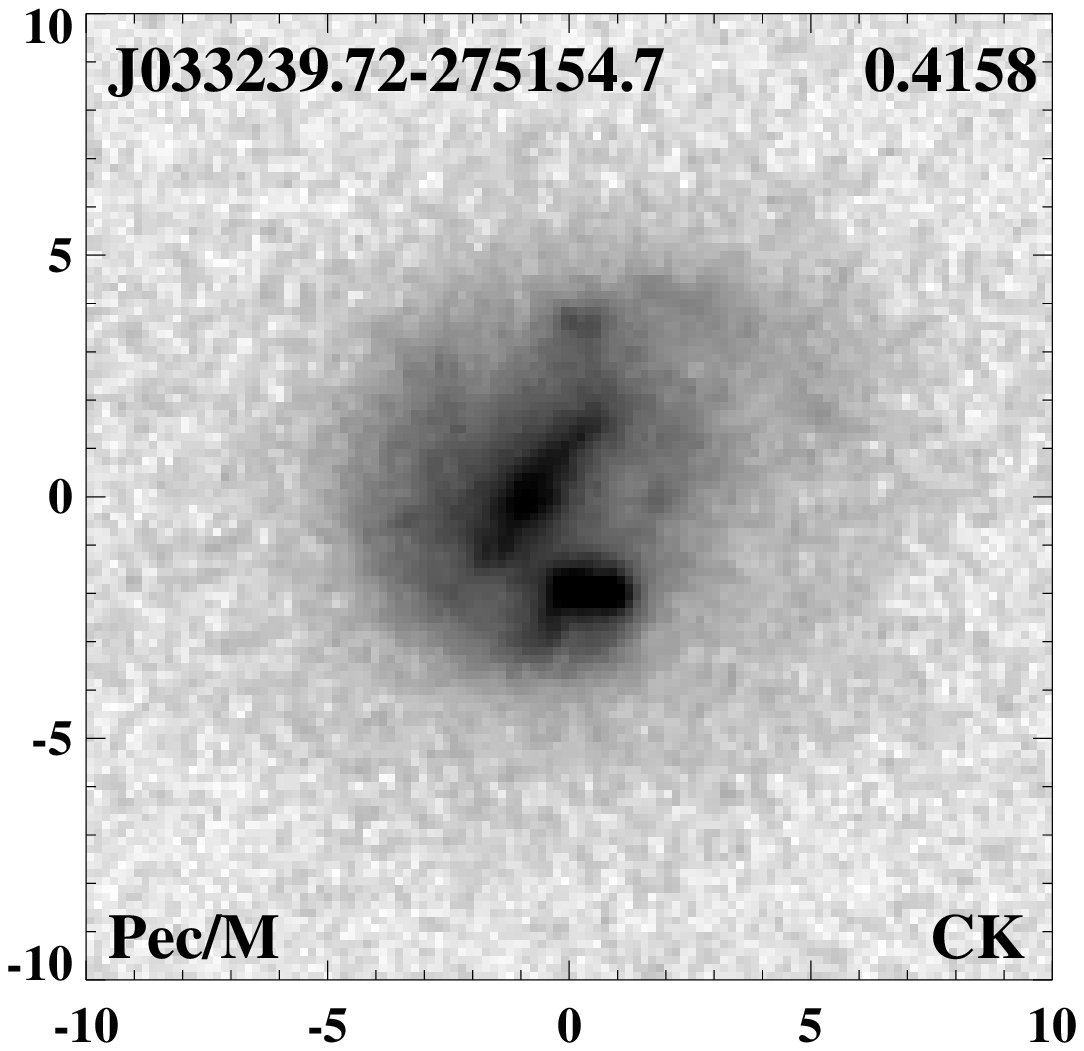} &
 \includegraphics[height=2.6cm]{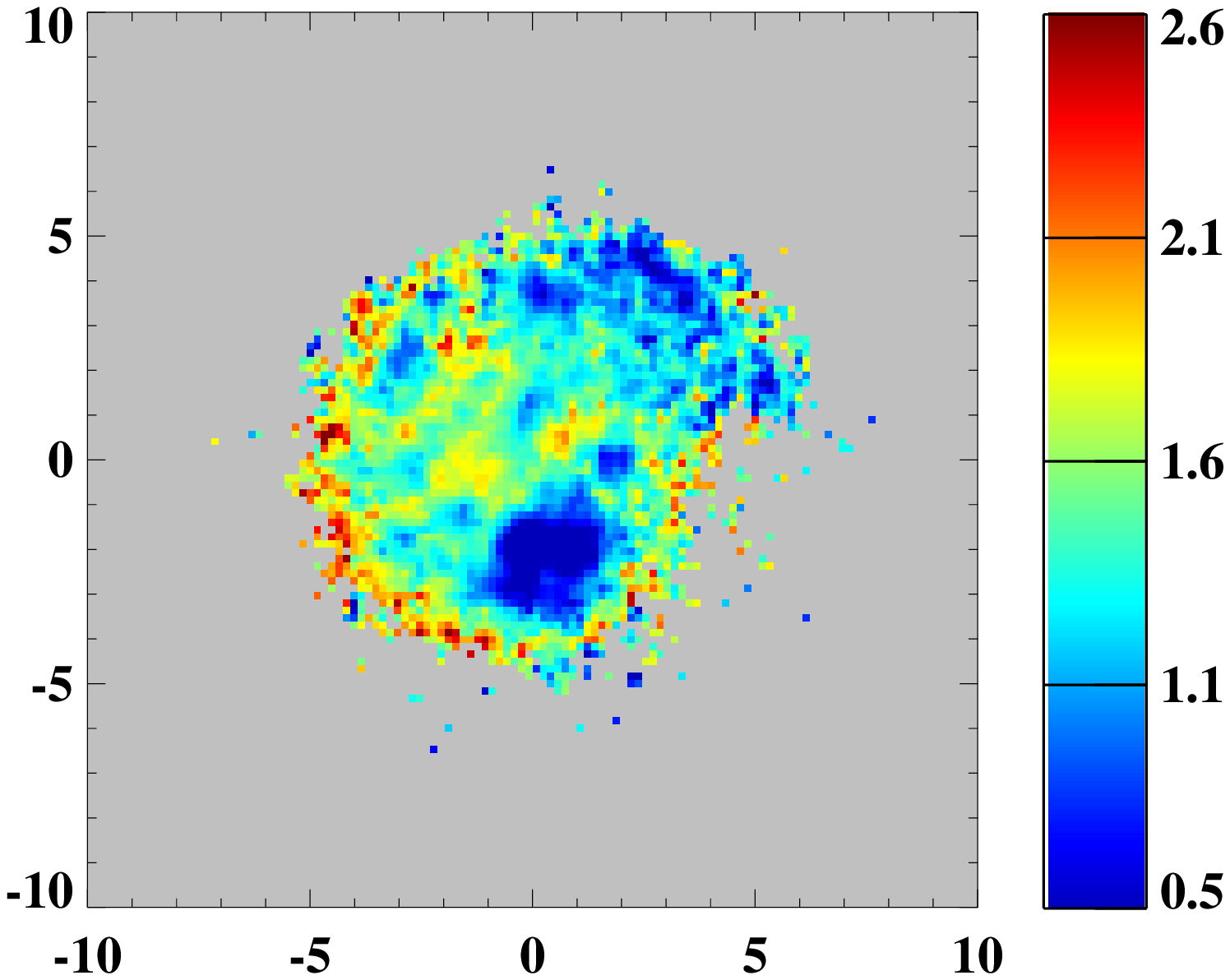} &
  \includegraphics[height=2.6cm]{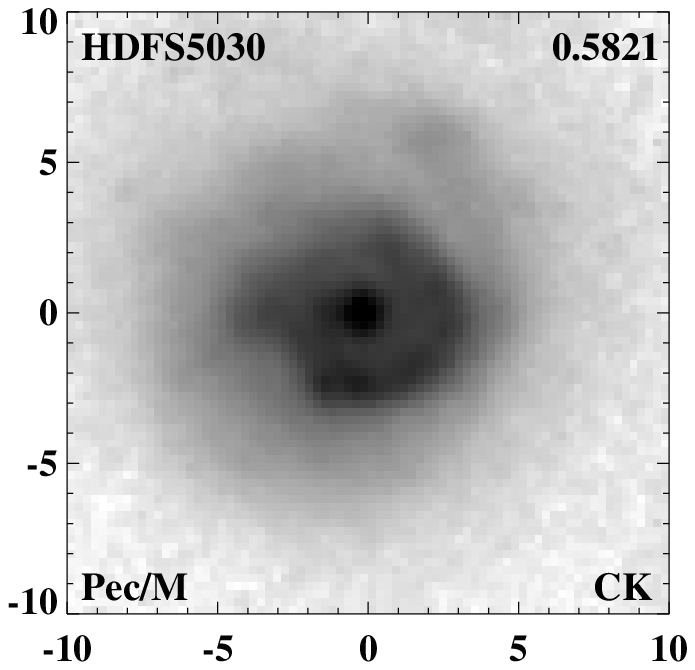} &
 \includegraphics[height=2.6cm]{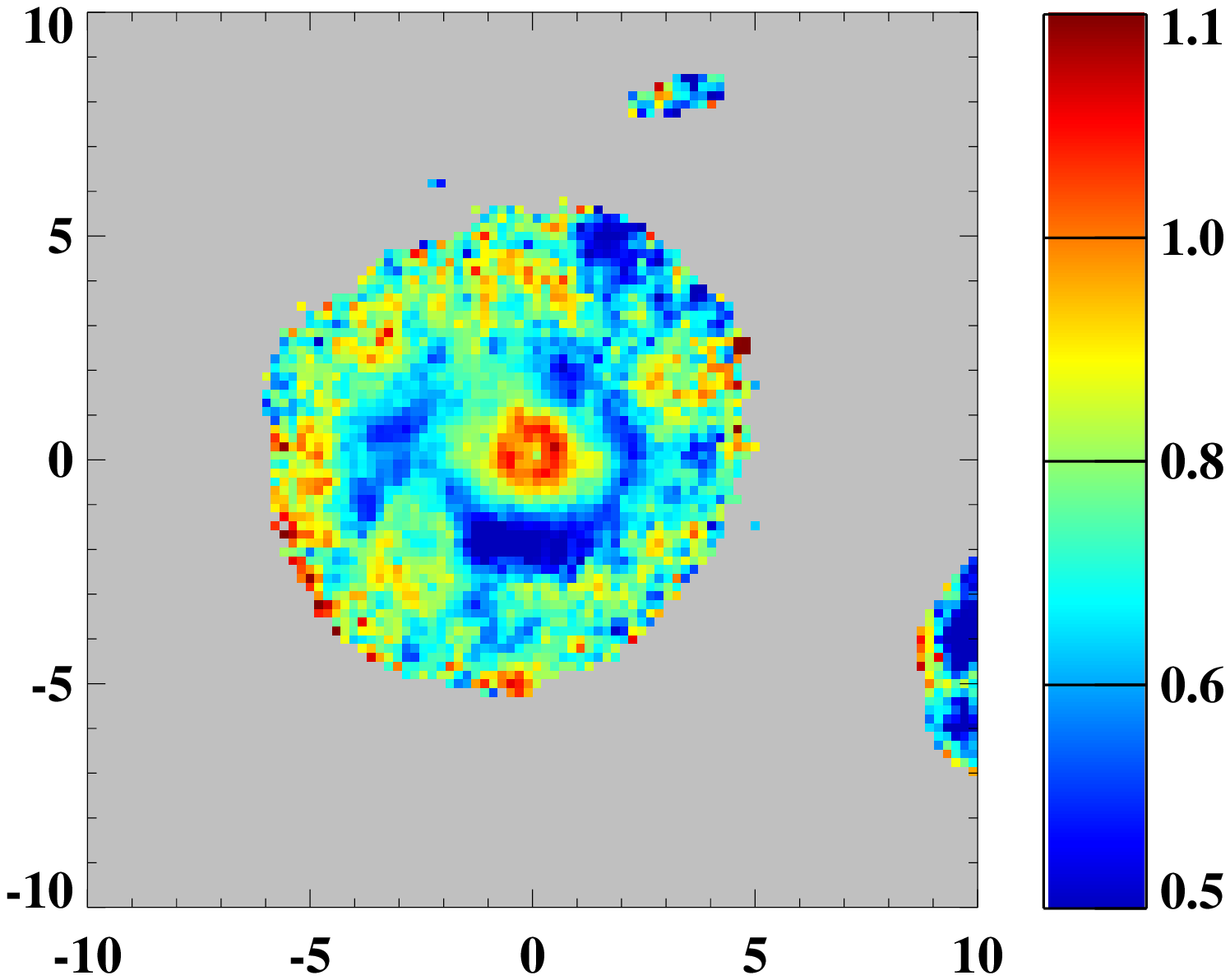} \\

      \includegraphics[height=2.6cm]{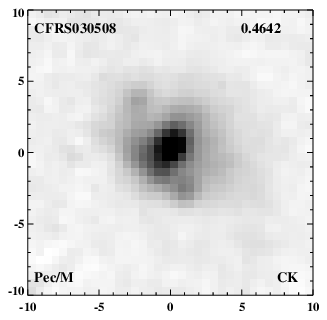} &
 \includegraphics[height=2.6cm]{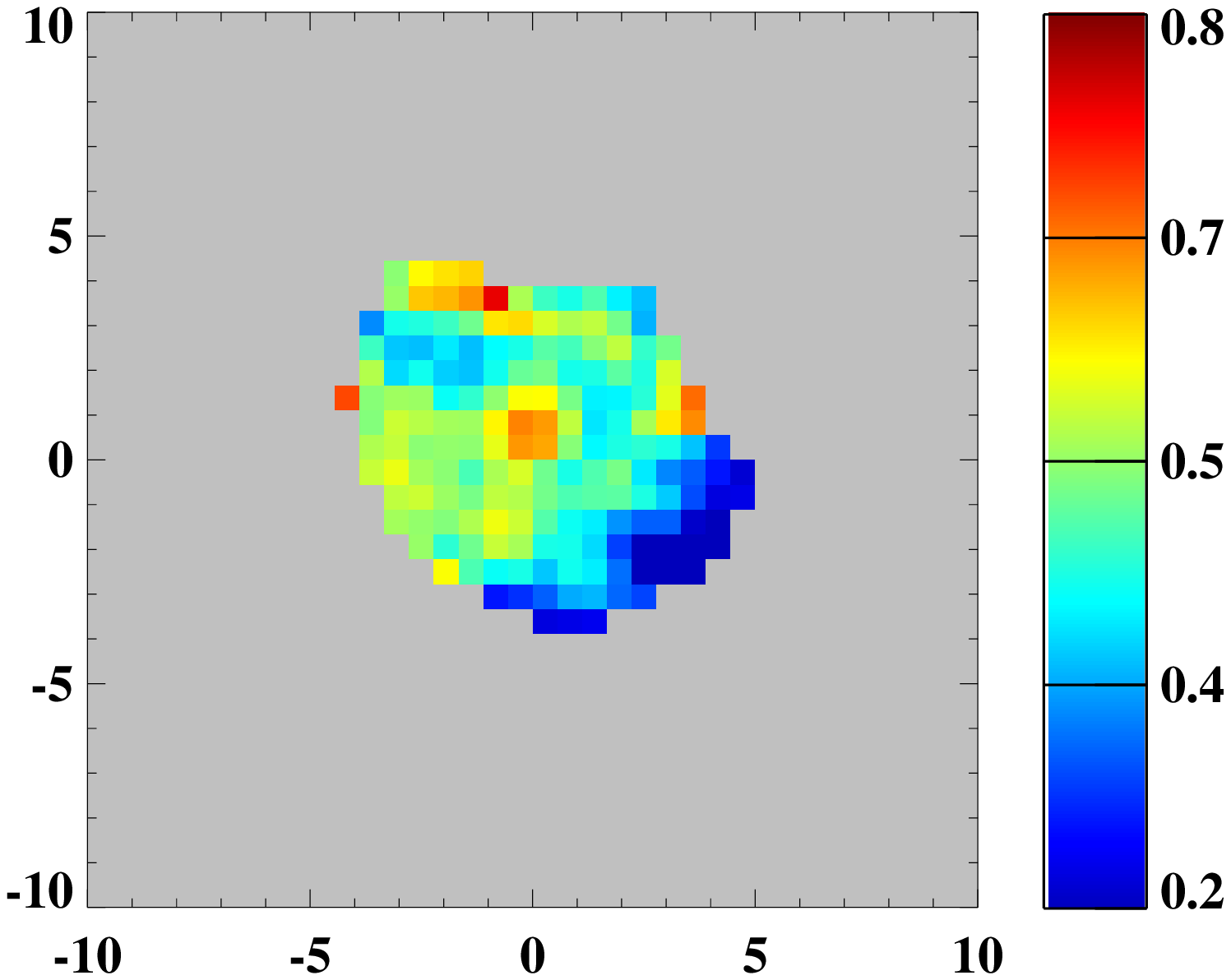} &
      \includegraphics[height=2.6cm]{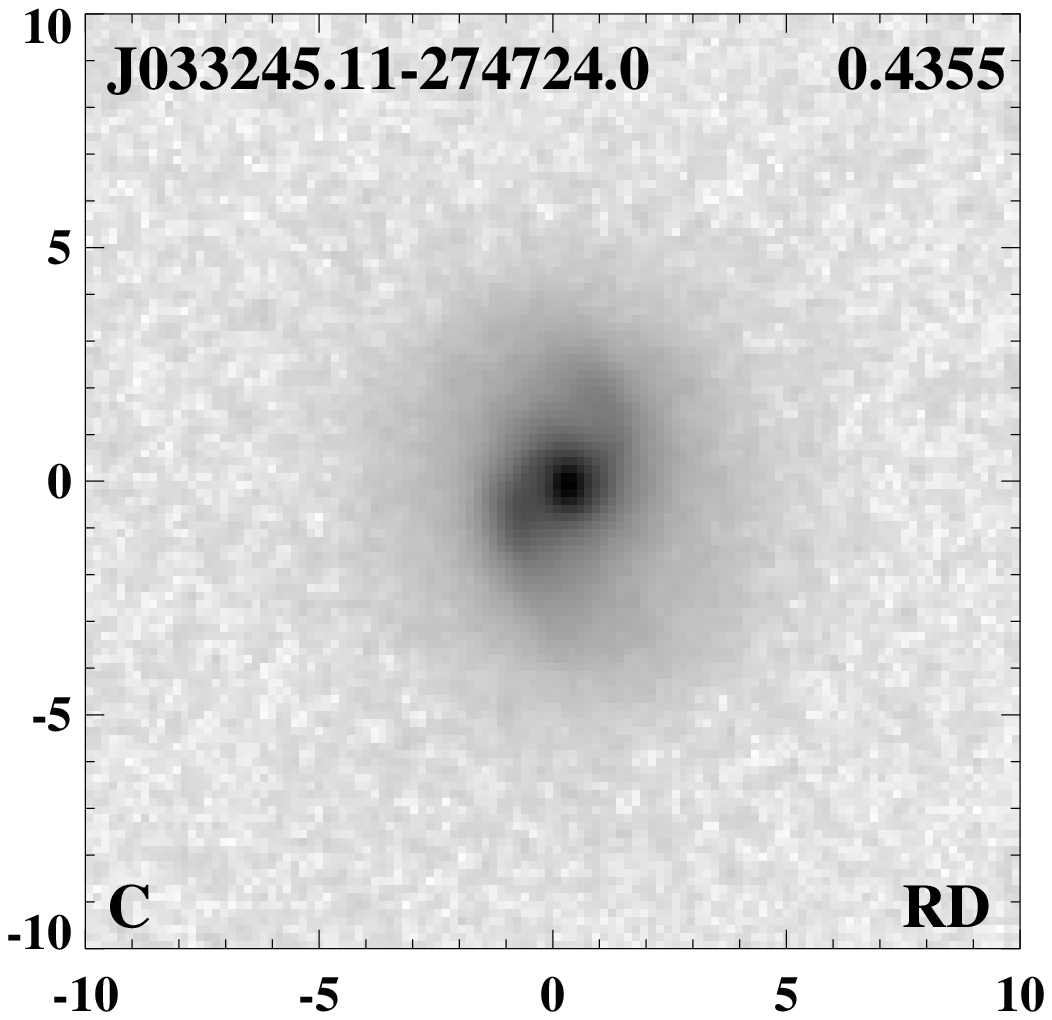} &
 \includegraphics[height=2.6cm]{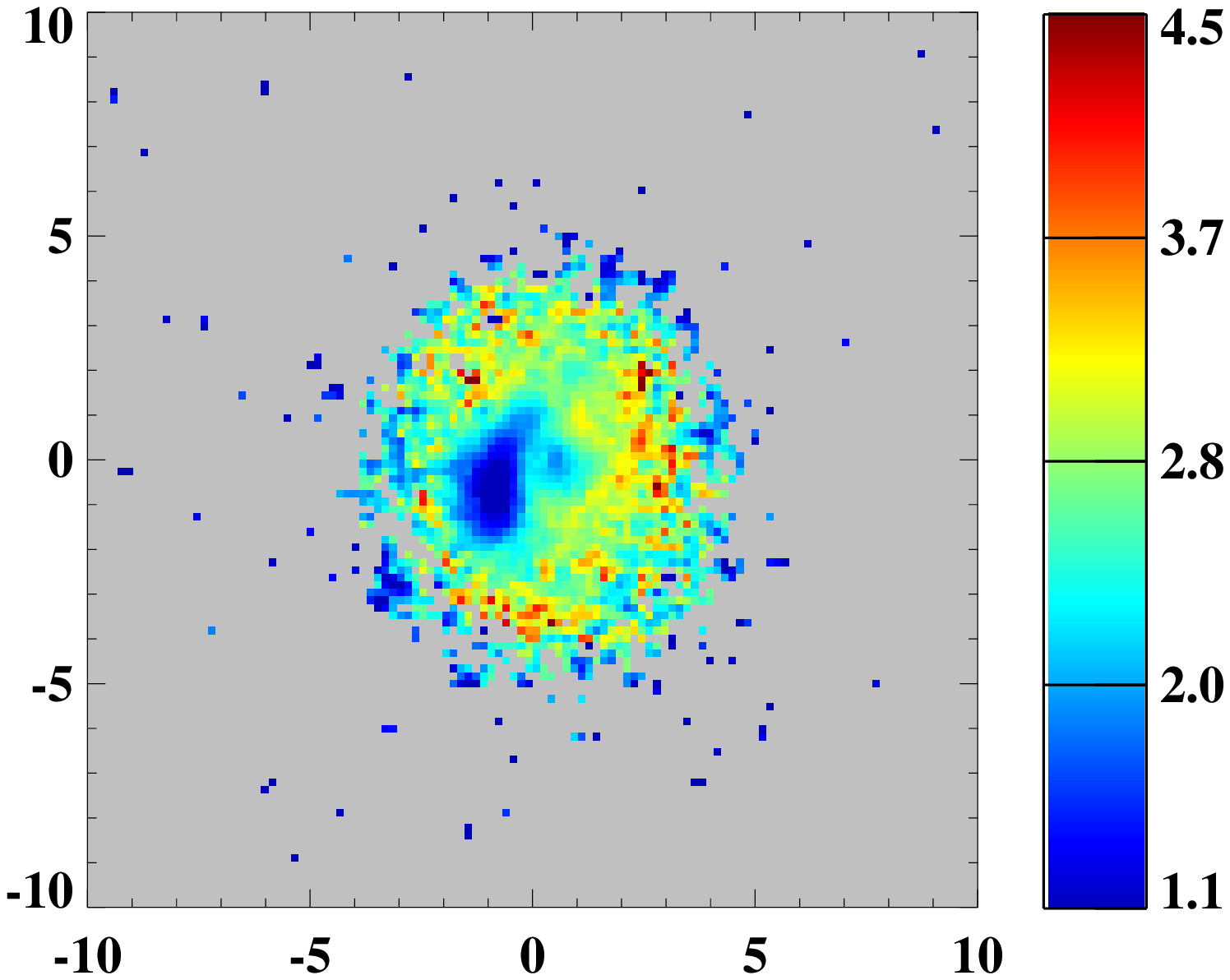} \\

  \includegraphics[height=2.6cm]{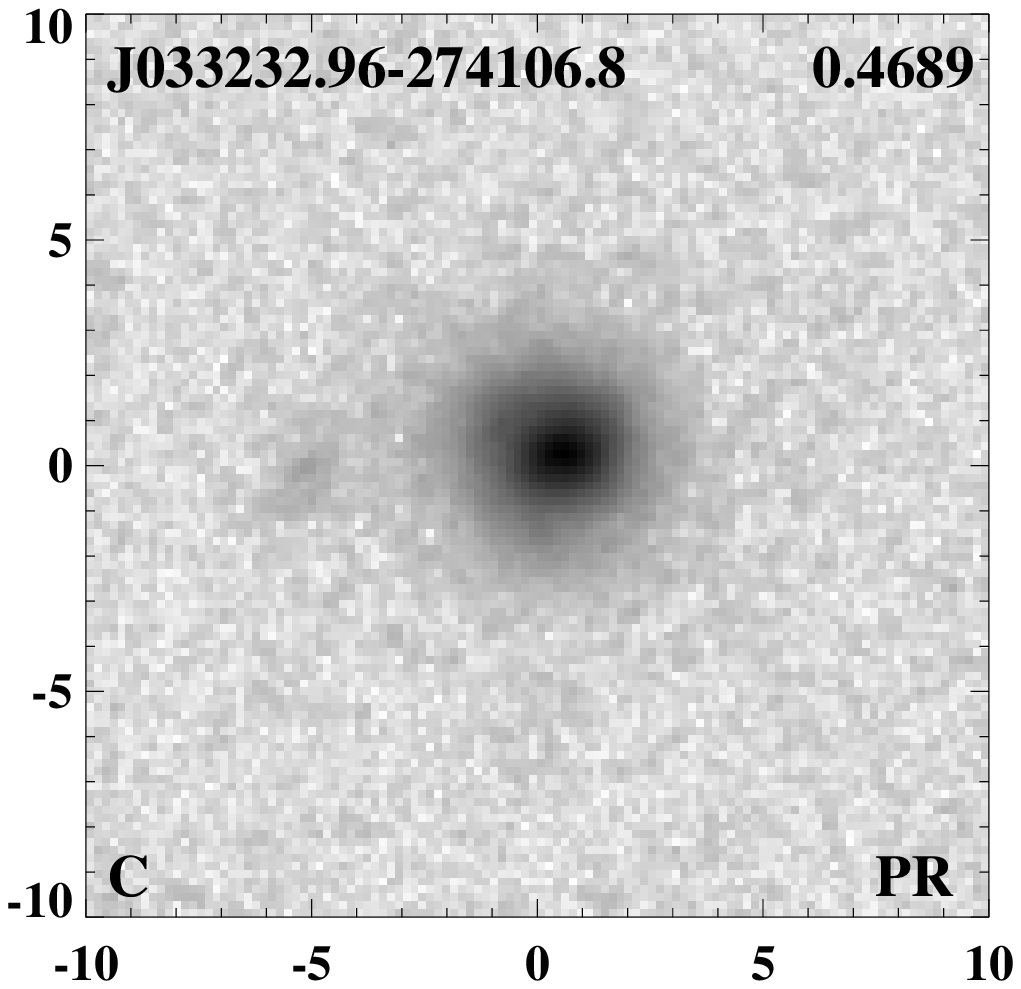} &
 \includegraphics[height=2.6cm]{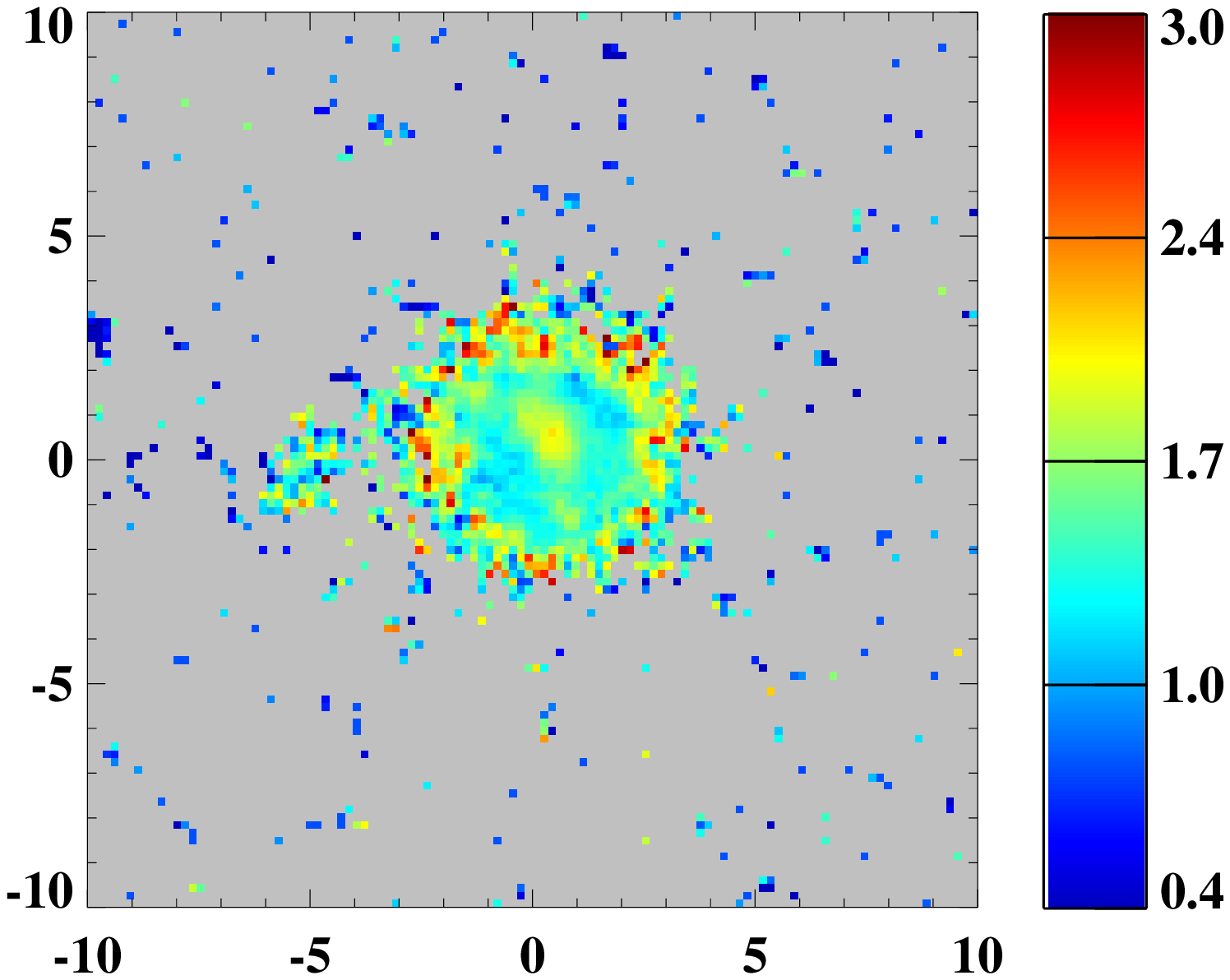} &   
  \includegraphics[height=2.6cm]{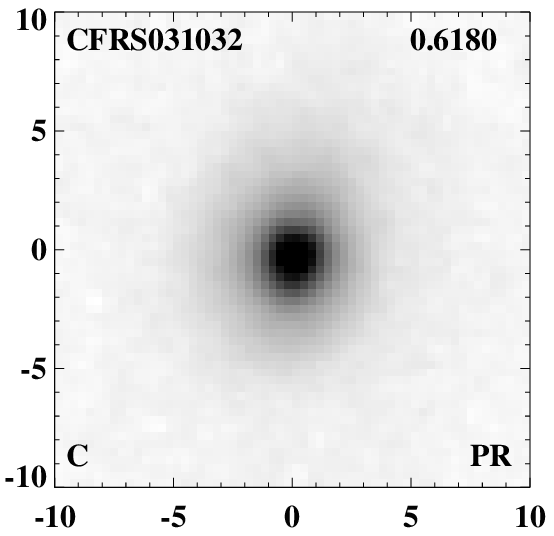} &
 \includegraphics[height=2.6cm]{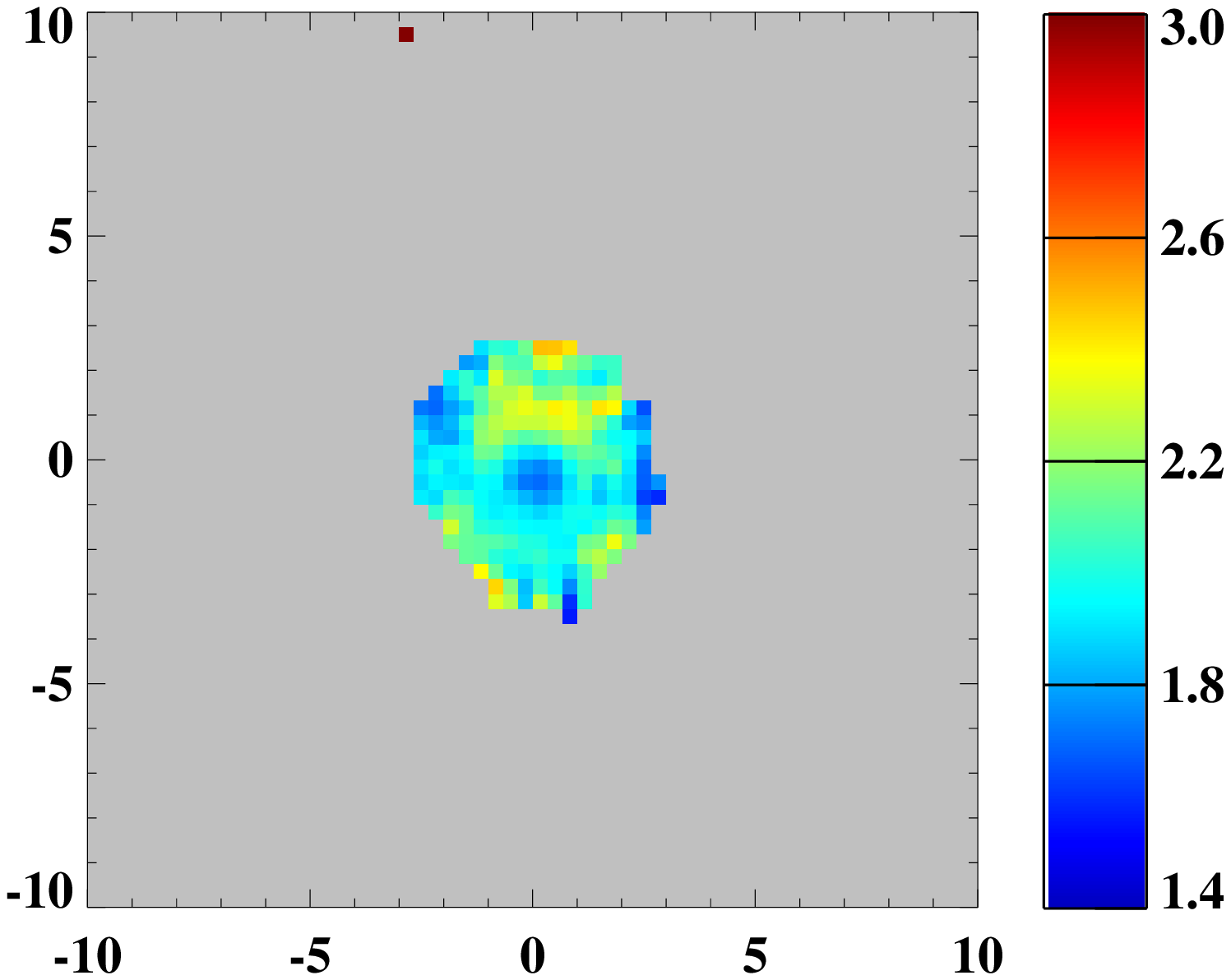} \\
  
  \includegraphics[height=2.6cm]{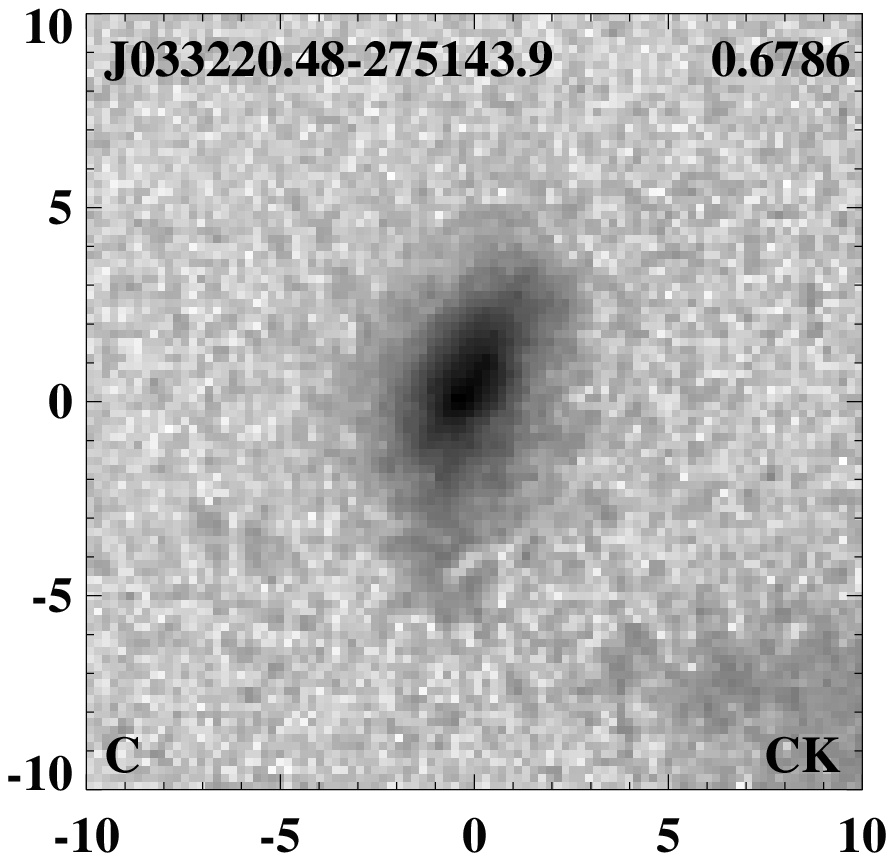} &
 \includegraphics[height=2.6cm]{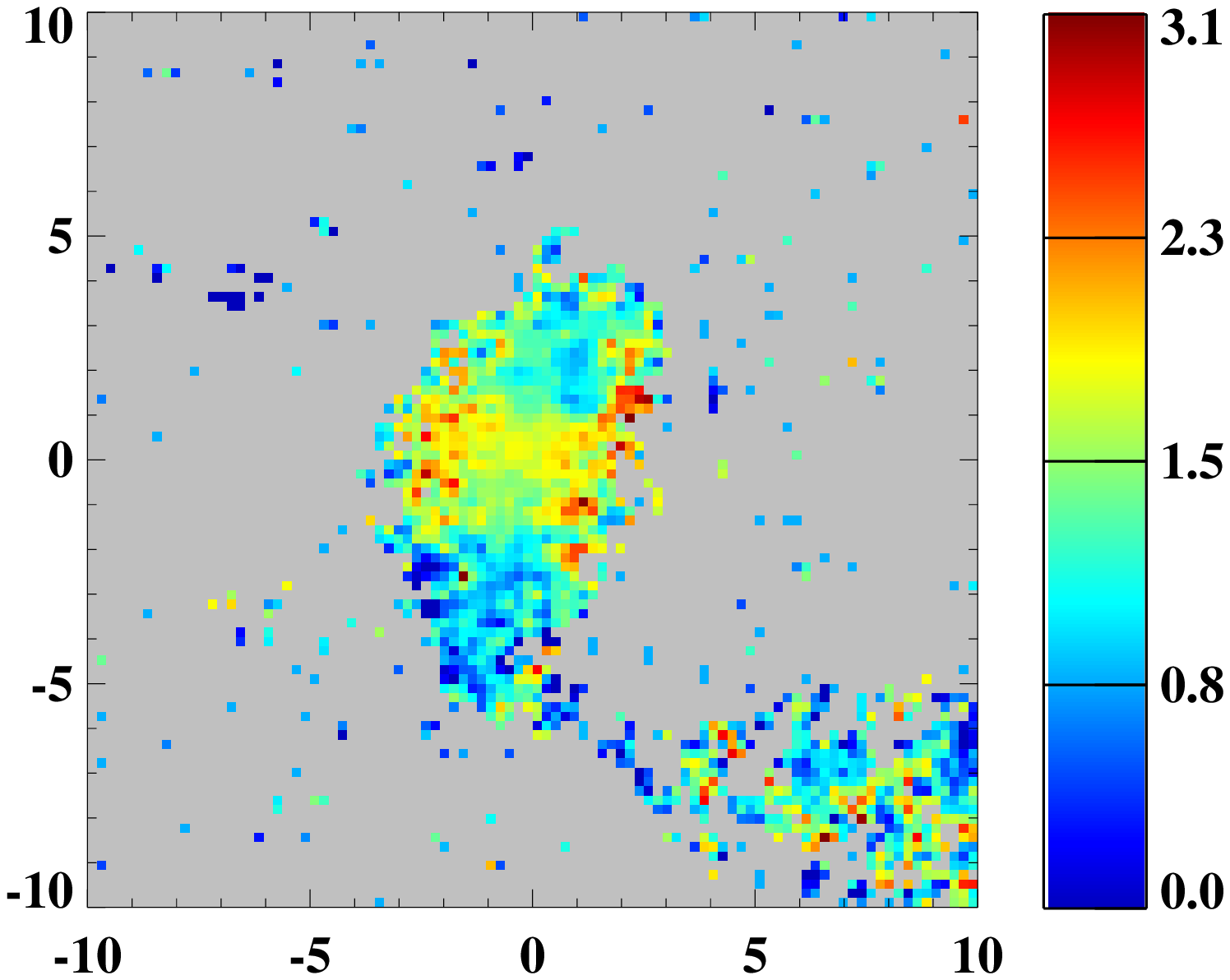} &
  \includegraphics[height=2.6cm]{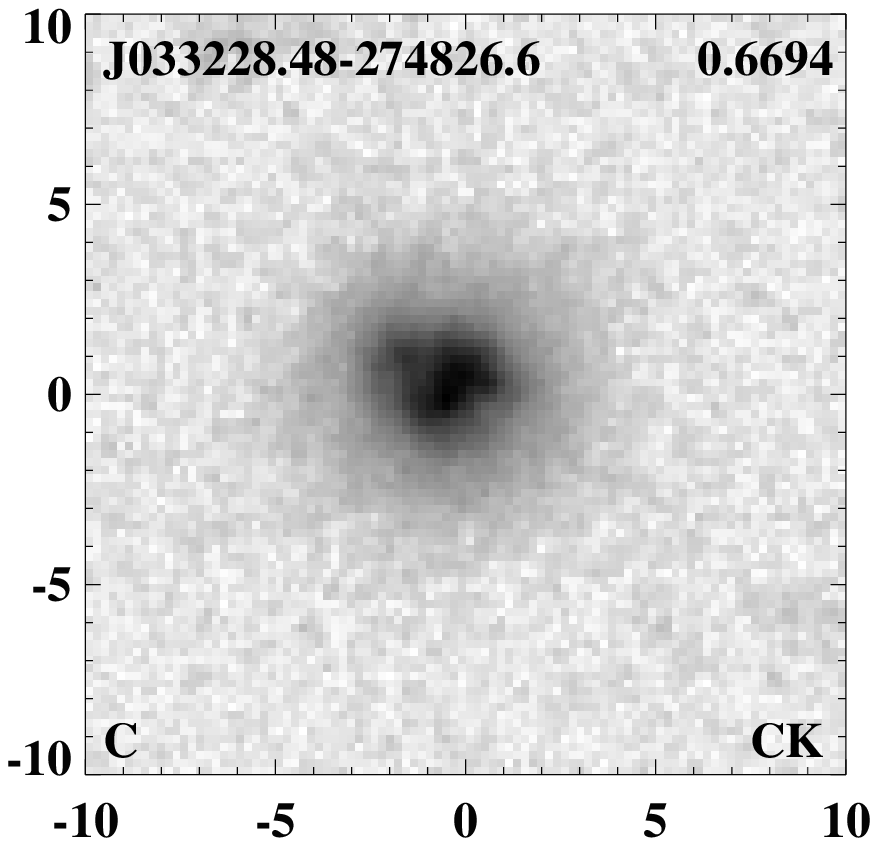} &
 \includegraphics[height=2.6cm]{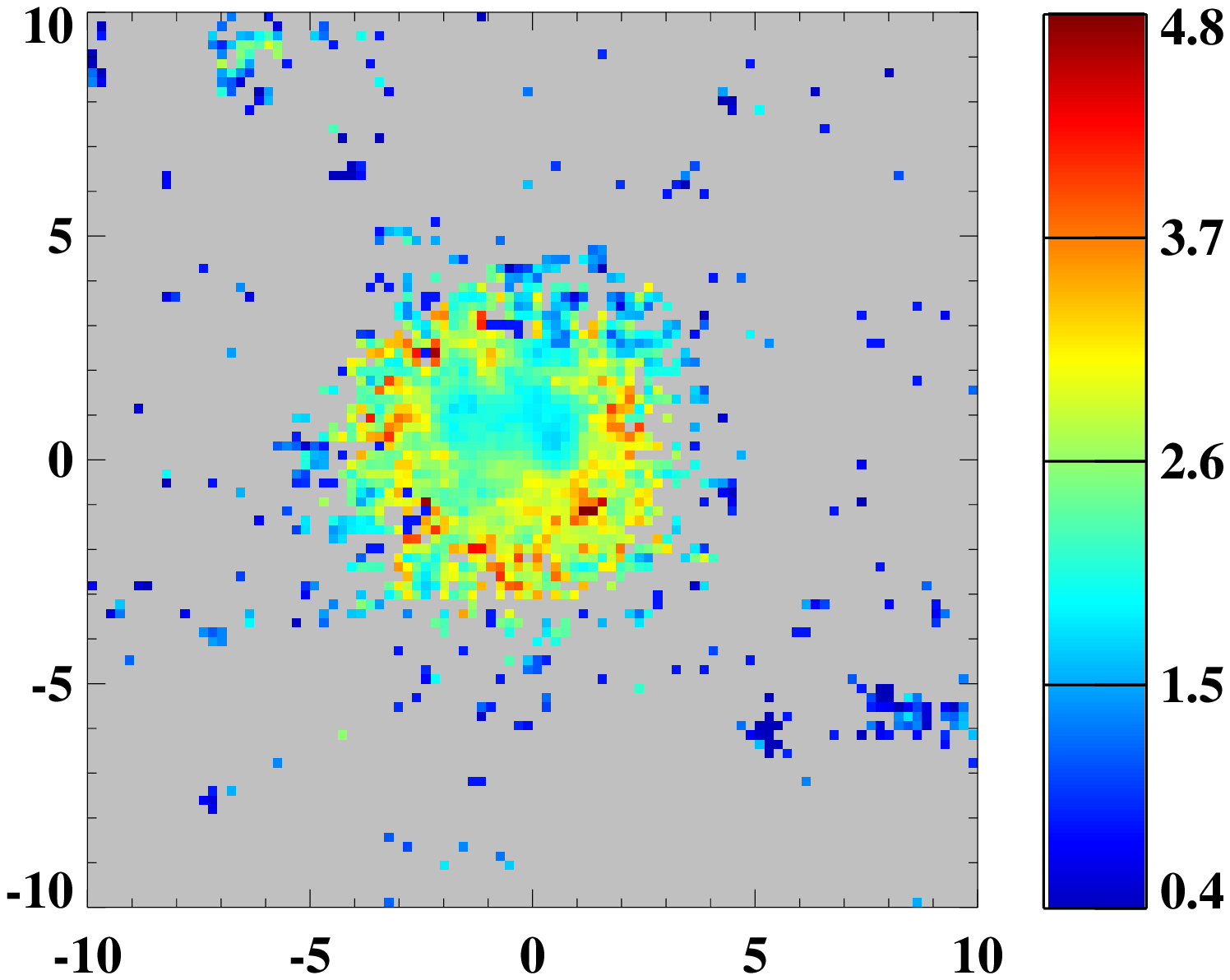} \\

 \includegraphics[height=2.6cm]{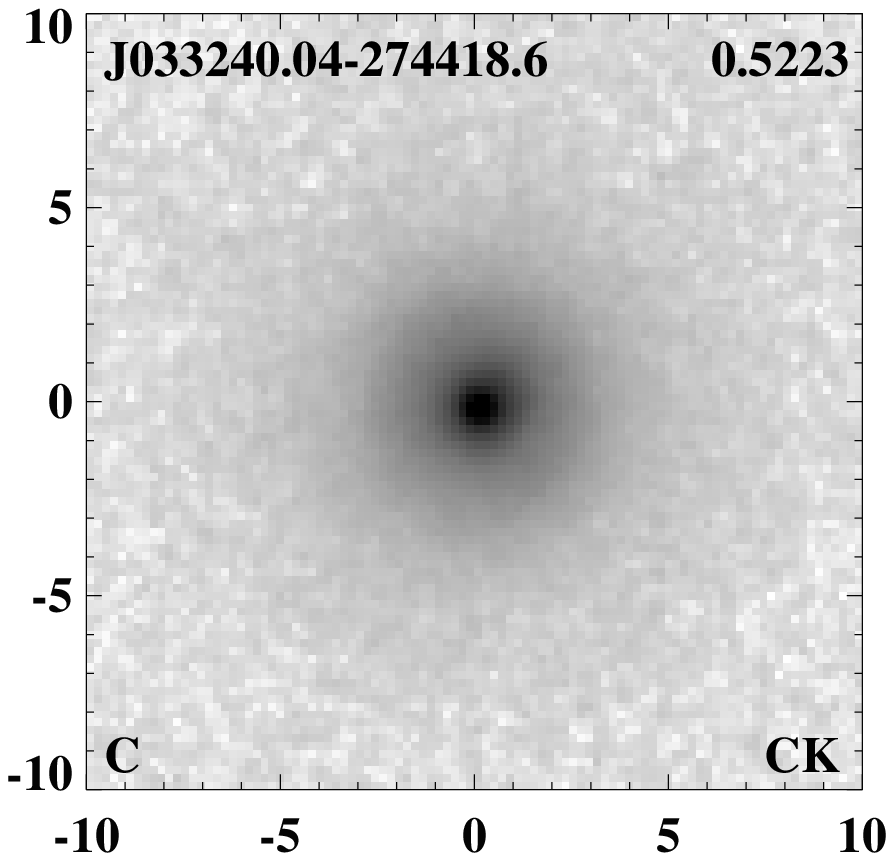} &
 \includegraphics[height=2.6cm]{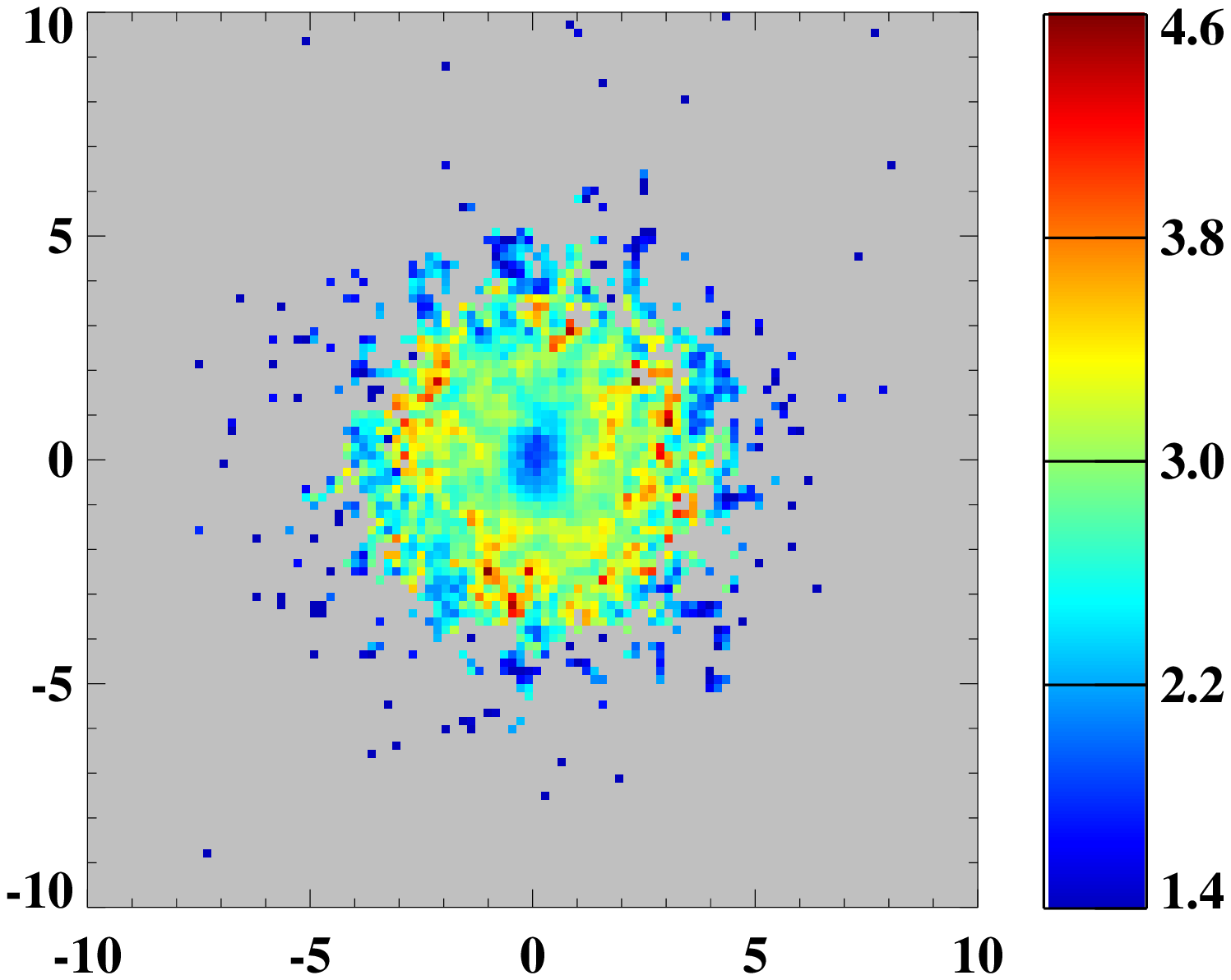} &
   \includegraphics[height=2.6cm]{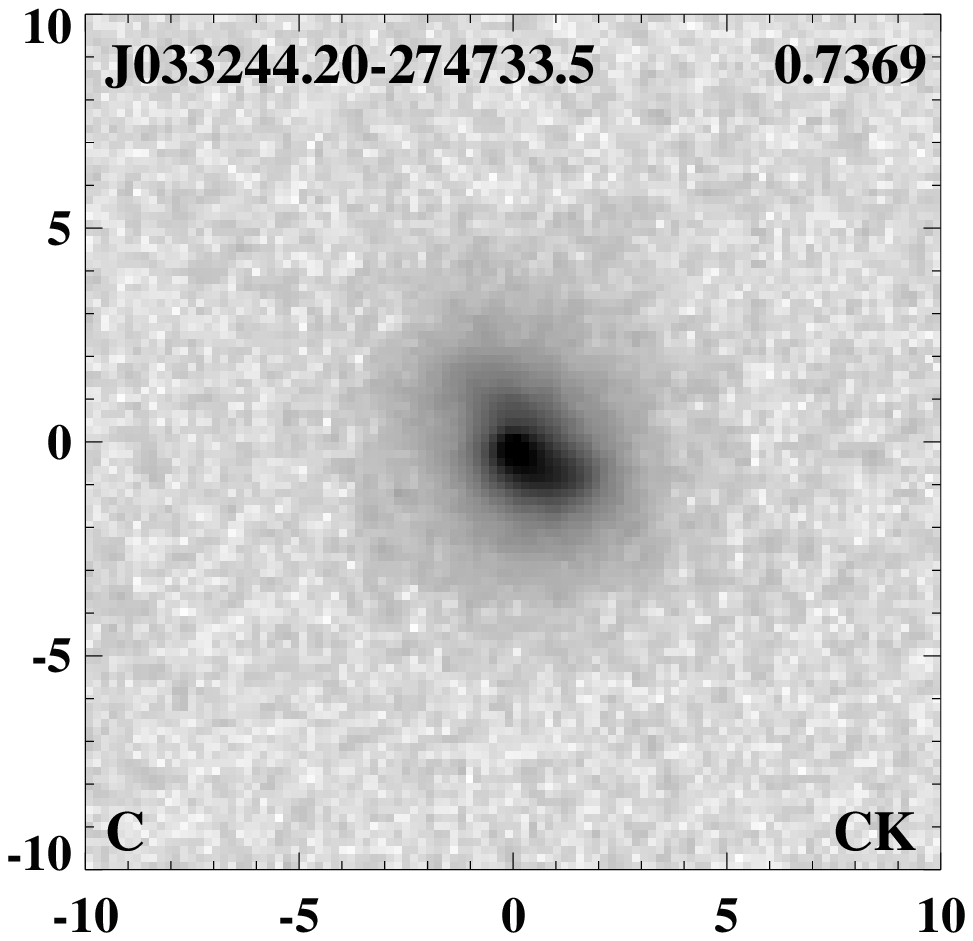} &
 \includegraphics[height=2.6cm]{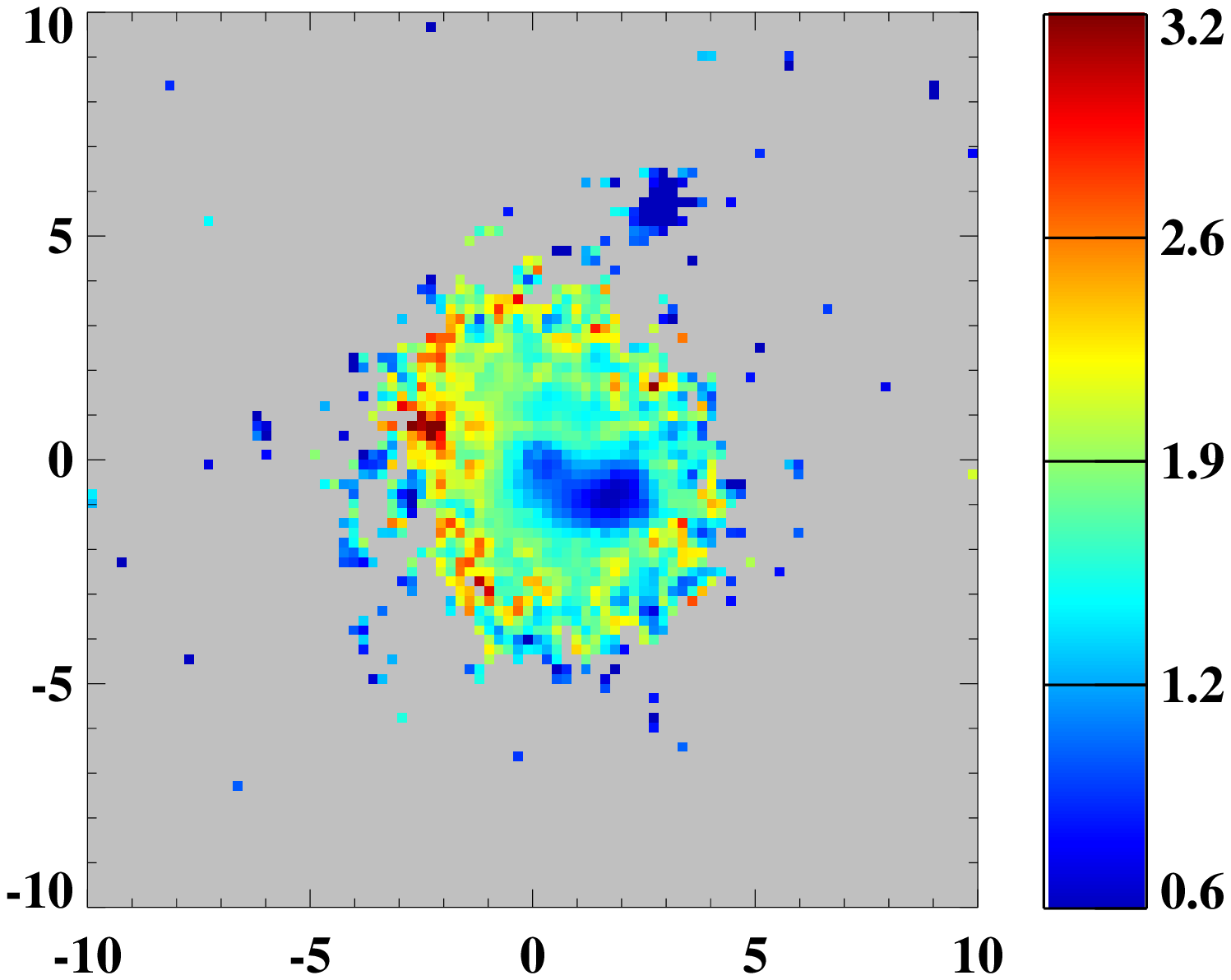} \\

   \includegraphics[height=2.6cm]{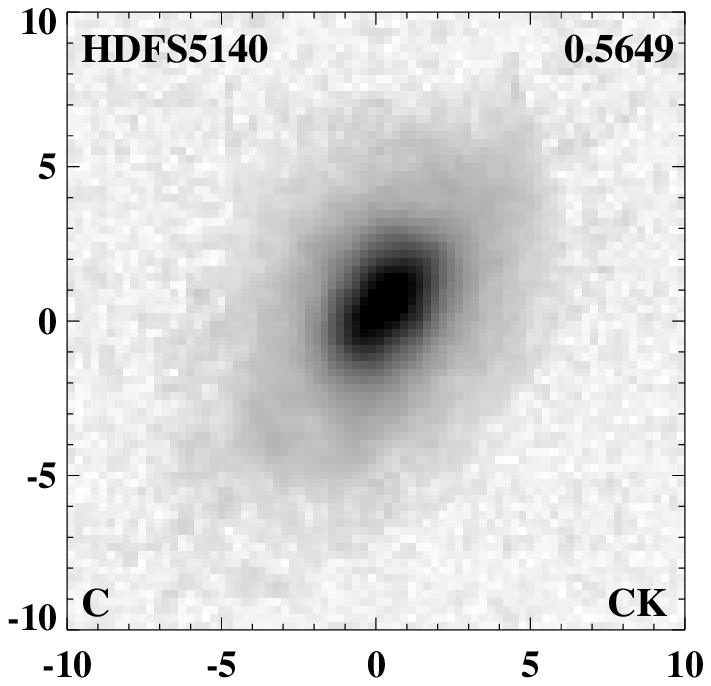} &
 \includegraphics[height=2.6cm]{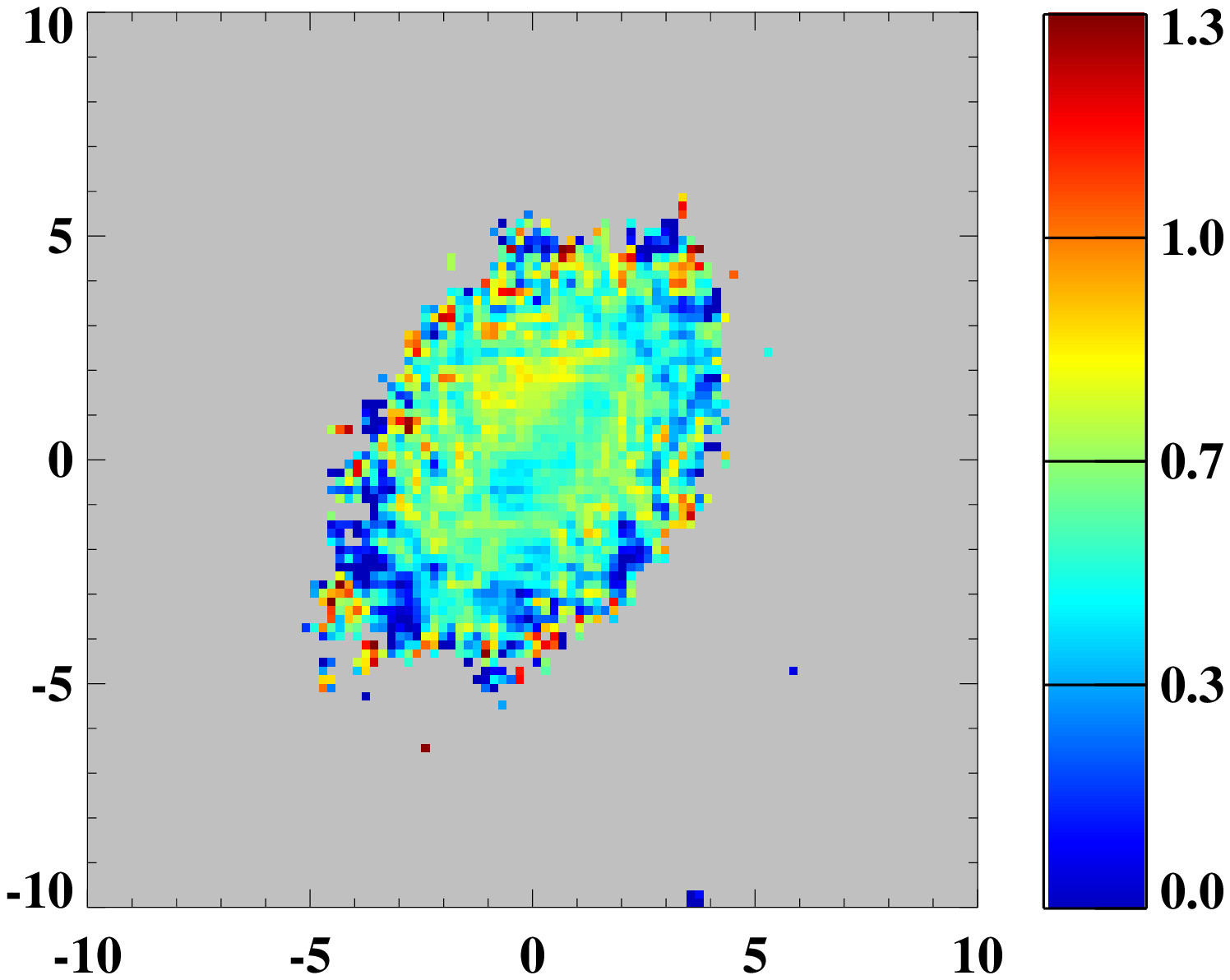} &
  \includegraphics[height=2.6cm]{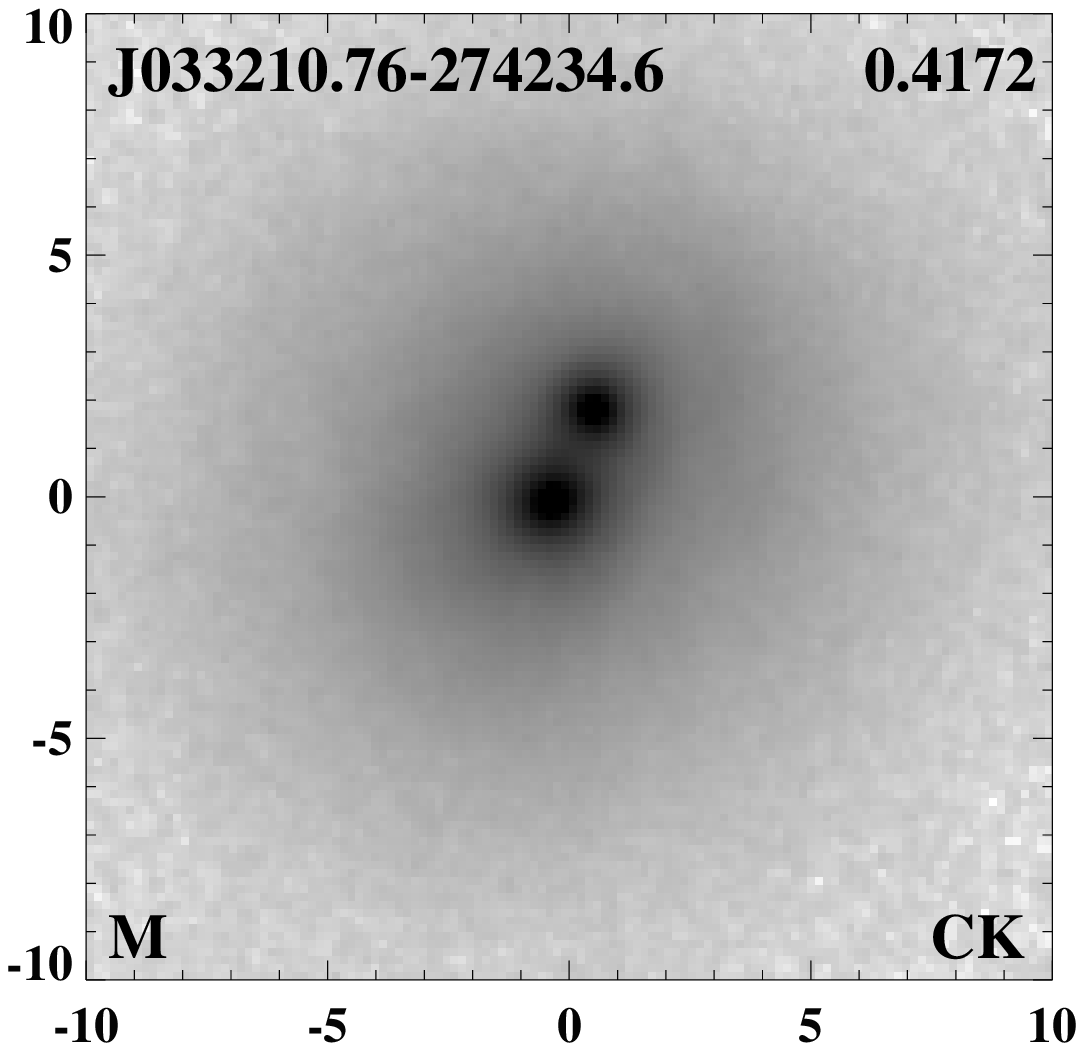} &
 \includegraphics[height=2.6cm]{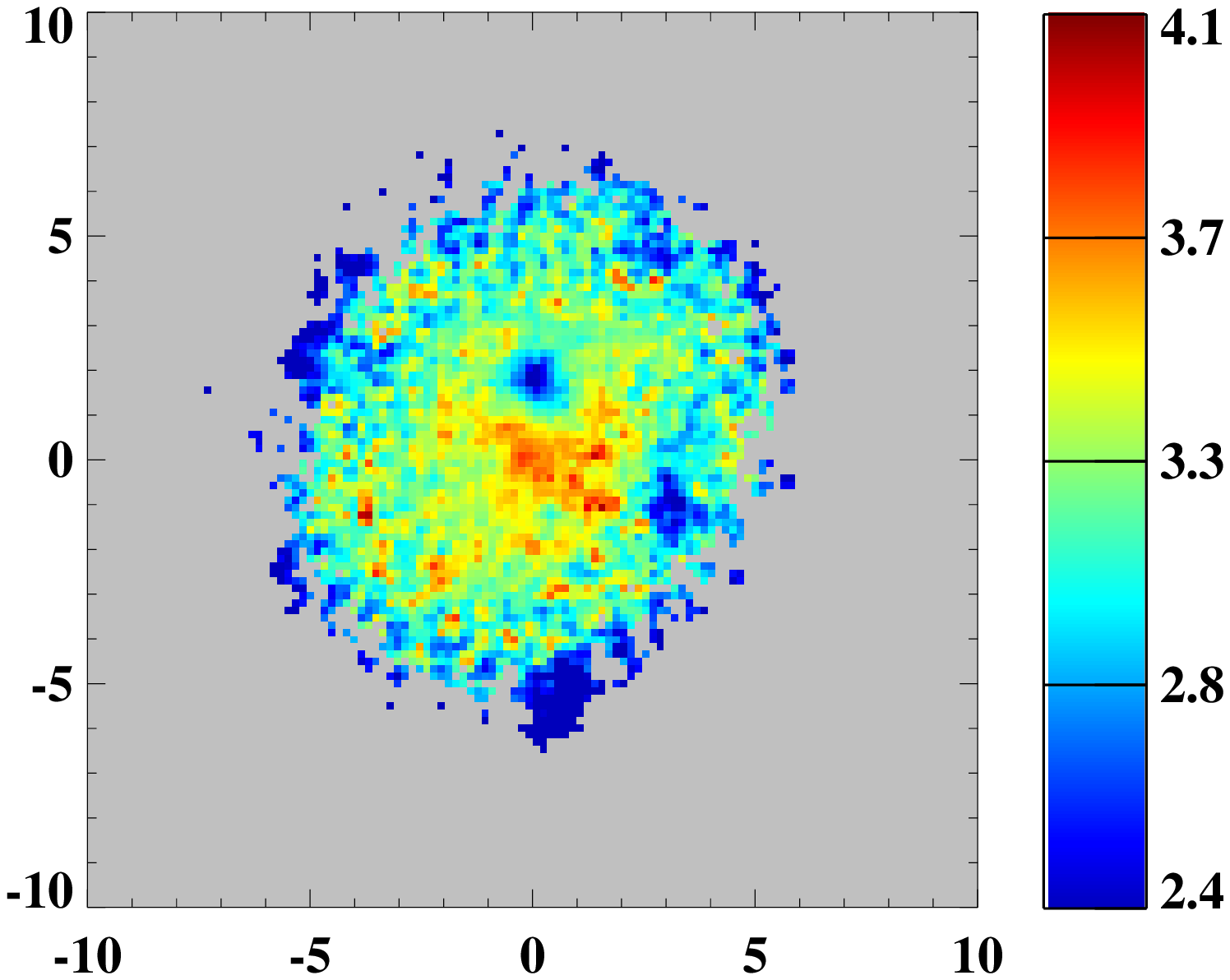} \\

\includegraphics[height=2.6cm]{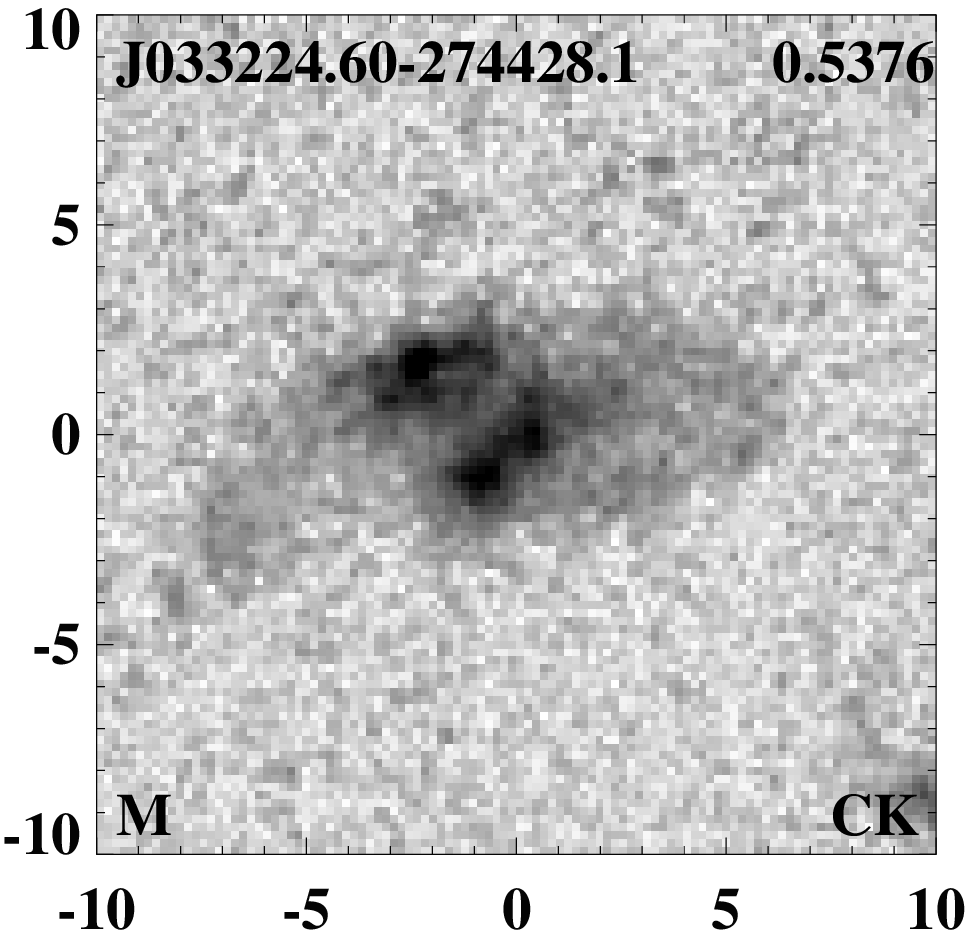} &
 \includegraphics[height=2.6cm]{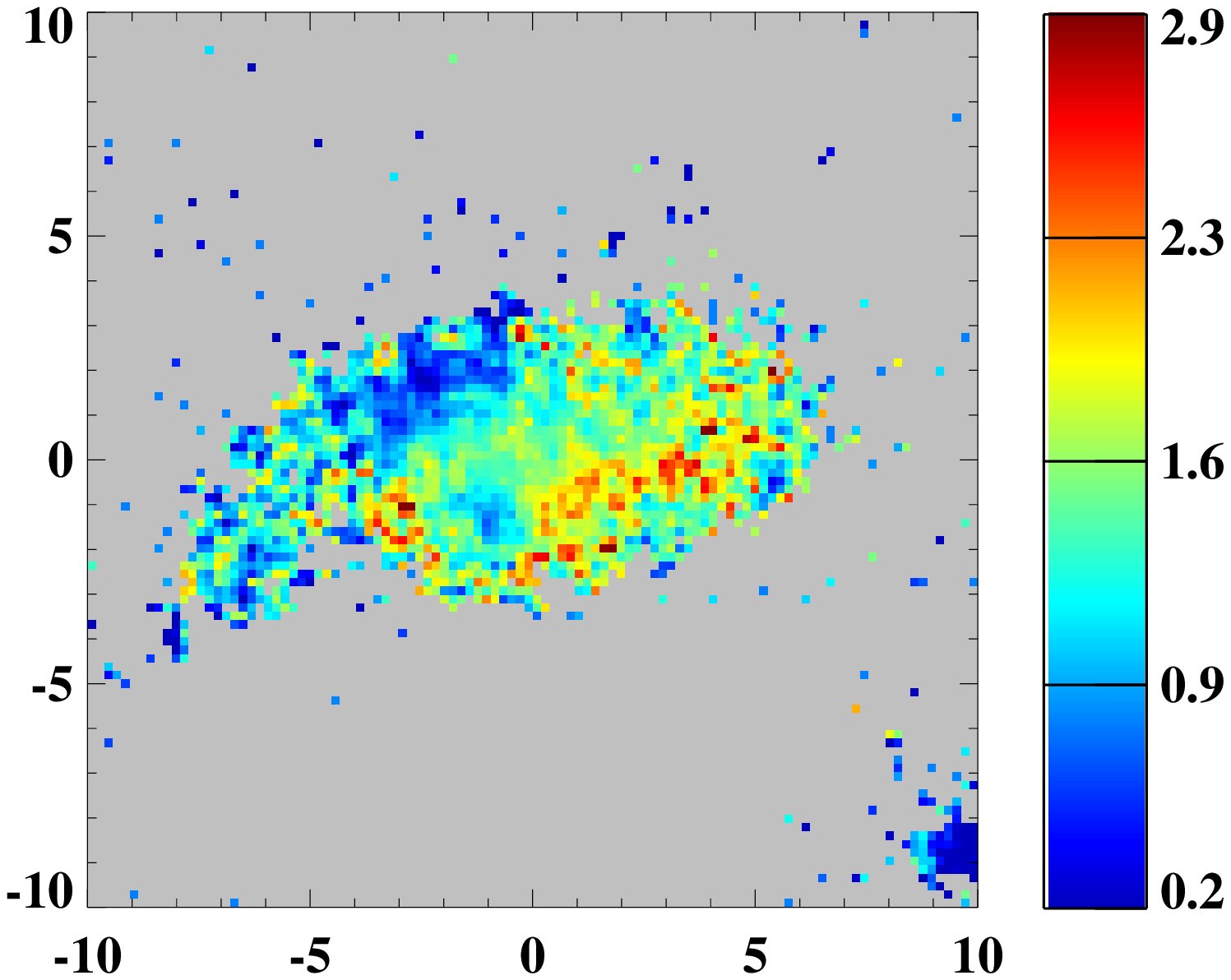} &
 \includegraphics[height=2.6cm]{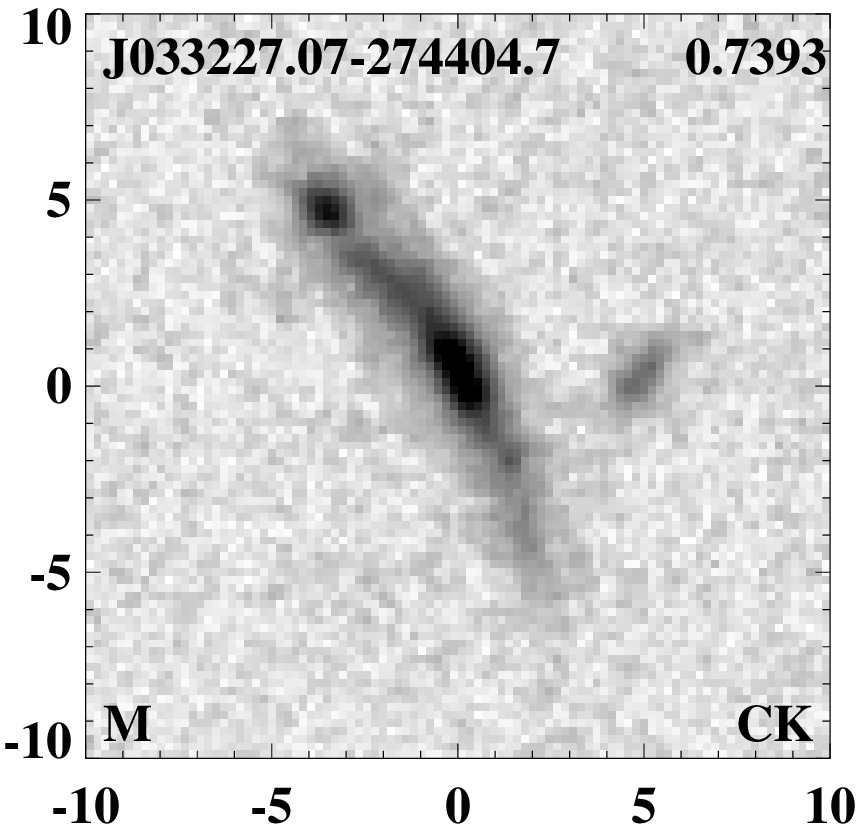} &
 \includegraphics[height=2.6cm]{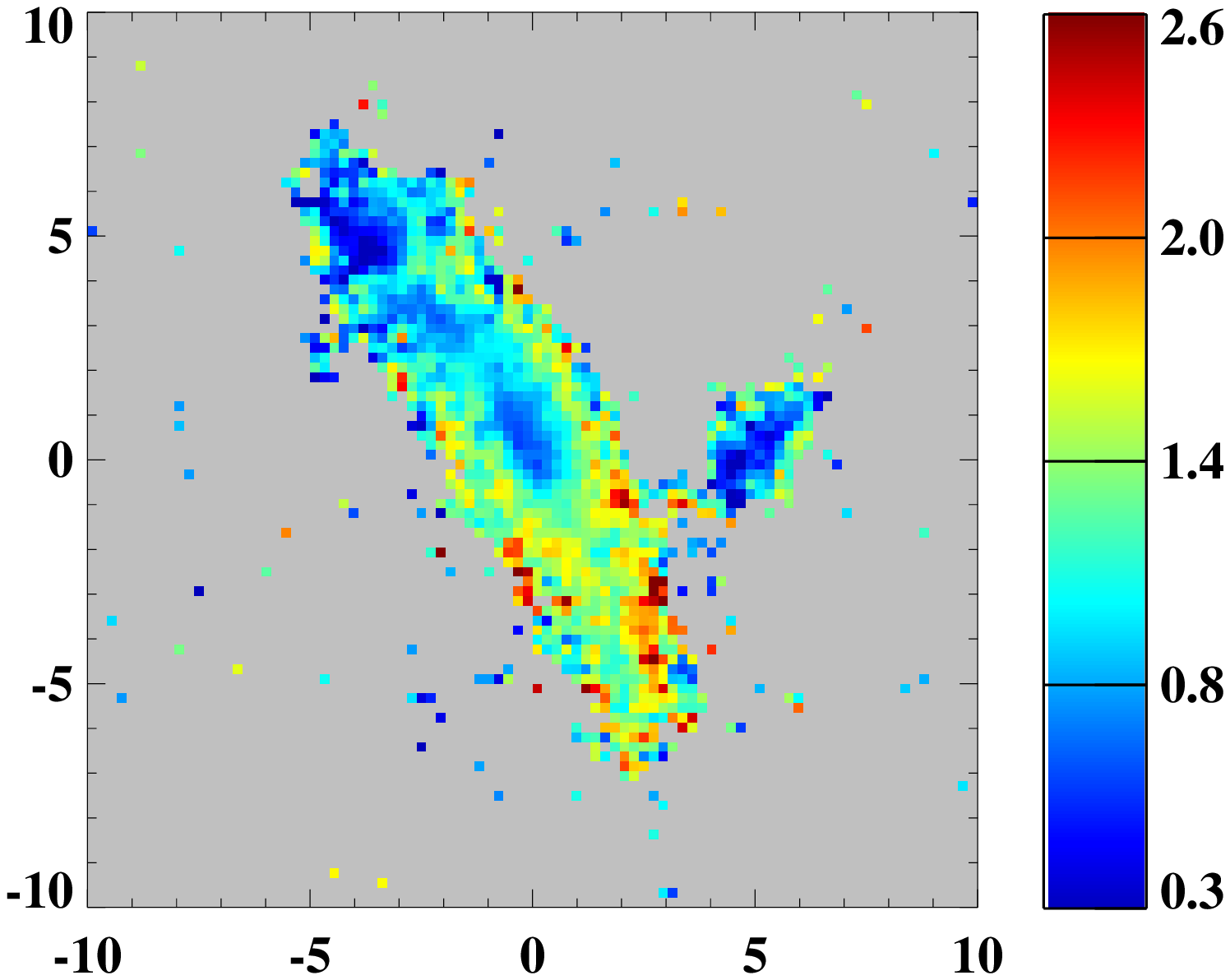} \\

  \includegraphics[height=2.6cm]{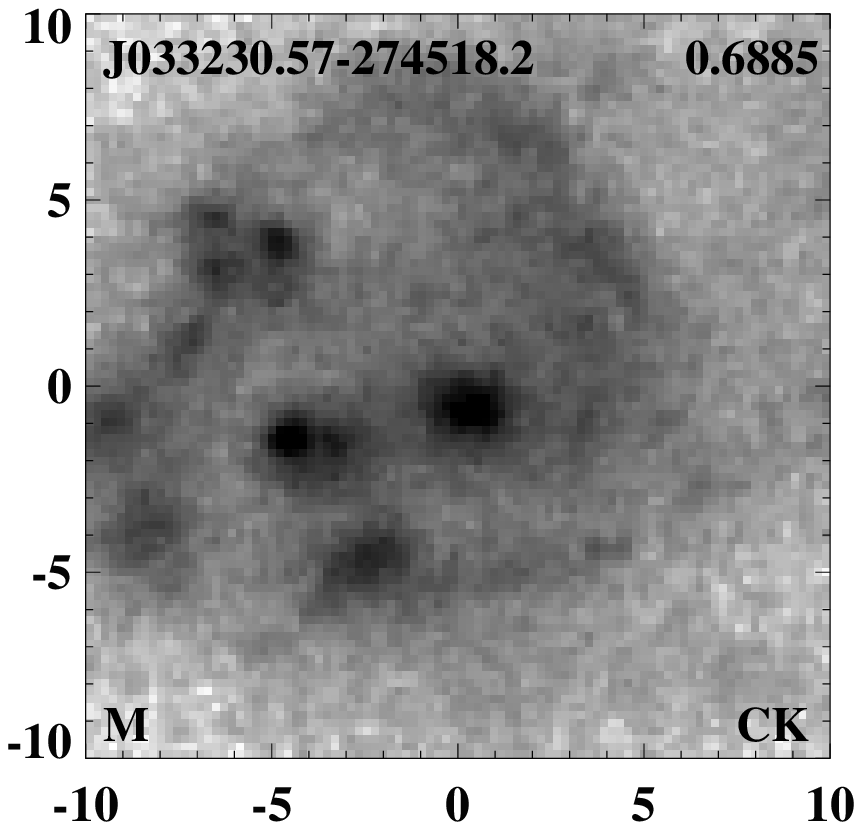} &
 \includegraphics[height=2.6cm]{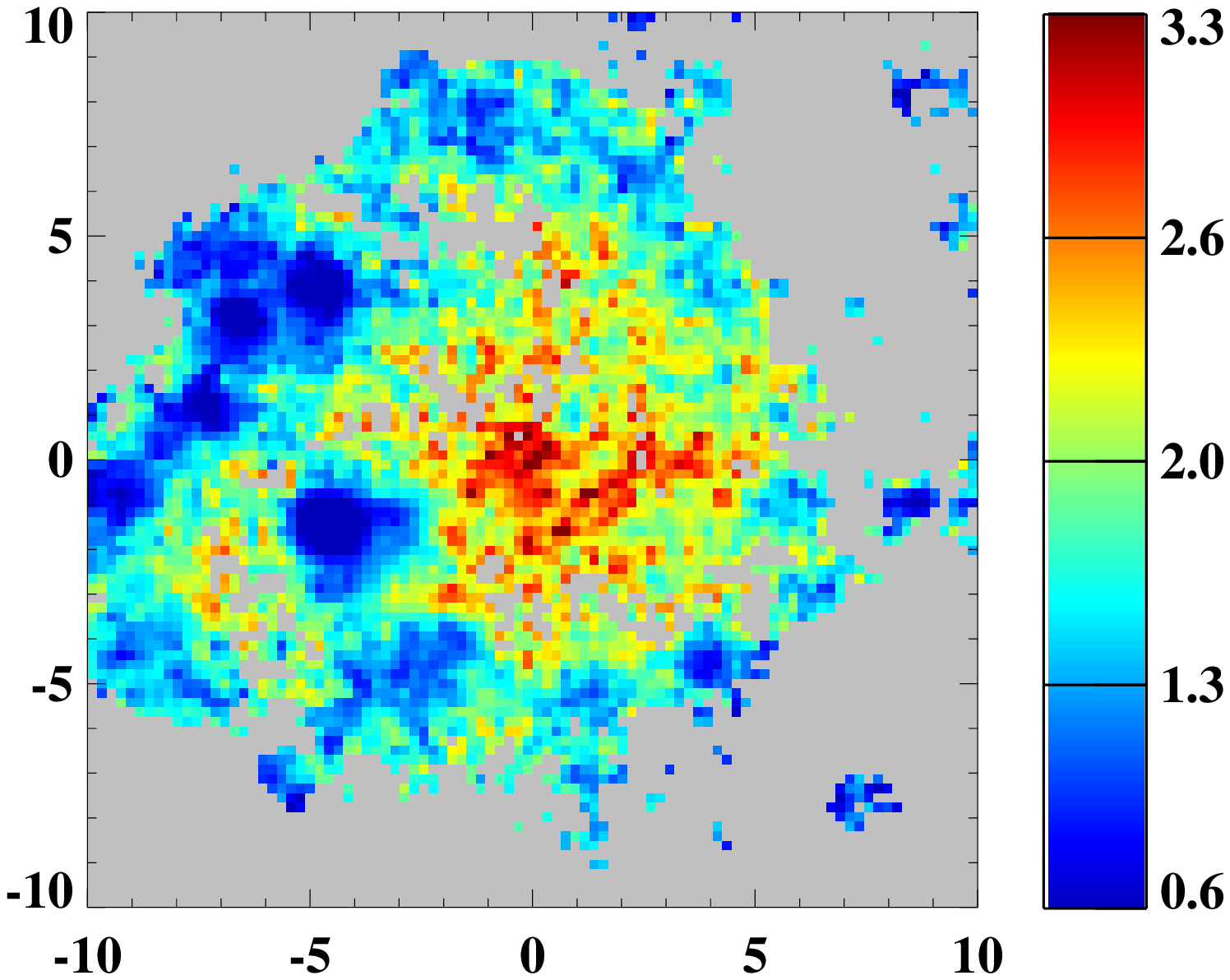} &
   \includegraphics[height=2.6cm]{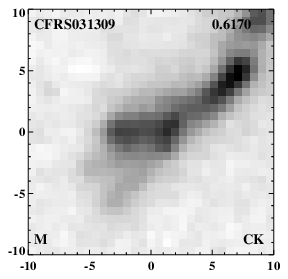} &
 \includegraphics[height=2.6cm]{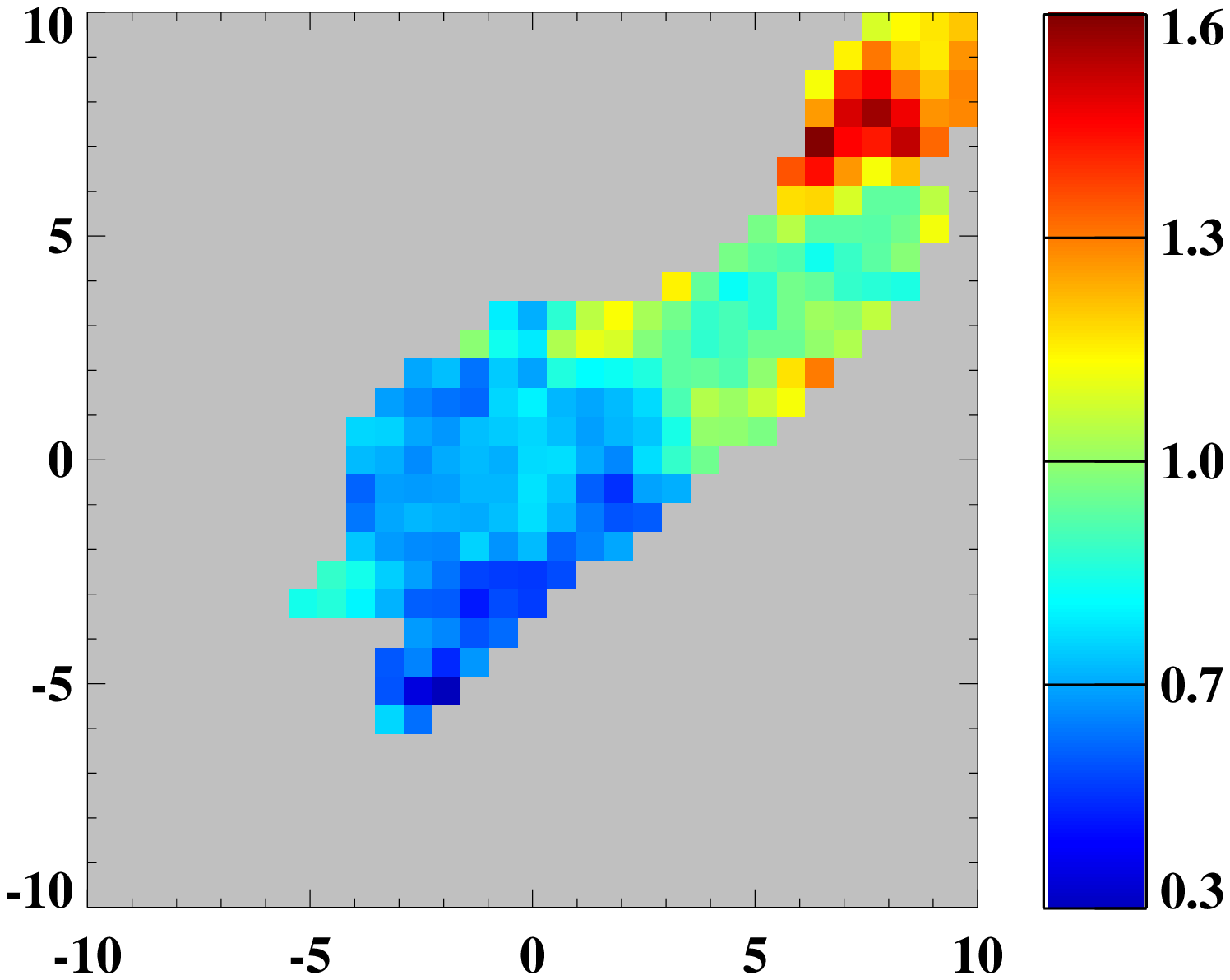} \\

    \end{tabular}
   \end{center}
   \caption{F850LP/F814W image, color map. The size of each image is 20x20 kpc. Galaxies observed with ACS have F850LP images, galaxies in CFRS and HDFS have F814W images. For each target, the name and redshift are labeled at the top-left and top right in the F850LP/F814W image. Rotating spiral disks are marked with an asterisk near their names. Morphological and kinematic classifications are indicated at the bottom-left and bottom right, respectively. Color maps are B$_{435}$-z$_{850}$ for CDFS objects and V$_{606}$-I$_{814}$ for objects in HDFS and CFRS (except CFRS039003, CFRS030488 with B$_{450}$-I$_{814}$ and CFRS031032 with V$_{555}$-I$_{814}$). All the galaxies are aligned with the GIRAFFE IFU (see Paper I).}
   \label{allimages} 
   \end{figure} 
 
\end{onecolumn}


\begin{thebibliography}{}


\bibitem[Abraham et al. 1994]{Abraham94}
     {Abraham, R.G., Valdes, F., Yee, H.K.C. \& van den Bergh, S.} 1994, ApJ, 432, 75 

\bibitem[Abraham et al. 1996]{Abraham96}
     {Abraham, R.G., Tanvir, N.R., Santiago, et al.} 1996, MNRAS, 279, 47 

\bibitem[Abraham et al. 1999]{Abraham99}
     {Abraham, R.G., Ellis, R.S., Fabian, et al.} 1999, MNRAS, 303, 641 
     
\bibitem[Abraham et al. 2003]{Abraham03}
     {Abraham, R.G.,  van den Bergh, S. \& Nair, P.} 2003, ApJ, 588, 218

\bibitem[Atkinson et al. 2007]{Atkinson07}
     {Atkinson, N.,  Conselice, C.J. \& Fox, N.} 2007, MNRAS, submitted, astro-ph/0712.1316


\bibitem[Barnes 2002]{Barnes02}
    {Barnes, J.E.} 2002, MNRAS, 333, 481 

\bibitem[Bell et al. 2003]{Bell03}
    {Bell, E. F., McIntosh, D. H., Katz, N. \& Weinberg, M. D.} 2003, ApJS, 149, 289

\bibitem[Bell et al. 2006]{Bell06}
    {Bell, E., Phleps, S., Somerville, et al.} 2006, ApJ, 652, 270 

\bibitem[Ben\'{i}tez et al. 2004]{Benitez04}
    {Ben\'{i}tez, N., Ford, H., Bouwens, R., et al.} 2004, ApJS, 150, 1    
   
   \bibitem[Bernardi et al. 2006]{Bernardi06}  
     {Bernardi, M., Nichol, R. C., Sheth, R. K., et al.} 2006, AJ, 131, 1288
     
\bibitem[Bershady et al. 2000]{Bershady00}
    {Bershady, M.A., Jangren, A., Conselice, C.J.} 2000, AJ, 119, 2645 
    
    
\bibitem[Birnboim et al. 2007]{Birnboim07}
    {Birnboim, Y., Dekel, A. \& Neistein, E.} 2007, MNRAS, 380, 339
     
\bibitem[Brinchmann et al. 1998]{Brinchmann98}
     {Brinchmann, J., Abraham, R., Schade, D., et al.} 1998, ApJ, 499, 112 



\bibitem[Bruzual \& Charlot, 2003]{Bruzual03}
     {Bruzual, G. \& Charlot, S.} 2003, MNRAS, 344, 1000 

\bibitem[Cassata et al. 2005]{Cassata05}
     {Cassata, P., Cimatti, A., Fransceschini, A. et al.} 2005, MNRAS, 357, 903

\bibitem[Casertano et al. 2000]{Casertano00}
     {Casertano, S., de Mello, D., Dickinson, M. et al.} 2000, AJ, 120, 2747 


\bibitem[Conselice et al. 2003]{Conselice03}
     {Conselice, C.J., Bershady, M.A., Dickinson, M. \& Papovich, C.} 2003, AJ, 126, 1183 

\bibitem[Conselice 2003]{Conselice03b}
     {Conselice, C.J.} 2003, ApJS, 147, 1

\bibitem[Conselice et al. 2005]{Conselice05}
    {Conselice, C.J., Blackburne, J.A. \& Papovich, C.} 2005, ApJ, 620, 564 

\bibitem[Dekel \& Birnboim 2006]{Dekel06}
     {Dekel, A. \& Birnboim, Y.} 2006, MNRAS, 368, 2 

\bibitem[Ellis et al. 2001]{Ellis01}
     {Ellis, R.S., Abraham, R.G. \& Dickinson, M.} 2001, ApJ, 551, 111 
   
\bibitem[Emsellem et al. 2004]{Emsellem04}
     {Emsellem, E., Cappellari, M., Peletier, R.F. et al.} 2004, MNRAS, 352, 721 
       
\bibitem[Flores et al. 2006]{Flores06}
     {Flores, H., Puech, M., Hammer, et al.} 2006, A\&A, 455, 107 

\bibitem[Governato et al. 2007]{Governato07}
     {Governato, F., Willman, B., Mayer, L. et al.} 2007, MNRAS, 374, 1479 

\bibitem[Giavalisco et al. 2004]{Giavalisco04}
     {Giavalisco, M., Ferguson, H.C., Koekemoer, A.M. et al.} 2004, ApJ, 600, 93 
     
\bibitem[Garrido et al. 2002]{Garrido02}
     {Garrido, O., Marcelin, M., Amram, P. et al.} 2002, A\&A, 387, 821     
     
\bibitem[Hammer et al. 1997]{Hammer97}
     {Hammer, F., Flores, H., Lilly, S. et al.} 1997, ApJ, 481, 49 
     
\bibitem[Hammer et al. 2001]{Hammer01}
     {Hammer, F., Gruel, N., Thuan, T.X., et al.} 2001, ApJ, 550, 570 

\bibitem[Hammer et al. 2005]{Hammer05}
     {Hammer, F., Flores, H., Elbaz, D., et al.} 2005, A\&A, 430, 115 

\bibitem[Hammer et al. 2007]{Hammer07}
     {Hammer, F., Puech, M., Chemin, L., et al.} 2007, ApJ, 662, 322



\bibitem[Jimenez et al. 2007]{Jimenez06}
     {Jimenez, R., Bernardi, M., Haiman, Z., et al.} 2007, ApJ, 669, 947

\bibitem[Kannappan \& Barton 2004]{Kannappan04}
     {Kannappan, S.J. \& Barton, E.J.} 2004, AJ, 127, 2694 

\bibitem[Kassin et al. 2007]{Kassin07}
     {Kassin, S.A., Weiner, B.J., Faber, S.M., et al.} 2007, ApJ, 660, 35 

\bibitem[Kent 1985]{Kent85}
     {Kent, S.M.} 1985, AJ, 59, 115 

\bibitem[Kere\u{s} et al. 2005]{Keres05}
    {Kere\u{s}, D., Katz, N., Weinberg, D. H. et al.} 2005, MNRAS, 363, 2


\bibitem[Law et al. 2007]{Law07}
     {Law, D.R., Steidel, C.C., Erb, D.K. et al.} 2007, ApJ, 669, 929

\bibitem[Le F\`{e}vre et al. 2000]{LeFevre00}
    {Le F\`{e}vre, O., Abraham, R., Lilly, S. J. et al.} 2000, MNRAS, 311, 565

\bibitem[Le F\`{e}vre et al. 2004]{LeFevre04}
    {Le F\`{e}vre, O., Vettolani, G., Paltani, S., et al.} 2004, A\&A, 428, 1043

\bibitem[Liang et al. 2006]{Liang06}
     {Liang, Y. C., Hammer, F. \& Flores, H.} 2006, A\&A, 447, 113

\bibitem[Lilly et al. 1998]{Lilly98}
     {Lilly, S., Schade, D., Ellis, R. et al.} 1998, ApJ, 500, 75 

\bibitem[Lucas et al. 2003]{Lucas03}
     {Lucas, R.A., Baum, S.A., Brown, T.M., Casertano, S. et al.} 2003, AJ, 125, 398 

\bibitem[Lotz et al. 2004]{Lotz04}
     {Lotz, J.M., Madau, P., Giavalisco, M. \& Primack, J.} 2004, AJ, 128, 163 
     
\bibitem[Lotz et al. 2006]{Lotz06a}
     {Lotz, J.M., Madau, P., Giavalisco, M. et al.} 2006, ApJ, 636, 592

\bibitem[Lotz et al. 2008]{Lotz08}
     {Lotz, J.M., Davis, M., Faber,S. et al.} 2008, ApJ, 672, 177

\bibitem[Melbourne et al. 2005]{Melbourne05}
     {Melbourne, J., Koo, D.C. \& Le Floc'h, E.} 2005, ApJ, 632, 65

\bibitem[Menanteau et al. 2001]{Menanteau01}
     {Menanteau, F., Abraham, R.G. \& Ellis, R.S.} 2001, MNRAS, 322, 1

\bibitem[Menanteau et al. 2004]{Menanteau04}
     {Menanteau, F., Ford, H.C., Illingworth, G.D., et al.} 2004, ApJ, 612, 202


\bibitem[Nakamura et al. 2004]{Nakamura04}
     {Nakamura, O., Fugukita, M., Brinkmann, J. et al.} 2004, AJ, 127, 2511


\bibitem[Peng et al. 2002]{Peng02}
     {Peng, C.Y., Ho, L.C., Impey, C.D. \& Rix, H.W. } 2002, AJ, 124, 266 

\bibitem[Peletier et al. 1999]{Peletier99}
     {Peletier, R.F., Balcells, M., Davies, R.L. et al.} 1999, MNRAS, 310, 703 
     
\bibitem[Persic et al. 1991]{Persic91}
     {Persic, M. \& Salucci, P.} 1991, ApJ, 368, 60
 
\bibitem[Pizagno et al. 2007]{Pizagno07}
     {Pizagno, J.; Prada, F.; Weinberg, D. H.; } 2007, ApJ, 134, 945
     
\bibitem[Puech et al. 2006a]{Puech06a}
     {Puech, M., Hammer, F., Flores, H., \"Osltin, G. \& Marquat, T.} 2006, A\&A, 455, 119 
     
\bibitem[Puech et al. 2006b]{Puech06b}
     {Puech, M., Flores, H., Hammer, F., Lehnert, M.D.} 2006, A\&A, 455, 131 
     
\bibitem[Puech et al. 2007]{Puech07} 
     {Puech, M., Hammer, Lehnert, M.D., F. \& Flores, H.} 2007, A\&A, 466, 83 
 
\bibitem[Puech et al. 2007b]{Puech07b} 
     {Puech, M., Hammer, F.,  Flores, H. et al.} 2007, A\&A, 476, 21 

\bibitem[Puech et al. 2008a]{Puech08} 
     {Puech, M., Hammer, F.,  Flores, H. et al.} 2007, A\&A, submitted.  

\bibitem[Puech et al. 2008b]{Puech08b} 
     {Puech, M., et al.} 2008, in preparation.  
     
\bibitem[Ravikumar et al. 2007]{Ravikumar07}
     {Ravikumar, C.D., Puech, M., Flores, et al.} 2007, A\&A, 465, 1099 

\bibitem[Ravindranath et al. 2004]{Ravindranath04}
     {Ravindranath, S., Ferguson, H.C., Conselice, C. et al.} 2004, ApJ, 604, 9 

\bibitem[Rawat et al. 2007]{Rawat07}
     {Rawat, A., Kembhavi, A.K., Hammer, F. et al.} 2007, A\&A, 469, 483

\bibitem[Rigopoulou et al. 2005]{Rigopoulou05}
     {Rigopoulou, D., Vacca, W. D., Berta, S. et al.} 2005, A\&A, 440, 61 

\bibitem[Roberts \& Haynes 1994]{Roberts94}
     {Roberts, Morton S. \& Haynes, M.P.} 1994, ARA\&A, 32, 115 


\bibitem[Sargent et al. 2007]{Sargent07}
     {Sargent, M. T.; Carollo, C. M.; Lilly, S. J.} 2007, ApJS, 172, 434

\bibitem[Schade et al. 1995]{Schade95}
     {Schade, D., Lilly, S.J., Crampton, D., et al.} 1995, ApJ, 451, 1

\bibitem[Schade et al. 1996]{Schade96}
     {Schade, D., Lilly, S.J., Le Fevre, O., et al.} 1996, ApJ, 464, 79

\bibitem[Semelin \& Combes 2005]{Semelin05}
     {Semelin, B. \& Combes, F.} 2005, A\&A, 441, 55 

\bibitem[Silk  2003]{Silk03}
    {Silk, J.} 2003, Ap\&SS, 284, 663
     

\bibitem[Somerville et al. 2001]{Somerville01}
     {Somerville, R.S., Primack, J.R. \& Faber, S.M.} 2001, MNRAS, 320, 504 

          

\bibitem[Thomas et al. 2006]{Thomas06}
     {Thomas, D. \& Davies, R.L.} 2006, MNRAS, 366, 510

     
\bibitem[Trujillo et al. 2005]{Trujillo05}
     {Trujillo, I. \& Pohlen, M.} 2005, ApJ, 630, 17

\bibitem[van den Bergh et al. 2001]{VandenBergh01}
     {van den Bergh, S., Cohen, J.G. \& Crabbe, C.} 2001, AJ, 122, 611

\bibitem[van der Kruit 1979]{vanderKruit79}
     {van der Kruit, P.C.} 1979, A\&AS, 38, 15

\bibitem[van der Wel 2008]{vanderWel08}
     {van der Wel, A.} 2008, ApJ, 675, 13


\bibitem[Vanzella et al. 2002]{Vanzella02}
     {Vanzella, E., Cristiani, S., Arnouts, S. et al.} 2002, A\&A, 396, 847

\bibitem[Vanzella et al. 2006]{Vanzella06}
     {Vanzella, E., Cristiani, S., Dickinson, M. et al.} 2006, A\&A, 454, 423

\bibitem[Werk et al. 2004]{Werk04}
     {Werk, J.K., Jangren, A. \& Salzer, J.J.} 2004, ApJ, 617, 1004

\bibitem[Williams et al. 2000]{Williams00}
     {Williams, R.E., Baum, S., Bergeron, L.E., Bernstein, N. et al.} 2000, AJ, 120, 2735

\bibitem[Yagi et al. 2006]{Yagi06}
     {Yagi, M., Nakamura, Y., Doi, M. et al.} 2006, MNRAS, 368, 211

\bibitem[Yang et al. 2007]{Yang07}
     {Yang, Y. et al.} 2007, A\&A, 477, 789
                        
\bibitem[Zheng et al. 2005]{Zheng05}
     {Zheng, X.Z.,  Hammer, F., Flores, et al.} 2005, A\&A, 435, 507  

\bibitem[Zheng et al. 2004]{Zheng04}
     {Zheng, X.Z.,  Hammer, F., Flores, Assemat, F. \& Pelat, D.} 2004, A\&A, 421, 847     

\end{thebibliography}
\end{document}